# Stock Performance Evaluation for Portfolio Design from Different Sectors of the Indian Stock Market

Capstone project report submitted in partial fulfillment of the requirements for the Post Graduate Program in Data Science at Praxis Business School

By

**ARPIT AWAD**

**AADITYA RAJ**

**GOURAV RAY**

**PUSPARNA CHAKRABORTY**

**SANKET DAS**

**SUBHASMITA MISHRA**

Under Supervision of

**Prof JAYDIP SEN**

**PRAXIS BUSINESS SCHOOL**

# ABSTRACT

Stock market offers platform where people buy and sell shares of publicly listed companies. Generally stock prices are quite volatile; hence predicting them is daunting task. There are still many researches going to develop more accuracy in stock price prediction. Portfolio construction refers to allocation of different sector stocks optimally to achieve maximum return by taking minimum risk. A good portfolio can help investor earn maximum profit by taking minimum risk. Beginning from Dow Jones Theory a lot of advancement has happened in area of building efficient portfolio. In this project we have tried to predict future value of few stocks from six important sectors of Indian economy and also built portfolio. As part of the project, our team has conducted a study of performance of various Time series, machine learning (regression and classification) and deep learning models in stock price prediction on selected stocks from the chosen six important sectors of the economy. As part of building an efficient portfolio we have studied multiple portfolio optimization theories beginning from MPT (Modern Portfolio theory). We have built minimum variance portfolio and optimal risk portfolio for all the six chosen sectors by using past five years' daily stock price as training data and have also conducted back testing to check the performance of the portfolio. We look forward to continue our study in area of stock price prediction and asset allocation and consider this project as first stepping stone.

# CONTENT

# CHAPTER-1

## Introduction

Stock market has become one of the money-spinning option for everyone because of its maximum return in minimum time. It is a secondary market where people buy and sell securities electronically. In India there are two stock exchanges- BSE and NSE. BSE is considered to be India's oldest stock exchange which came to existence in 1875 which has SENSEX as its flagship index comprising of top 30 stocks which are largest most liquid and financially stable; later NSE came to existence and started trading in 1994. NSE has a flagship index named as NIFTY 50 comprising of top 50 companies based on its trading volume and market capitalization. Securities Exchange Board of India (SEBI) is the regulator of Indian Stock Market. A company first launches its IPO in primary market and then on the day of listing it becomes a part of secondary market and investors as well as traders buys shares of company. All orders in the trading systems needs to be done electronically through brokers from Monday to Friday within 9.00 AM- 3.30PM. Equity markets follow T+2 settlements. An investor earns by investing optimally in different sectors hence by building an optimal portfolio. Before building portfolio investor should first decide his/her financial goal and have a look on his/her current asset, liabilities and other investments. Then he should allocate asset in diversified manner in order to get maximum return by taking minimum risks. Once portfolio is constructed its crucial to monitor investment and reevaluate goals annually and make changes semi-annually or annually. Many new methods and concepts have emerged in financial portfolio construction, risk management and performance evaluation. Markowitz who used variance of returns as a measure of risk was one of the pioneers in proposing a quantitative methodology for portfolio construction, his work along with Sharpe and Lintner generated discussions that formed modern theory of portfolio. The pandemic saw a large number of surges in Trading and Demat Account. All stock trading data are available in NSE websites. In this project we have taken six important sectors of Indian Economy and top four sectors from six sectors. In this project we have done stock prediction using various techniques like time series, machine learning and deep learning. In Time Series we have used Simple Exponential Smoothing, Holt Winter's Model, Rolling Window Method, and Sliding Window Method for stock price prediction. In machine learning we used various Regression and Classification techniques. In deep learning we used LSTM and CNN method to predict various stock prices.

Building an efficient portfolio is the process of allocating weights to a collection of stocks in such a way that the risk and return are optimized. Markowitz's Minimum Variance Portfolio is considered as the foundation of all the later works in the field of portfolio optimization. A detailed comparative study of various statistical, machine learning and deep learning models has been done for stock price prediction. Machine Learning models have also been employed for building classification models to predict future price of stock. Building an optimal portfolio along with the ROI computation helps an investor to make investment decisions wisely. The investor can equally divide the fund in different sectors to maximize return in minimum risk. Generally stock price fluctuate in news flow, we have also tried to show price fluctuation based on news flow. We shall discuss in details in subsequent chapters.

# CHAPTER-2

## Methodology

The first step towards achieving the goal of stock price prediction is selection of stocks. We took top four stocks from six important sectors of Indian Economy. These four stocks were chosen from top four stock of fact sheet published at the end of every month in NSE website. We chose Metal sector, IT sector, Banking sector, FMCG sector, Auto sector and Pharma sector. For the same, the latest monthly published sectoral Index report was referred to and the top 5 contributors to the index of each sector were chosen. The lists of four stocks chosen from six sectors for stock price prediction are as follows:-

| Sl No | METAL | IT | BANK | AUTO | FMCG | PHARMA |
|-------|-------|----|----|------|------|--------|
| 1 | TATA STEEL | TCS | HDFC BANK | BAJAJ AUTO | HINDUSTAN UNILEVER | SUN PHARMA |
| 2 | HINDALCO | INFY | ICICI BANK | M&M | ITC | CIPLA |
| 3 | JSW STEEL | WIPRO | AXIS BANK | MARUTI | NESTLE | DR REDDY |
| 4 | VEDANTA | HCL TECH | KOTAK BANK | TATA MOTORS | TATA CONSUMER | DIVISLAB |

The daily data from 1st Jan 2016 to 31st Dec 2021 was fetched using Yahoo finance API. The four years data was which is from 1st Jan 2016 to 31st Dec 2020 was used as training data and the one year data which is from 1st Jan 2021 to 31st December 2021 is used as testing data. Usually, stock markets work 5 days a week and will be off on Saturday and Sunday. But in the dataset, some of the weekday trading data was missing due to holidays. The missing days were identified and imputed using forward fill. After the imputation, there are 1482 data points. Before going deep into the topic let us discuss few terms associated with stock market:

i)    Open- The price at which stock opens on a trading day.
ii)   High- The highest price that the stock touches on a trading day.
iii)  Low- The lowest price that the stock touches on a trading day.
iv)   Volume- The number of stocks traded on a particular day.
v)    Close- The price of the stock during closing of market.
vi)   Average Traded Price- This is what buyers have paid for one share on average over the course of a specific time period.
vii)  RSI- This measures the magnitude of recent price changes to analyze overbought and oversold condition.
viii) Moving Averages- These features capture average changes in series of data over a period of time

The models used for prediction are as follows:-

TIME SERIES MODELS:  Simple Exponential Smoothing

Holt's Trend Method

ARIMA

Sliding Window Method

Rolling Window Method

MACHINE LEARNING MODELS: Linear Regression

Random Forest

Gradient Boost

Naïve Bayes

KNN

DEEP LEARNING MODELS: LSTM (Long Short-Term Memory)

CNN (Convolutional NeuralNetwork)

Python libraries are used for building time series model and machine learning models and keras for building deep learning models.

The validation method used for statistical, econometric and machine learning models is walk-forward validation. There are two variants of walk-forward validation, rolling window and sliding window walk-forward validation. In rolling window validation, with every iteration both the training set and test set moves forward by a fixed number of data points, the size of the training set would keep on increasing whereas the size of the test set would remain constant. Similarly, in sliding window validation with every iteration both training set and test set moves forward, but unlike rolling window the size of the training set and test set remains constant. That is, as the training window moves forward, it would leave the past values. The walk-forward validation method allows us to train the model with the recent values.

For stock price validation, the sliding window method is considered to be more appropriate compared to the rolling window method as it leaves the past values as the window moves forward. For stock price prediction more than the amount of the data with which a model has trained the recency of the data is important.

# CHAPTER-3

## TIME SERIES MODEL

## Simple Exponential Smoothing

A simple exponential smoothing is one of the simplest ways to forecast a time series. The idea of this model is future will be almost similar to the past. The intention of exponential smoothing is to smooth original series like moving average does and then to use smoothed series in forecasting future asset value. Exponential smoothing is a simple and realistic approach to forecasting whereby forecast is constructed from an exponentially weighted average of past observations. Simple Exponential Smoothing is widely used forecasting technique which requires little computation. This method is used when data pattern is approximately horizontal (i.e., there is no neither cyclic variation nor pronounced trend in the historical data).

Let an observed time series be y1, y2, …. yn. Formally, the simple exponential smoothing equation takes the form of

$$S_{t+1} = \alpha y_t + (1-\alpha) S_t$$

$S_i$-The smoothed value of time series at time

$Y_i$-Actual value of time series at time

$\alpha$ -Smoothing constant

 In case of simple exponential smoothing, the smoothed statistic is the Forecasted value.

$$F_{t+1} = \alpha y_t + (1-\alpha) F_t$$

$F_{t+1}$ -Forecasted value of time series at time t+1

$F_t$ -Forecasted value of time series at time t

This means:

$$F_t = \alpha y_{t-1} + (1-\alpha) F_{t-1}$$

$$F_{t-1} = \alpha y_{t-1} + (1-\alpha) F_{t-2}$$

$$F_{t-2} = \alpha y_{t-2} + (1-\alpha) F_{t-3}$$

$$F_{t-3} = \alpha y_{t-3} + (1-\alpha) F_{t-4}$$

Substituting, $F_{t+1} = \alpha y_t + (1-\alpha) F_t =$

$\alpha y_t + (1-\alpha)(\alpha y_{t-1} + (1-\alpha)F_{t-1}) =$

$= \alpha y_t + \alpha (1-\alpha) y_{t-1} + (1-\alpha)^2 F_{t-1} =$

$= \alpha y_t + \alpha (1-\alpha) y_{t-1} + \alpha (1-\alpha)^2 y_{t-2} + (1-\alpha)^3 F_{t-2}$

$= \alpha y_t + \alpha (1-\alpha) y_{t-1} + \alpha (1-\alpha)^2 y_{t-2} + \alpha(1-\alpha)^3 y_{t-3} + (1-\alpha)^4 F_{t-3}$

Generalizing,

$$F_{t+1} = \sum^{t-1}_{i=0} \alpha (1-\alpha) i \, y_{t-i} + (1-\alpha)^t F_1$$

The series of weights used in producing the forecast Ft are $\alpha$ , $\alpha (1-\alpha)$ , $\alpha(1-\alpha)^2$ , $\alpha(1-\alpha)^3$….

These weights decline toward zero in an exponential fashion; thus, as we go back in the series, each value has a smaller weight in terms of its effect on the forecast. The exponential decline of the weights towards zero is evident.

Choosing α:

After the model is specified, its important to validate its performance characteristics by comparison of its forecast with historical data for the process it was designed to forecast. We can use the error measures such as MAPE (Mean absolute percentage error), MSE (Mean square error) or RMSE (Root mean square error) and α is chosen such that the error is minimum. Usually the MSE or RMSE can be used as the criterion for selecting an appropriate smoothing constant. For instance, by assigning a value from 0.1 to 0.99, we select the value that produces the smallest MSE or RMSE.

This method is call exponential smoothing because weight given to each observation is exponentially reduced. This is quite better than moving average model. But still it has few limitations:

- ❖ It does not project trend
- ❖ It does not recognize its seasonal pattern
- ❖ It cannot use any external information like pricing or marketing expenses

Lastly we can say Simple Exponential Smoothing will help in getting good results and can be foundation block to build complex models later.

# Holt-Winters Trend Method

This method is named after researchers who proposed this model. An early form of exponential smoothing forecast was initially proposed by R.G. Brown in 1956. His equations were refined by Charles C.Holt in 1957. These smoothing models were again improved by Peter Winters. On their name this method named to be Holt-Winters Method. Both proposed different exponential smoothing models that also can understand and project a trend or seasonality. This method is very common time series forecasting method capable of including both trend and seasonality. It is combination of Simple Exponential Smoothing, Holt's Exponential Smoothing, Winter's Exponential Smoothing. It's therefore otherwise referred to as triple exponential smoothing.

**Holt-Winters' additive method**

The component form for the additive method is:

$$y_{t+h|t} = \ell_t + hb_t + s_{t+h-m(k+1)}$$
$$\ell_t = \alpha(y_t - s_{t-m}) + (1-\alpha)(\ell_{t-1} + b_{t-1})$$
$$b_t = \beta*(\ell_t - \ell_{t-1}) + (1-\beta*)b_{t-1}$$
$$s_t = \gamma(y_t - \ell_{t-1} - b_{t-1}) + (1-\gamma)s_{t-m},$$

where k is the integer part of $(h-1)/m$, which ensures that the estimates of the seasonal indices used for forecasting come from the final year of the sample.

The level equation shows a weighted average between the seasonally adjusted observation $(y_t - s_{t-m})(y_t - s_{t-m})$ and non-seasonal forecast $(\ell_{t-1} + b_{t-1})(\ell_{t-1} + b_{t-1})$ for time t. The trend equation is identical to Holt's linear method. The seasonal equation shows a weighted average between the current seasonal index, $(y_t - \ell_{t-1} - b_{t-1})(y_t - \ell_{t-1} - b_{t-1})$, and the seasonal index of the same season last year (i.e., m time periods ago).

The equation for the seasonal component is often expressed as

$$s_t = \gamma*(y_t - \ell_t) + (1-\gamma*)s_{t-m}$$

If we substitute $\ell_t$ from the smoothing equation for the level of the component form above, we get

$$s_t = \gamma*(1-\alpha)(y_t - \ell_{t-1} - b_{t-1}) + [1 - \gamma*(1-\alpha)]s_{t-m},$$

which is identical to the smoothing equation for the seasonal component we specify here, with $\gamma = \gamma*(1-\alpha)$. The usual parameter restriction is $0 \leq \gamma* \leq 1$, which translates to $0 \leq \gamma \leq 1-\alpha$

**Holt-Winters' multiplicative method**

The component form for the multiplicative method is:

$$y_{t+h|t}=(\ell_t+hb_t)s_{t+h-m(k+1)}$$
$$\ell_t=\alpha[y_t/_{t-m}]+(1-\alpha)(\ell_{t-1}+b_{t-1})$$
$$b_t=\beta*(\ell_t-\ell_{t-1})+(1-\beta*)b_{t-1}$$
$$s_t=\gamma[y_t(\ell_{t-1}+b_{t-1})]+(1-\gamma)s_{t-m}$$

This method is capable of capturing level, trend and seasonality component and promptly utilizes them in a forecast. It is incredibly intuitive and relatively simple forecasting procedure capable of modeling plethora of time series.

# ARIMA

It stands for Auto Regressive Integrated Moving Average. It is a linear regression model that uses its own lags as predictors. It helps to get better insight into data and predict future trend. A dataset is stationary if it has constant mean, variance and covariance over time.

AR terms are the lags of stationary series
MA terms are the lags of forecast errors.

It is mainly of two types- Non-Seasonal Arima and Seasonal Arima. These models are applied to stationary data. If data is not stationary then process of differencing is followed. The ACF and PACF patterns are studied to study the presence of lags in data. Then model is fitted and checked for residuals. A series is made stationary by differencing the time series with its lag value. After each differencing, the Augmented Dickey-Fuller (ADF) test is conducted to check the stationarity of the series, and the process is repeated until the series passes the ADF test. The Auto Regression parameter (p), the Difference parameter (d), and the Moving Average parameter (q) are required to fit the ARIMA model to a time series and to perform the univariate forecasting. Python has the auto_arima() function which finds the appropriate p, d, and q value of a series.

p otherwise known as lag order

d otherwise known as degree of differencing

q otherwise known as moving average

ARIMA is a method for forecasting future outcomes based on a historical time series. It is mainly based on statistical concept of serial correlation where past data influence future trends. These models take into account trends, cycles, seasonality, and other non-static types of data while making forecast.

# SECTOR WISE RESULTS AND ANALYSIS

## <u>METAL SECTOR</u>

The performance metric used is RMSE/mean percentage. It checks what percentage of the mean of the test value is RMSE (Root Mean Square Error). RMSE/mean help to compare across stocks as the value range of the variable in consideration won't affect the metric.

1. TATASTEEL

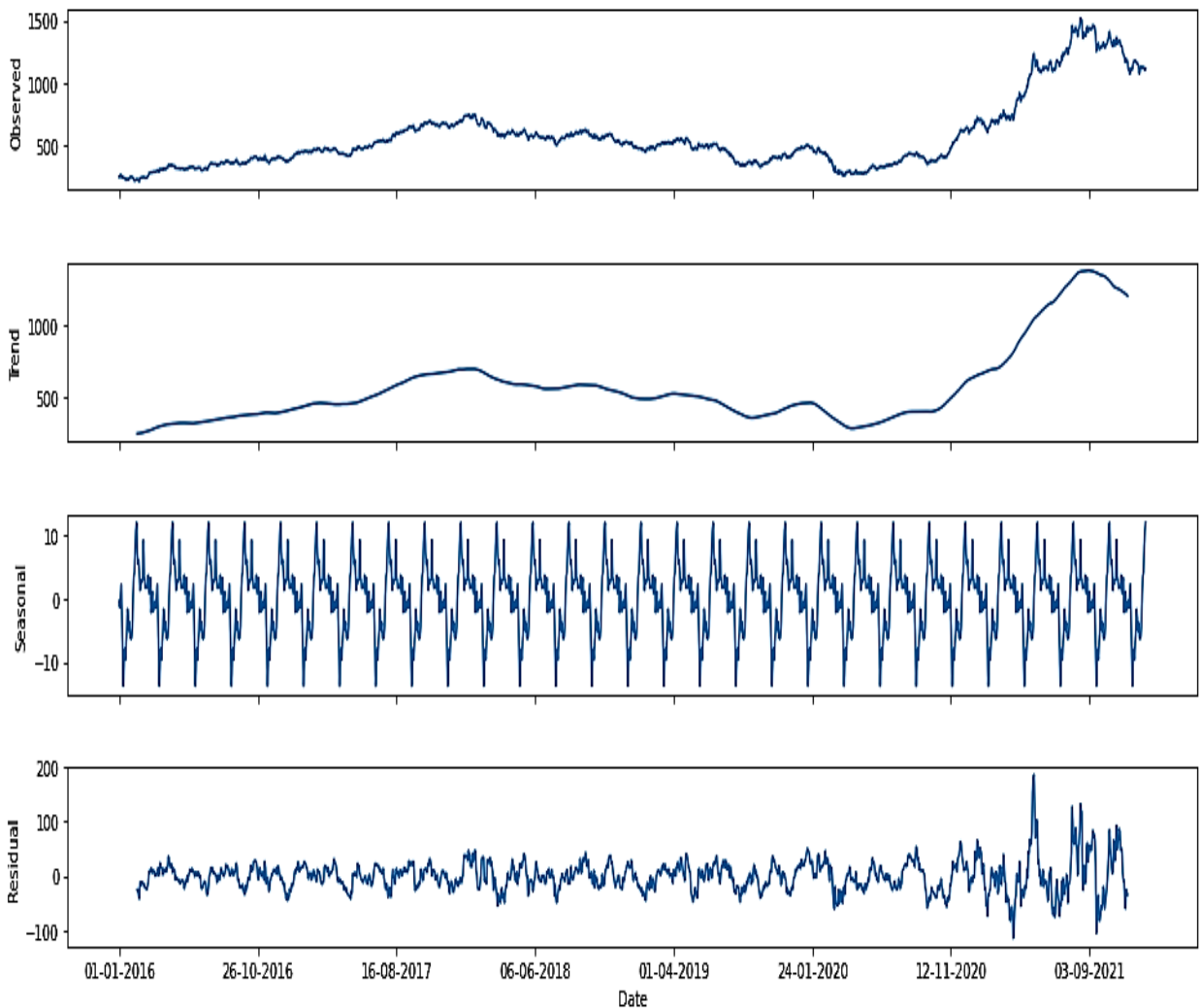

This is the seasonal decomposition of TATASTEEL

## Simple Exponential Smoothing

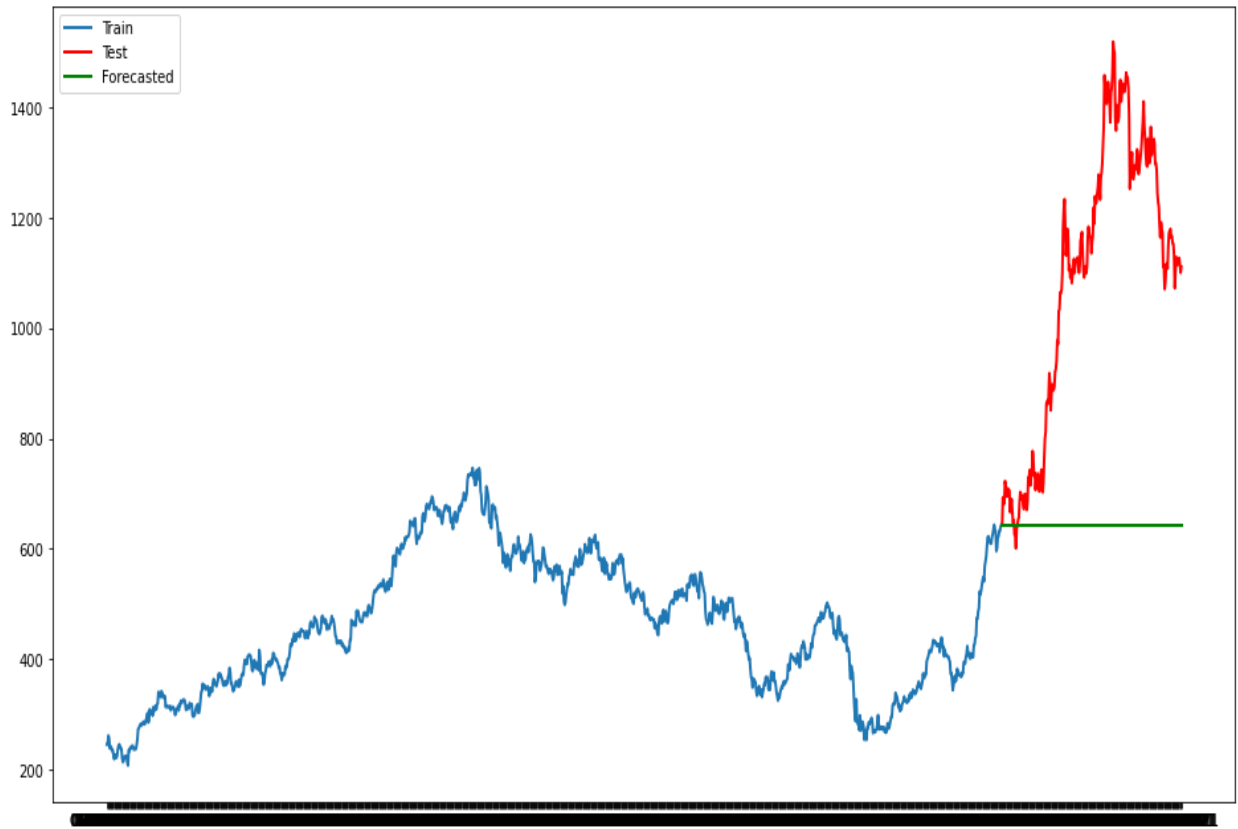

Here is graphical representation test, train and forecast graph. Training data is from 1$^{st}$ January 2016 to 31$^{st}$ December 2020, Test data is from 1st January 2021 to 31$^{st}$ December 2021. Green line shows the forecast value

# Holt Winter Trend Method

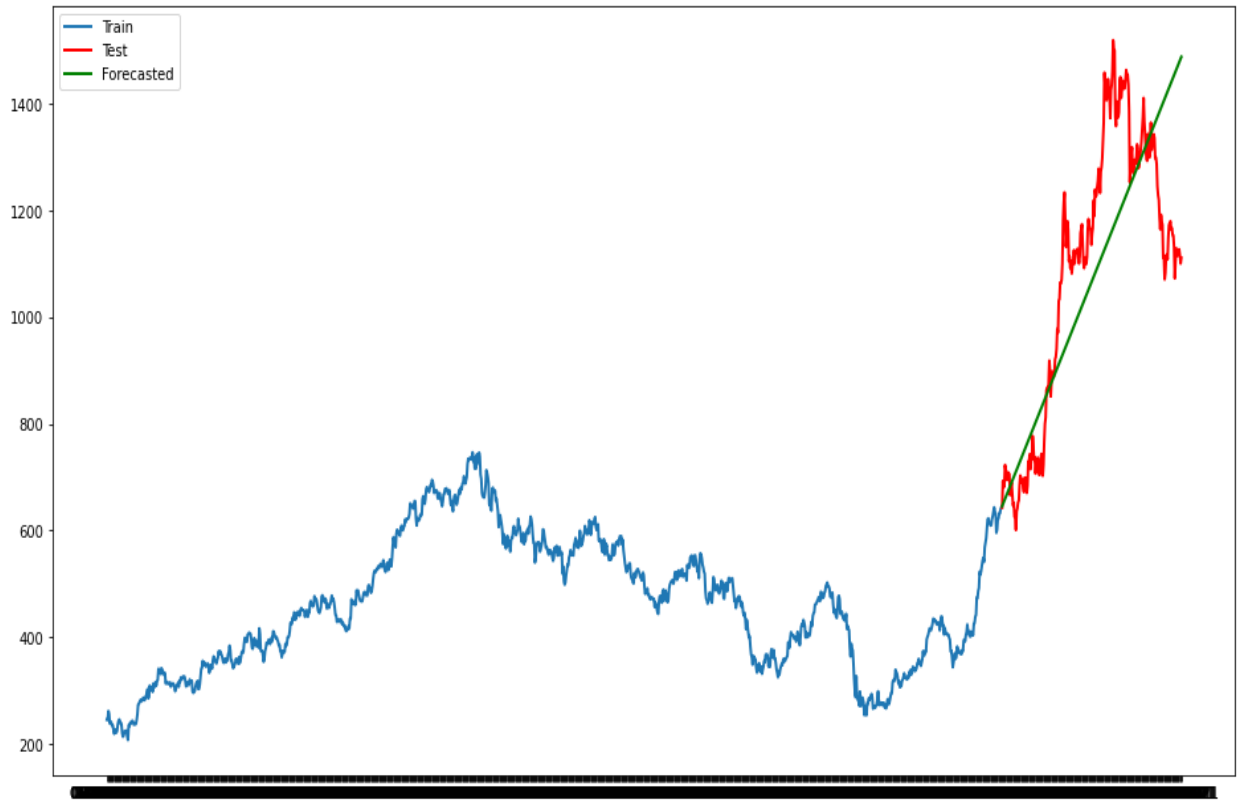

# Arima

Below is log return and Auto correlation plot

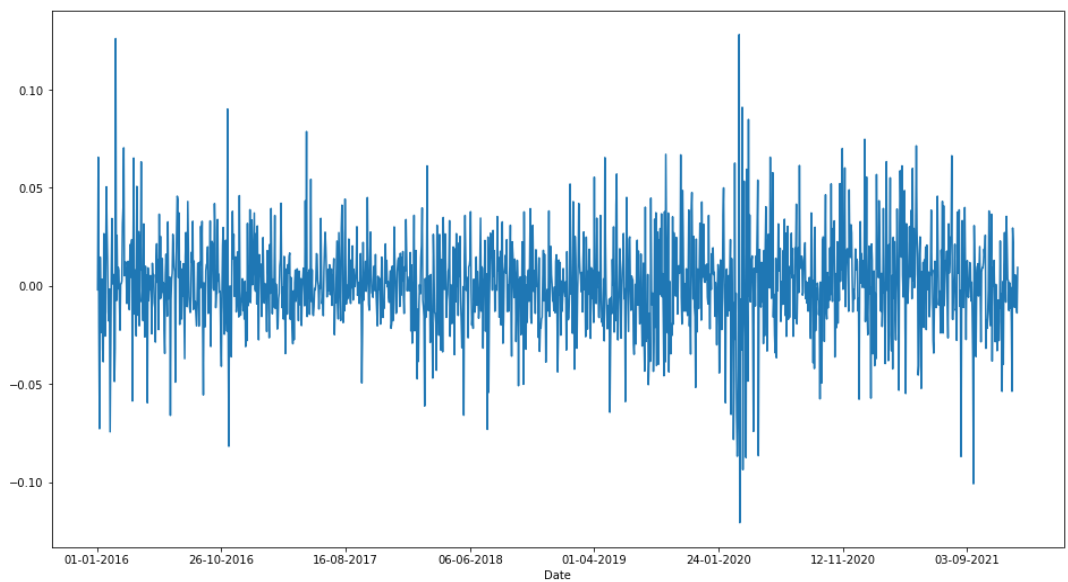

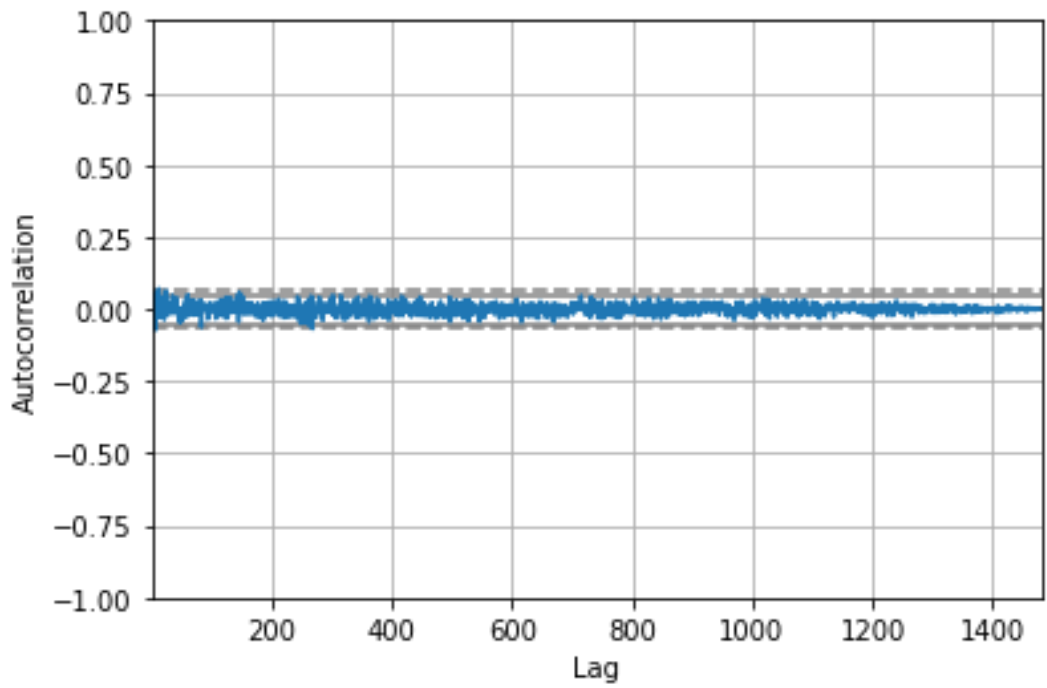

Below is the ACF and PACF

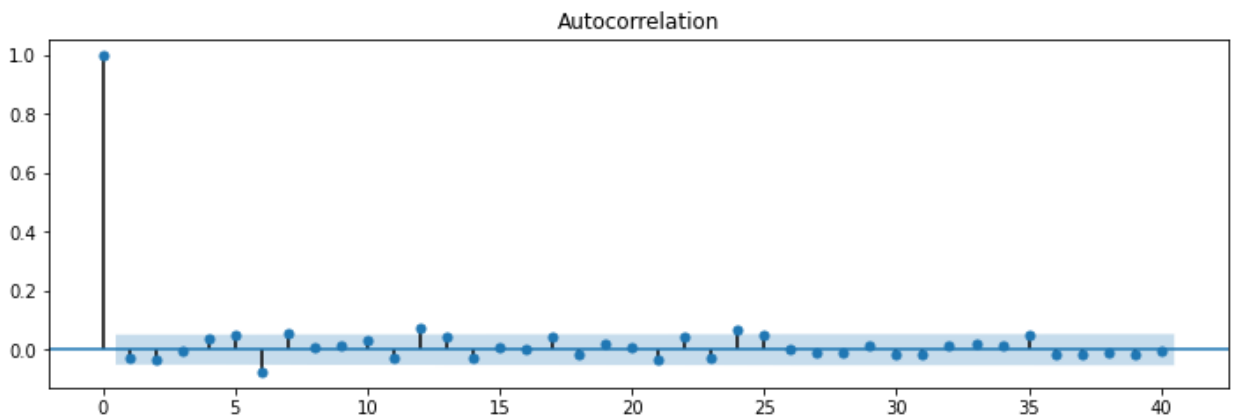

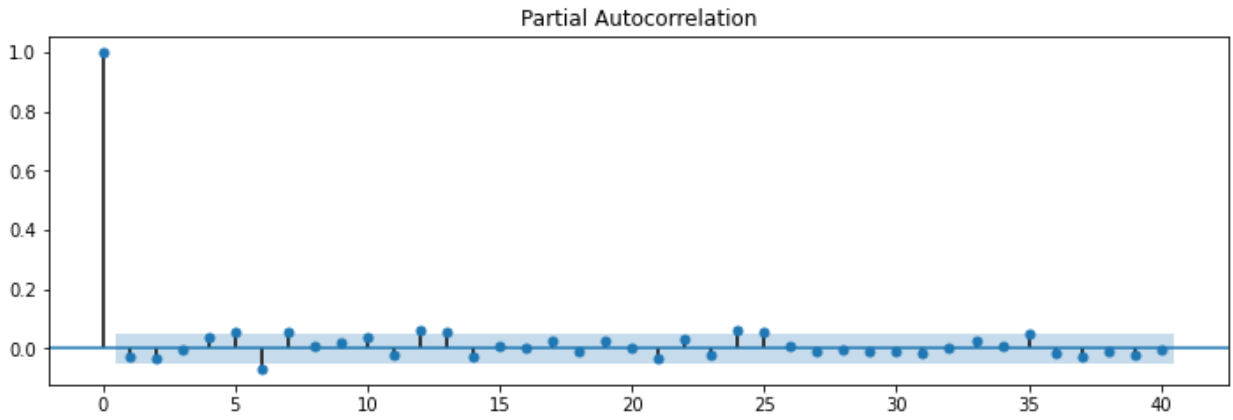

## Sliding Window Method

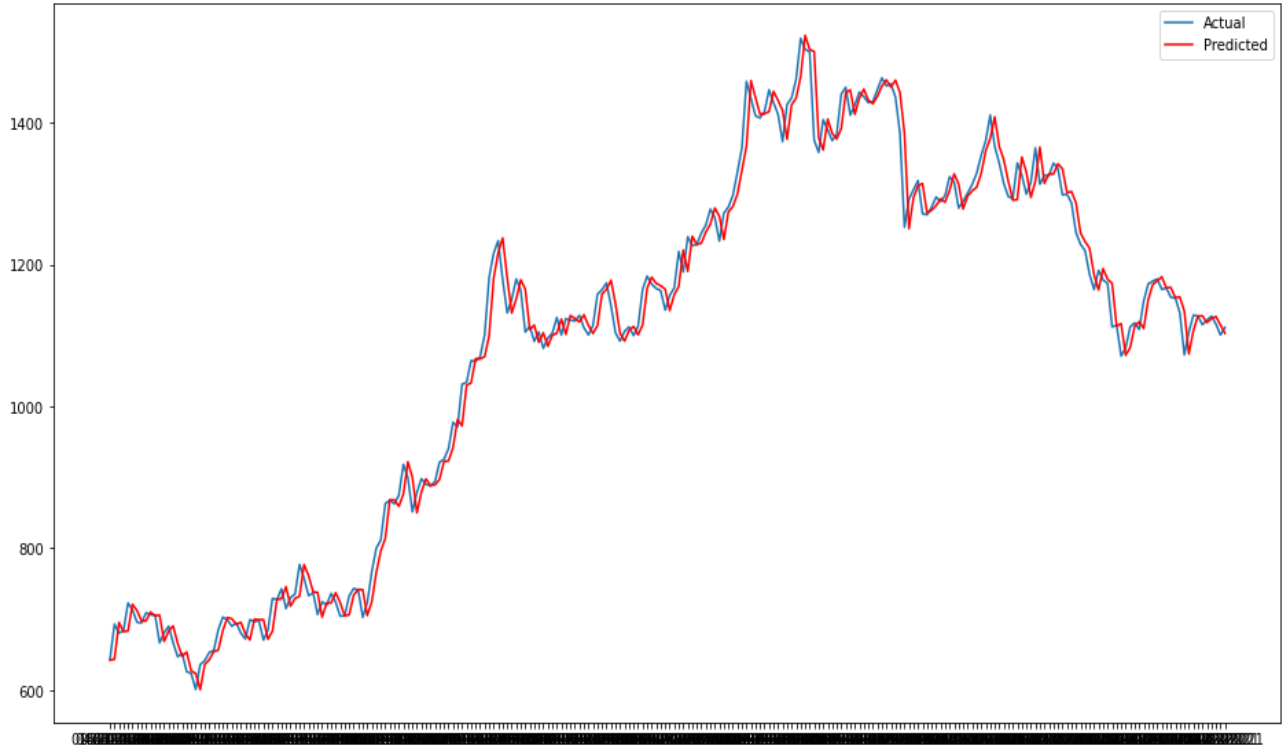

## Rolling Window Method

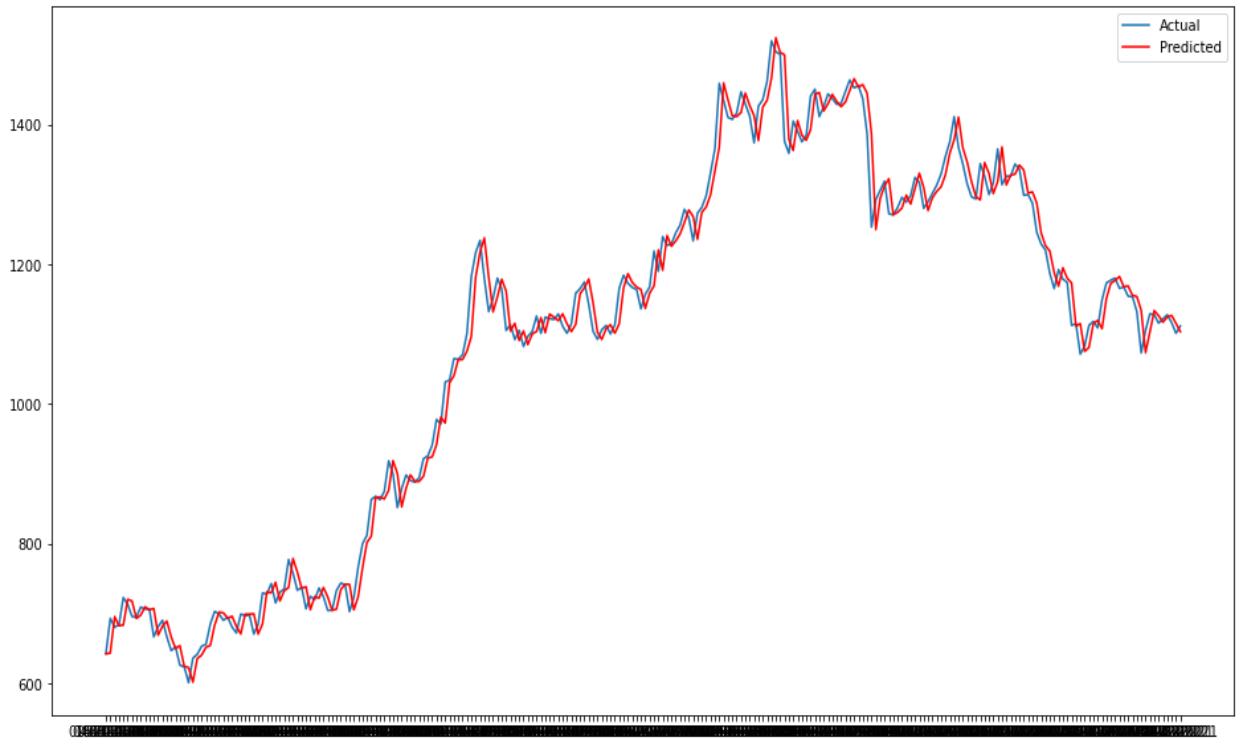

## 2. Hindalco

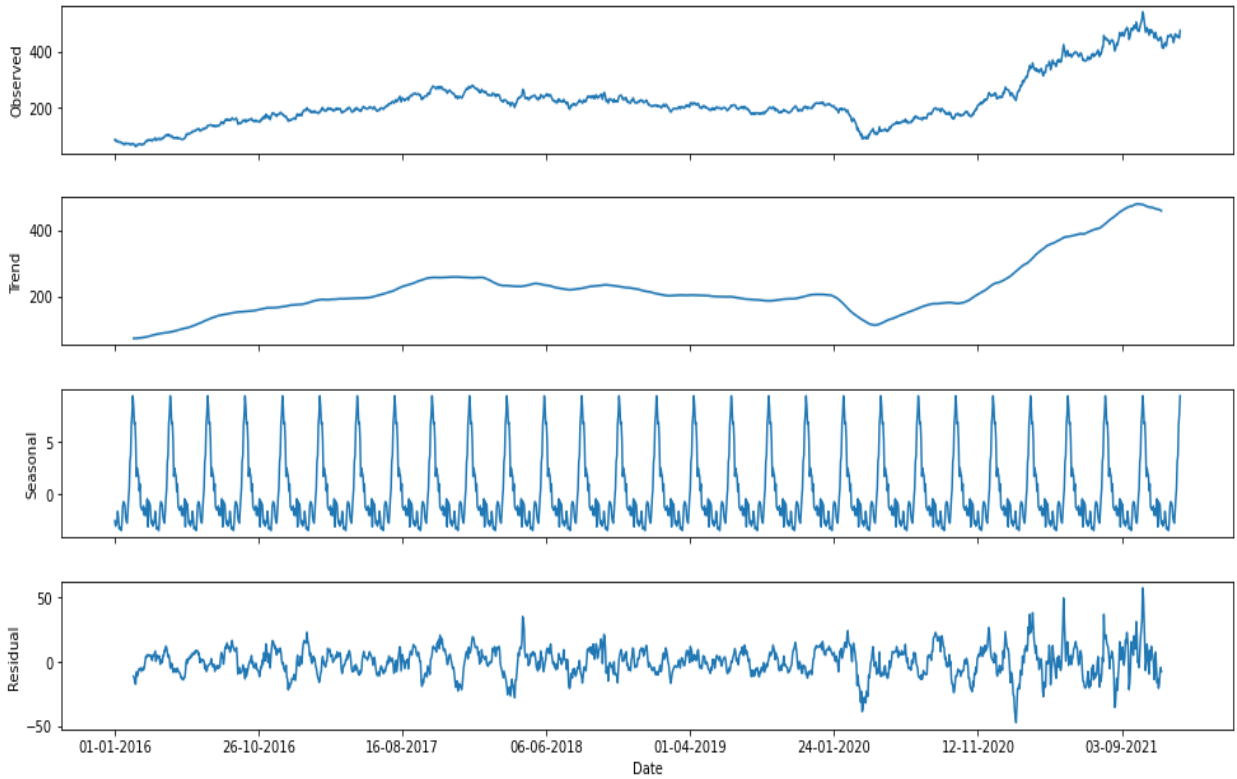

# Simple Exponential Smoothing

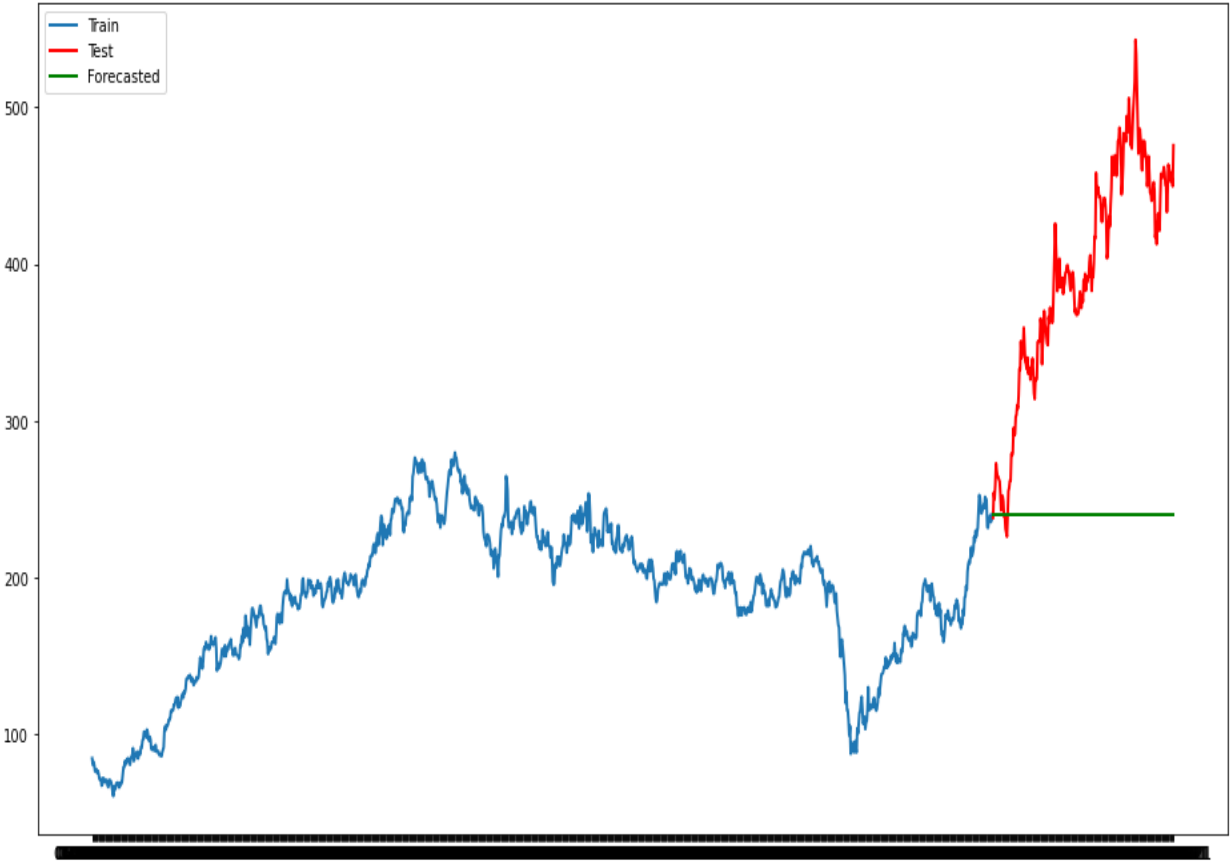

## Holt Winter Trend Method

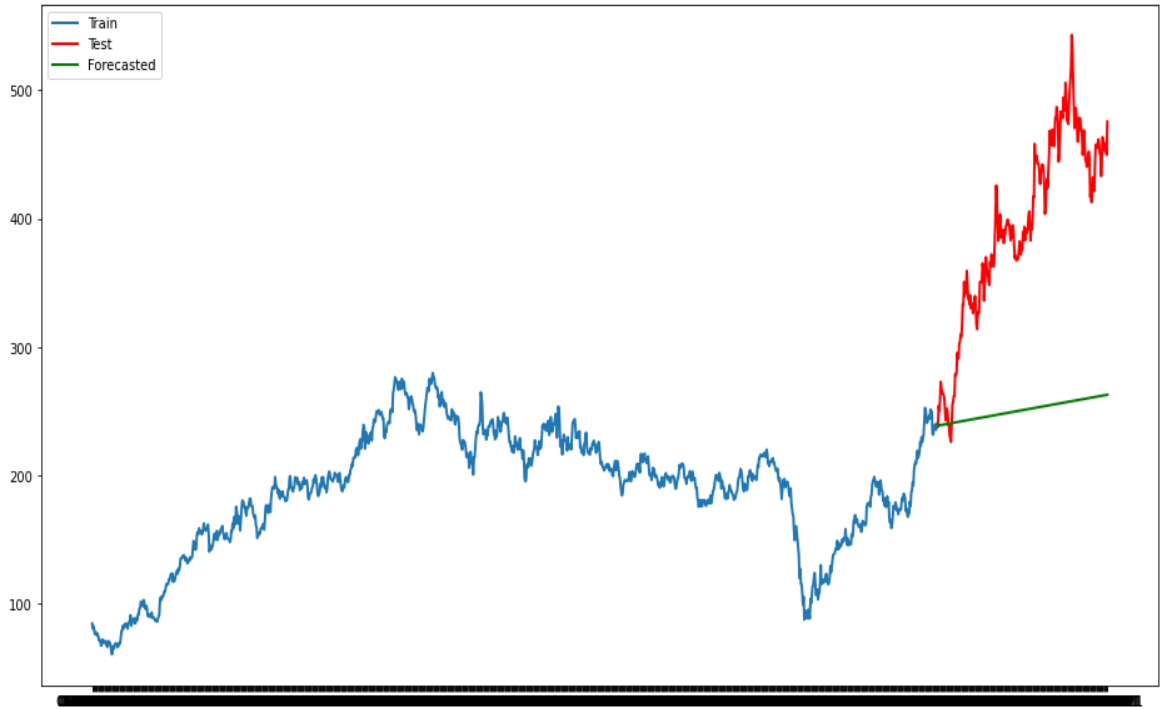

## Sliding Window Method

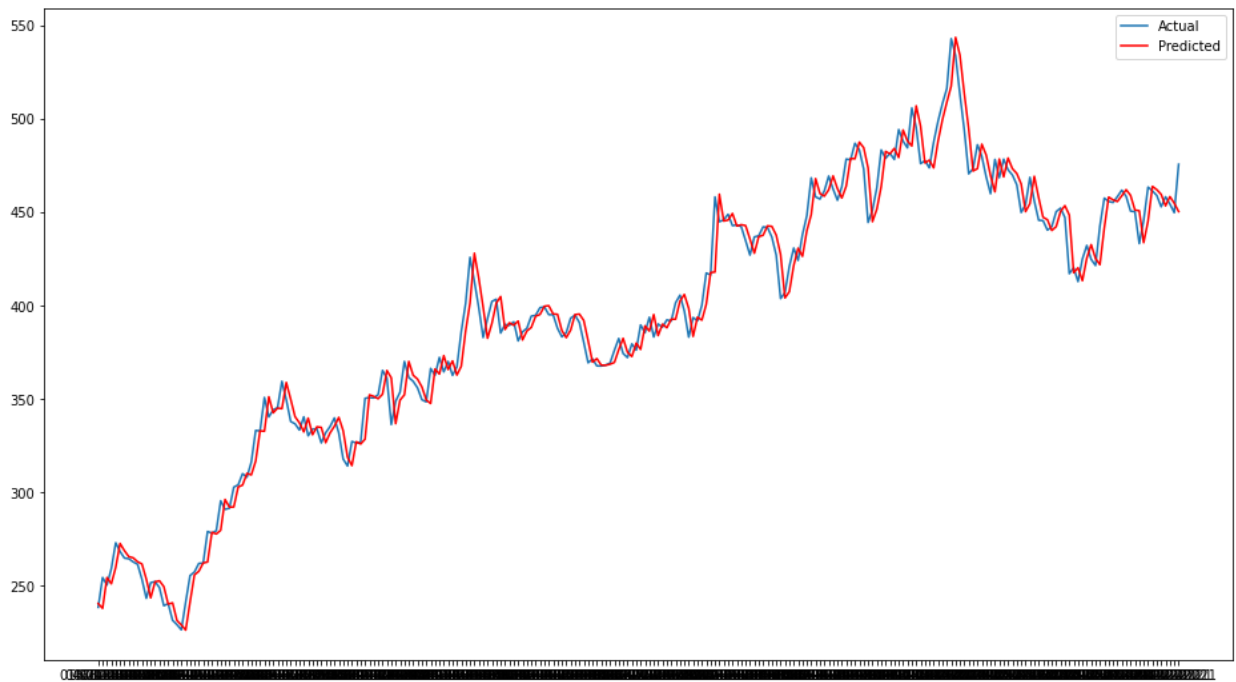

Rolling Window Method

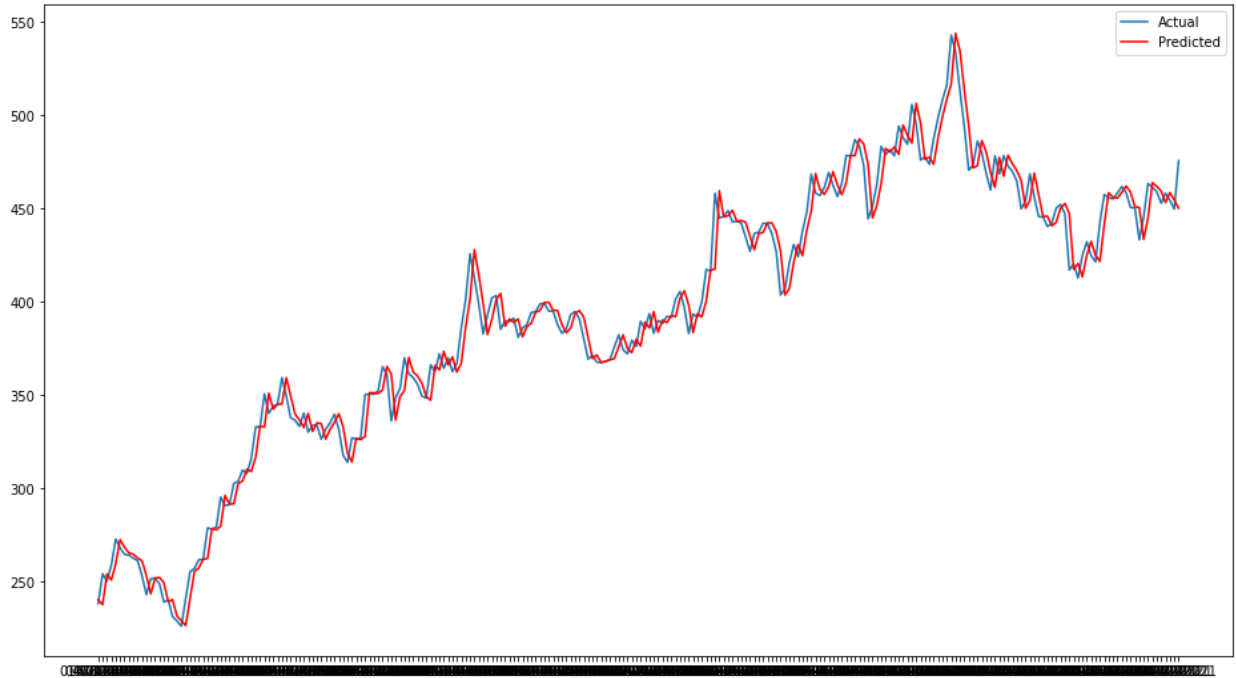

# **IT SECTOR**

1. TCS

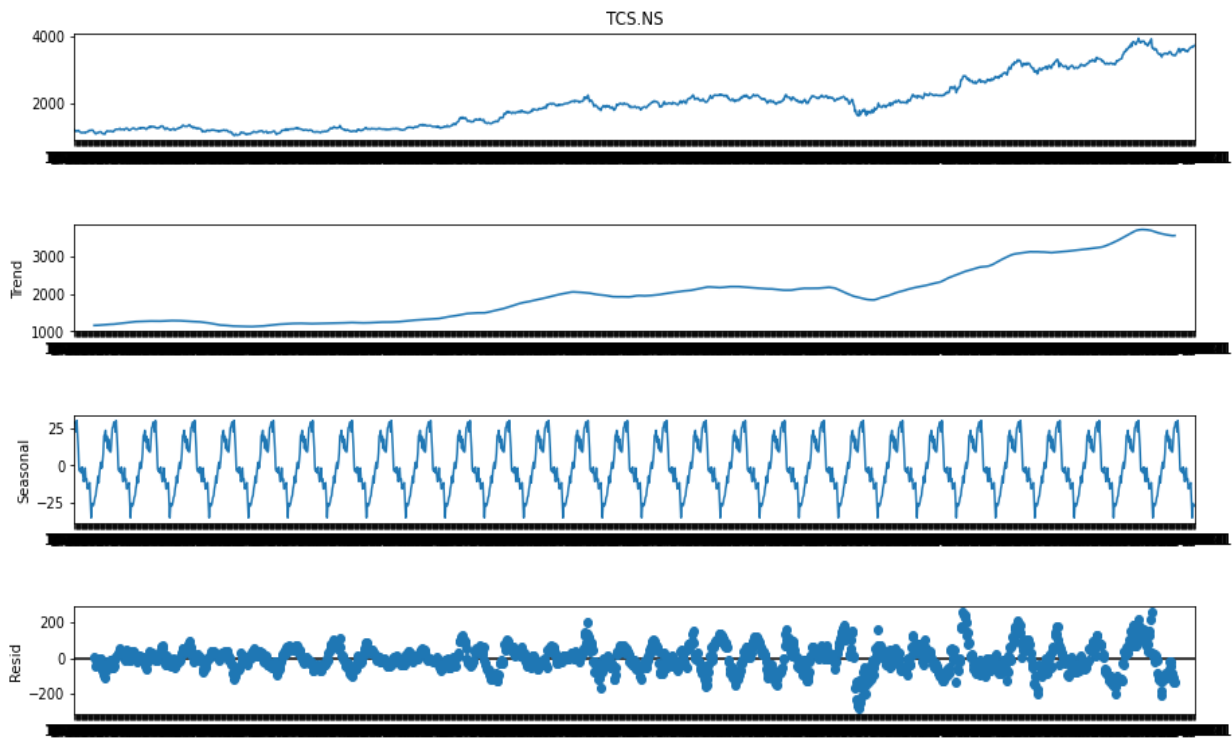

## SIMPLE EXPONENTIAL SMOOTHING

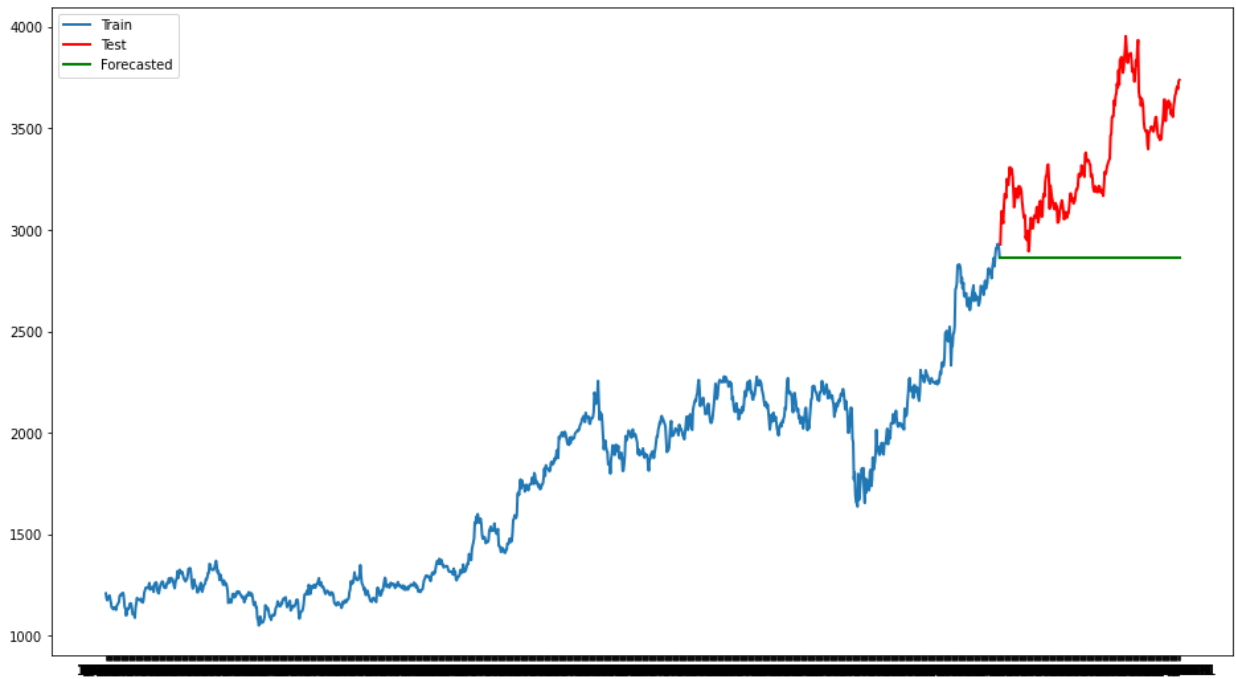

## HOLT WINTER TREND METHOD

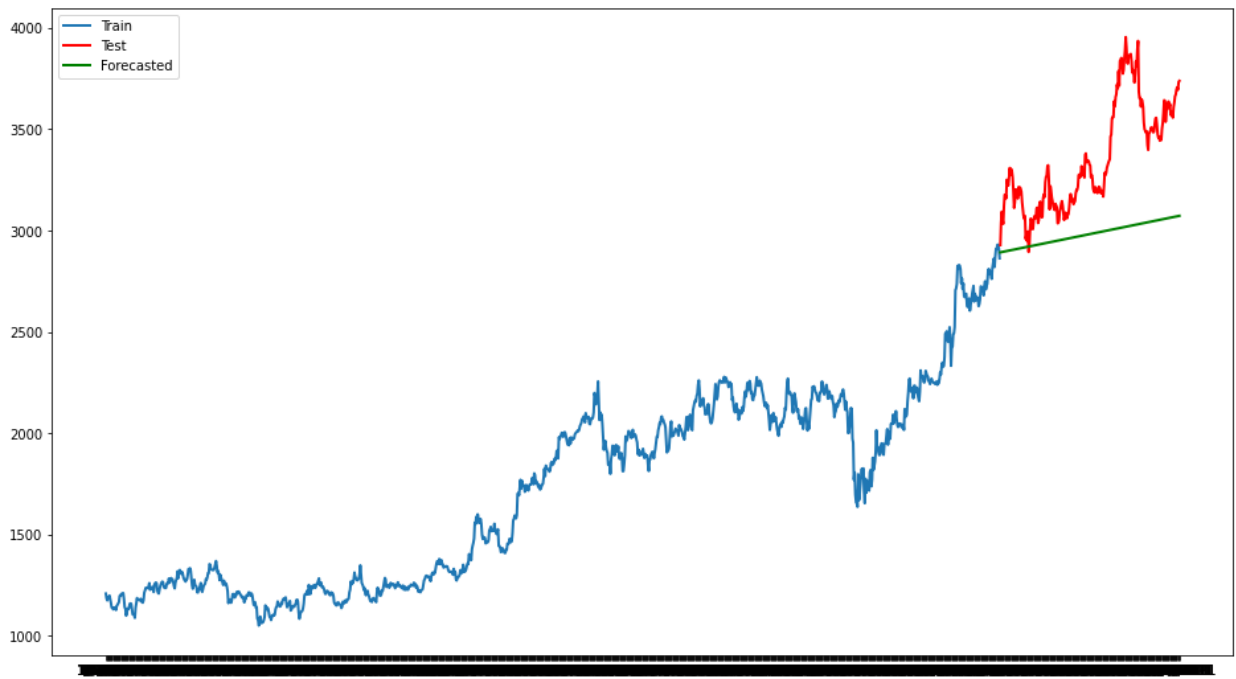

# ARIMA

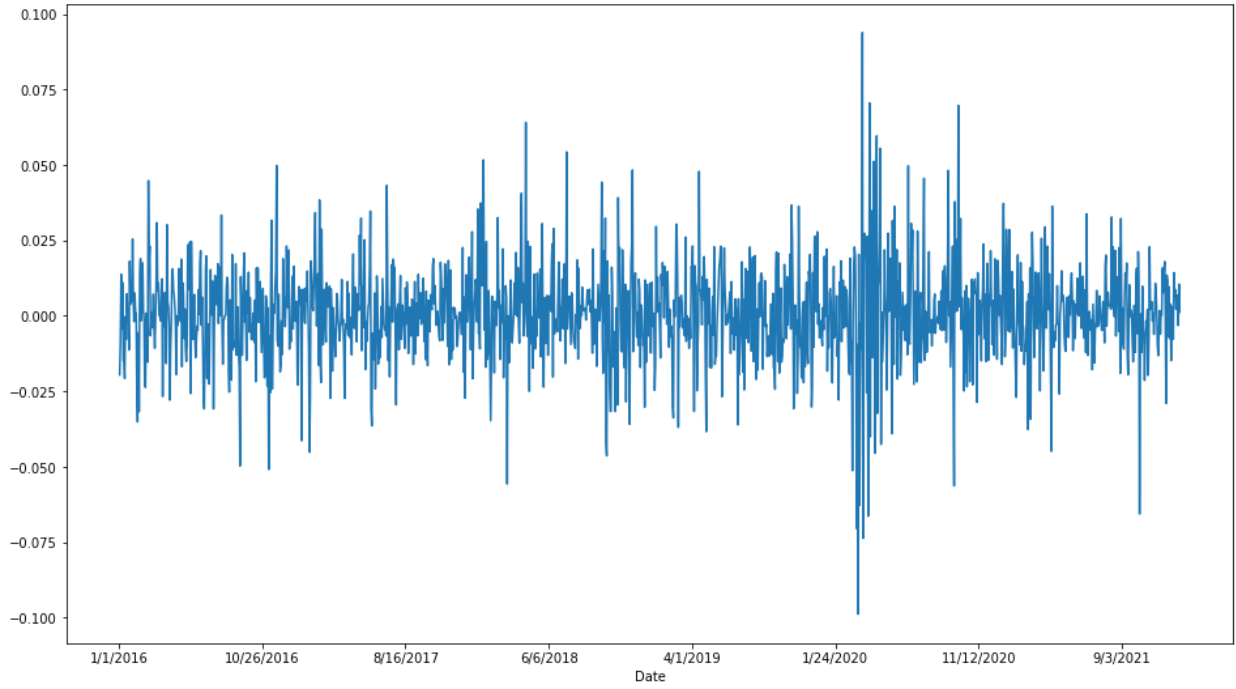

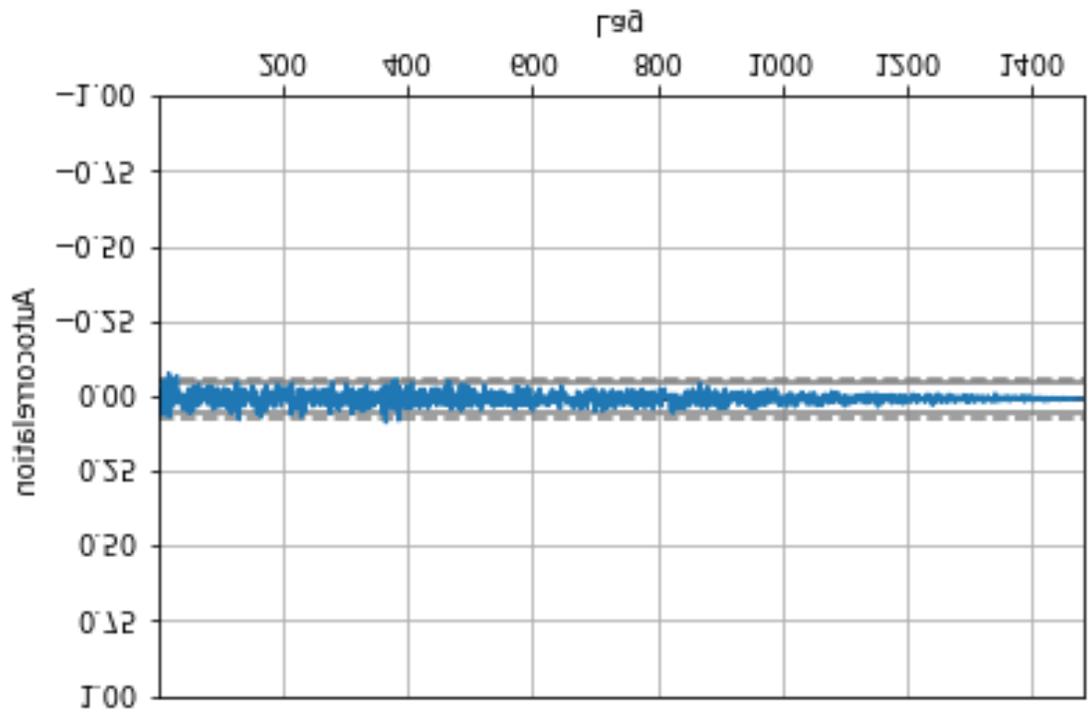

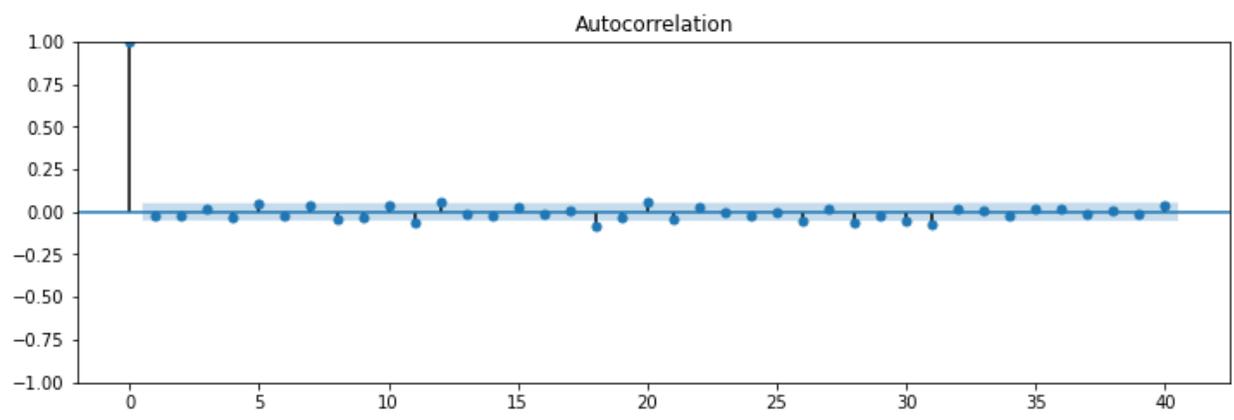

Autocorrelation

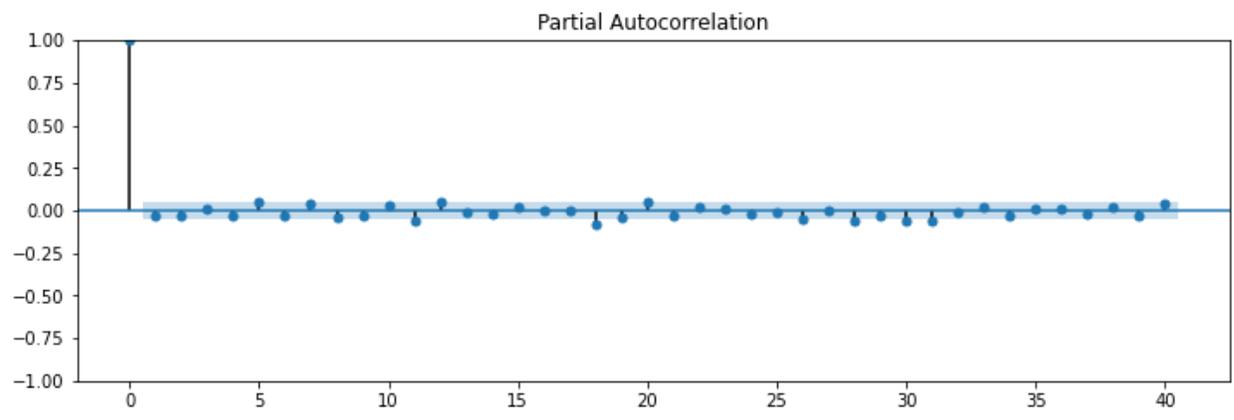

Partial Autocorrelation

## SLIDING WINDOW

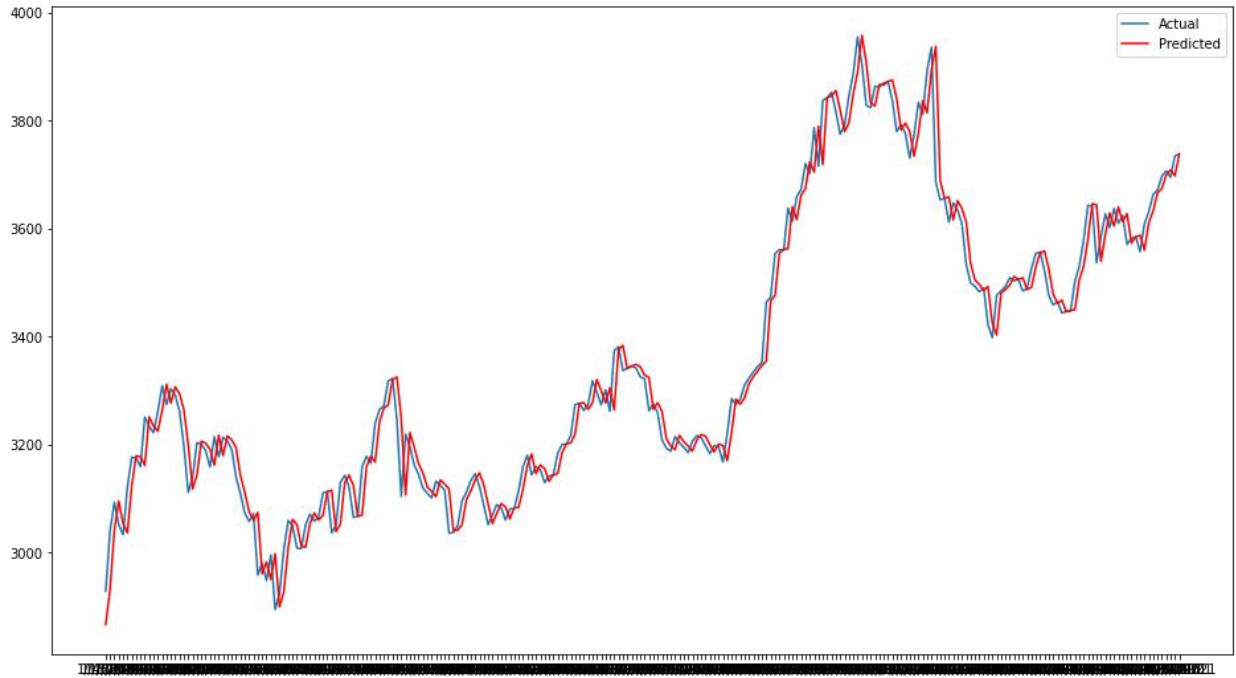

## ROLLING WINDOW

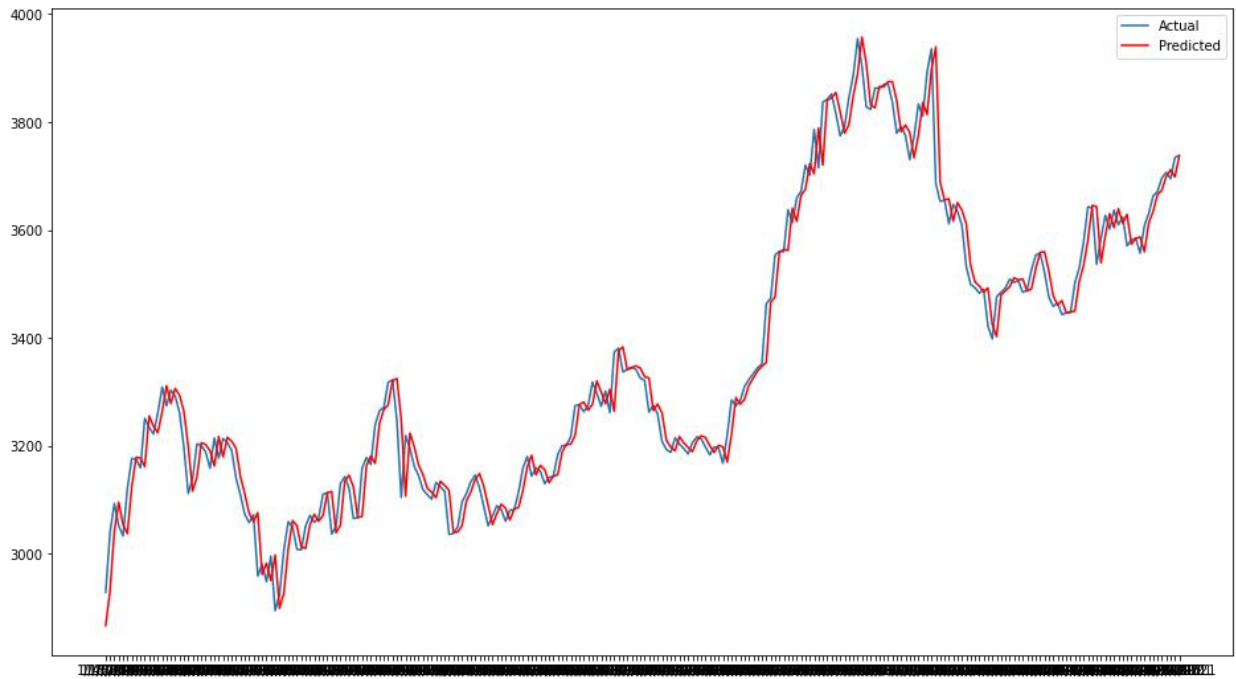

## 2. WIPRO

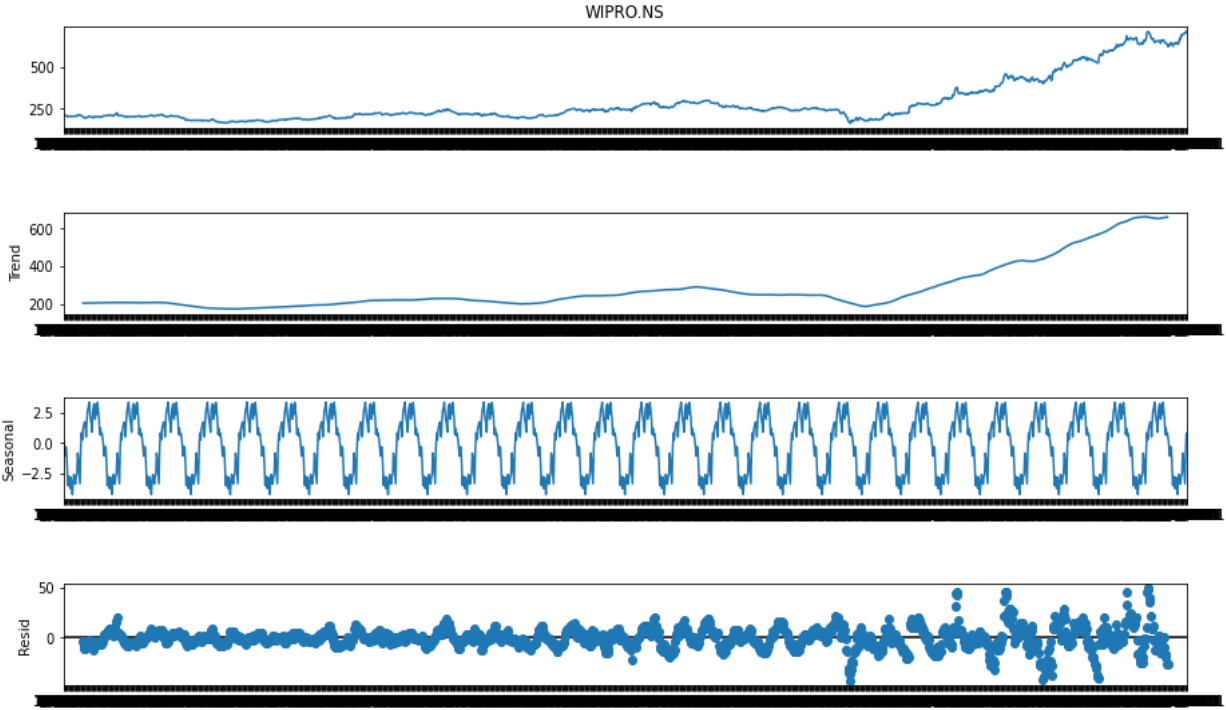

## SIMPLE EXPONENTIAL SMOOTHING

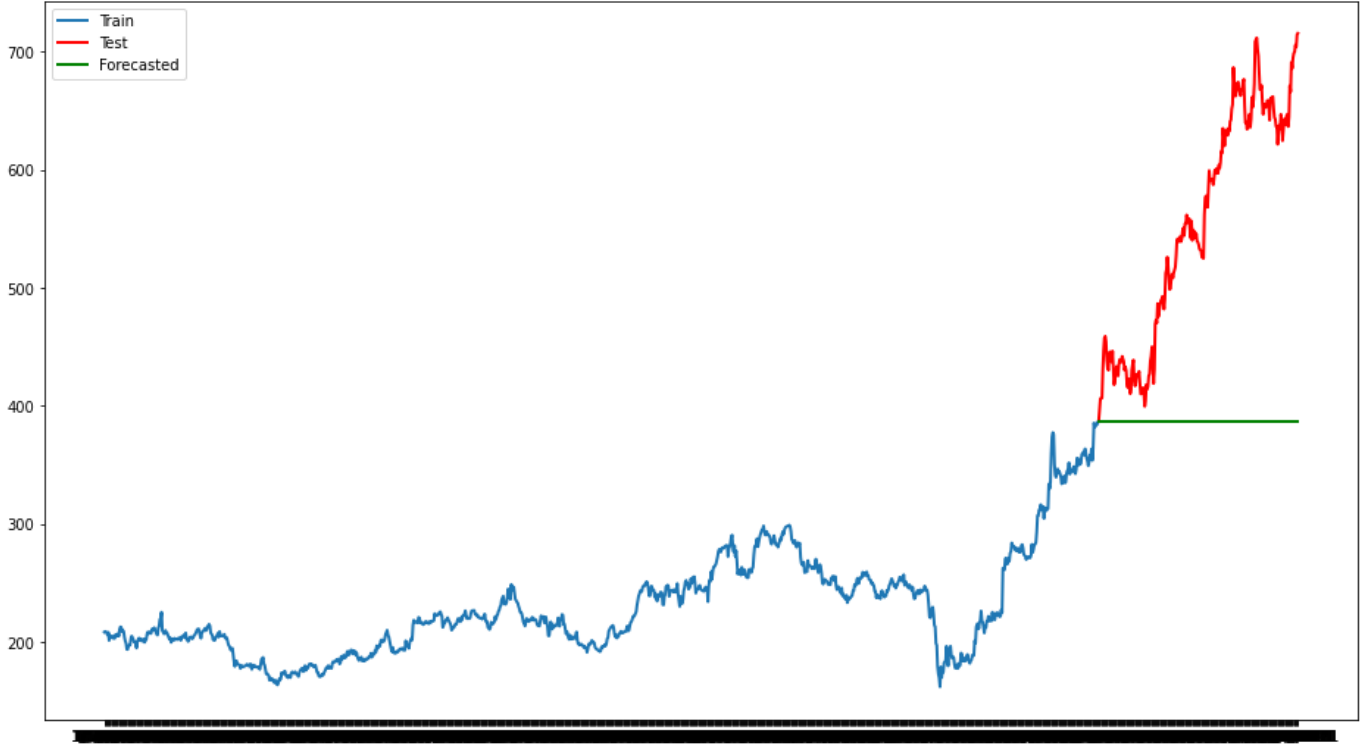

## HOLT WINTER METHOD

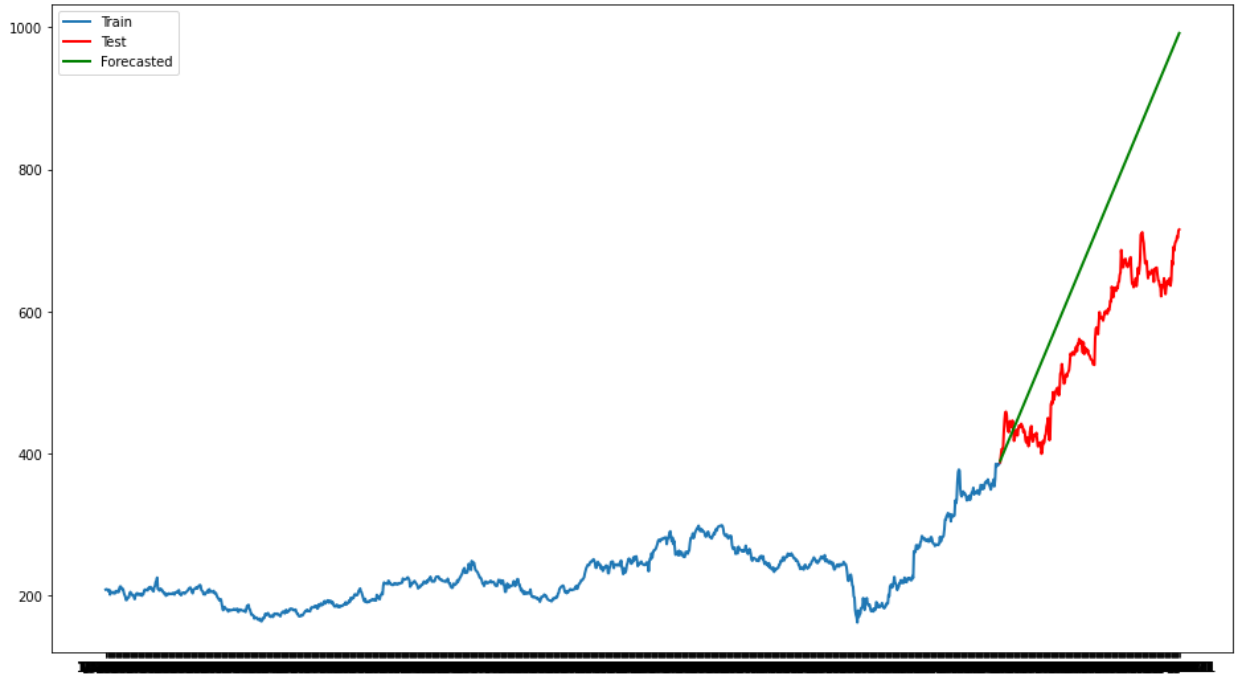

## ARIMA

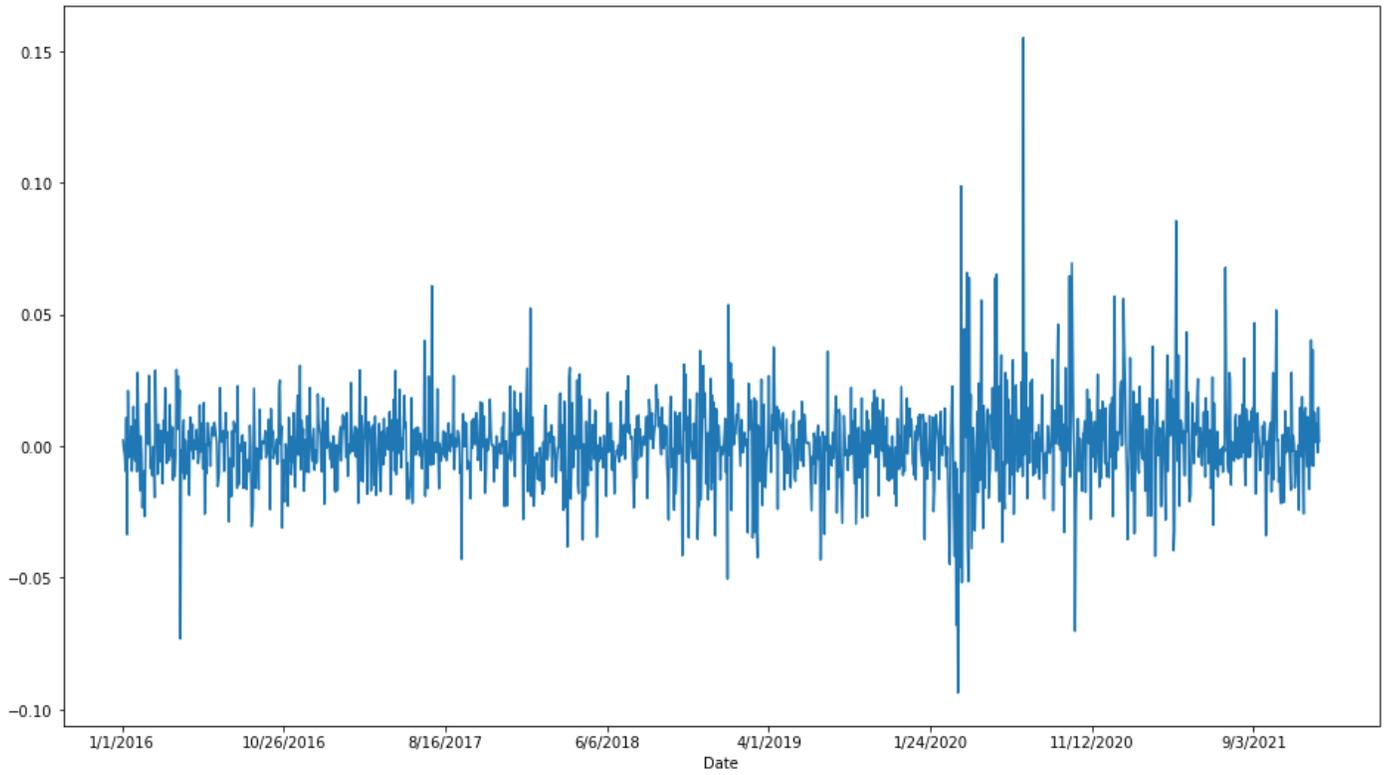

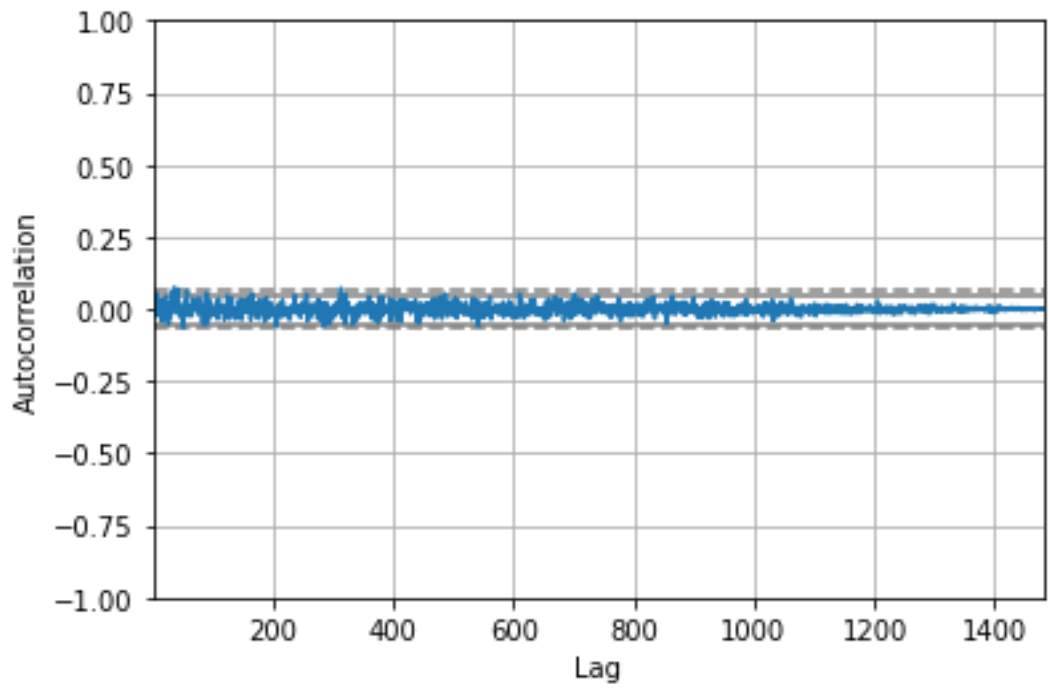

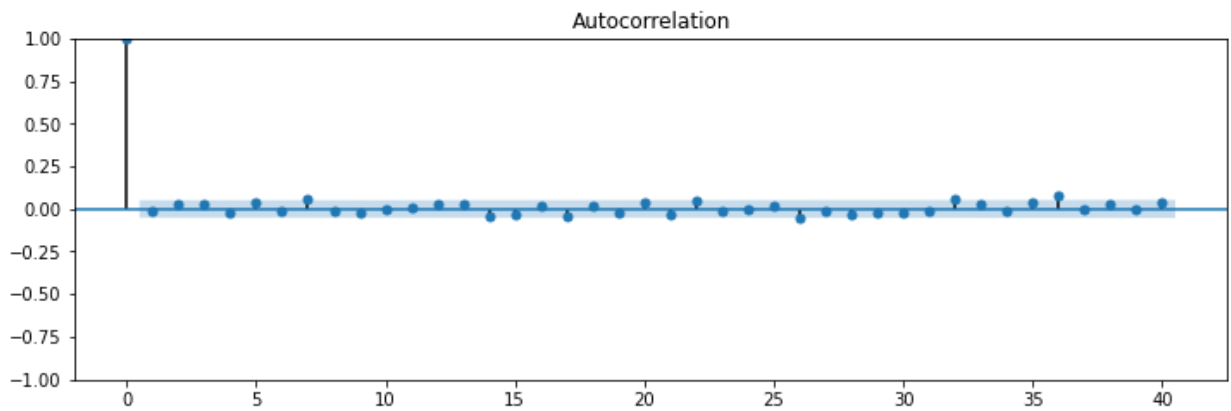

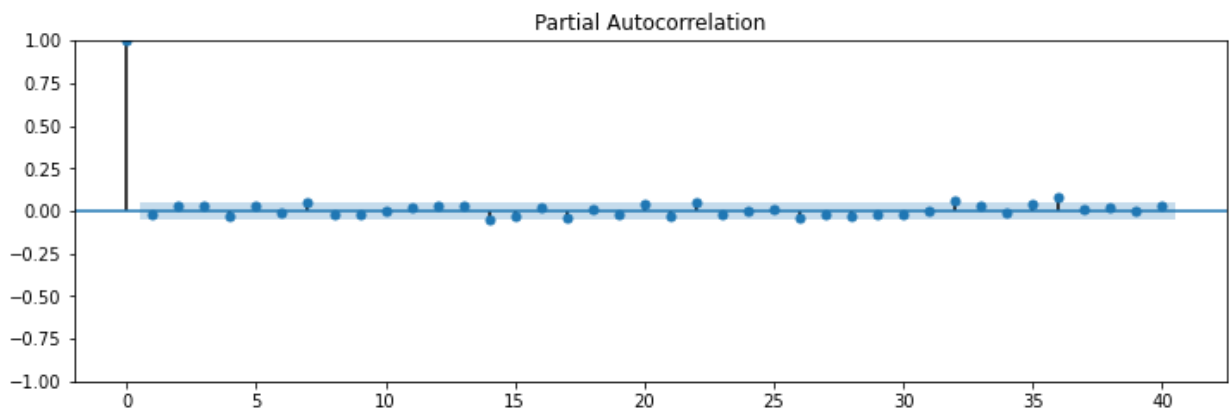

## SLIDING WINDOW

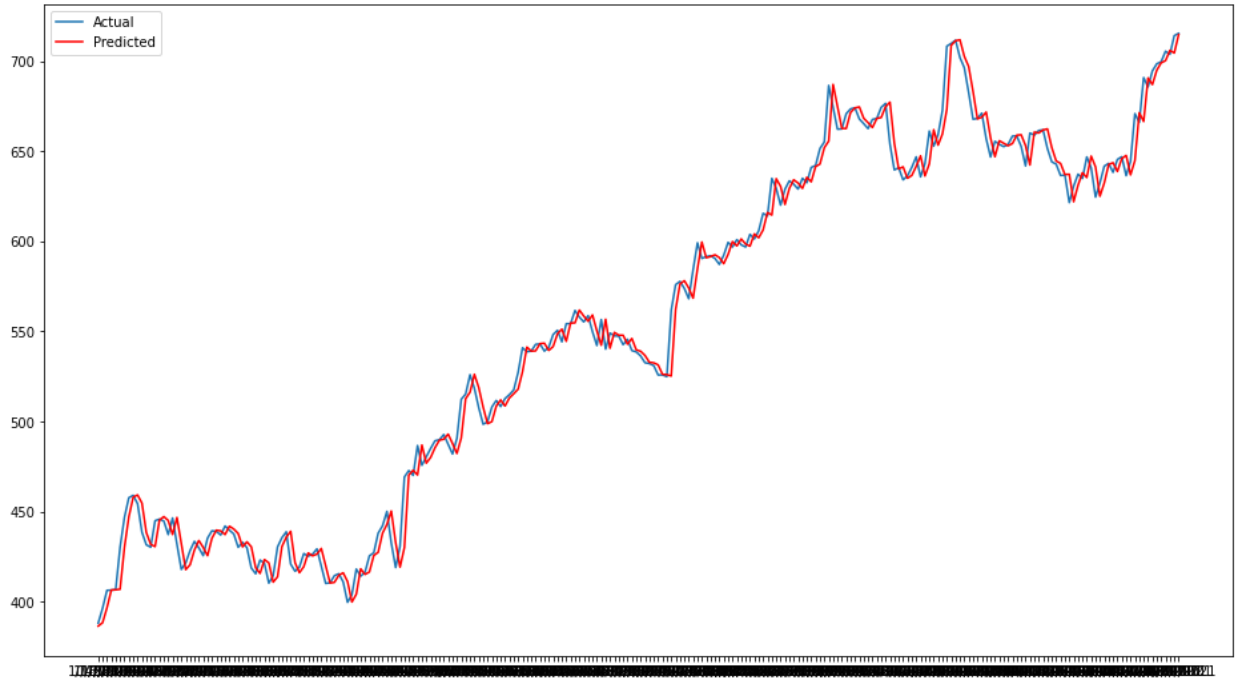

## ROLLING WINDOW

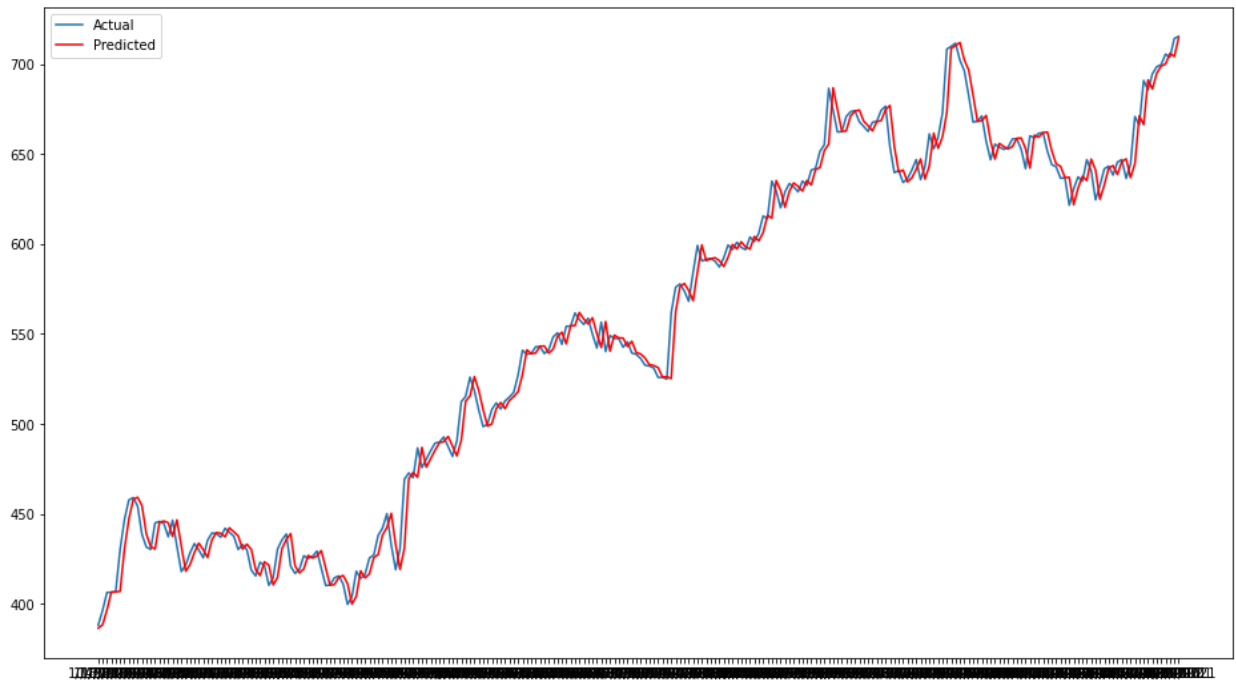

# BANKING SECTOR

1. HDFC BANK

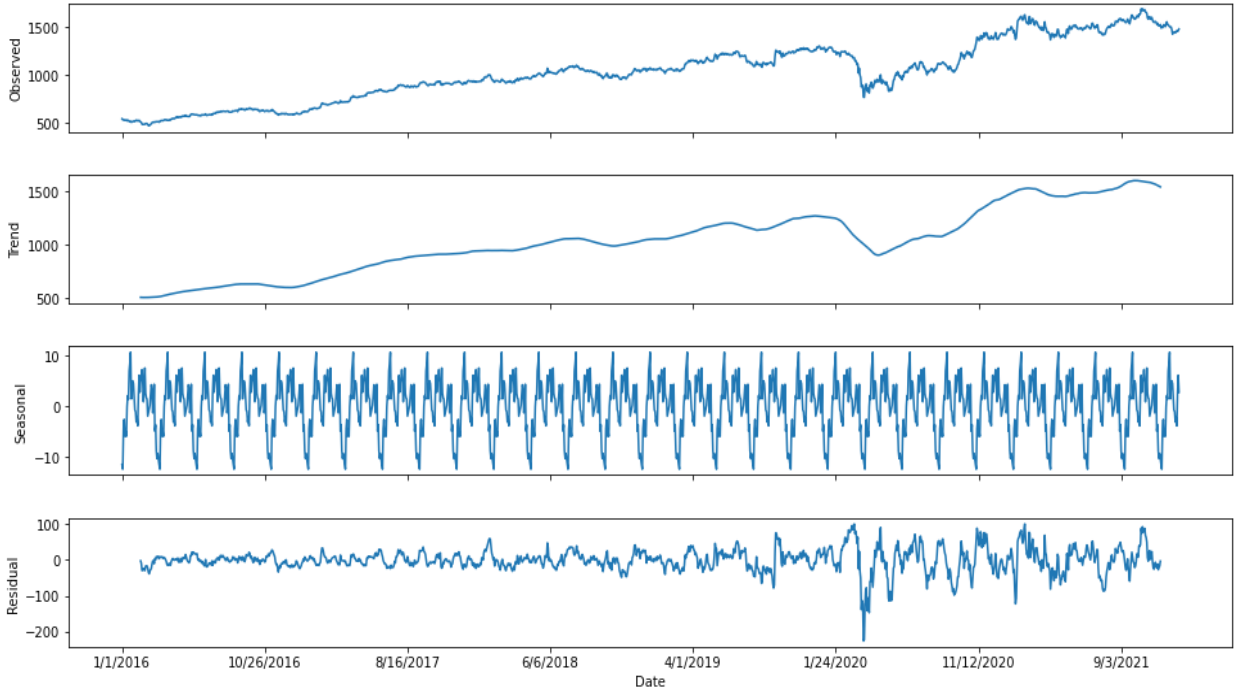

SIMPLE EXPONENTIAL SMOOTHING

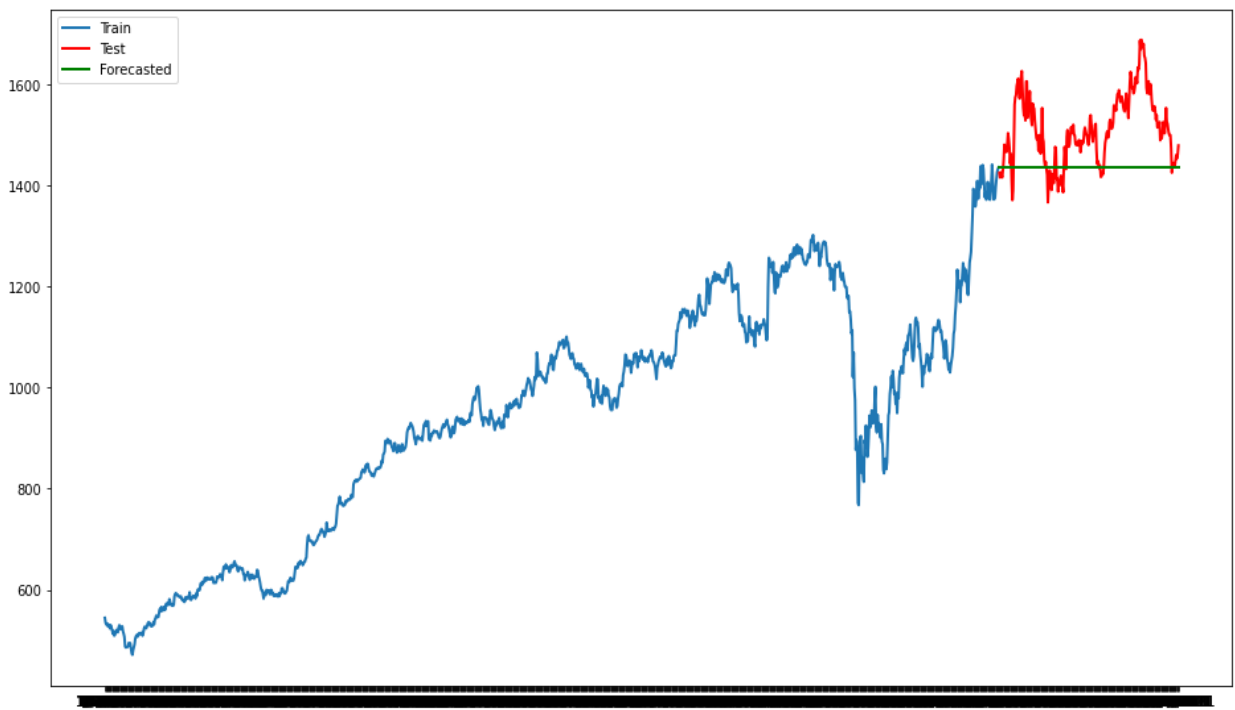

# HOLT WINTER TREND METHOD

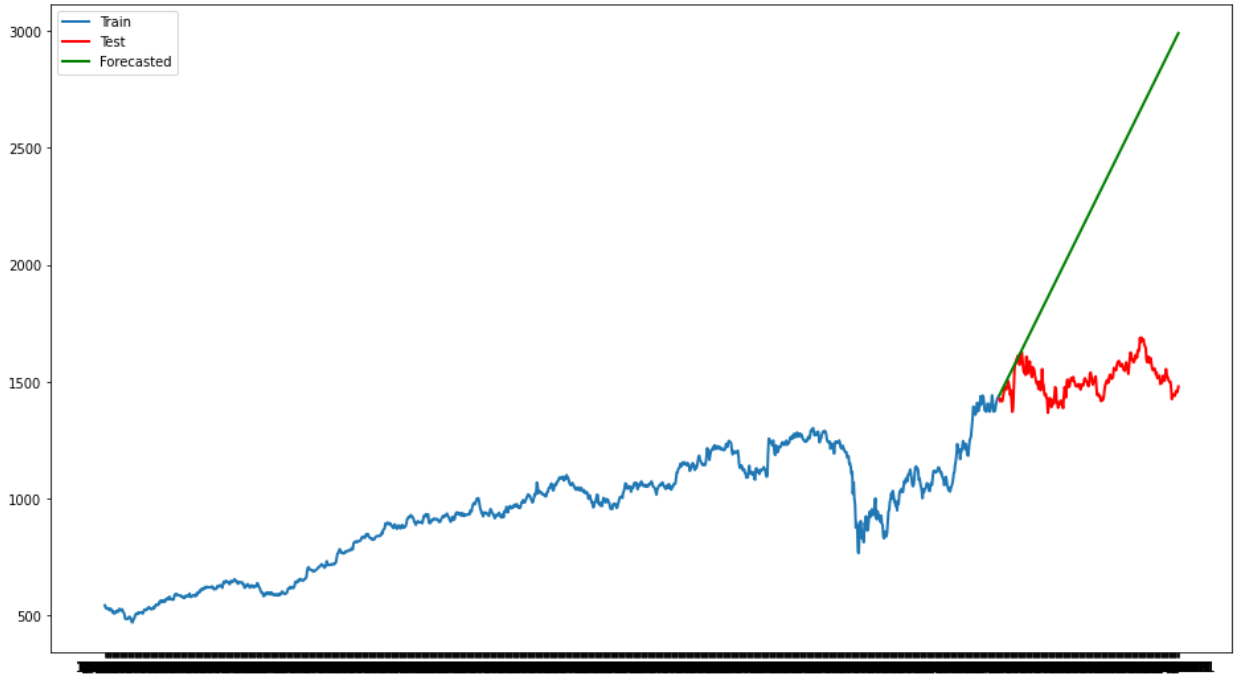

# ARIMA

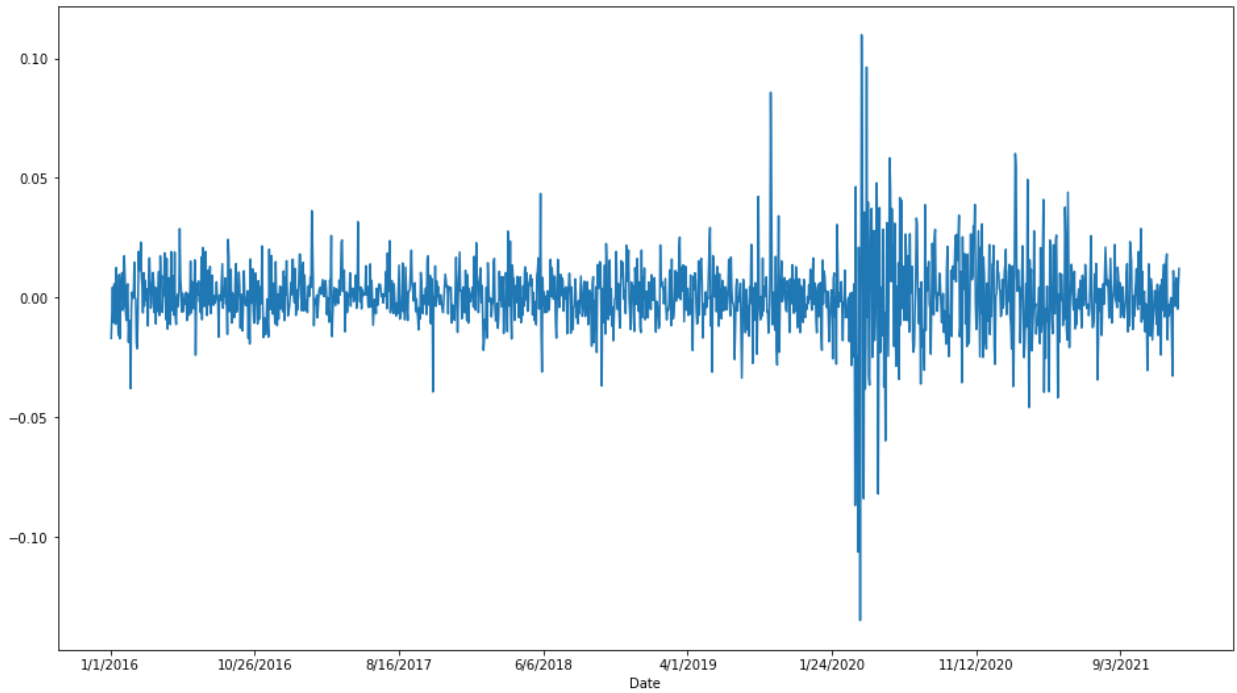

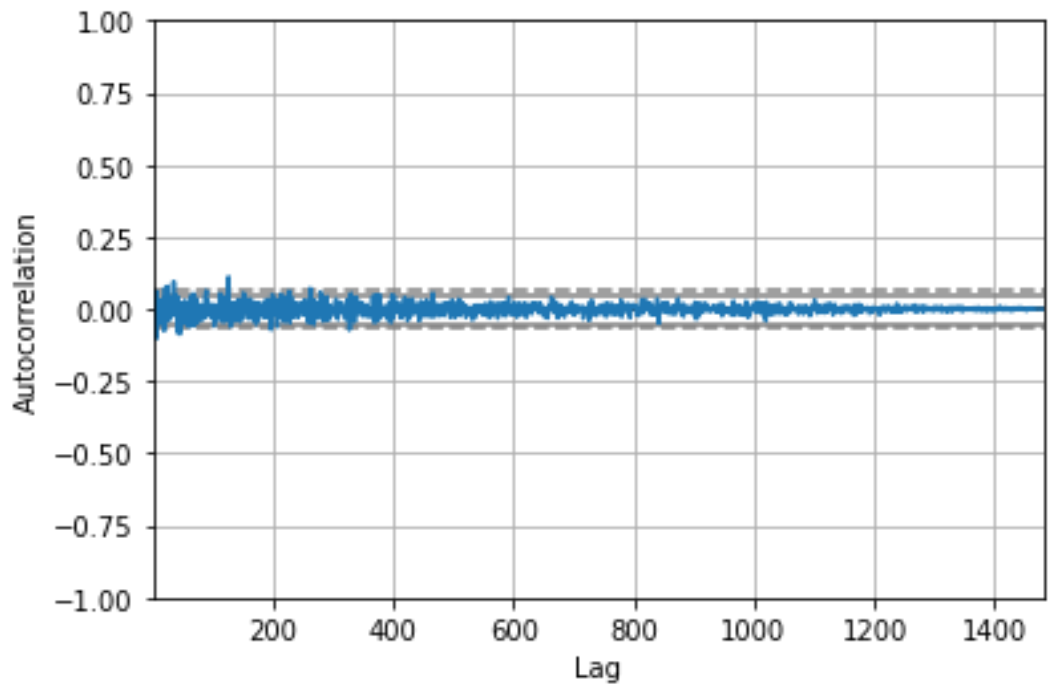

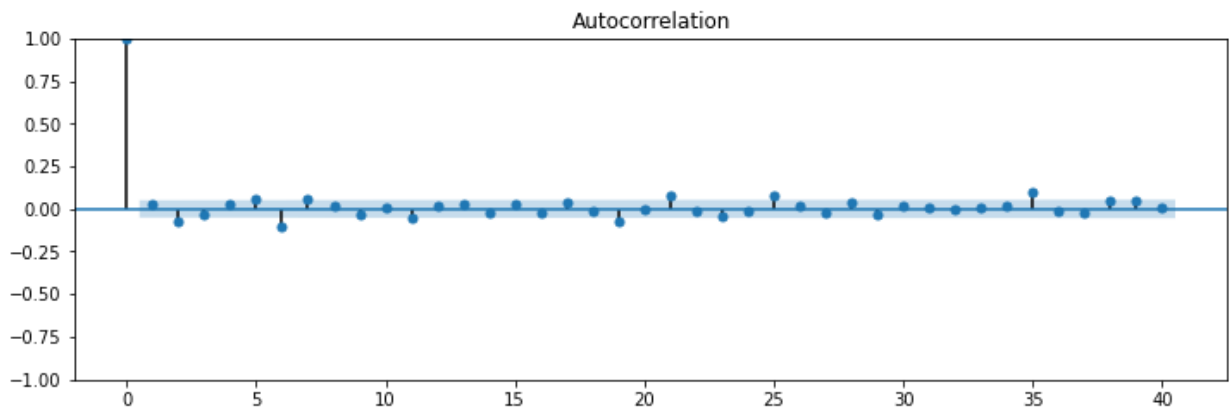

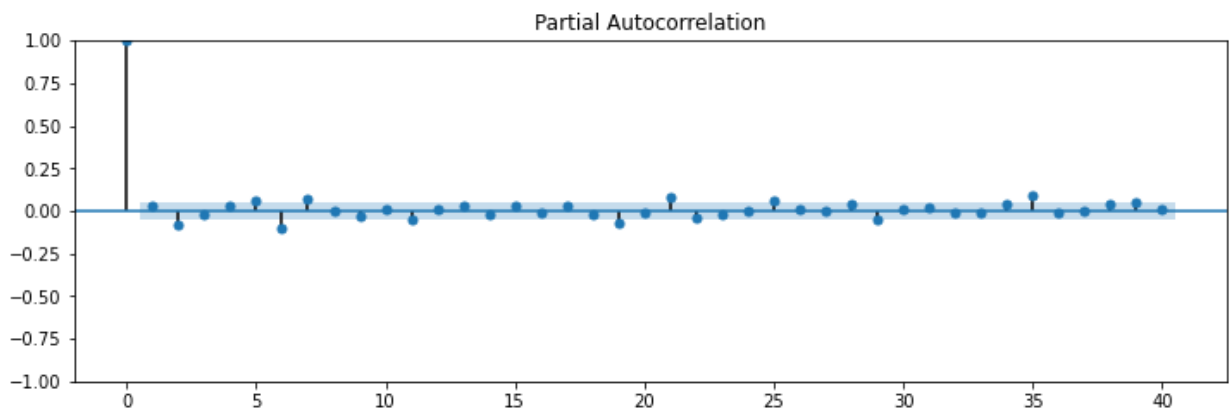

## SLIDING WINDOW METHOD

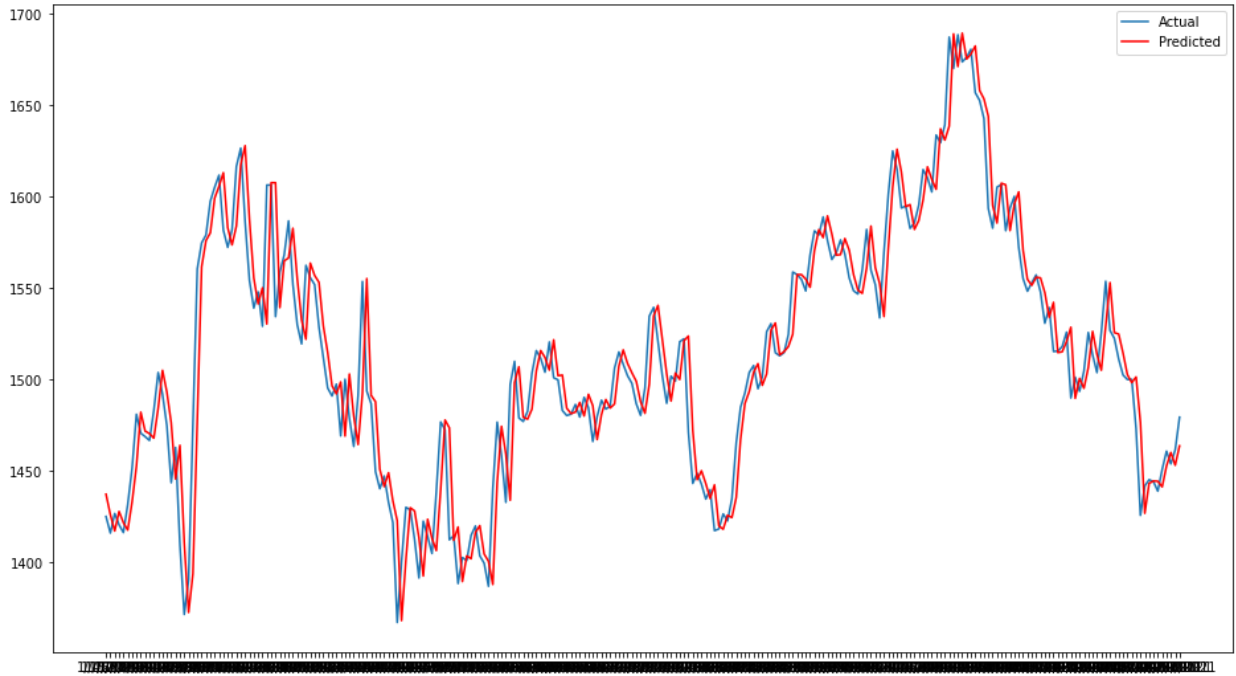

## ROLLING WINDOW METHOD

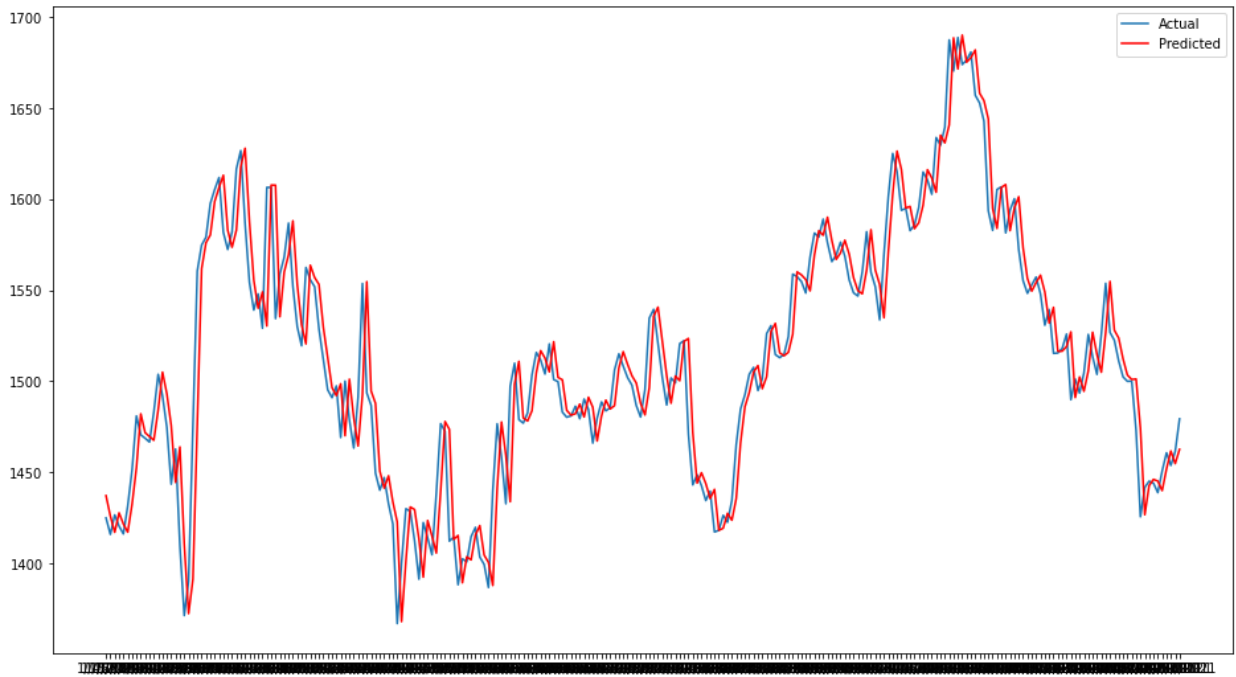

## 2. ICICI BANK

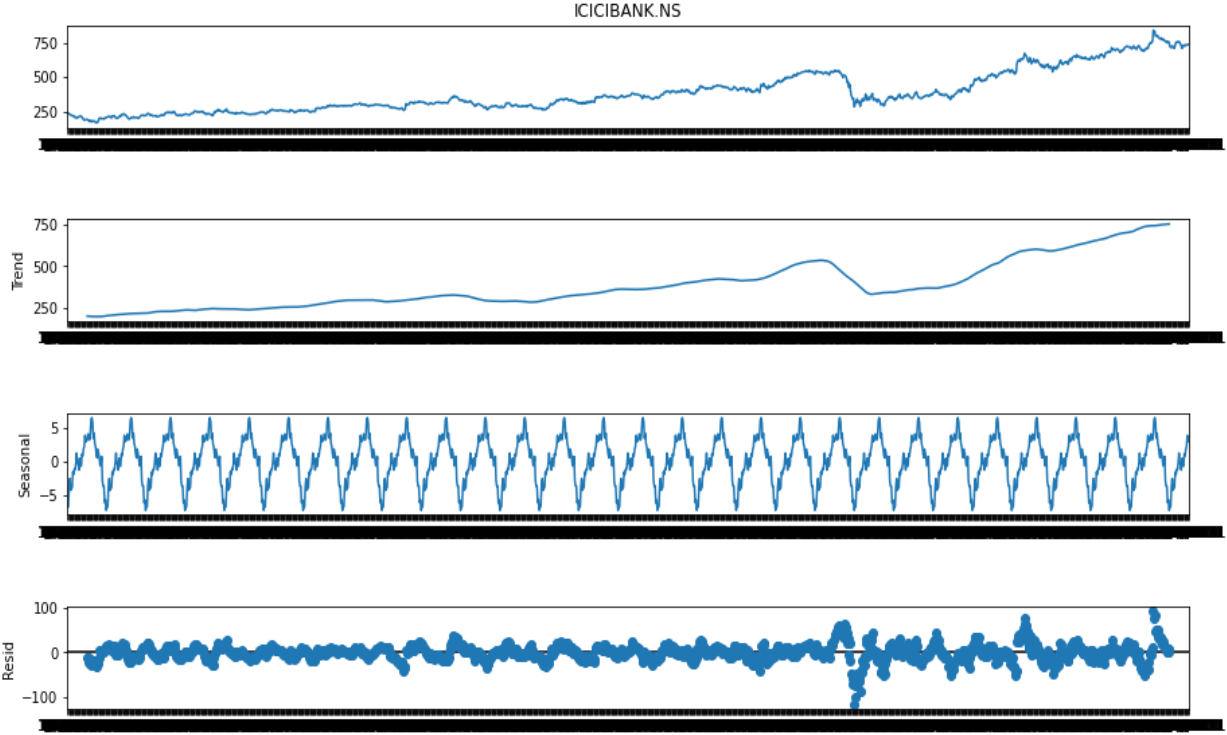

## SIMPLE EXPONENTIAL SMOOTHING

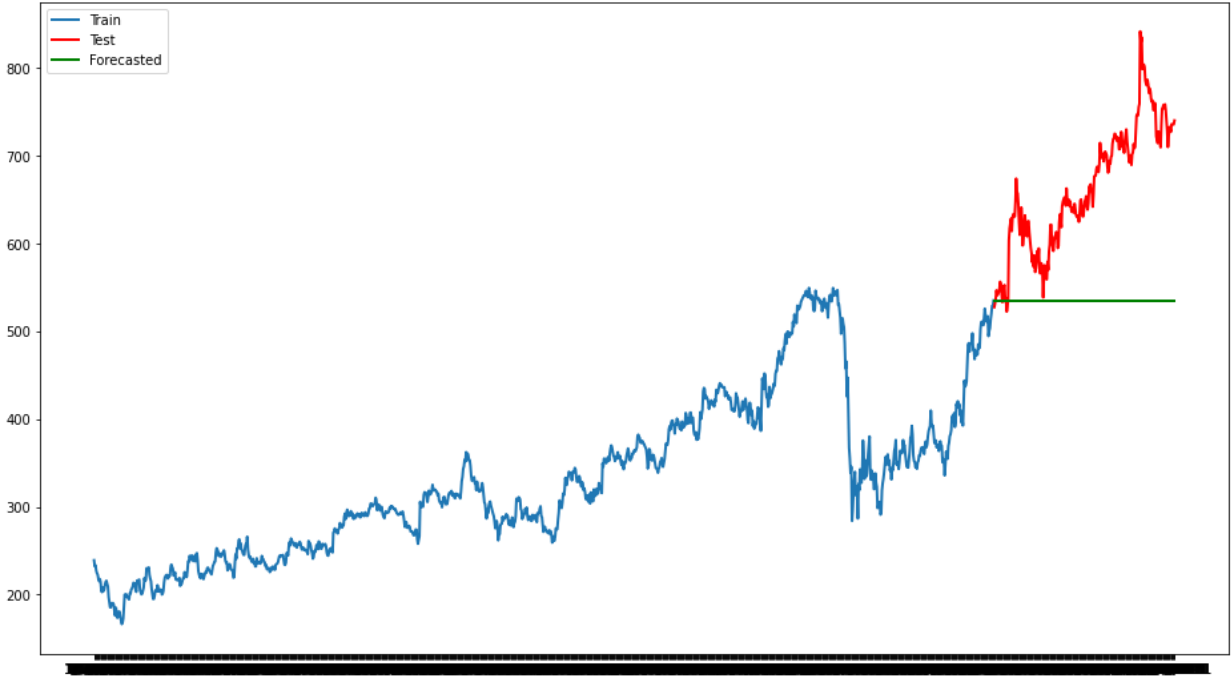

# HOLT WINTER TREND METHOD

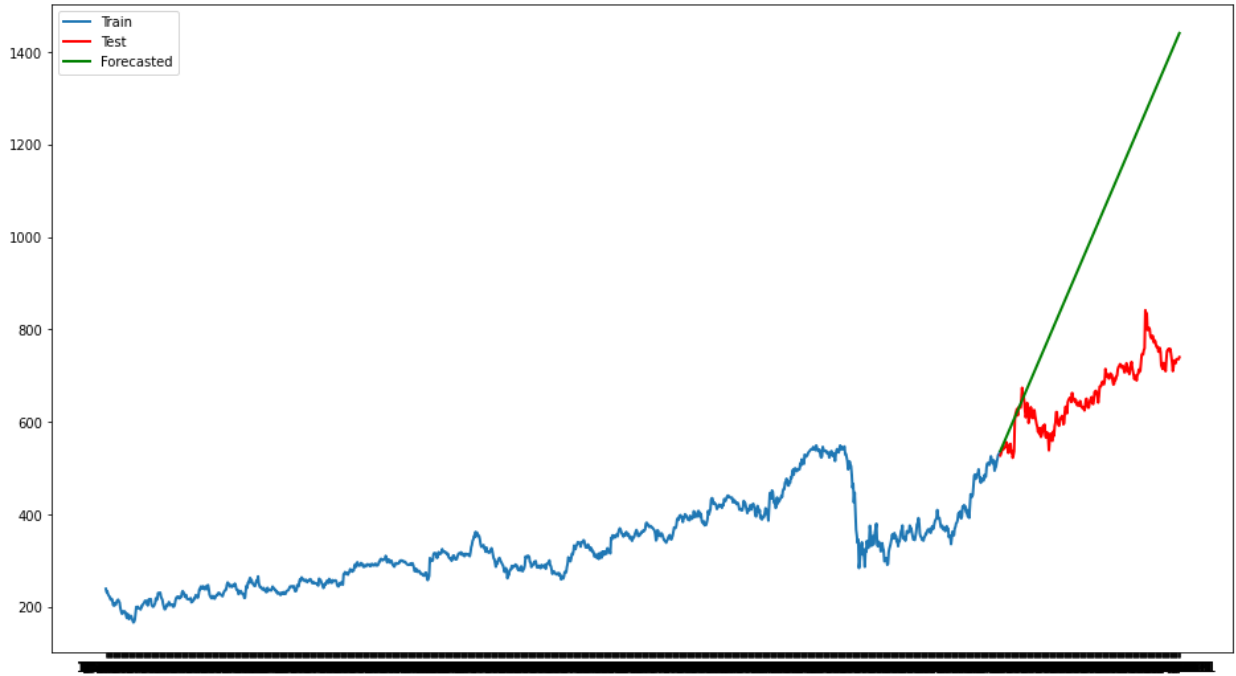

# ARIMA

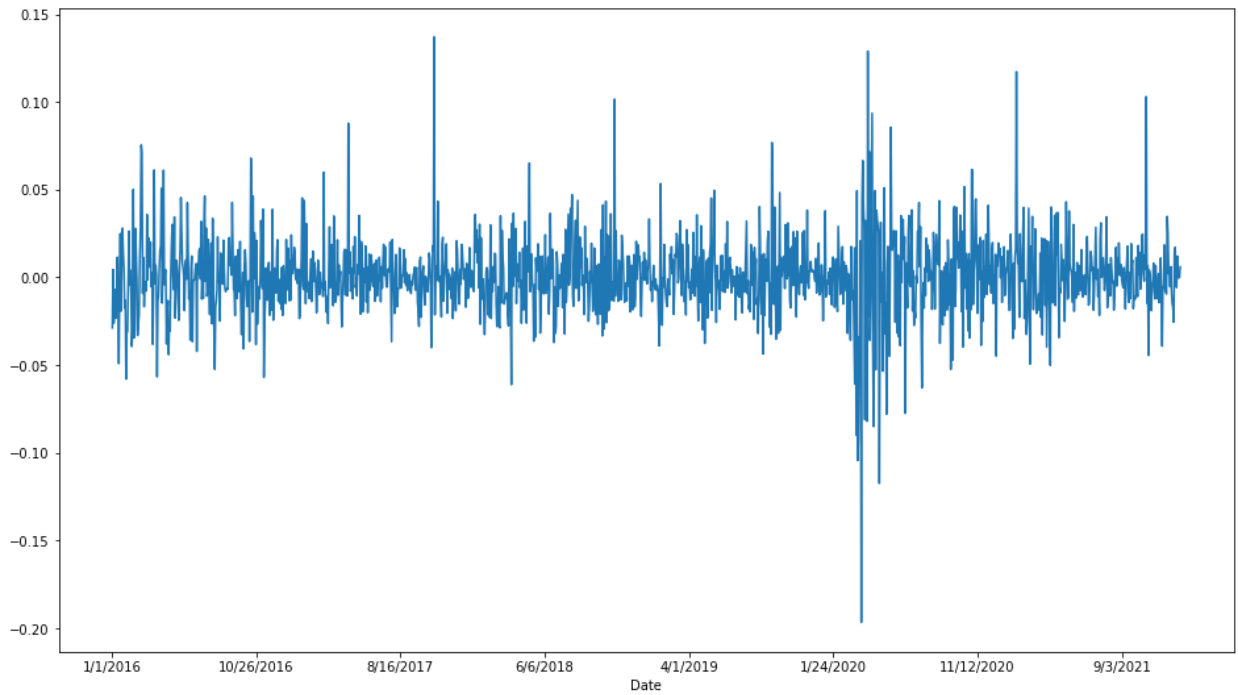

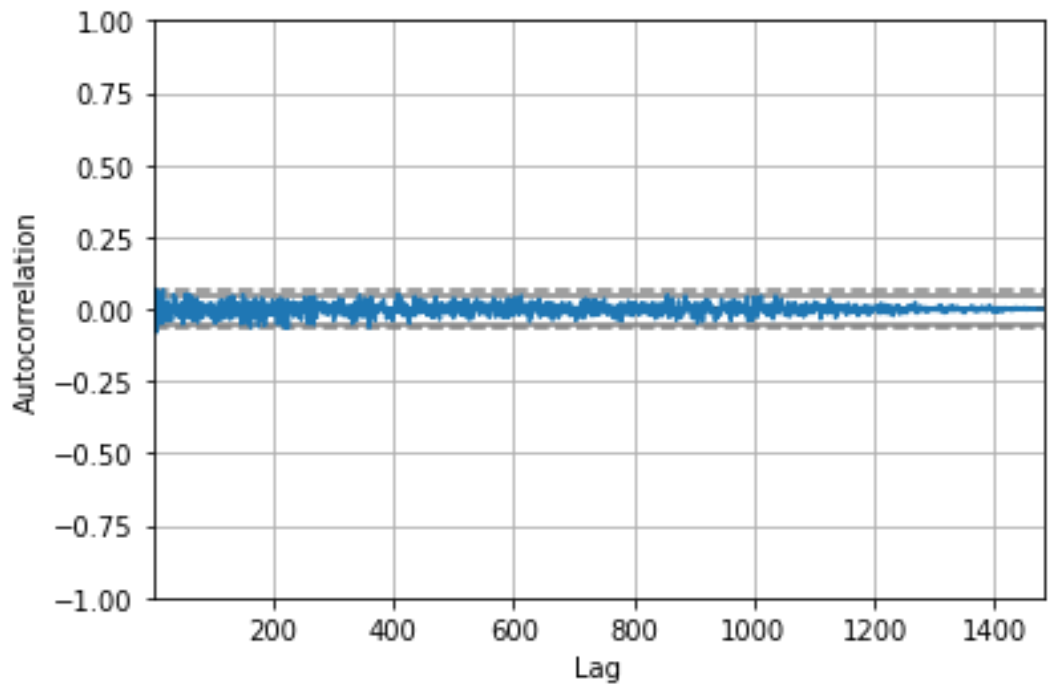

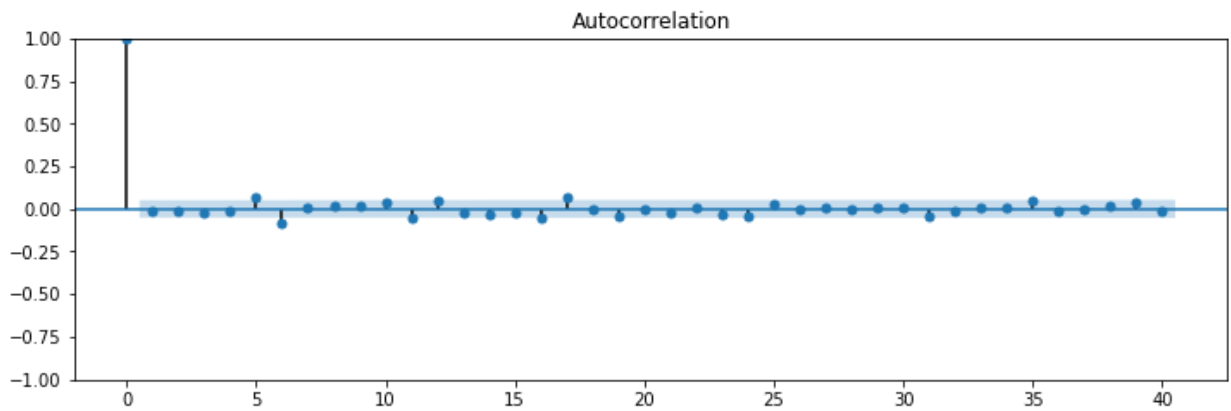

Autocorrelation

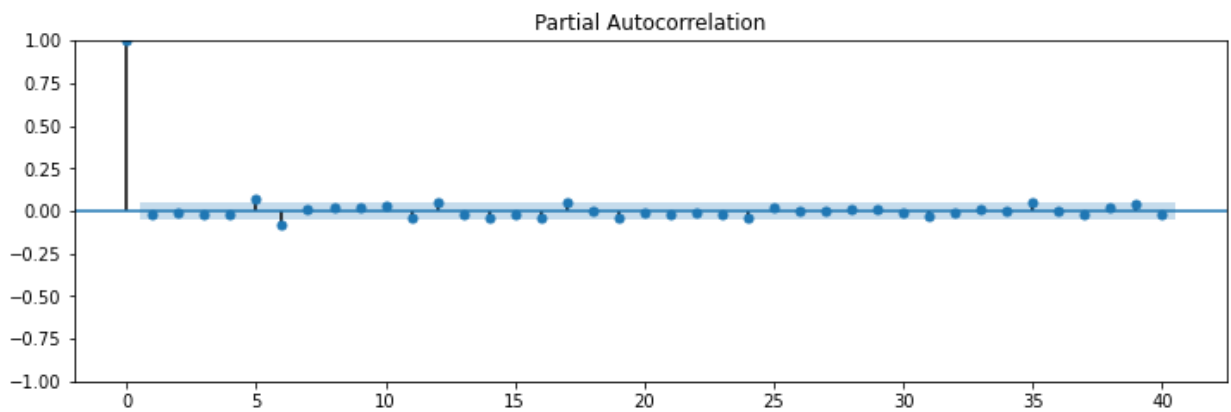

Partial Autocorrelation

## SLIDING WINDOW

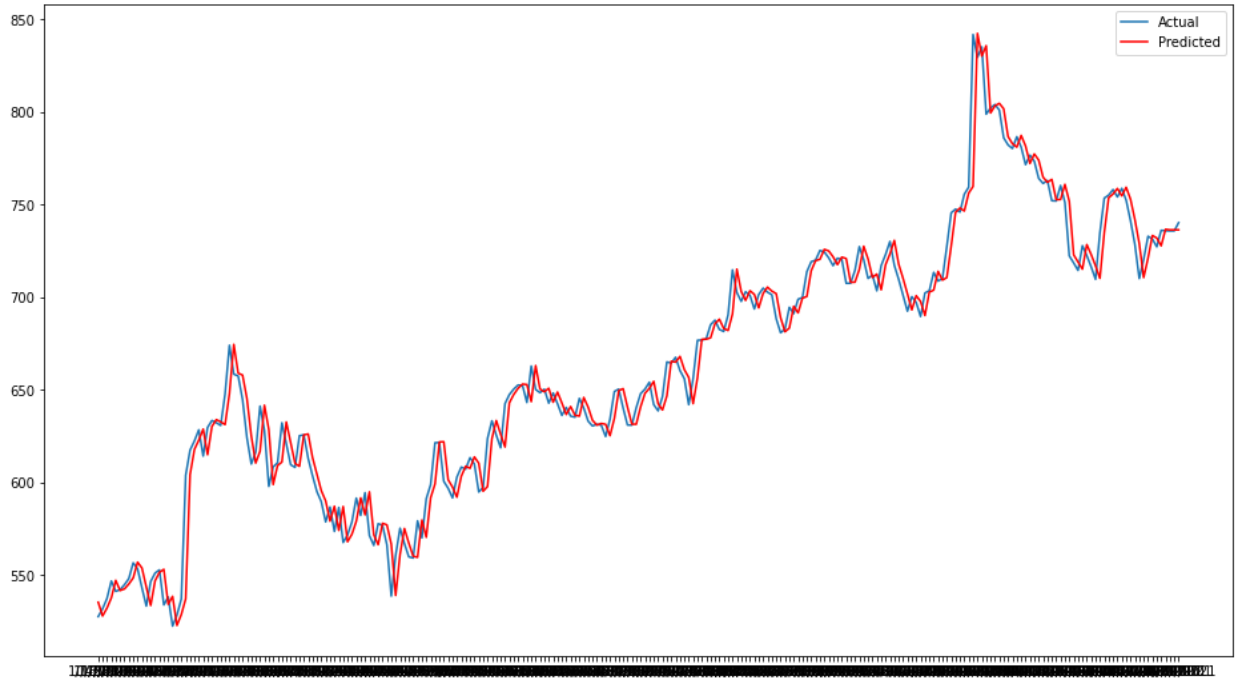

## ROLLING WINDOW

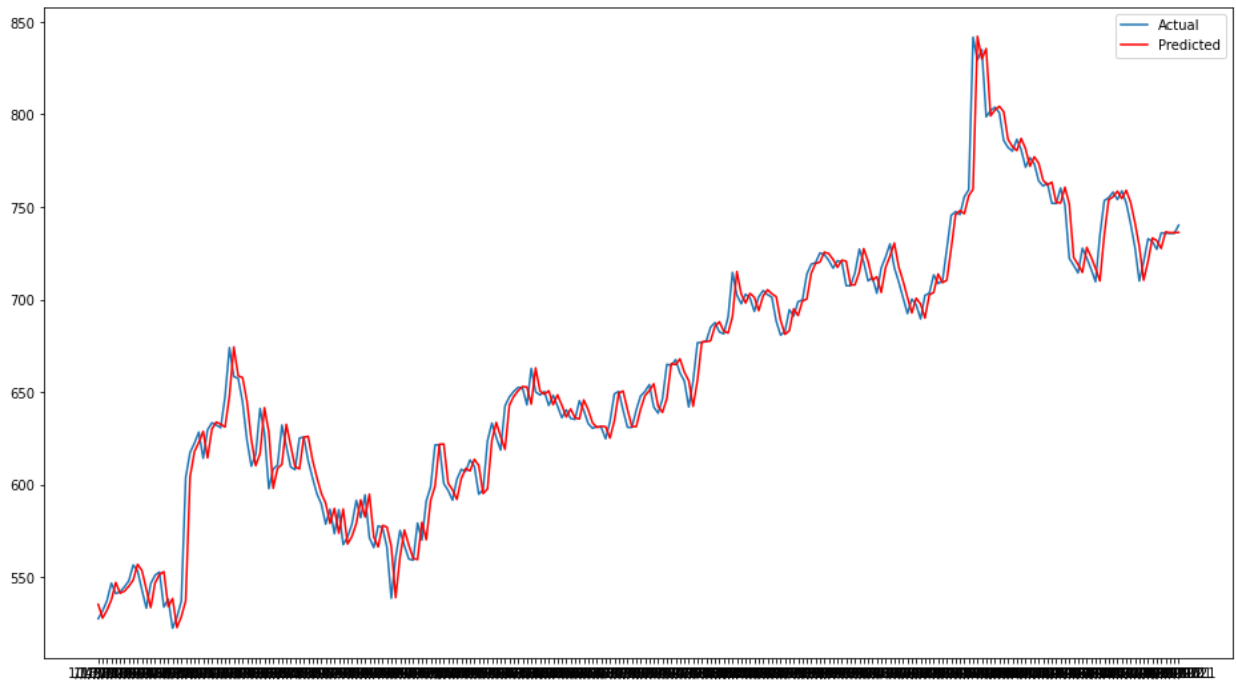

# FMCG SECTOR

1. HINDUSTAN UNILEVER

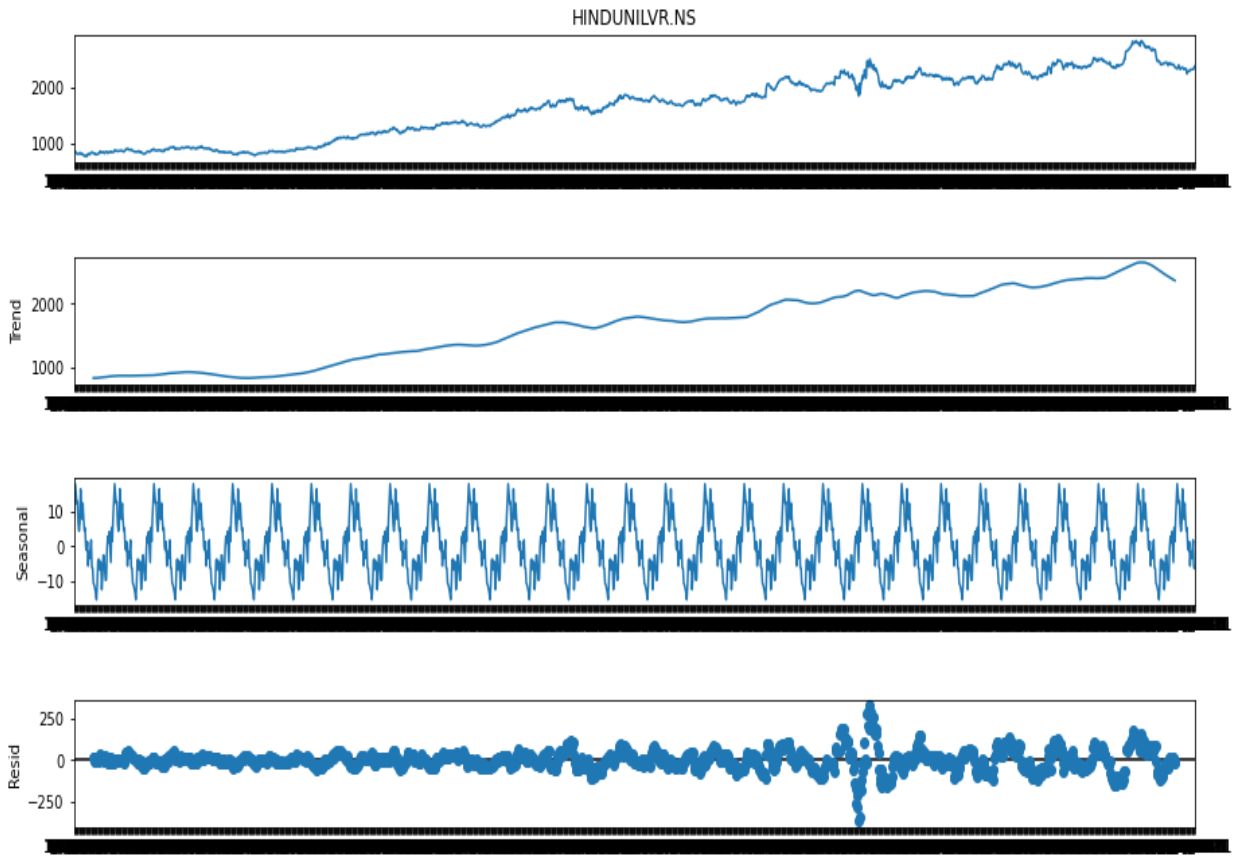

SIMPLE EXPONENTIAL SMOOTHING

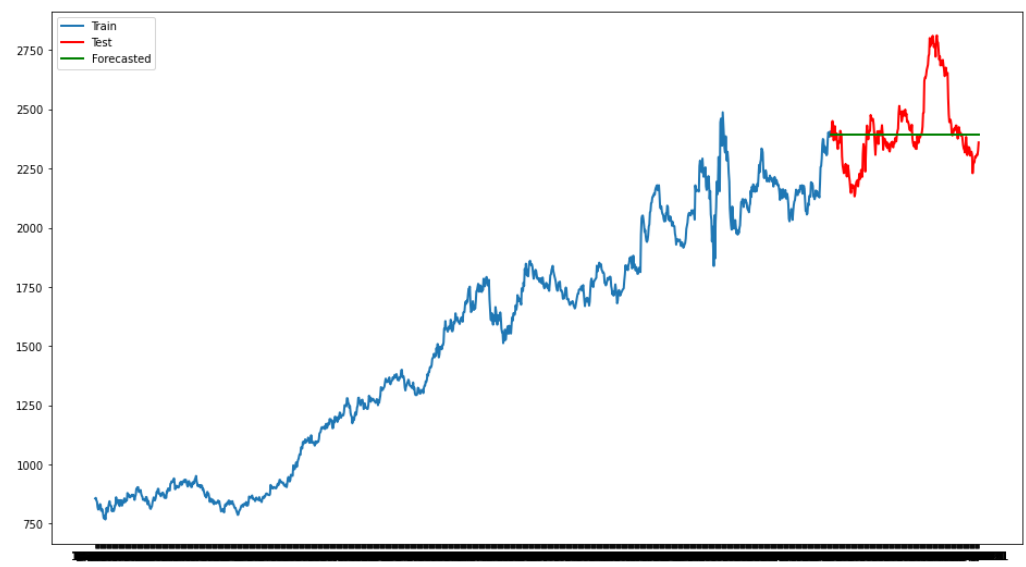

# HOLT WINTER TREND METHOD

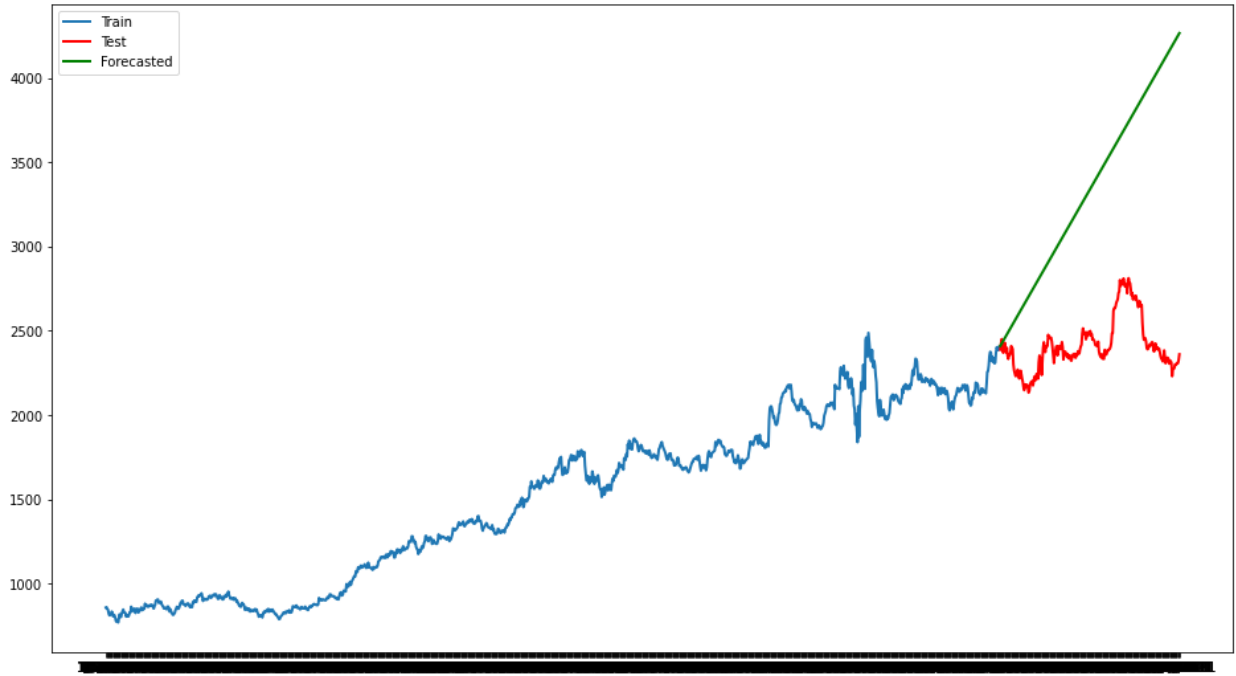

# ARIMA

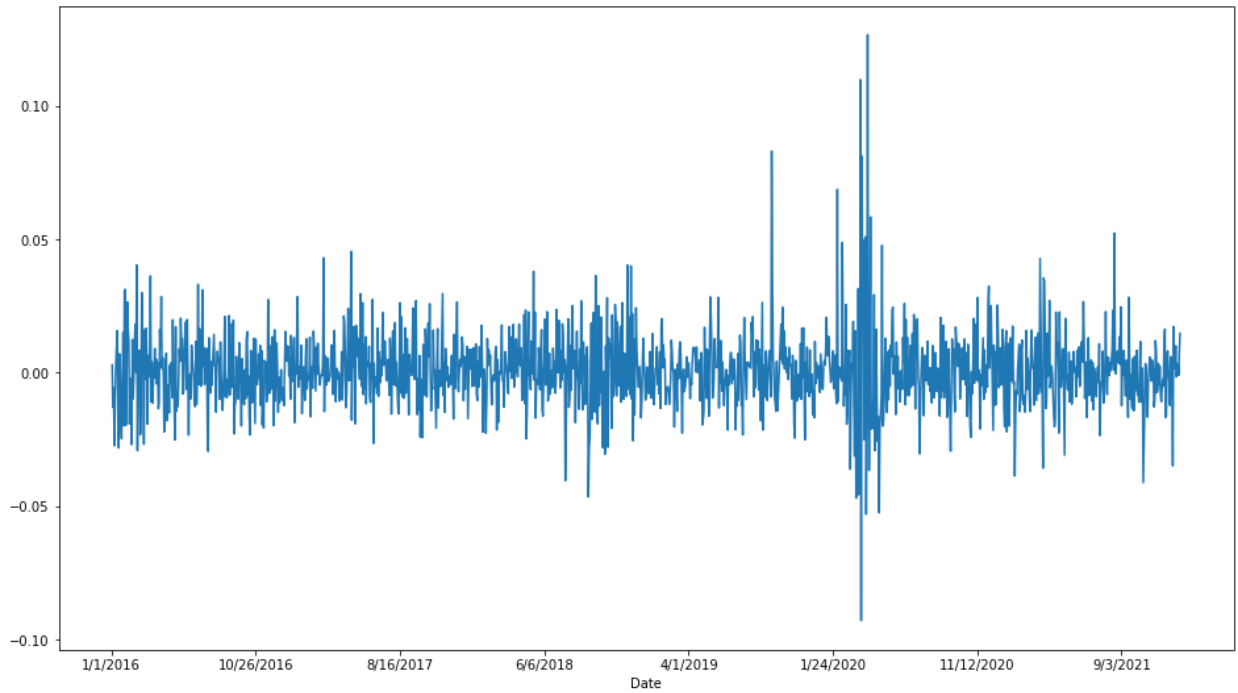

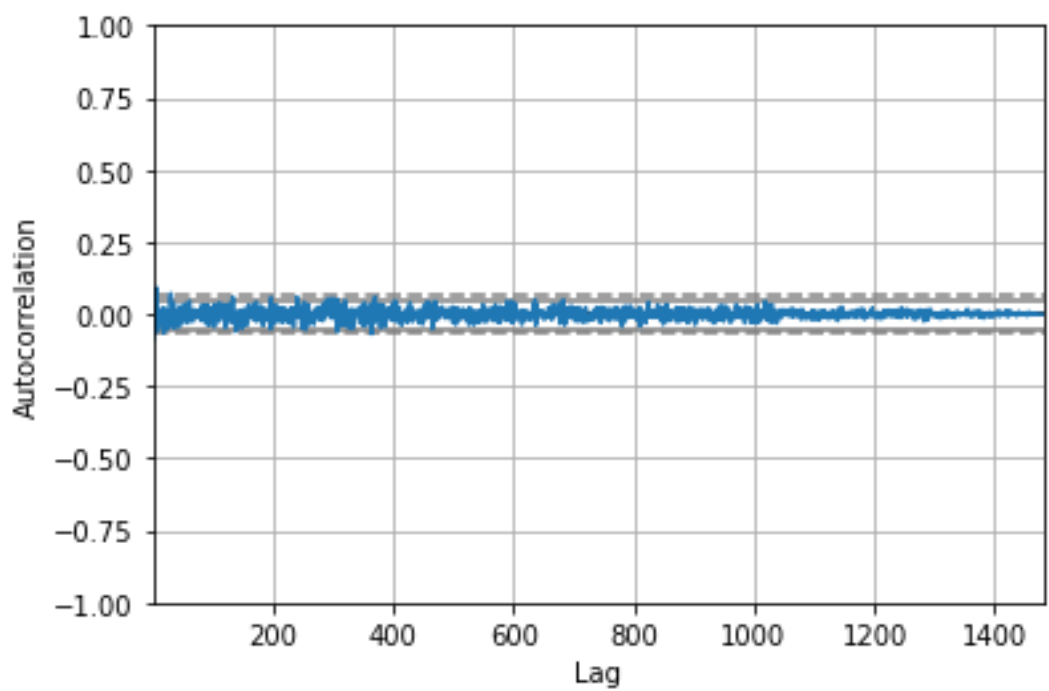

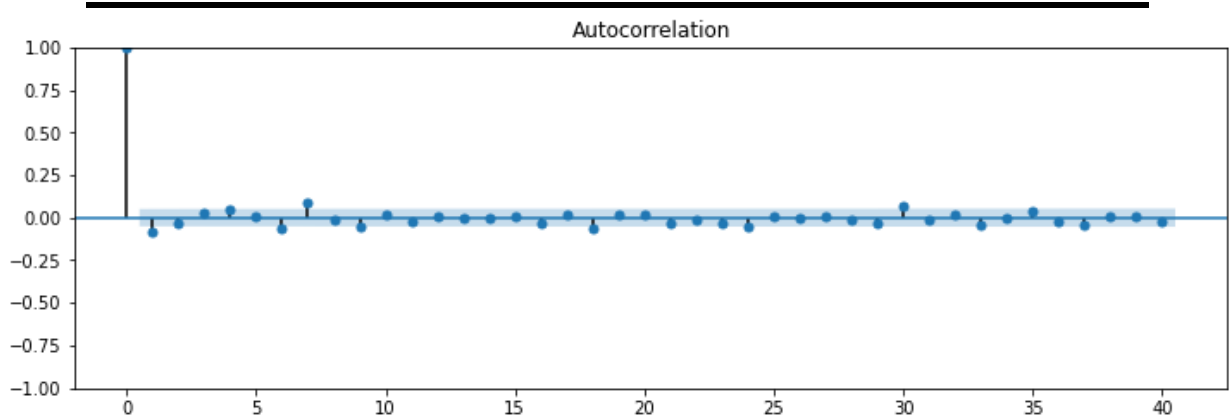

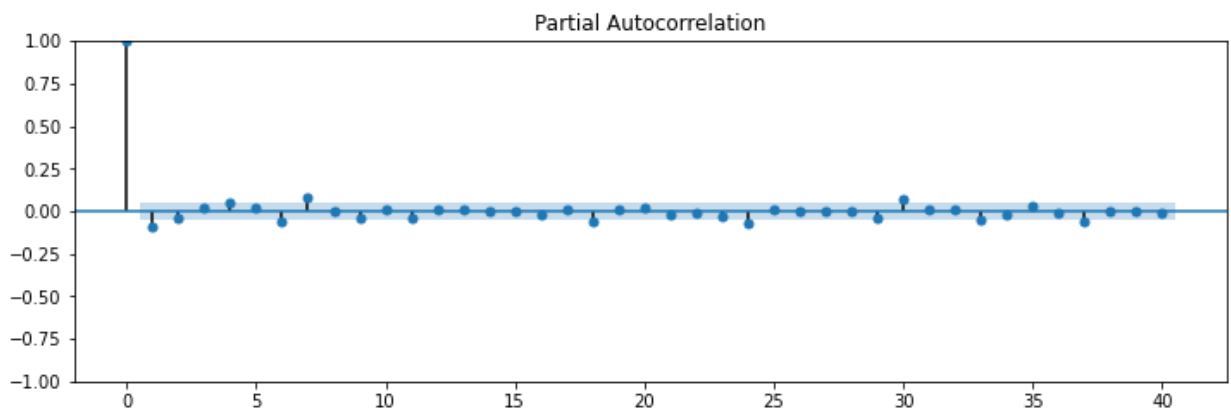

## SLIDING WINDOW METHOD

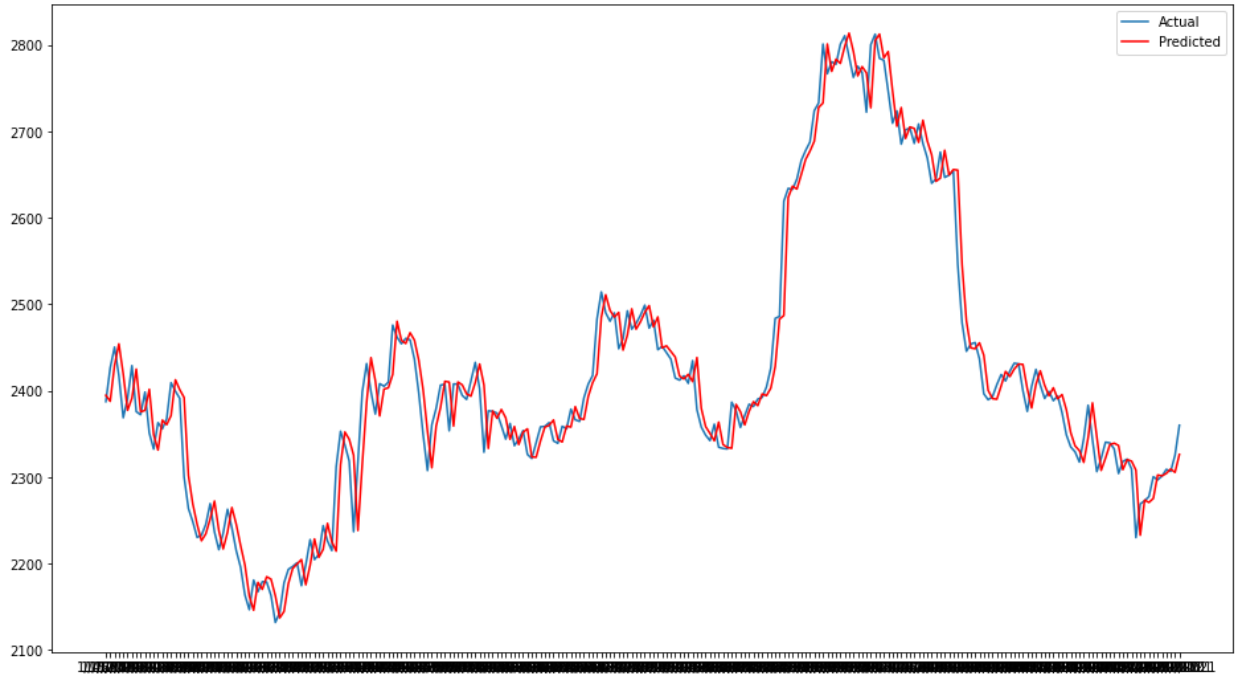

## ROLLING WINDOW METHOD

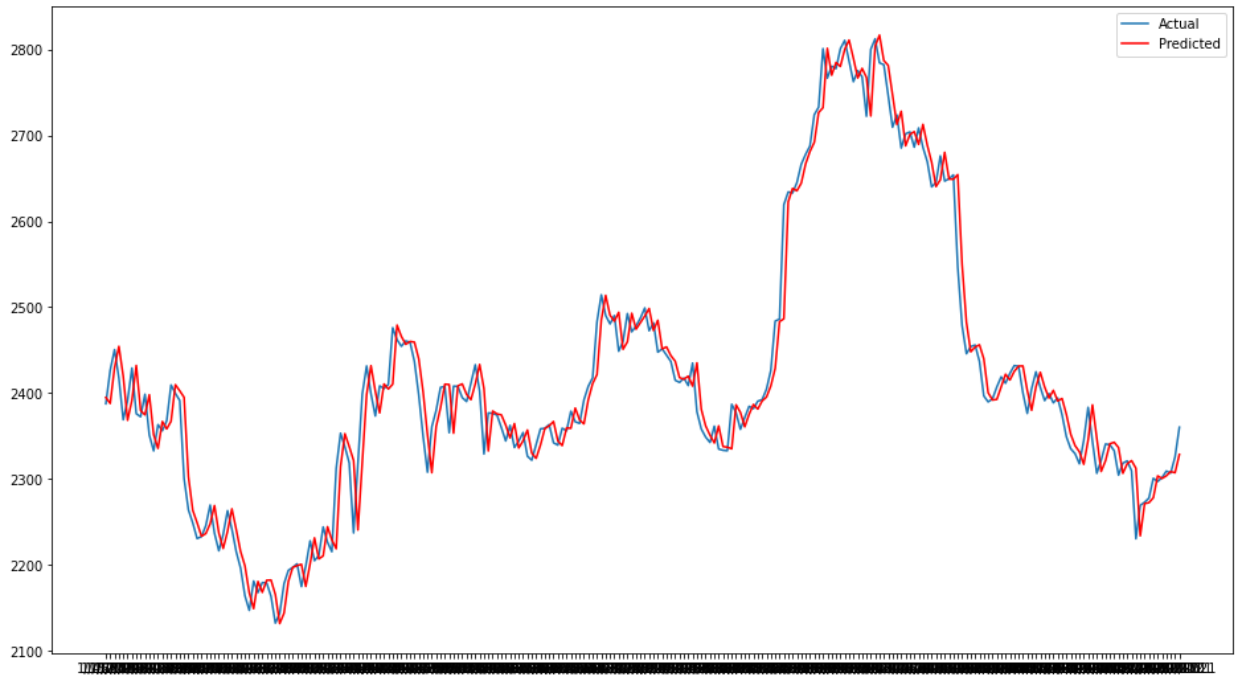

## 2. ITC

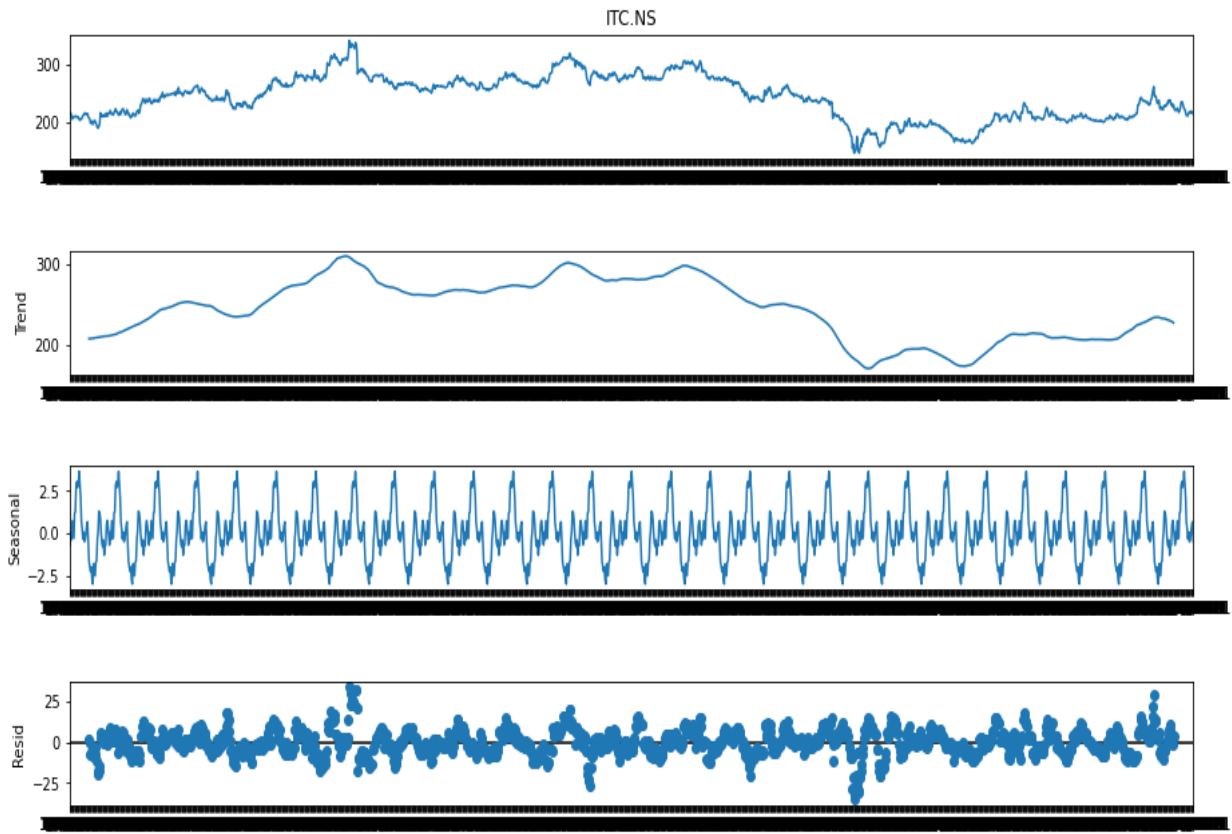

## SIMPLE EXPONENTIAL SMOOTHING

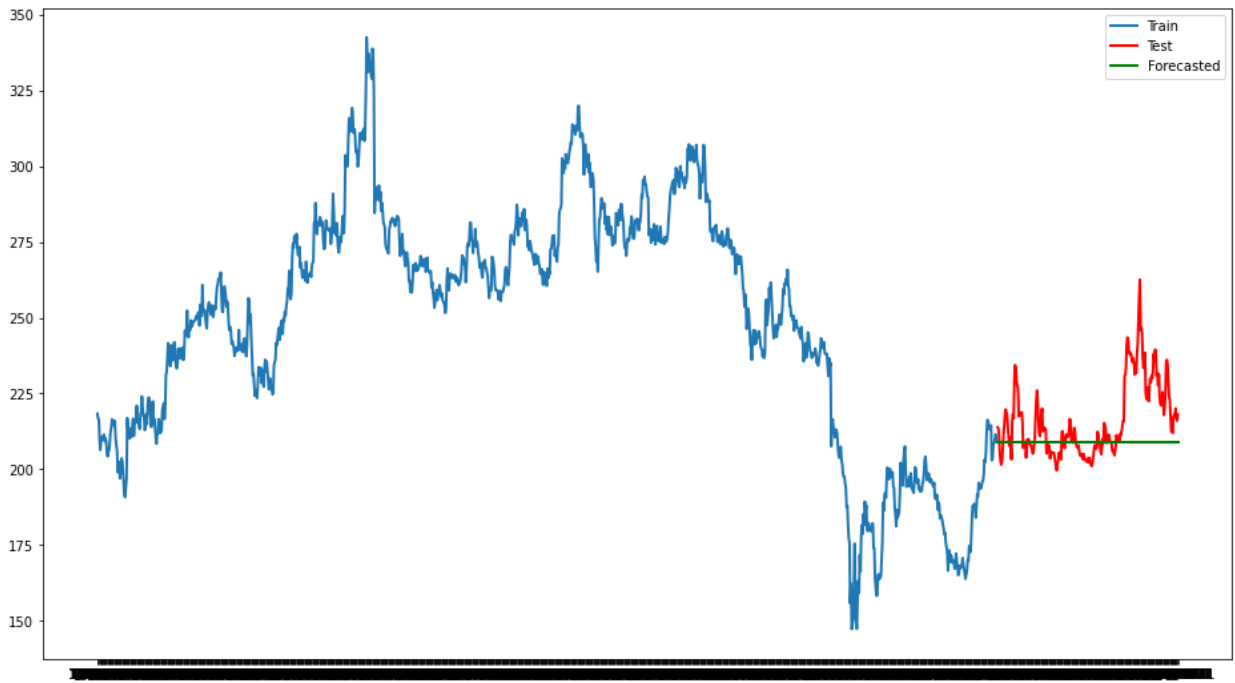

# HOLT WINTER TREND METHOD

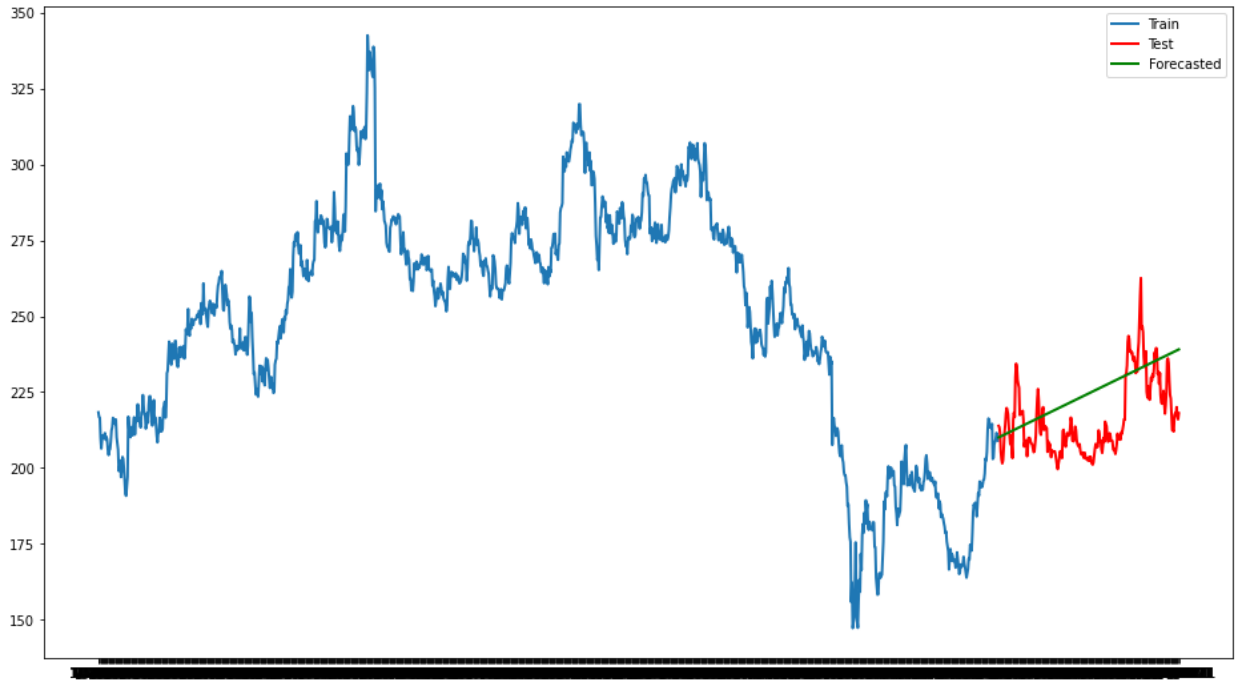

# ARIMA

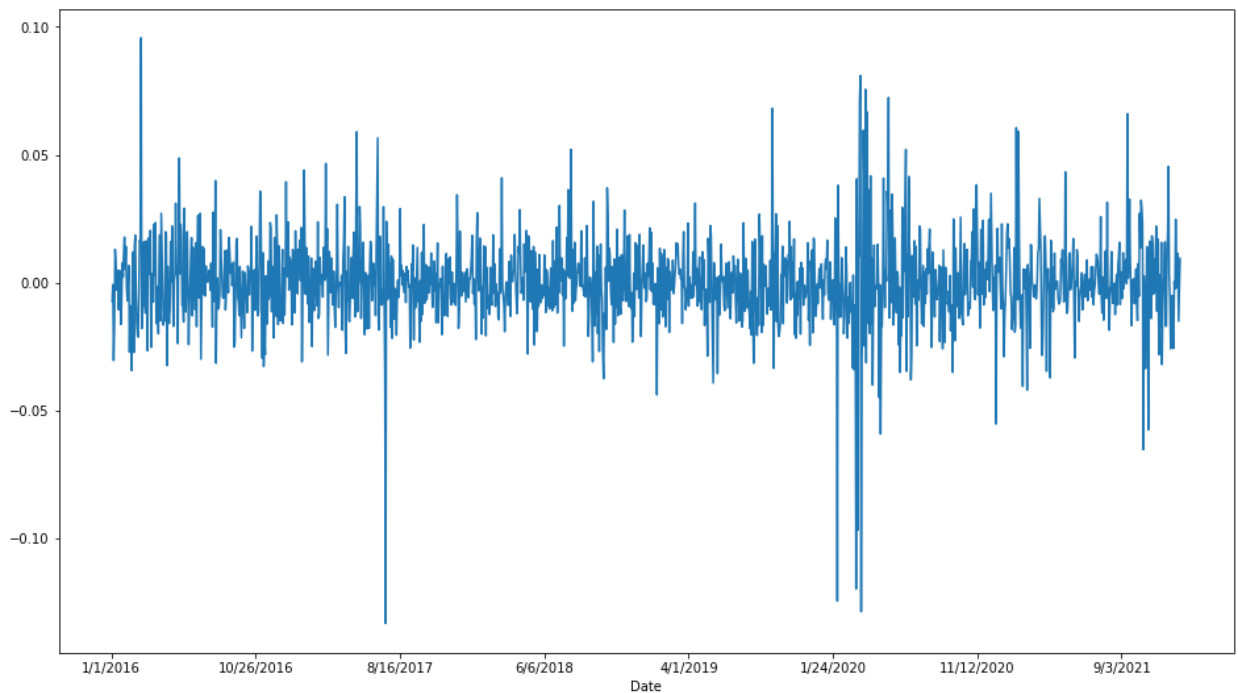

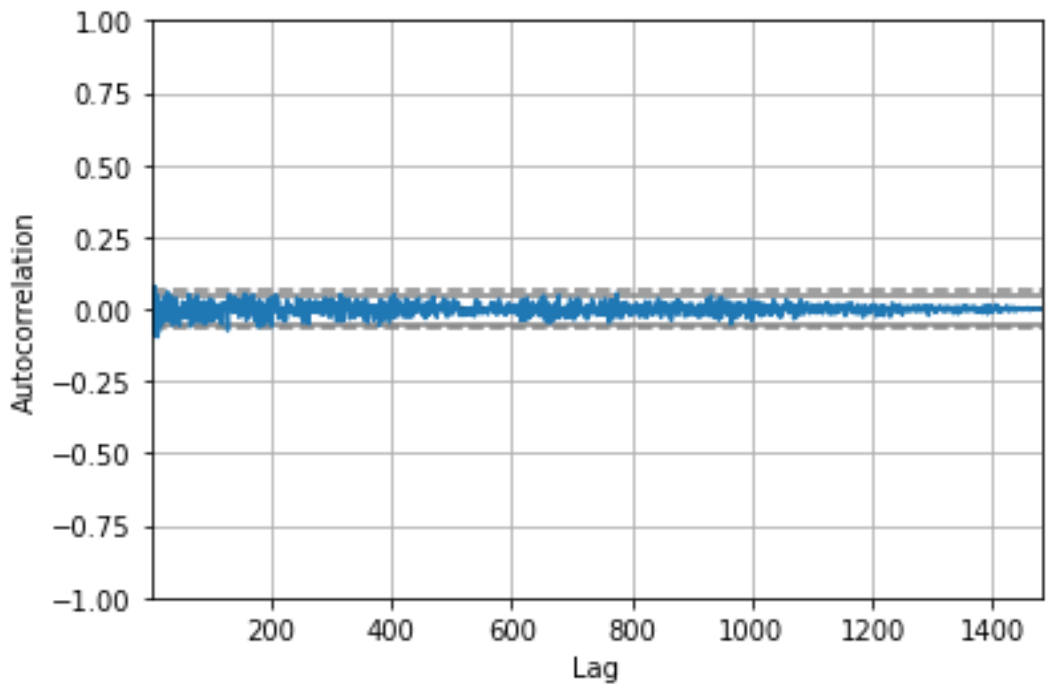

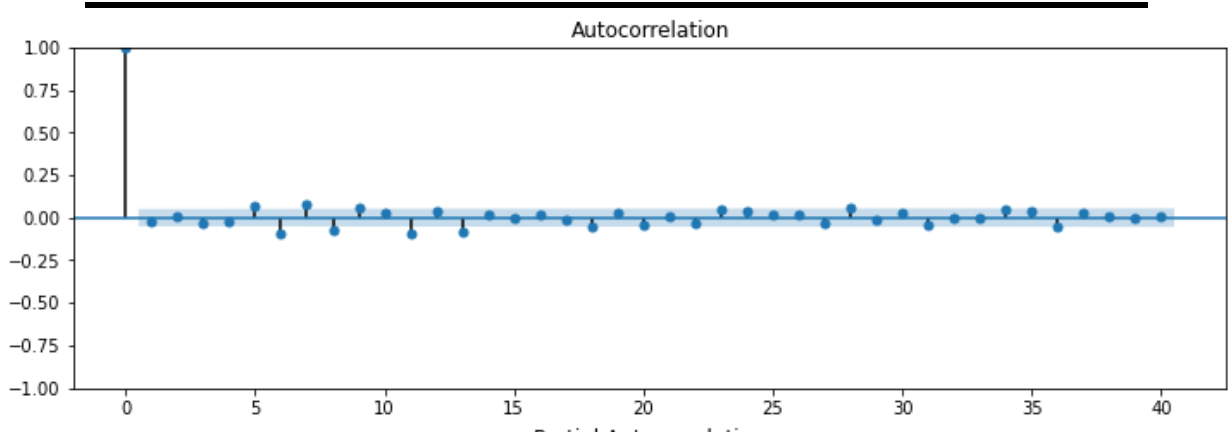

Autocorrelation

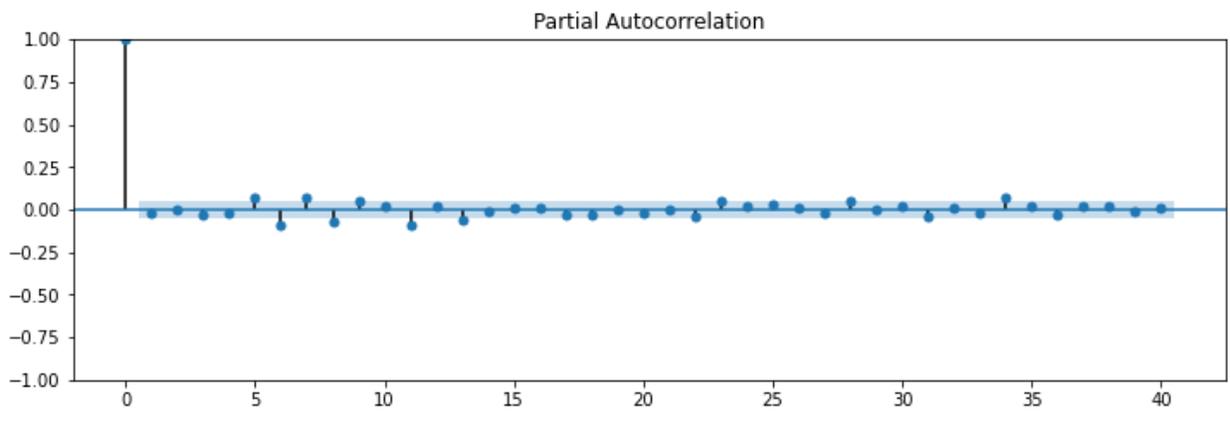

Partial Autocorrelation

## SLIDING WINDOW METHOD

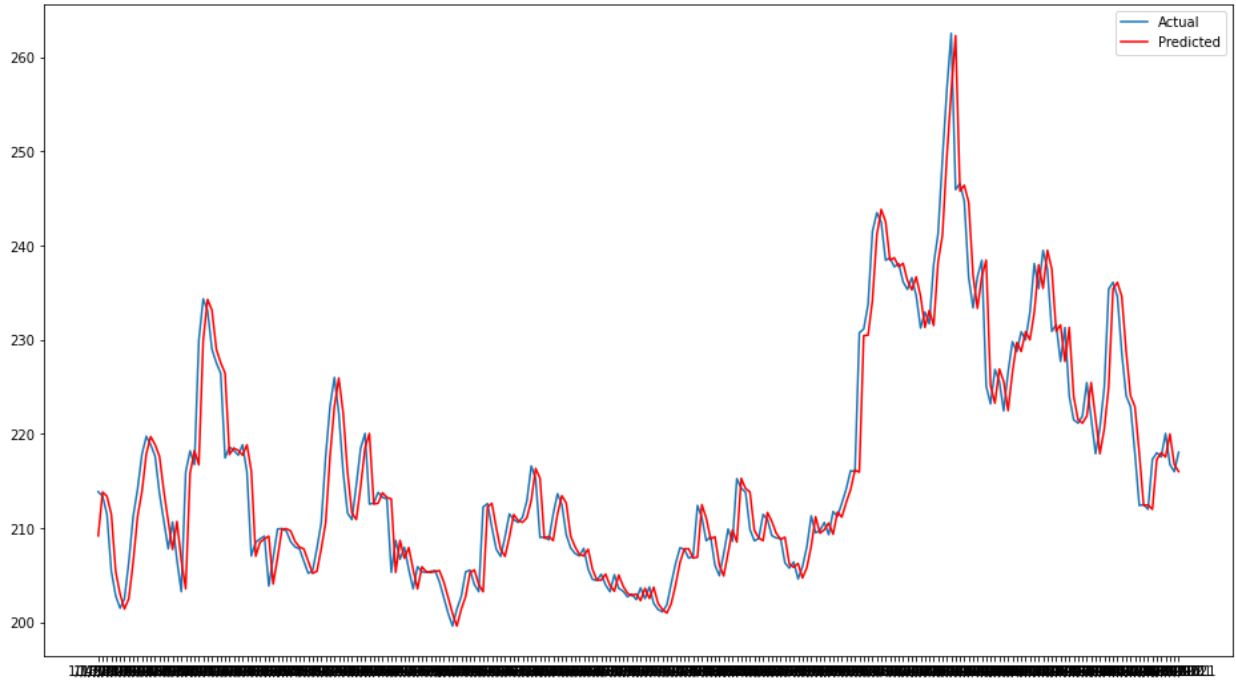

## ROLLING WINDOW METHOD

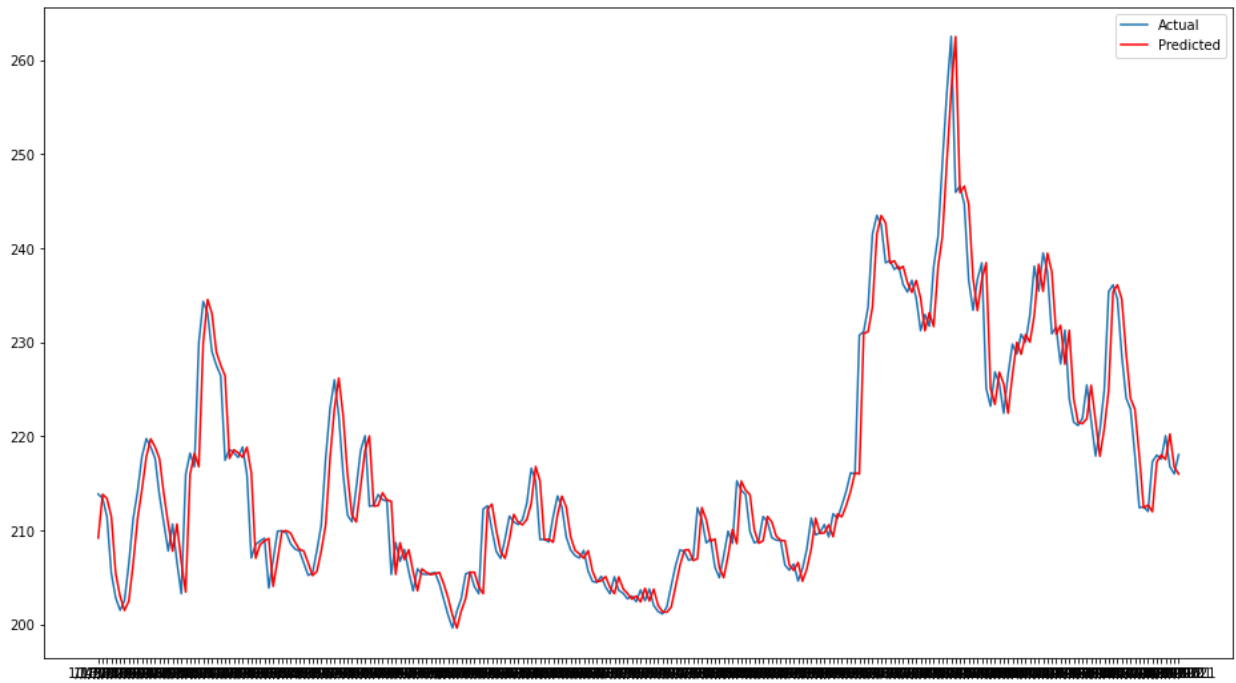

# AUTO SECTOR

## 1. TATA MOTORS

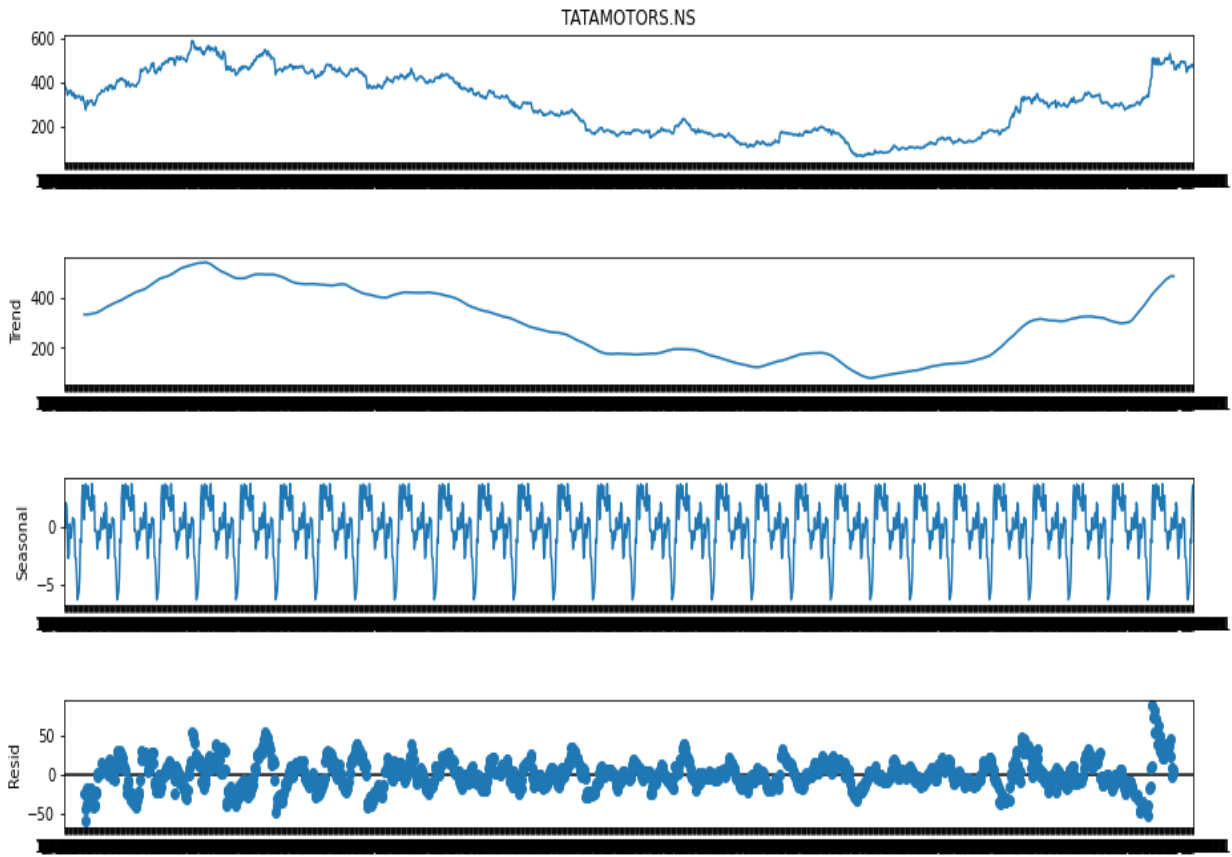

## SIMPLE EXPONENTIAL SMOOTHING

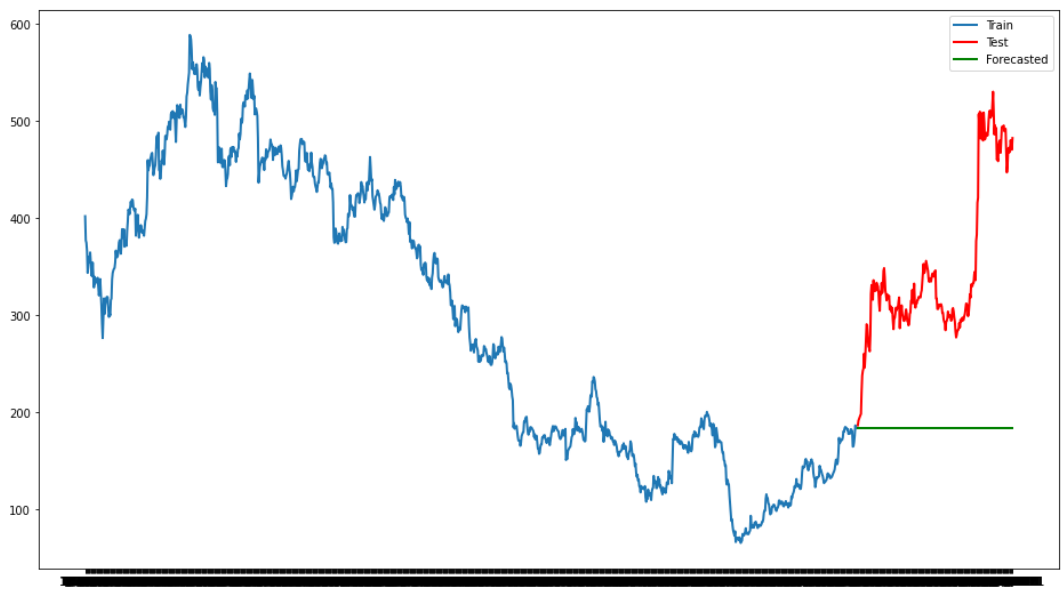

# HOLT WINTER TREND METHOD

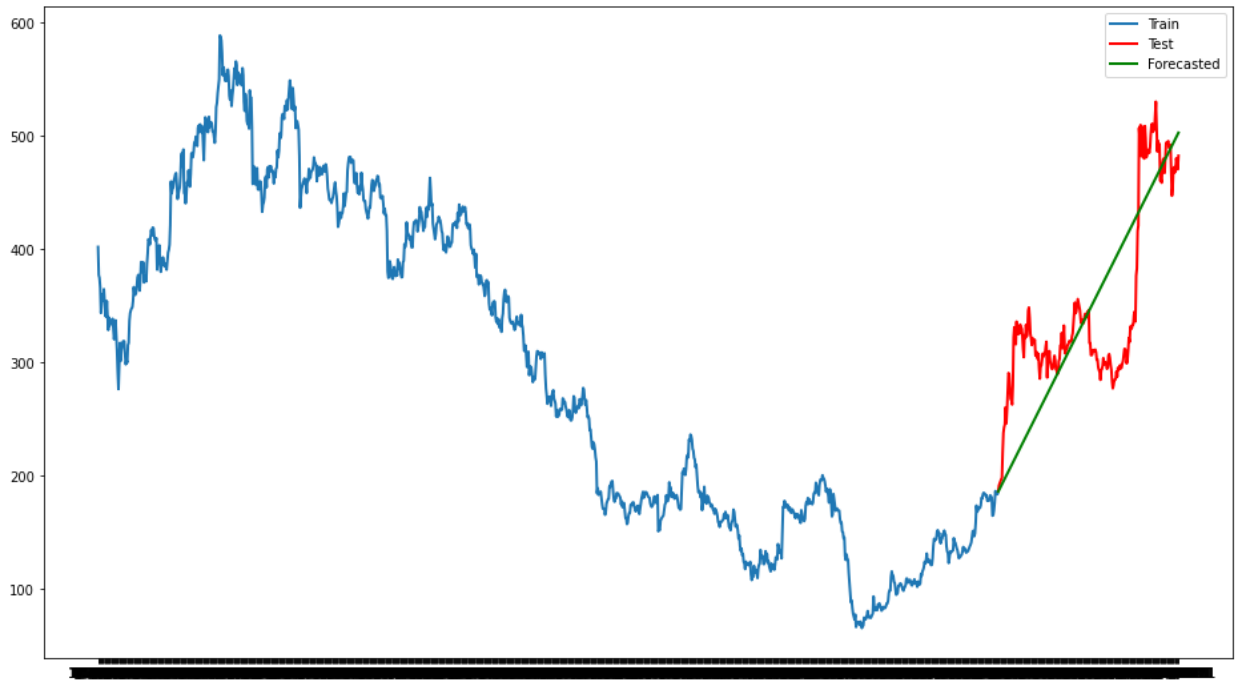

# ARIMA

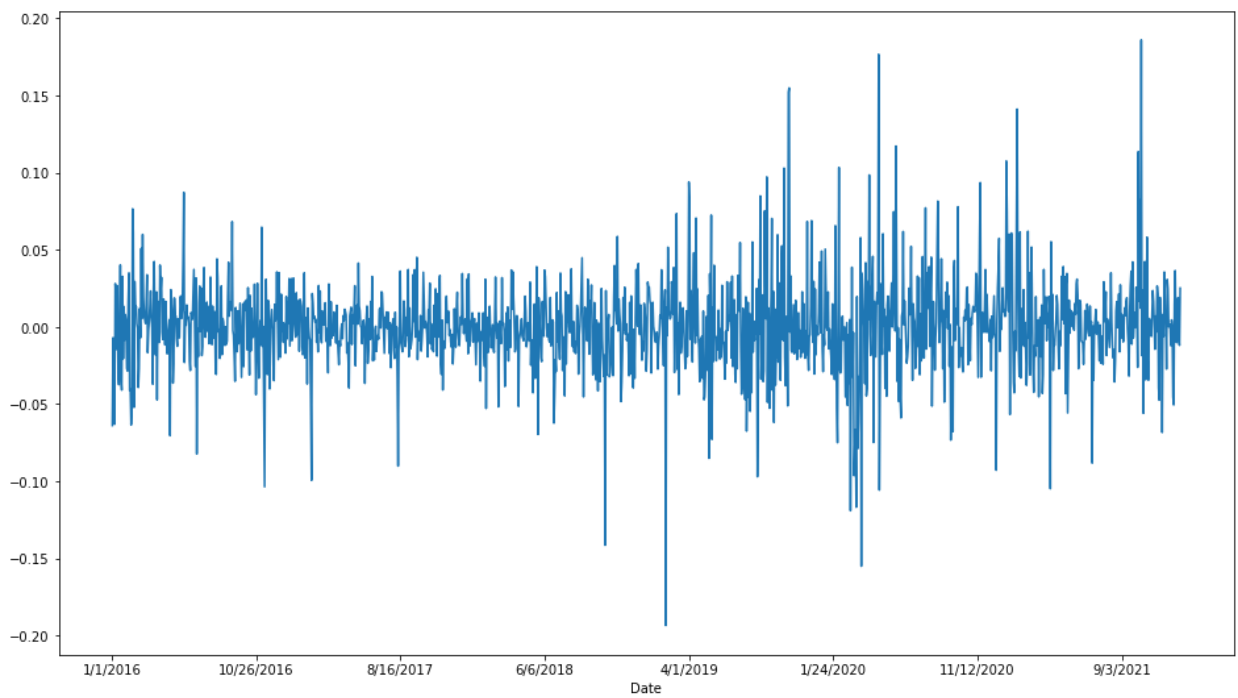

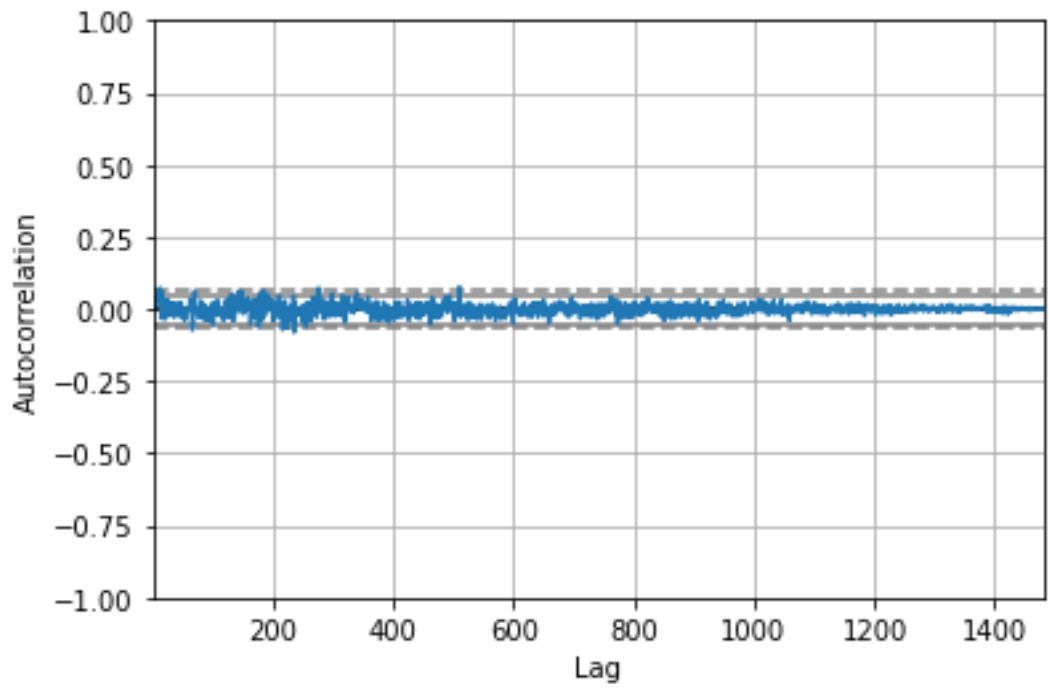

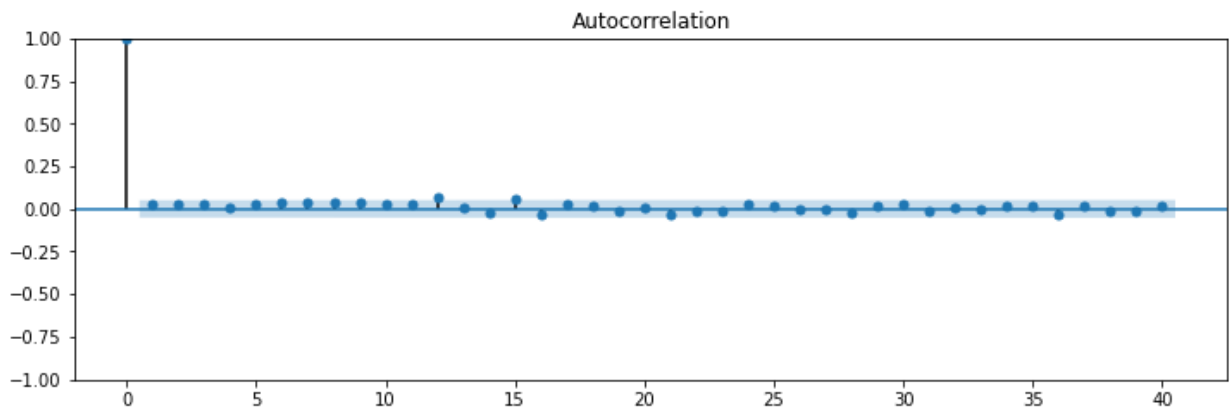

Autocorrelation

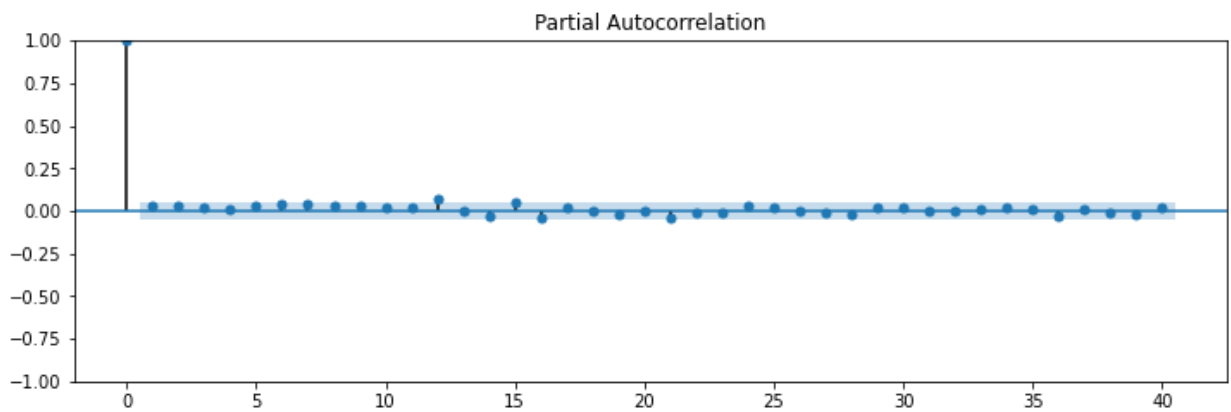

Partial Autocorrelation

## SLIDING WINDOW

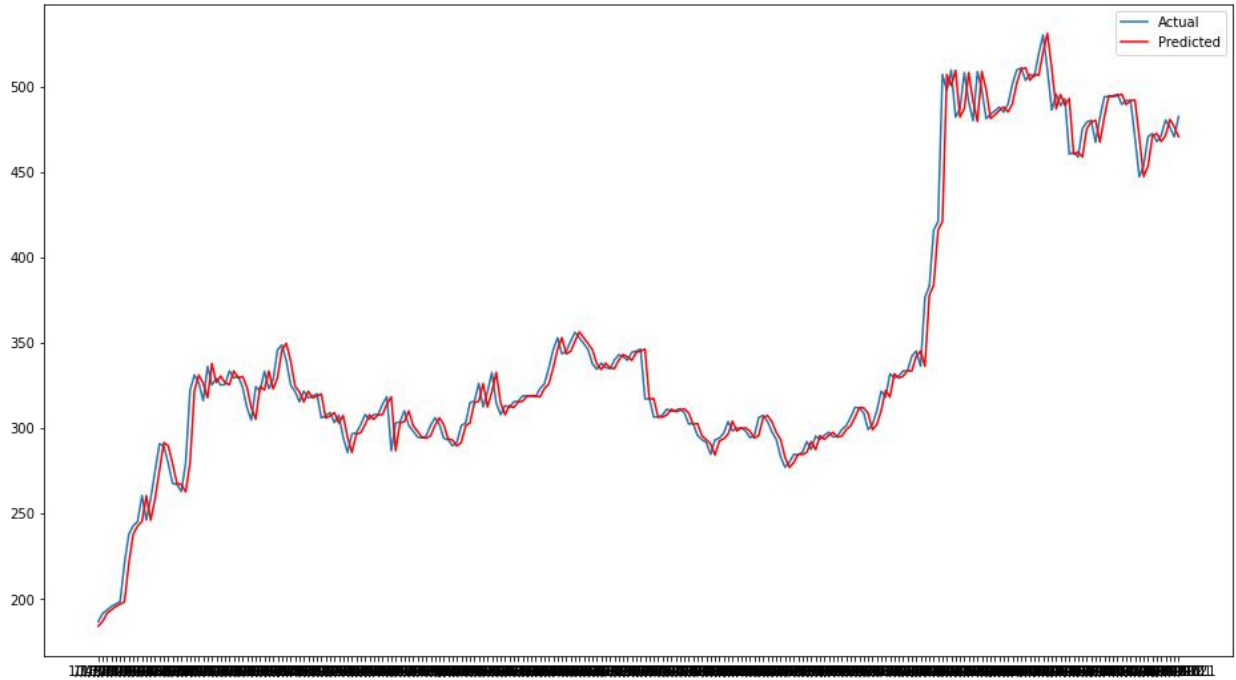

## ROLLING WINDOW

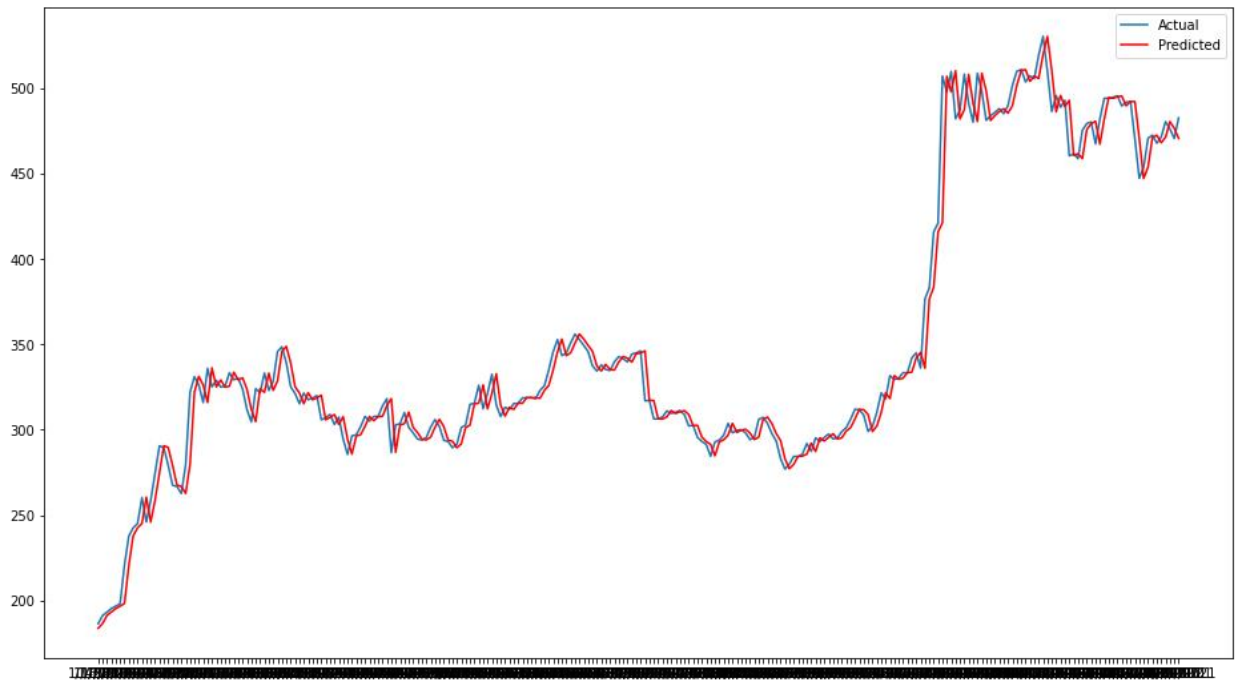

## 2. MARUTI

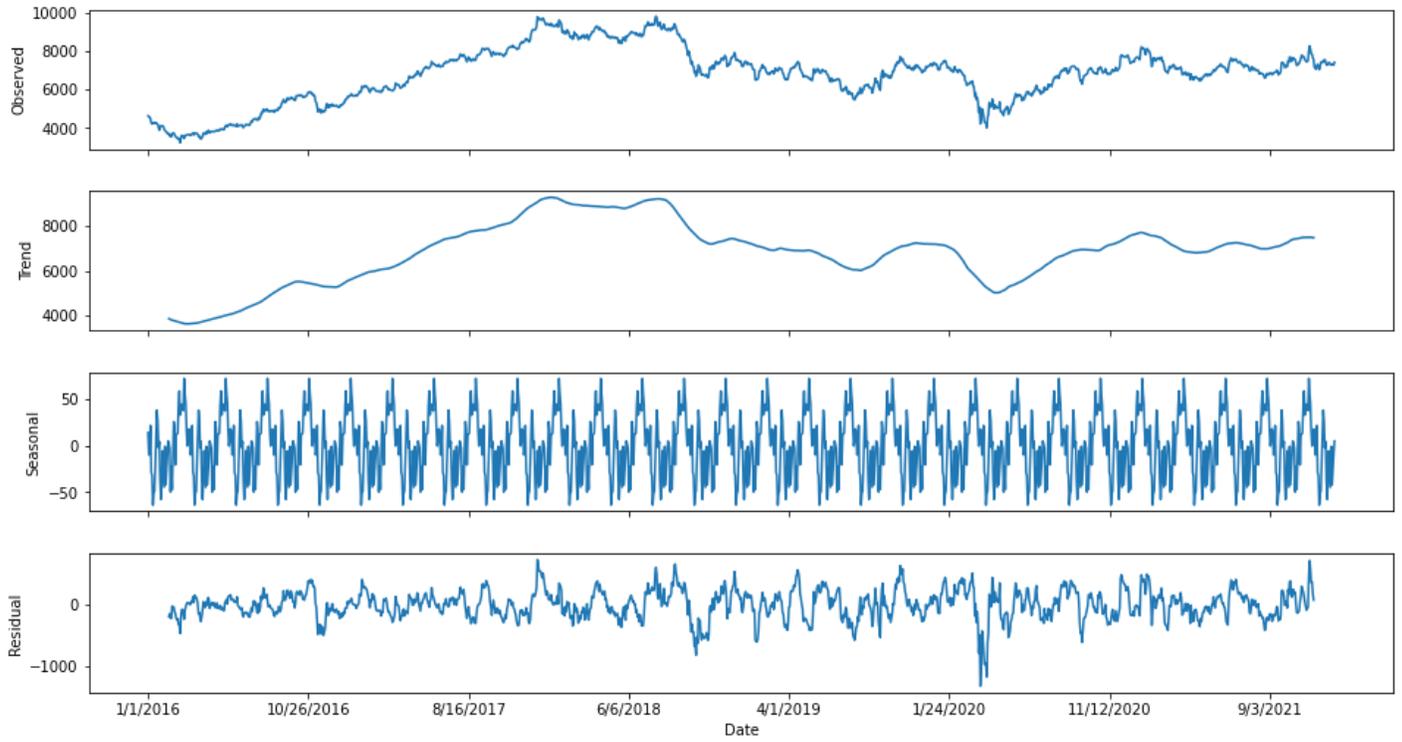

## SIMPLE EXPONENTIAL SMOOTHING

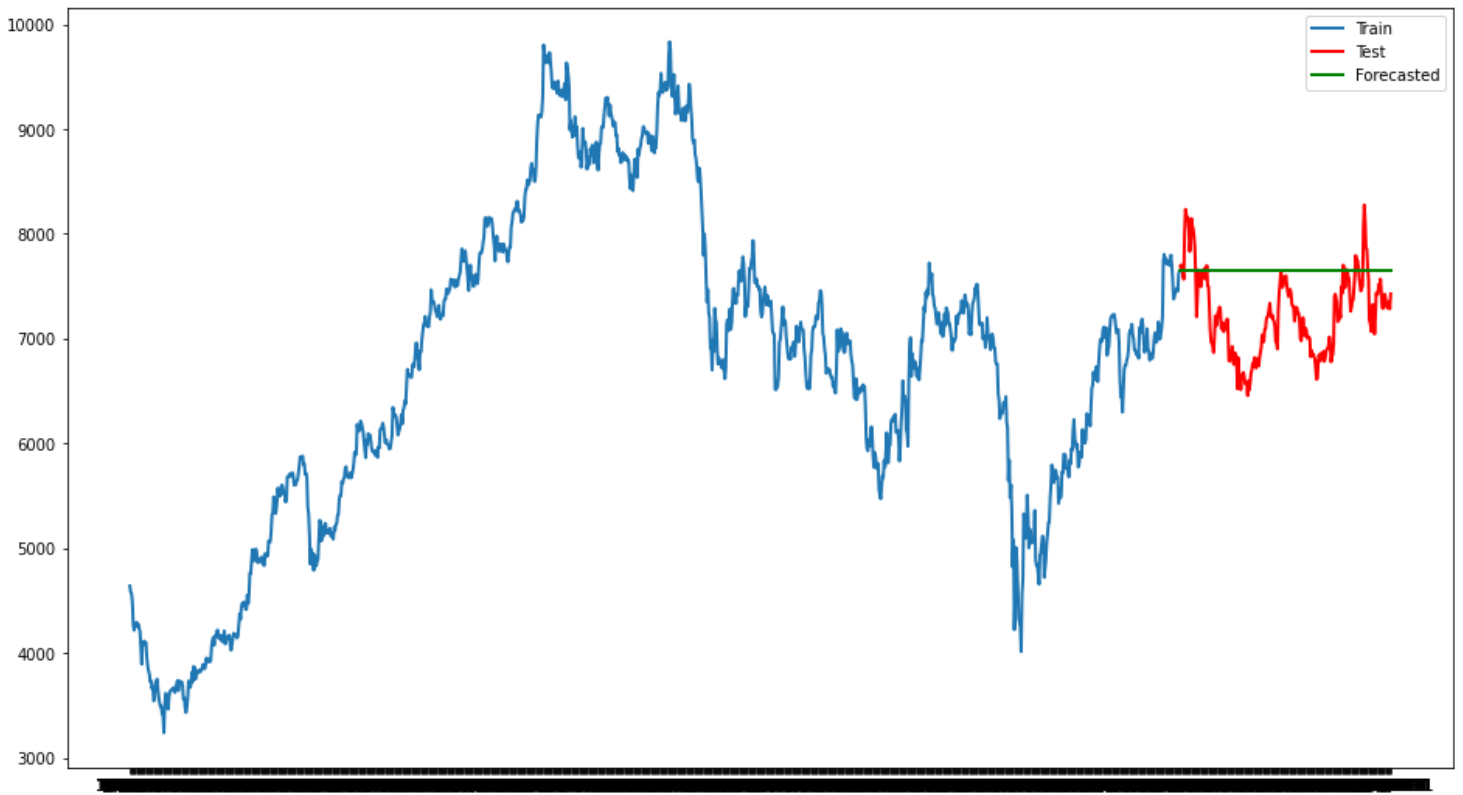

## HOLT WINTER TREND METHOD

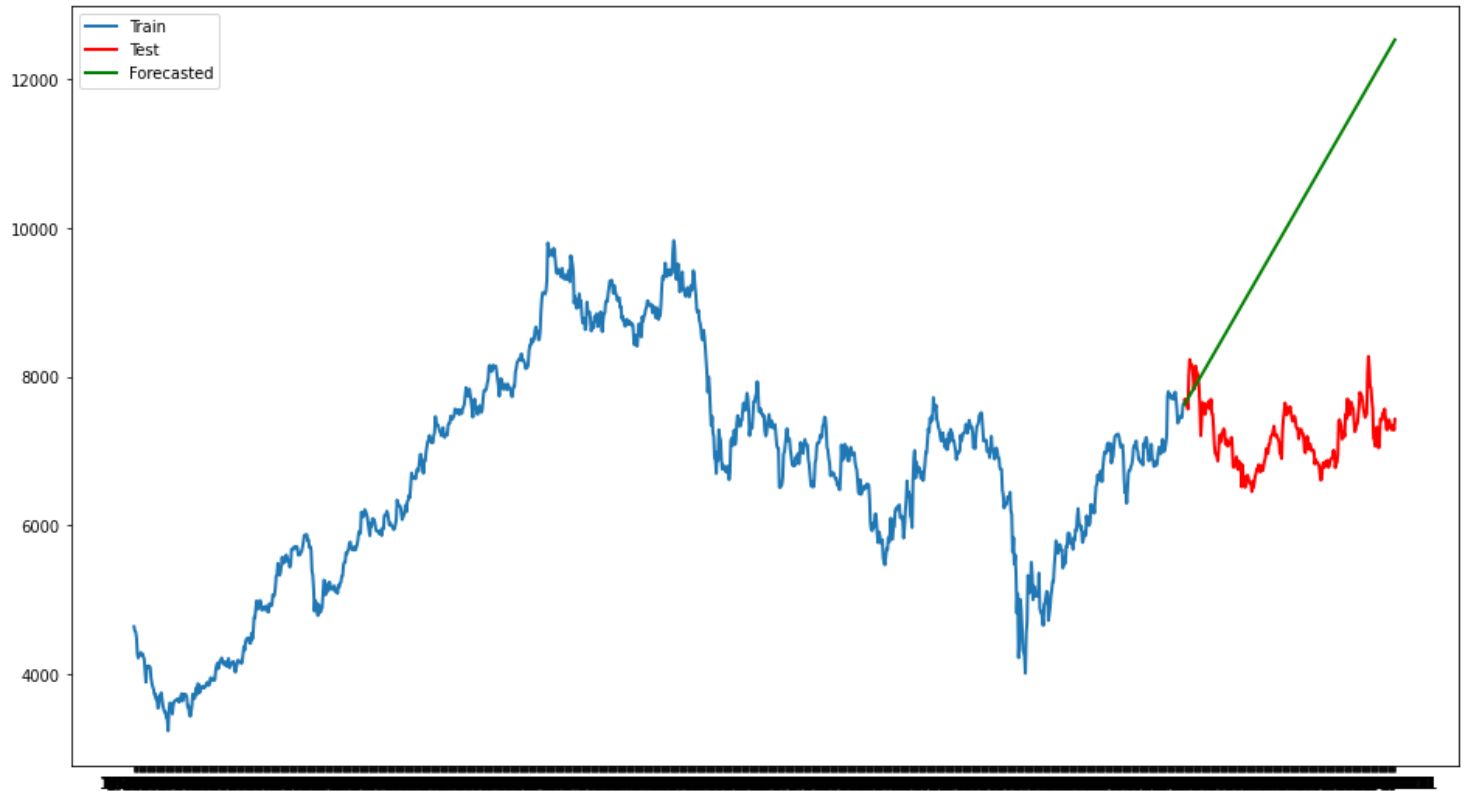

## ARIMA

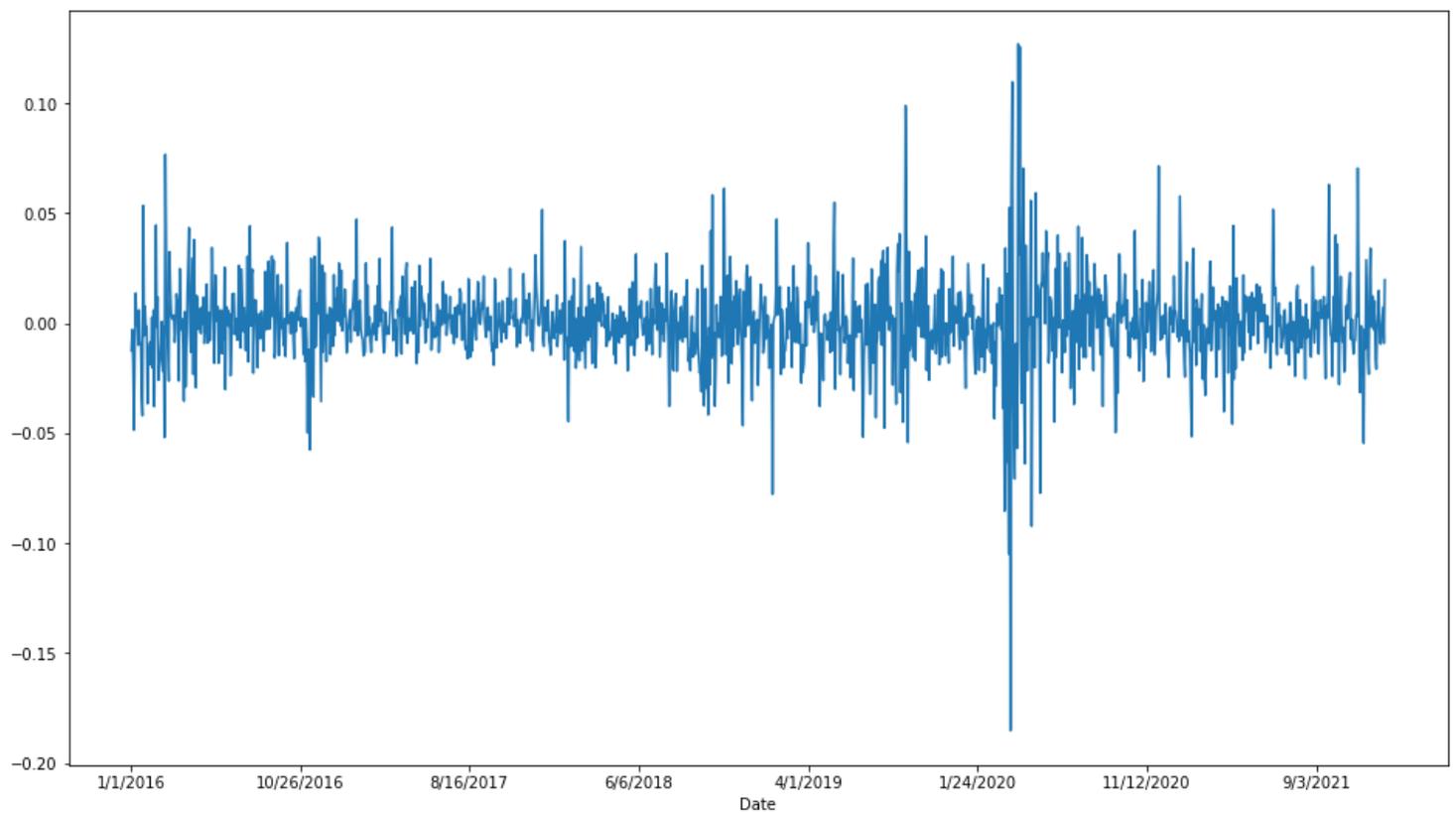

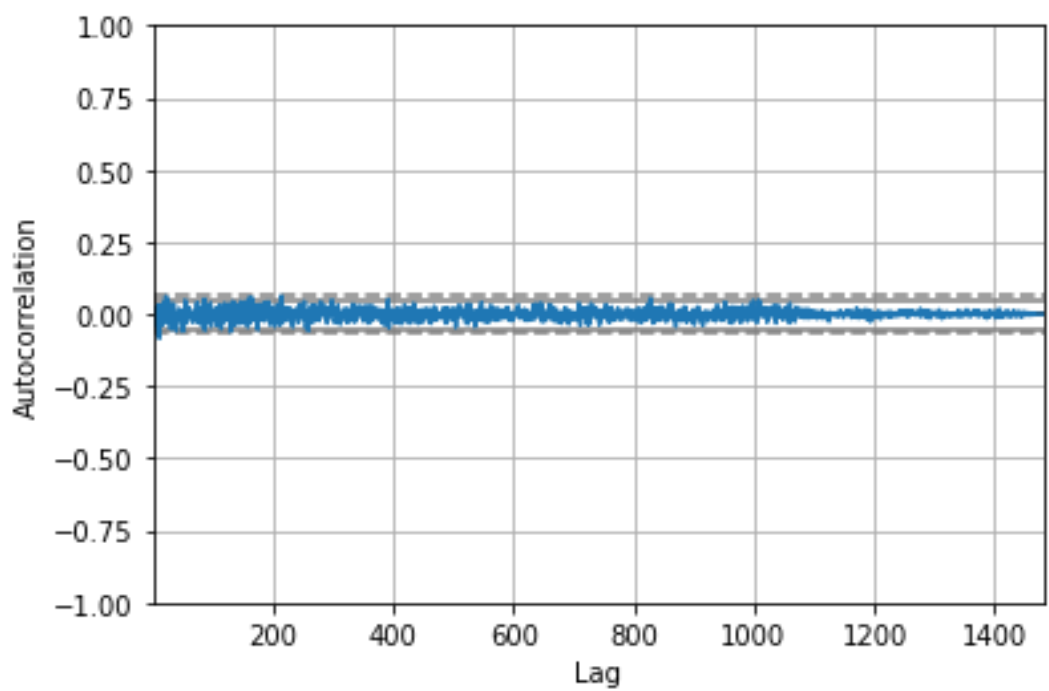

## Autocorrelation

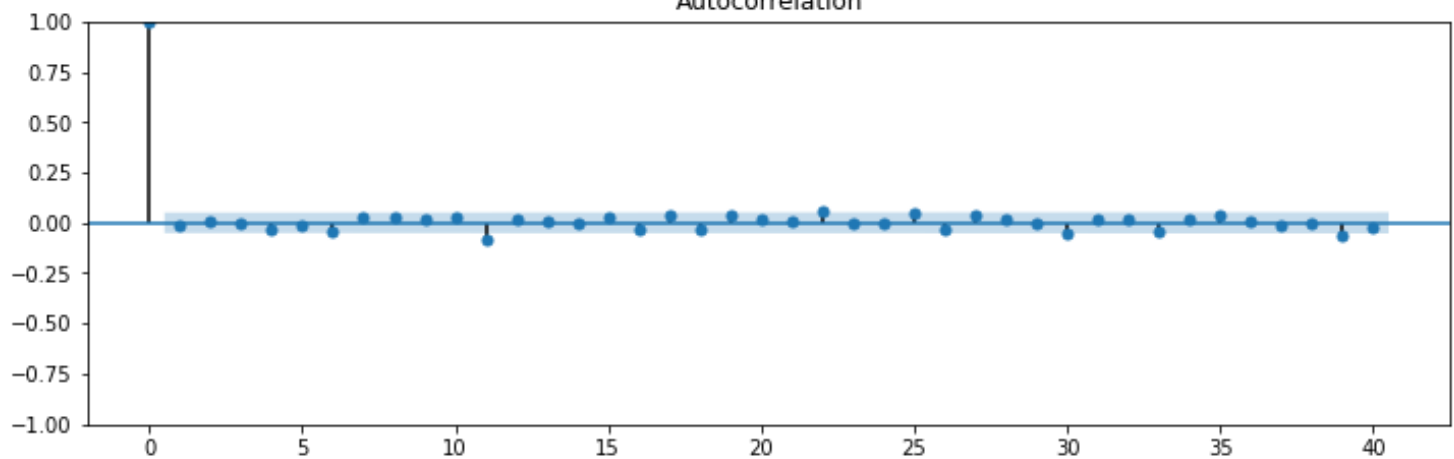

## Partial Autocorrelation

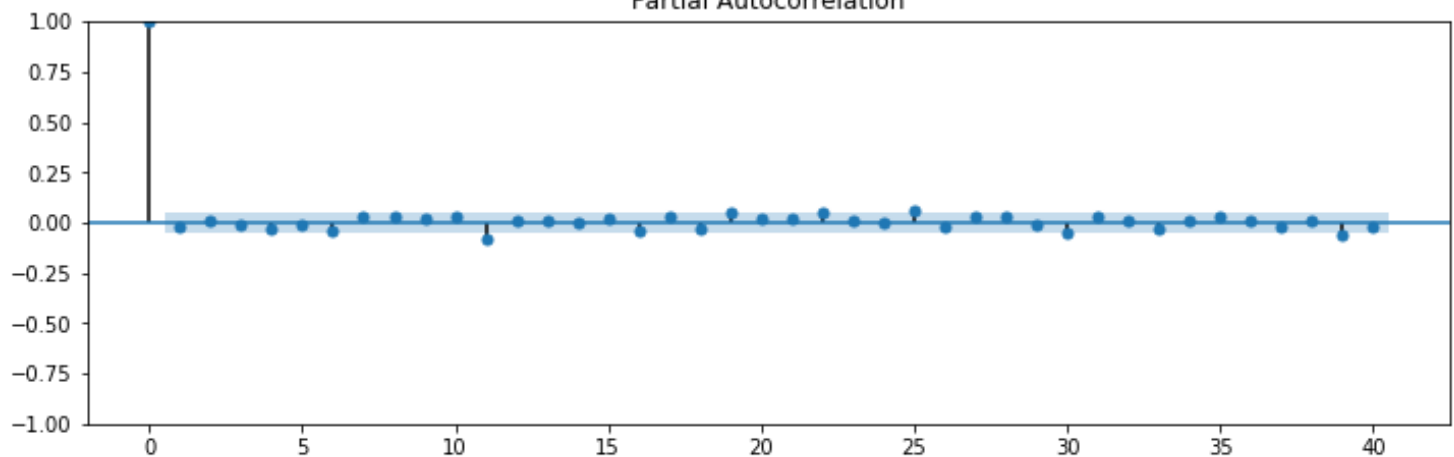

## SLIDING WINDOW METHOD

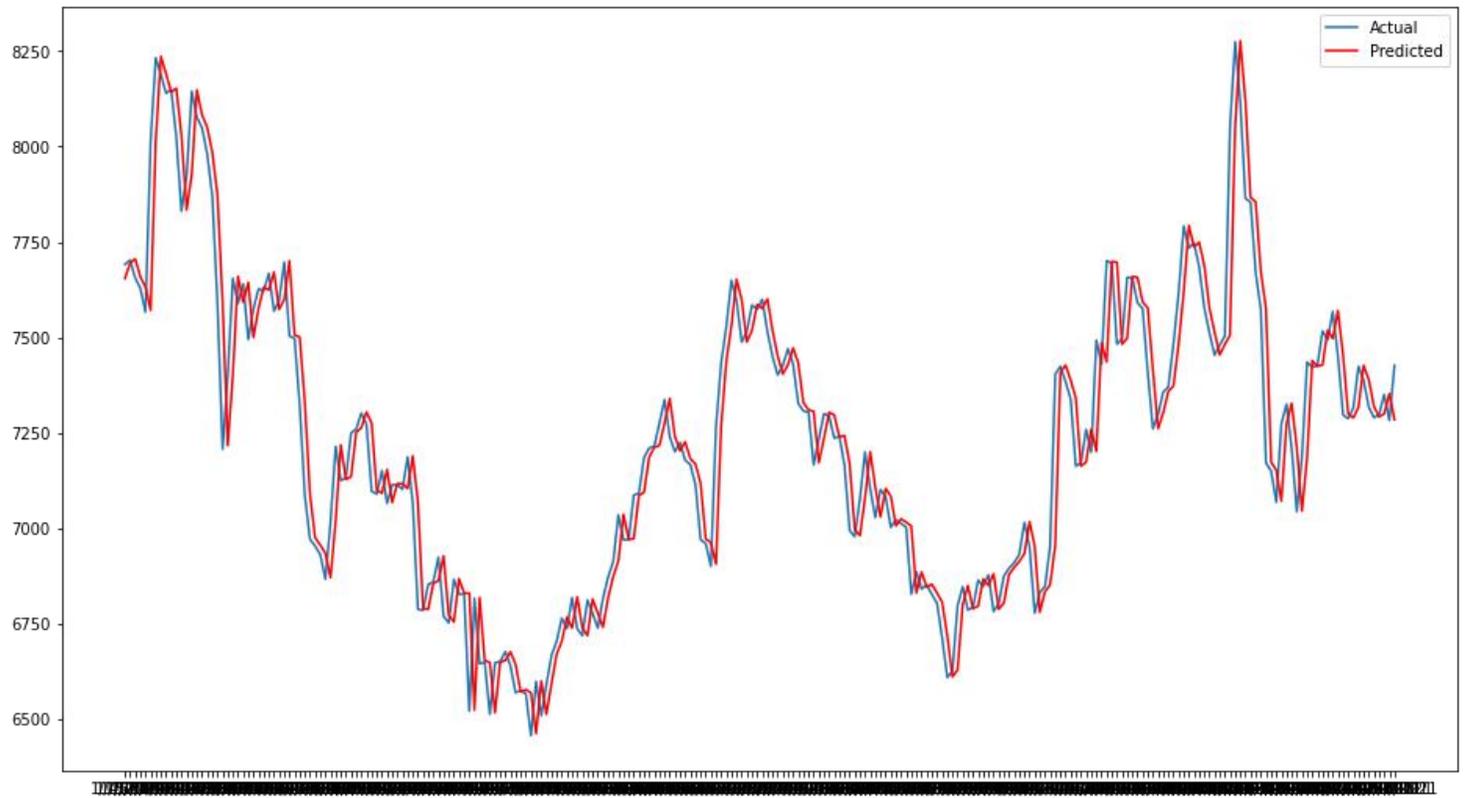

## ROLLING WINDOW METHOD

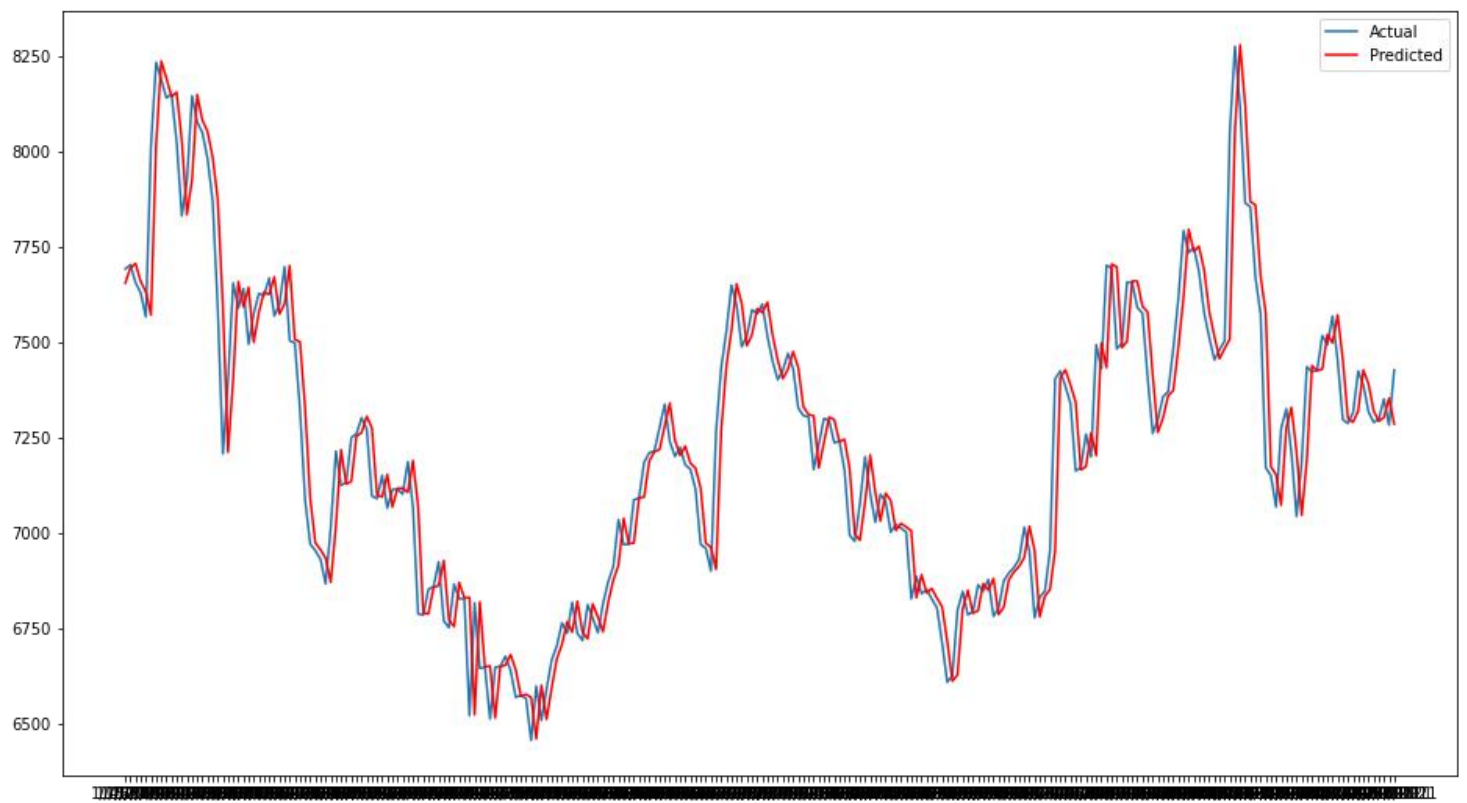

# PHARMA SECTOR

1. SUN PHARMA

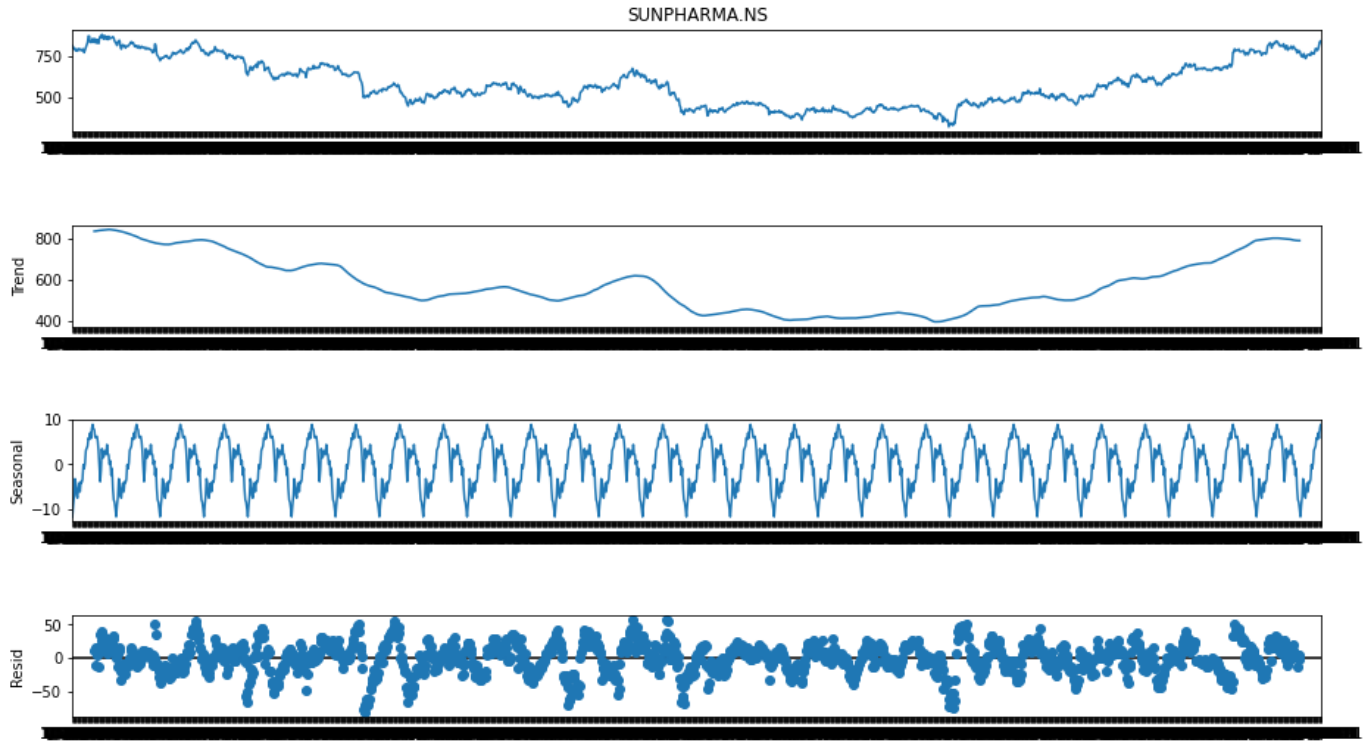

SIMPLE EXPONENTIAL SMOOTHING

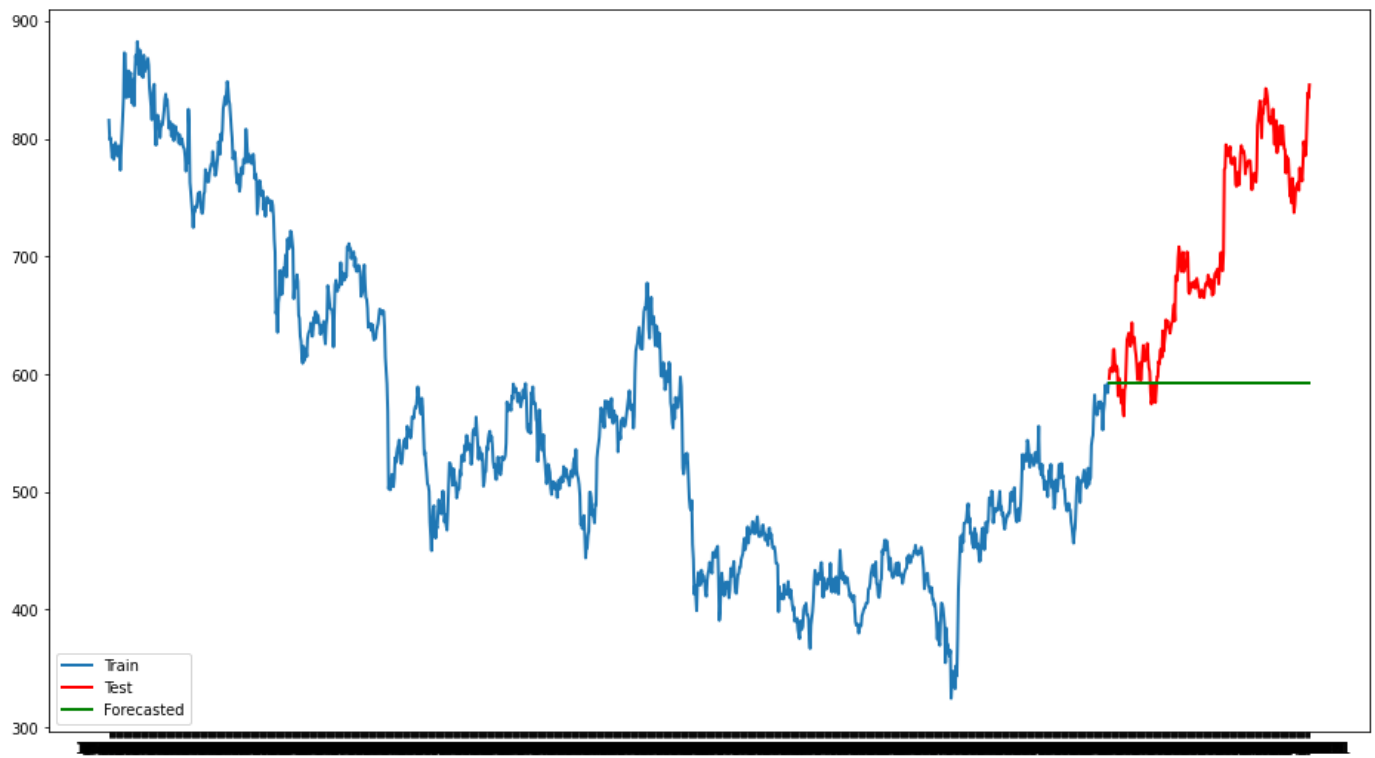

## HOLT WINTER TREND METHOD

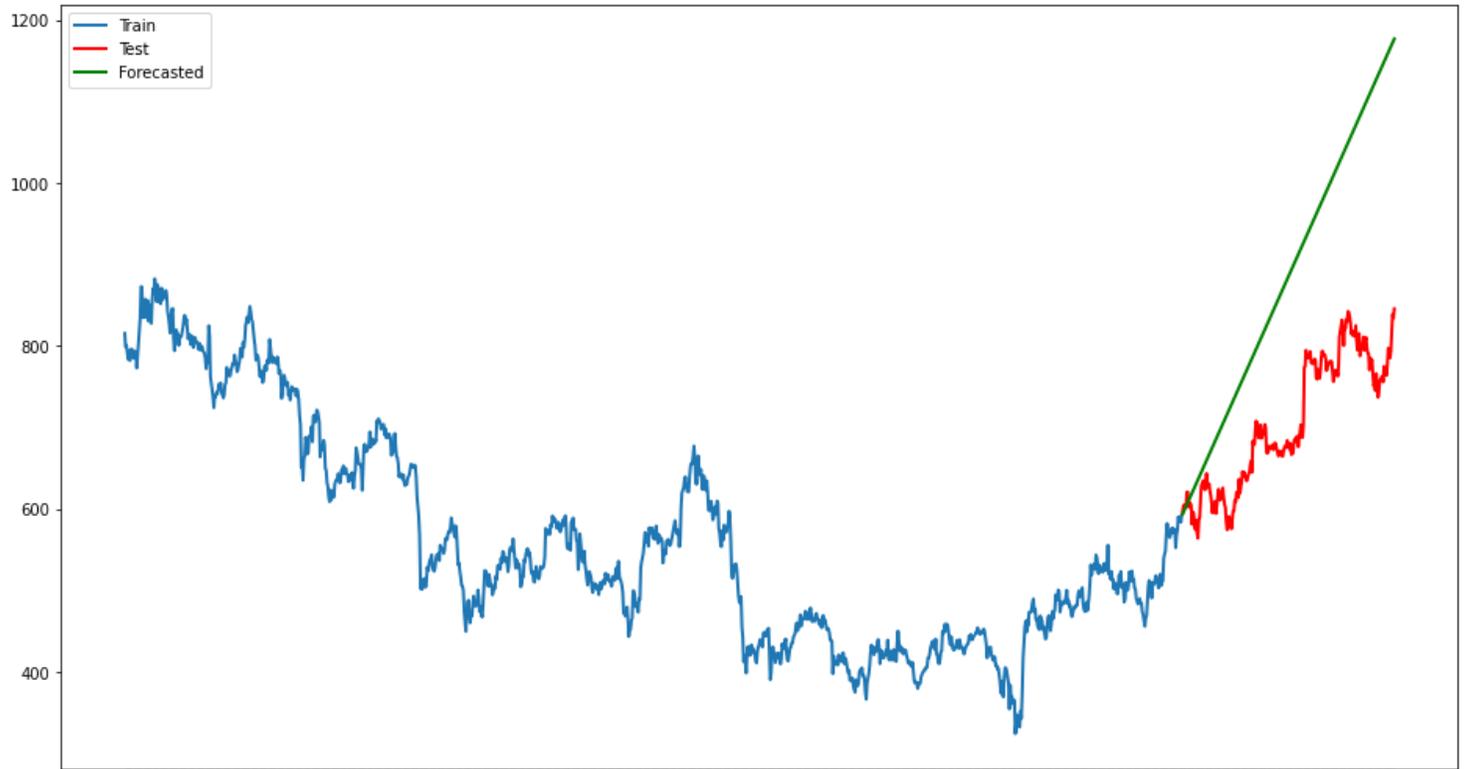

## ARIMA

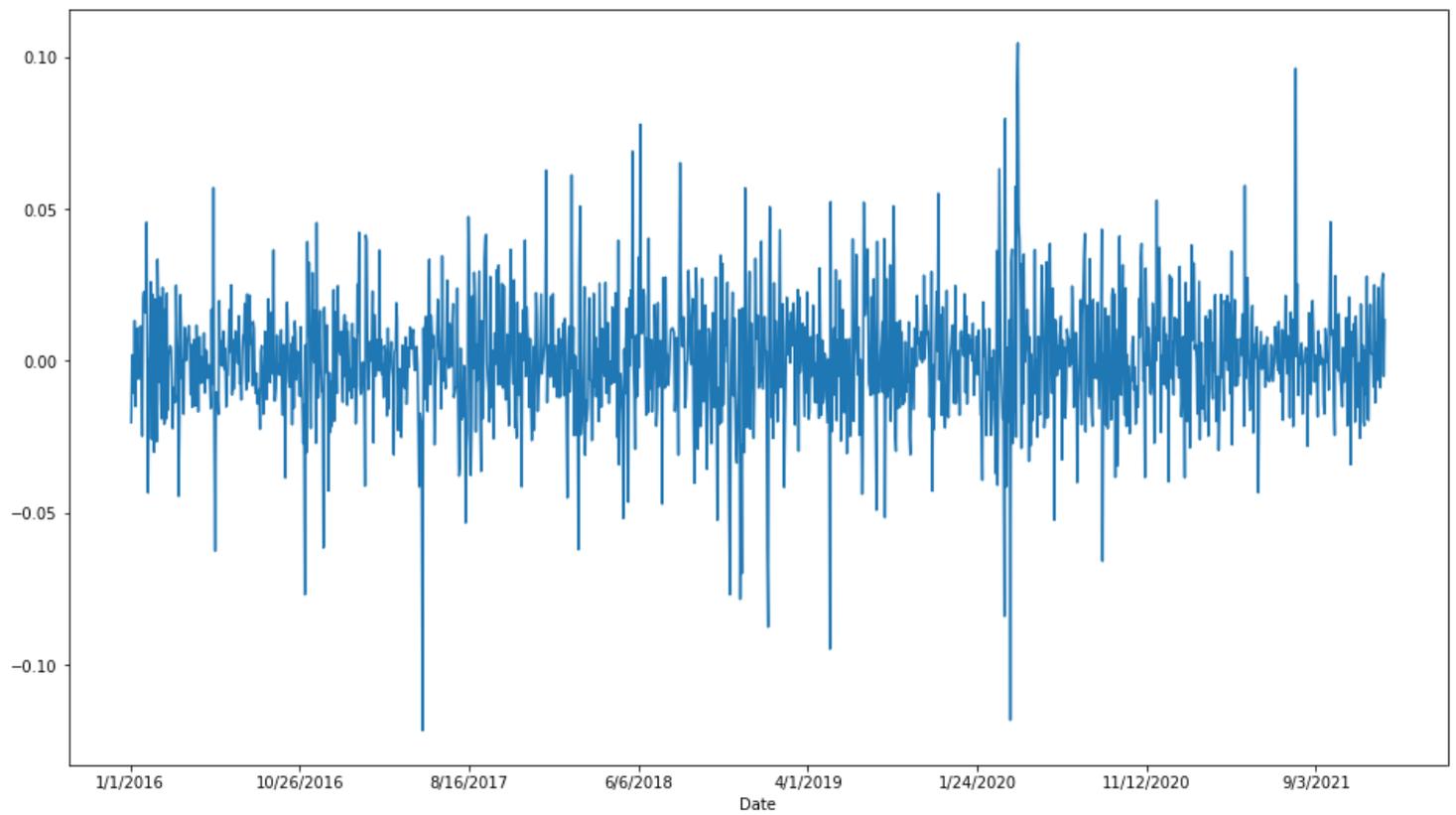

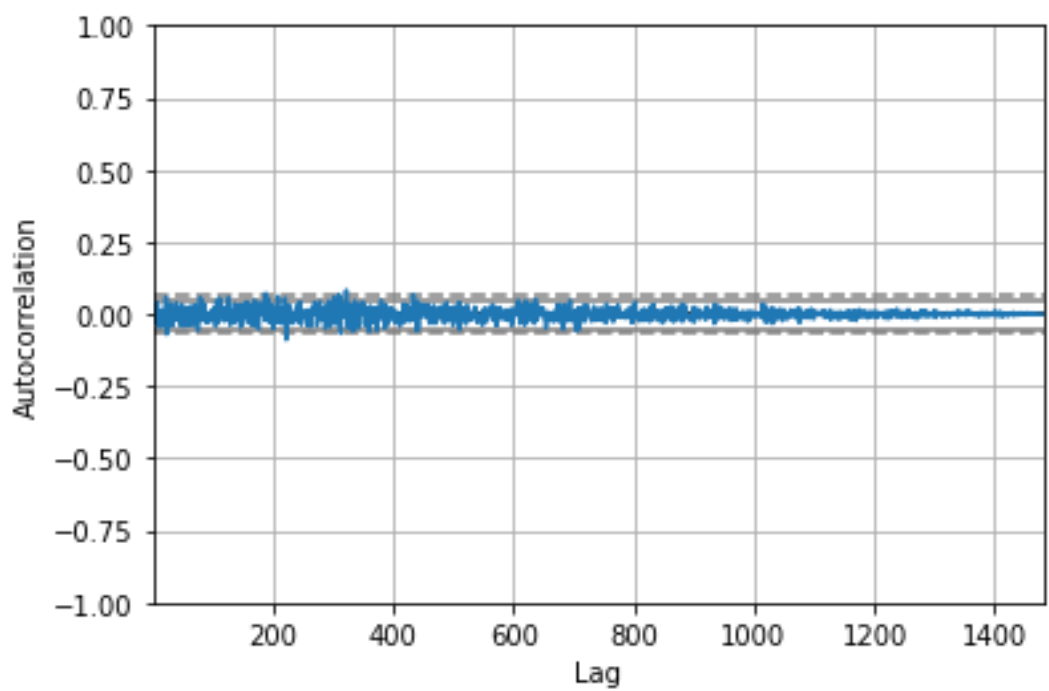

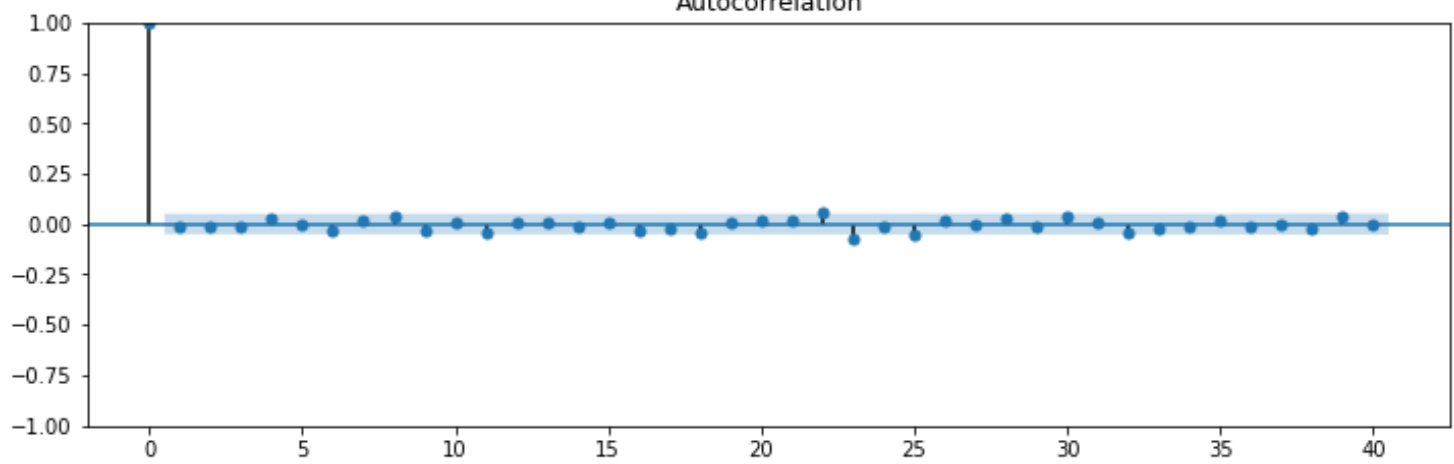

Autocorrelation

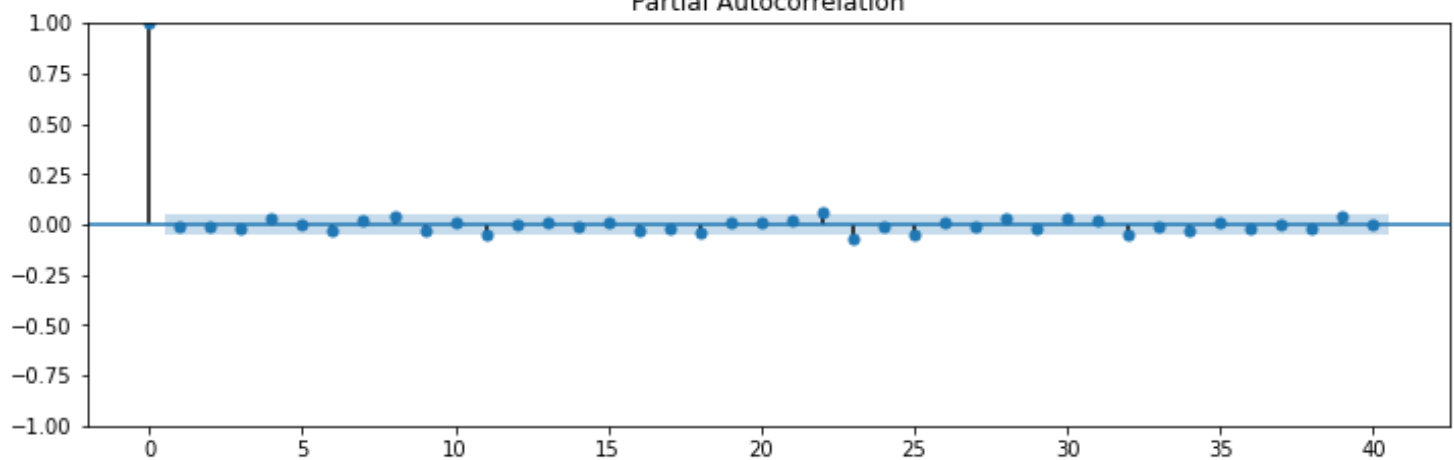

Partial Autocorrelation

## SLIDING WINDOW METHOD

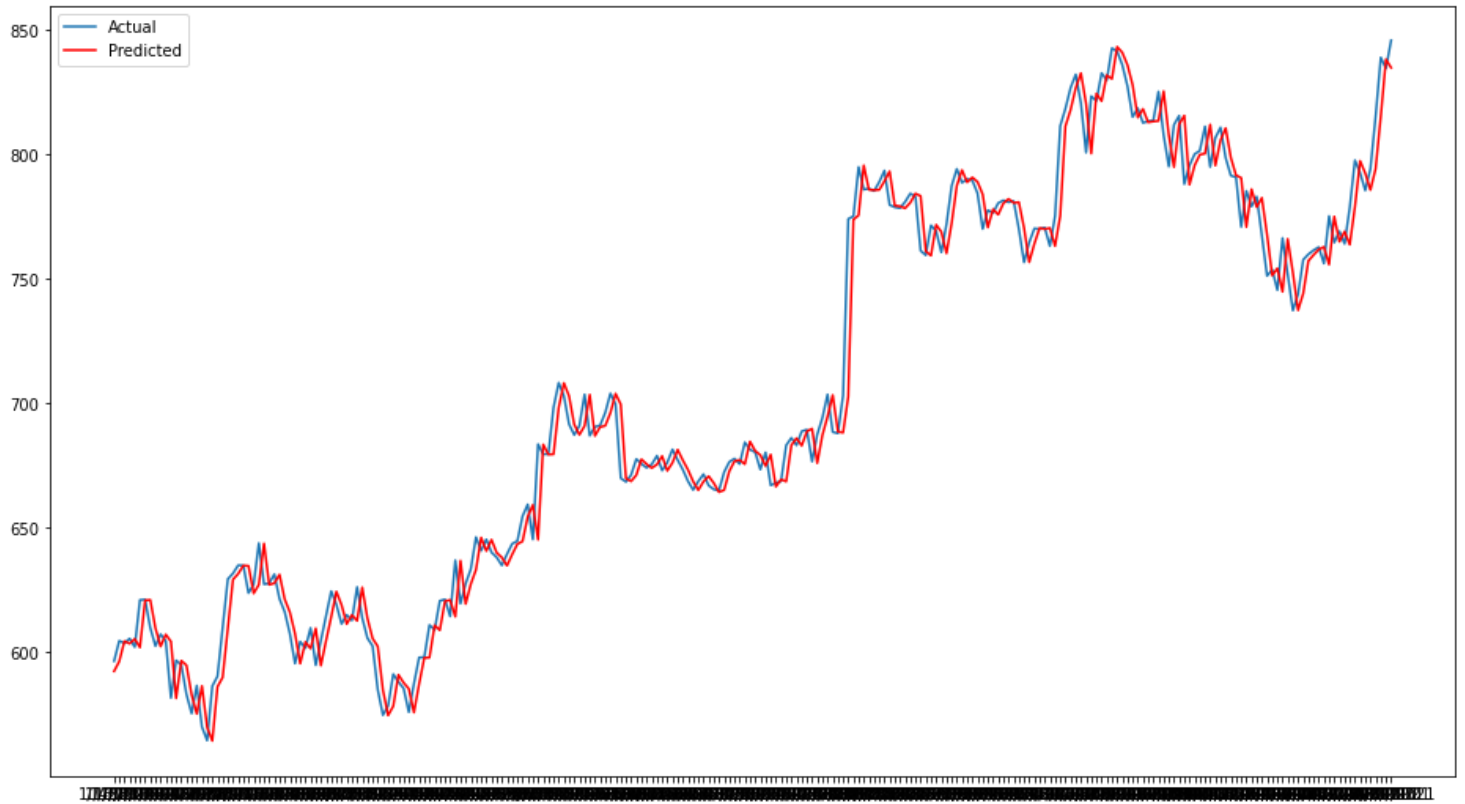

## ROLLING WINDOW METHOD

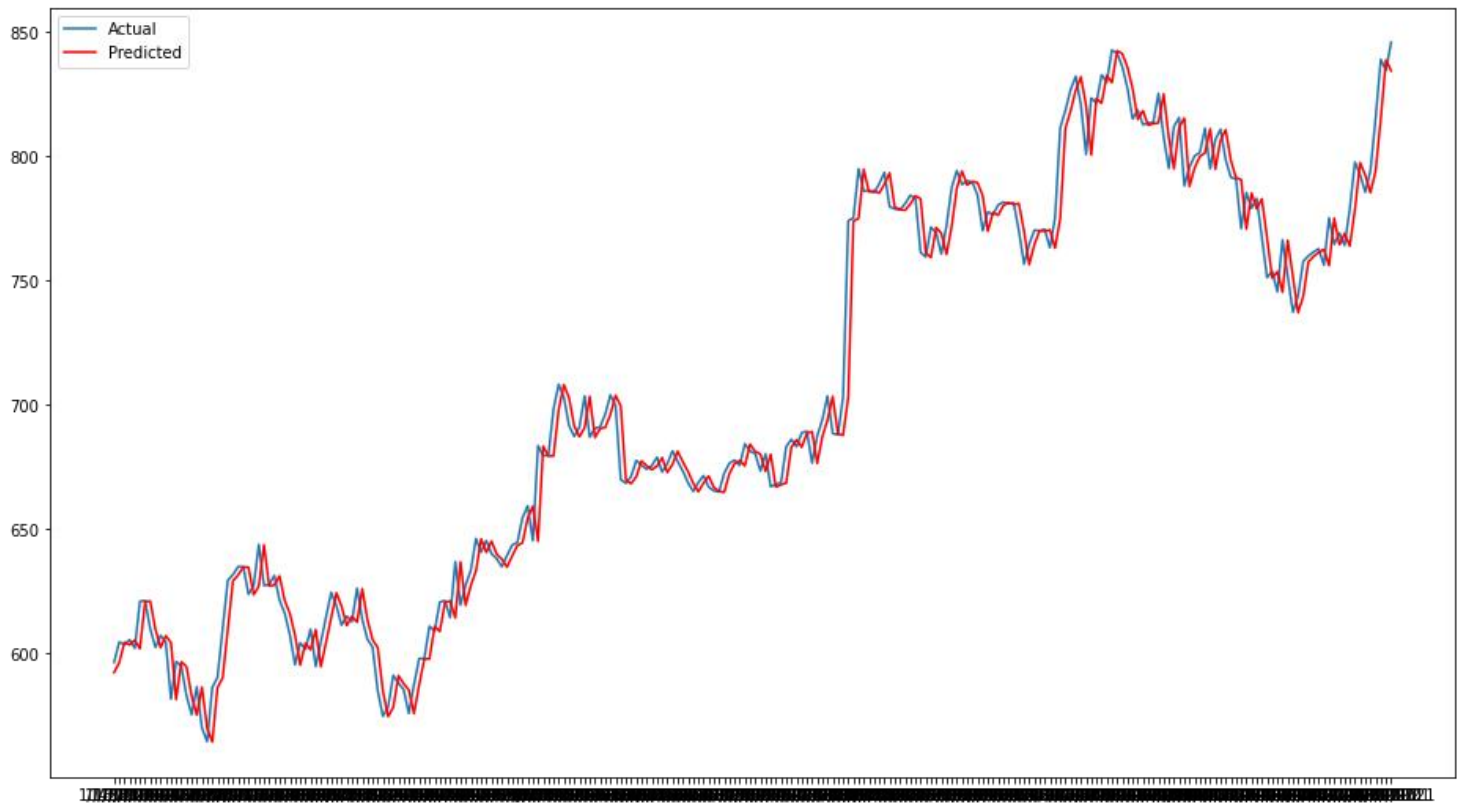

## 2. CIPLA

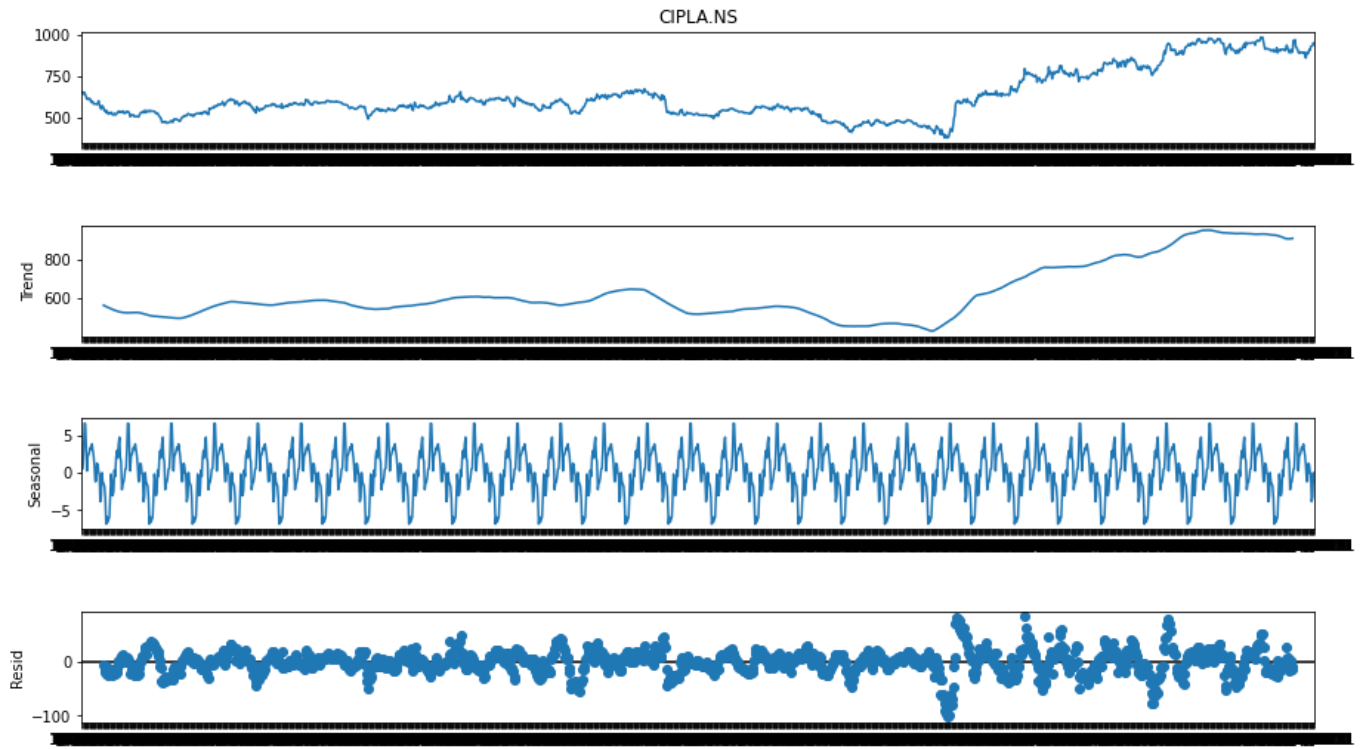

## SIMPLE EXPONENTIAL SMOOTHING

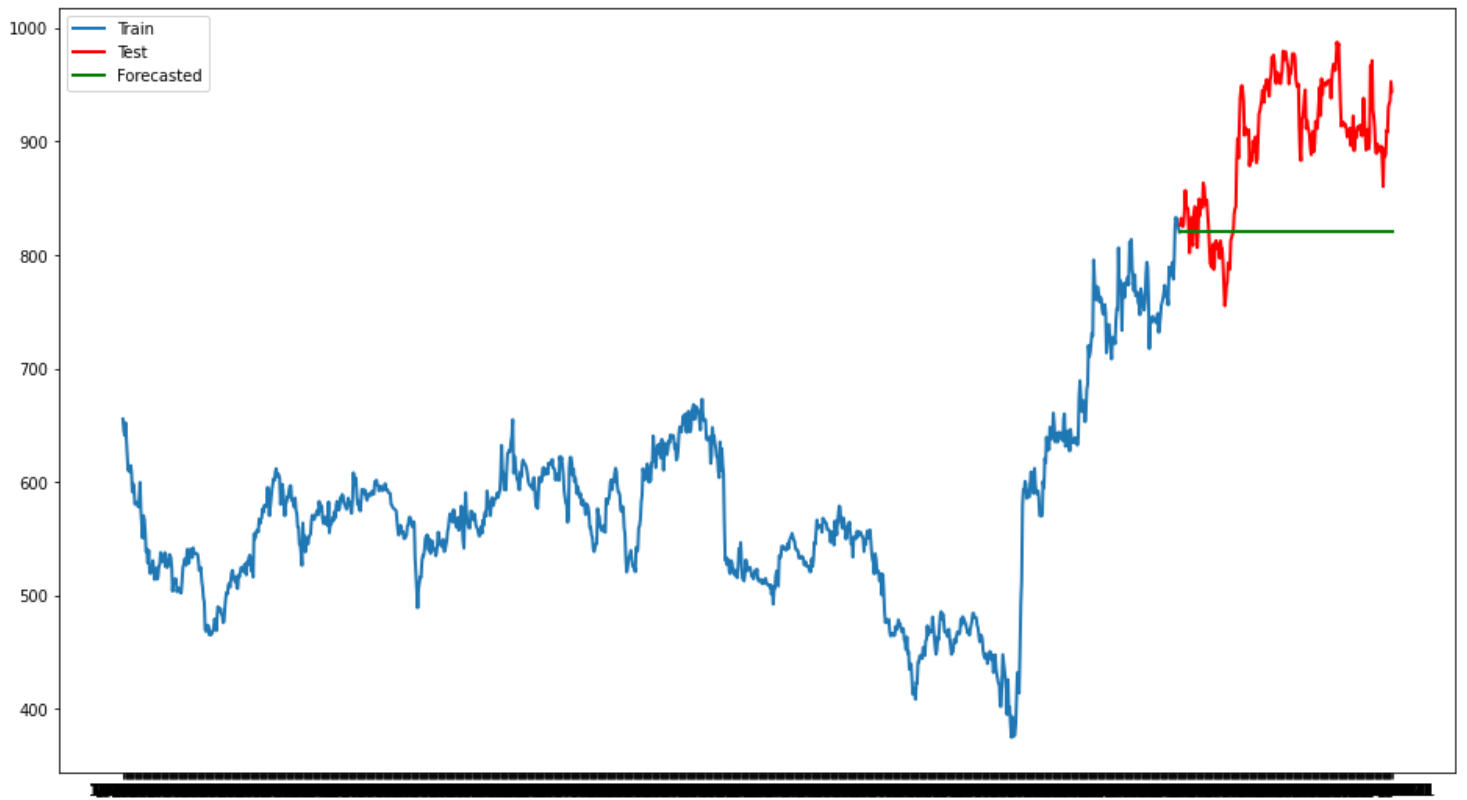

## HOLT WINTER TREND METHOD

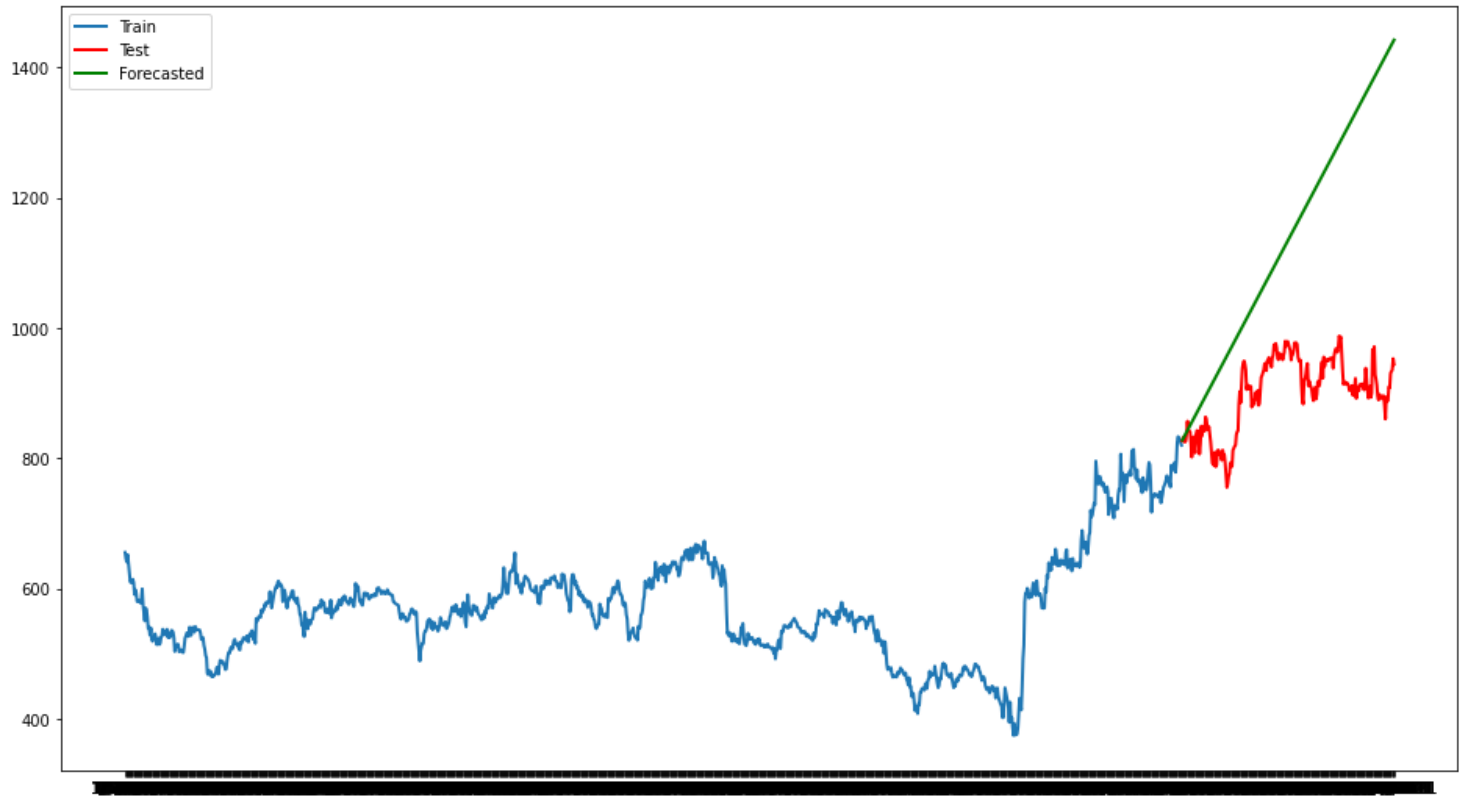

## ARIMA

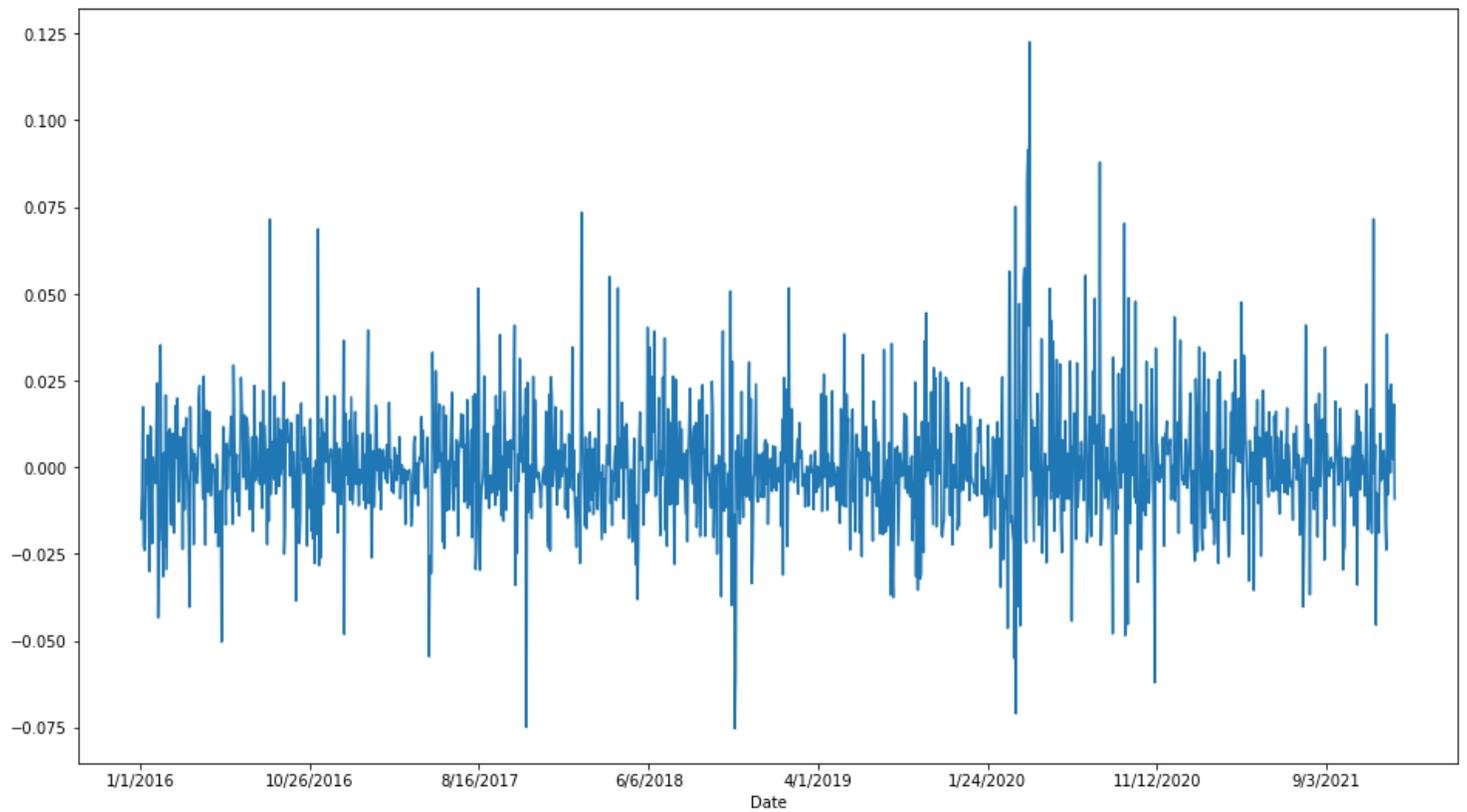

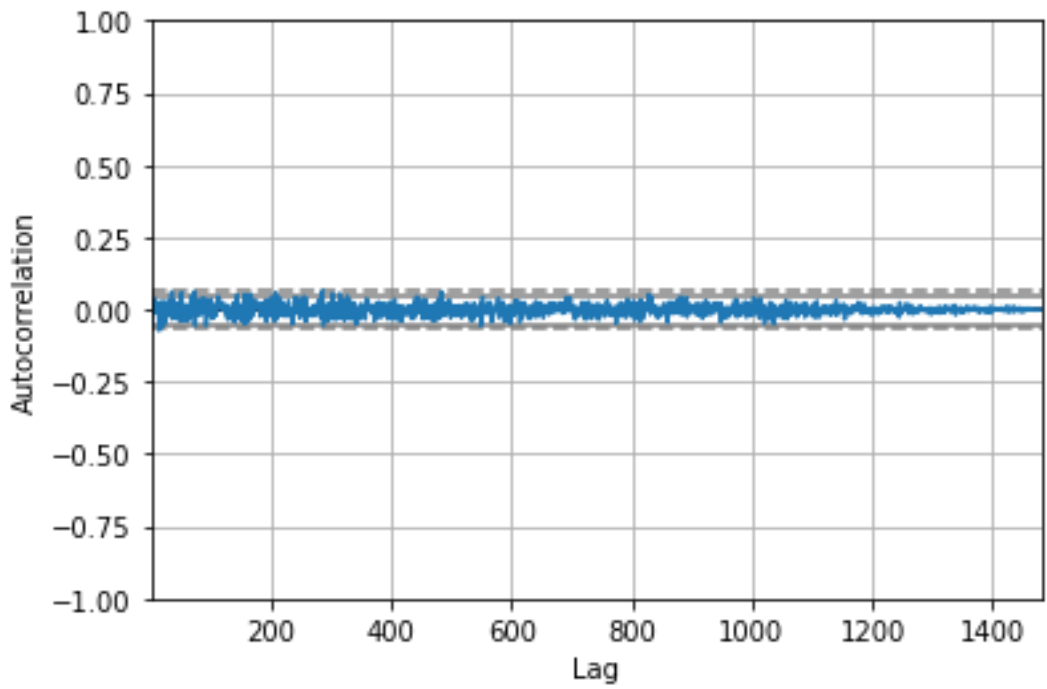

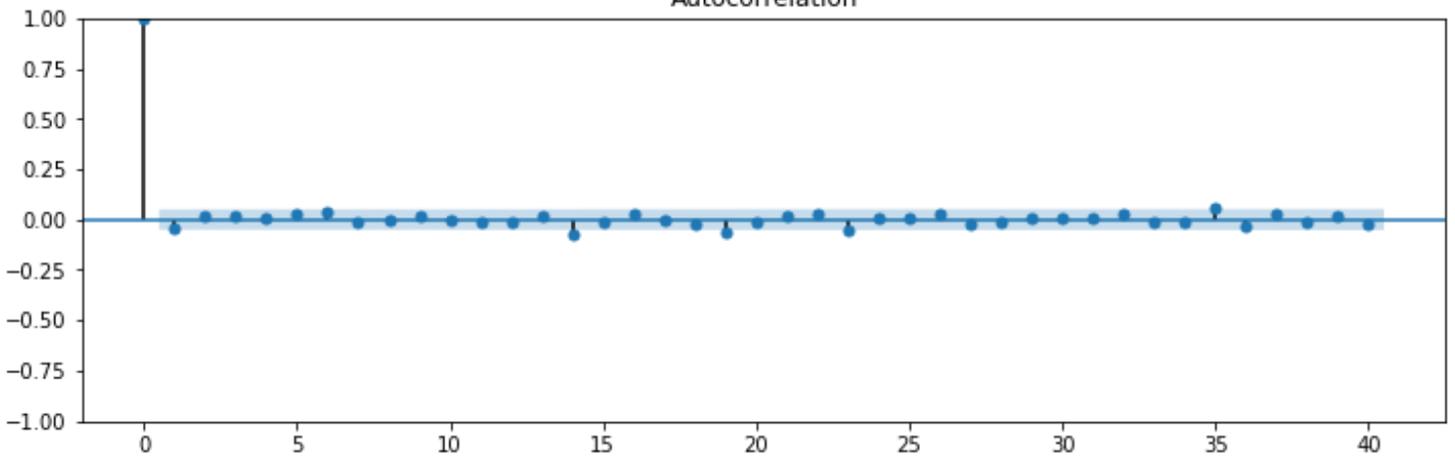

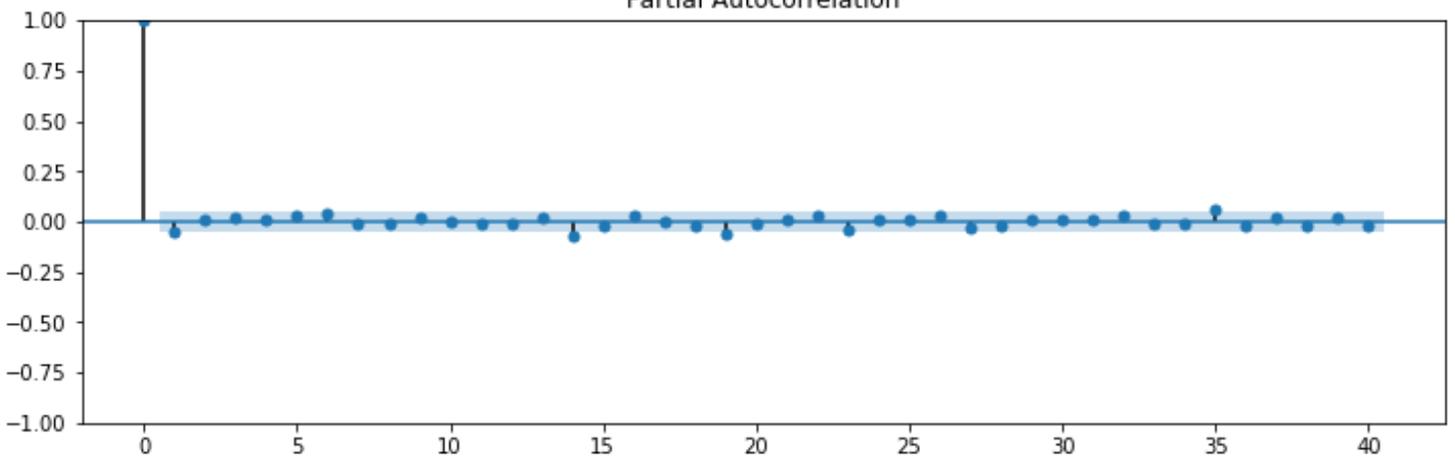

## SLIDING WINDOW METHOD

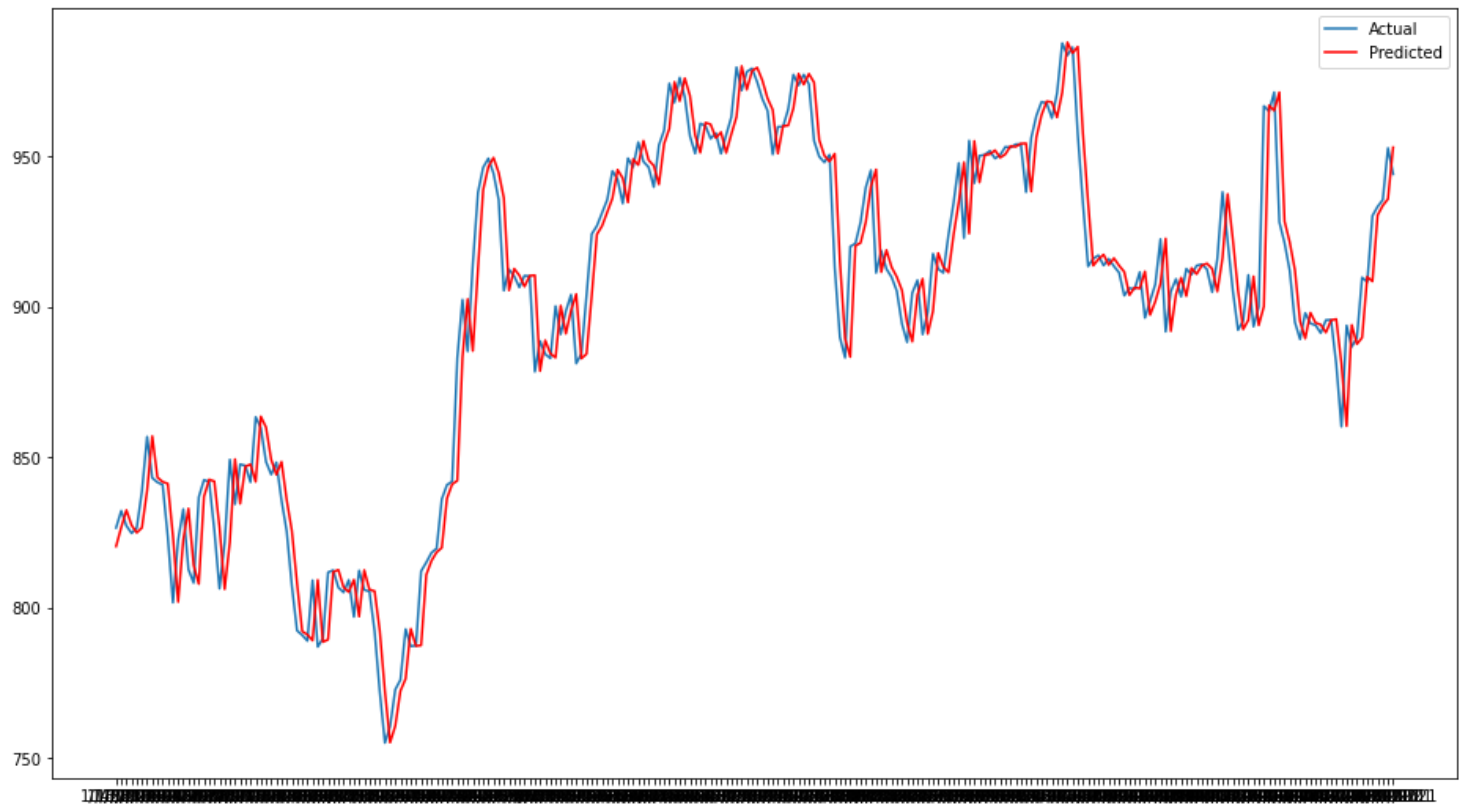

## ROLLING WINDOW METHOD

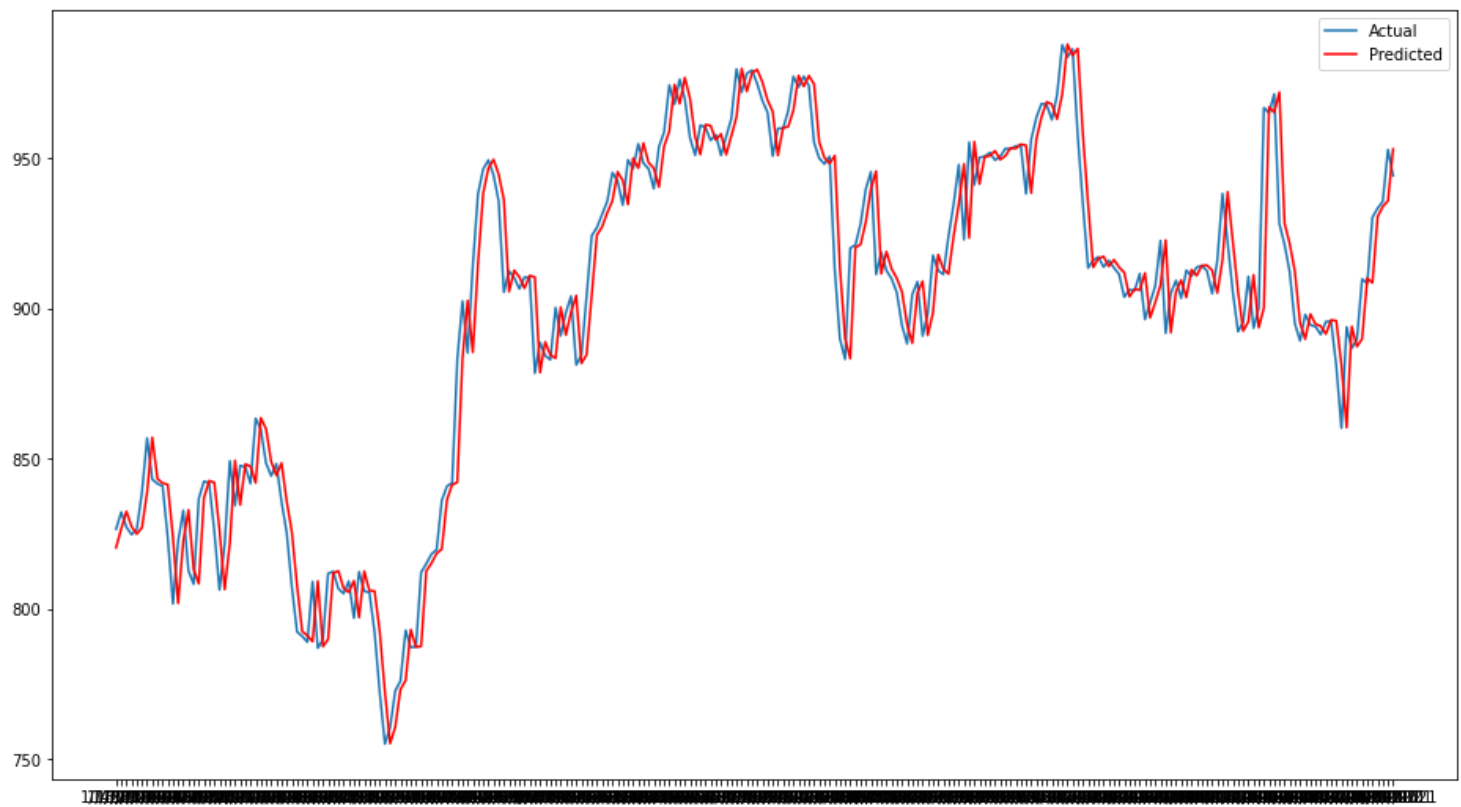

# CHAPTER-4

## MACHINE LEARNING MODELS

## Linear Regression

It is mathematical representation of patterns hidden in data. When ML model is trained to the training data it discovers some structures within it. If a model is trained on a training data and then apply that model to new data, the model would be able to infer some relationship within it.

Below are few steps for process of training a ML model:

- ❖ Get input data
- ❖ Split data into training and test data
- ❖ Fit algorithm to train data
- ❖ Evaluate model
- ❖ Use model in real life

Linear Regression is one of the best supervised machine learning algorithms. It is used to predict a numeric value based on a set of inputs. It tries to achieve a straight line of best fit between input variable and output variable. Linear regression attempts to model the relationship between two variables by fitting a linear equation to observed data. One variable is considered to be an explanatory variable, and the other is considered to be a dependent variable. Linear-regression models are relatively simple and provide an easy-to-interpret mathematical formula that can generate predictions. For each value of the independent variable, the distribution of the dependent variable must be normal. The variance of the distribution of the dependent variable should be constant for all values of the independent variable. The relationship between the dependent variable and each independent variable should be linear and all observations should be independent.

# Random Forest

Random forest is a Supervised Machine Learning Algorithm that is used widely in Classification and Regression problems. It builds decision trees on different samples and takes their majority vote for classification and average in case of regression. It can handle the dataset containing continuous variables and categorical variable It performs better results for classification problems. It consists of multiple random decision trees. First each tree is built on a random sample from original data, and then at each node a subset of features are randomly selected to generate best split. It is not only used for classification/regression but also for outlier detection, clustering and interpreting a dataset. It is based on the concept of ensemble learning, which is a process of combining multiple classifiers to solve a complex problem and to improve the performance of the model.

One of the main advantage of using random forest is it takes less time as compared to other algorithms. It predicts output with high accuracy and also maintains accuracy when large proportion of data is missing. Apart from these there are few assumptions of Random Forest. There should be some actual values in the feature variable of the dataset so that the classifier can predict accurate results rather than a guessed result. The predictions from each tree must have very low correlations. Random Forest is mainly applied in banking, pharma and marketing sectors. It is capable of performing both classification and regression tasks and can also handle large dataset with high dimensionality.

# Gradient Boost

Gradient Boosting is a supervised machine learning algorithm used for classification and regression problems. It is an ensemble technique which uses multiple weak learners to produce a strong model for regression and classification. It requires few inputs such as a loss function to optimize, a weal learner to make prediction and an additive model to add weak learners to minimize the loss function. Gradient descent is a first-order iterative optimization algorithm for finding a local minimum of a differentiable function. As gradient boosting is based on minimizing a loss function, different types of loss functions can be used resulting in a flexible technique that can be applied to regression, multi-class classification

Some advantages of Gradient Boosting are as follows:

1. Most of the time predictive accuracy of gradient boosting algorithm on higher side.

2. It provides lots of flexibility and can optimize on different loss functions and provides several hyper parameter tuning options that make the function fit very flexible.

3. Most of the time no data pre-processing required.

4. Gradient Boosting algorithm works great with categorical and numerical data.

5. Handles missing data — missing value imputation not required.

# XGBoost

XGBoost stands for "Extreme Gradient Boosting". XGBoost is one of the most popular variants of gradient boosting. It is a decision-tree-based ensemble Machine Learning algorithm that uses a gradient boosting framework. XGBoost is basically designed to enhance the performance and speed of a Machine Learning model. In prediction problems involving unstructured data (images, text, etc.), artificial neural networks tend to outperform all other algorithms or frameworks. However, when it comes to small-to-medium structured/tabular data, decision tree-based algorithms are considered best-in-class right now. XGBoost is used for supervised learning problems, where we use the training data to predict a target variable. XGBoost is an optimized distributed gradient boosting library designed to be highly efficient, flexible, and portable. It implements Machine Learning algorithms under the Gradient Boosting framework. It provides a parallel tree boosting to solve many data science problems in a fast and accurate way.

XGBoost uses pre-sorted algorithm & histogram-based algorithm for computing the best split. The histogram-based algorithm splits all the data points for a feature into discrete bins and uses these bins to find the split value of the histogram. Also, in XGBoost, the trees can have a varying number of terminal nodes and left weights of the trees that are calculated with less evidence is shrunk more heavily. XGBoost is a faster algorithm when compared to other algorithms because of its parallel and distributed computing. XGBoost is developed with both deep considerations in terms of systems optimization and principles in machine learning.

# Gaussian NB

Naïve Bayes method is a type of supervised machine learning model. This namedto be "naïve" because the classifier assumes that the input features that go into the model are independent of each other. Hence, changing one input feature won't affect any of the others. Naive Bayes are a group of supervised machine learning classification algorithms based on the Bayes theorem. It is a simple classification technique, but has high functionality. They find use when the dimensionality of the inputs is high. Complex classification problems can also be implemented by using Naive Bayes Classifier. Gaussian Naive Bayes is a variant of Naive Bayes that follows Gaussian normal distribution and supports continuous data. Naive Bayes Classifiers are based on the Bayes Theorem. One assumption taken is the strong independence assumptions between the features. These classifiers assume that the value of a particular feature is independent of the value of any other feature. In a supervised learning situation, Naive Bayes Classifiers are trained very efficiently. Naive Bayed classifiers need a small training data to estimate the parameters needed for classification. Naive Bayes Classifiers have simple design and implementation and they can applied to many real life situations.

Some Application of Naïve Bayes are as follows:-

- ❖ Real-time prediction
- ❖ Multi-class prediction
- ❖ Text Classification
- ❖ Recommendation System

Some Advantages of Naïve Bayes are:

- ❑ Naive Bayes is easy to grasp and works quickly to predict class labels. It also performs well on multi-class prediction.
- ❑ When the assumption of independence holds, a Naive Bayes classifier performs better compared to other models like logistic regression, and you would also need less training data.
- ❑ It performs well when the input values are categorical rather than numeric.

# Logistic Regression

Logistic Regression comes under the umbrella of supervised machine learning. It is used when data is binary in nature and linearly separable. It is mainly used for binary classification problem. Logistic model is a statistical model that models the probability of one event taking place by having the log-odds for the event be a linear combination of one or more independent variables. In regression analysis, logistic regression is estimating the parameters of a logistic model (the coefficients in the linear combination). Formally, in binary logistic regression there is a single binary dependent variable, coded by a indicator variable, where the two values are labeled "0" and "1", while the independent variables can each be a binary variable or a continuous variable.

**Key Assumptions of Effective Logistic Regression:**

Multinomial can be used to classify subjects into groups based on a categorical range of variables to predict behavior. For example, you can conduct a survey in which participants are asked to select one of several competing products as their favorite. You can create profiles of people who are most likely to be interested in your product, and plan your advertising strategy accordingly.

Binary is most useful when you want to model the event probability for a categorical response variable with two outcomes. A loan officer wants to know whether the next customer is likely to default or not default on a loan. Binary analysis can help assess the risk of extending credit to a particular customer.

The goal of Logistic Regression is to discover a link between characteristics and the likelihood of a specific outcome. A Logistic Regression model is similar to a Linear Regression model, except that the Logistic Regression utilizes a more sophisticated cost function, which is known as the "Sigmoid function" or "logistic function" instead of a linear function.

# KNN Model

The k-nearest neighbors (KNN) algorithm is a simple, supervised machine learning algorithm that can be used to solve both classification and regression problems. It's easy to implement and understand, but has a major drawback of becoming significantly slows as the size of that data in use grows.

KNN works by finding the distances between a query and all the examples in the data, selecting the specified number examples (K) closest to the query, then votes for the most frequent label or averages the labels. Knn model is often called lazy learner because model is not learned using training data prior and the learning process is postponed to a time when prediction is requested on the new instance. Some of its application are in data preprocessing, recommendation engines, financial sector, health care sector and pattern recognition problems.

# SECTOR WISE RESULTS AND ANALYSIS

## <u>METAL SECTOR</u>

1. TATA STEEL

LINEAR REGRESSION

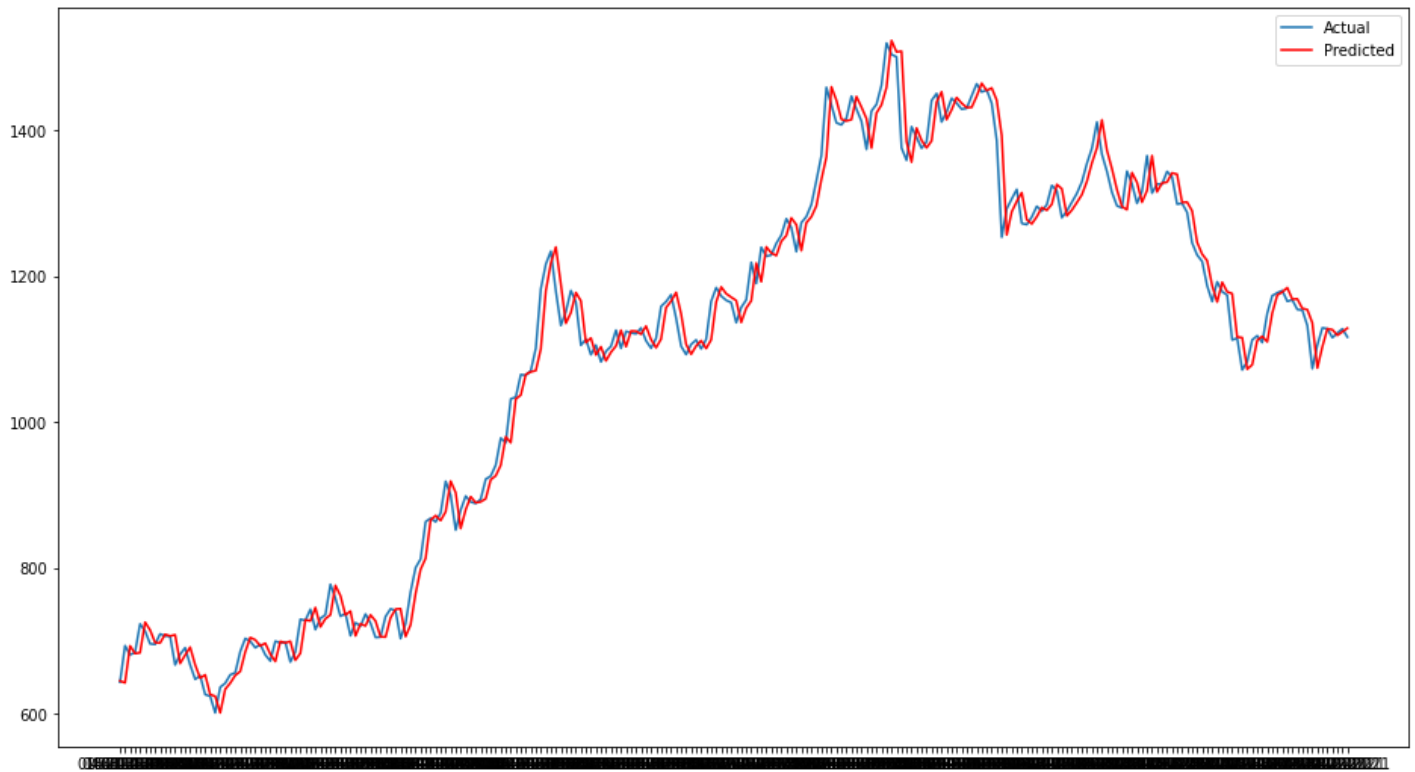

## RANDOM FOREST

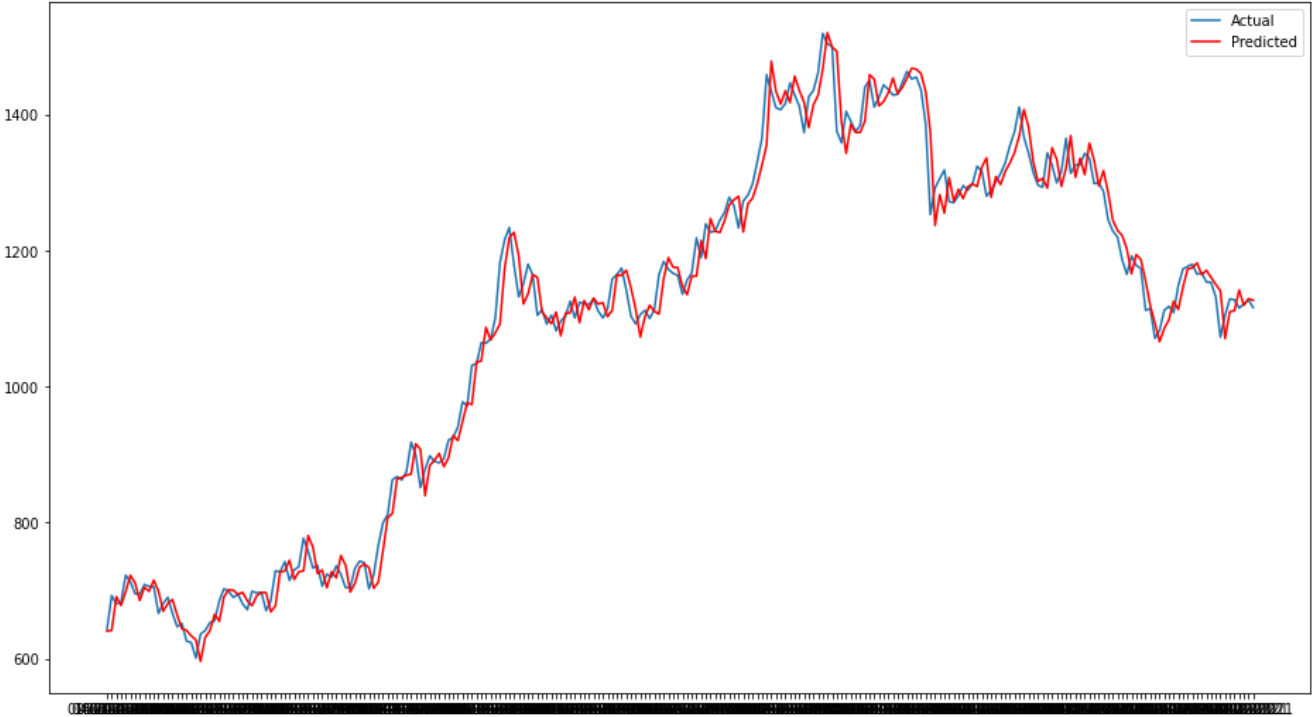

## GRADIENT BOOST

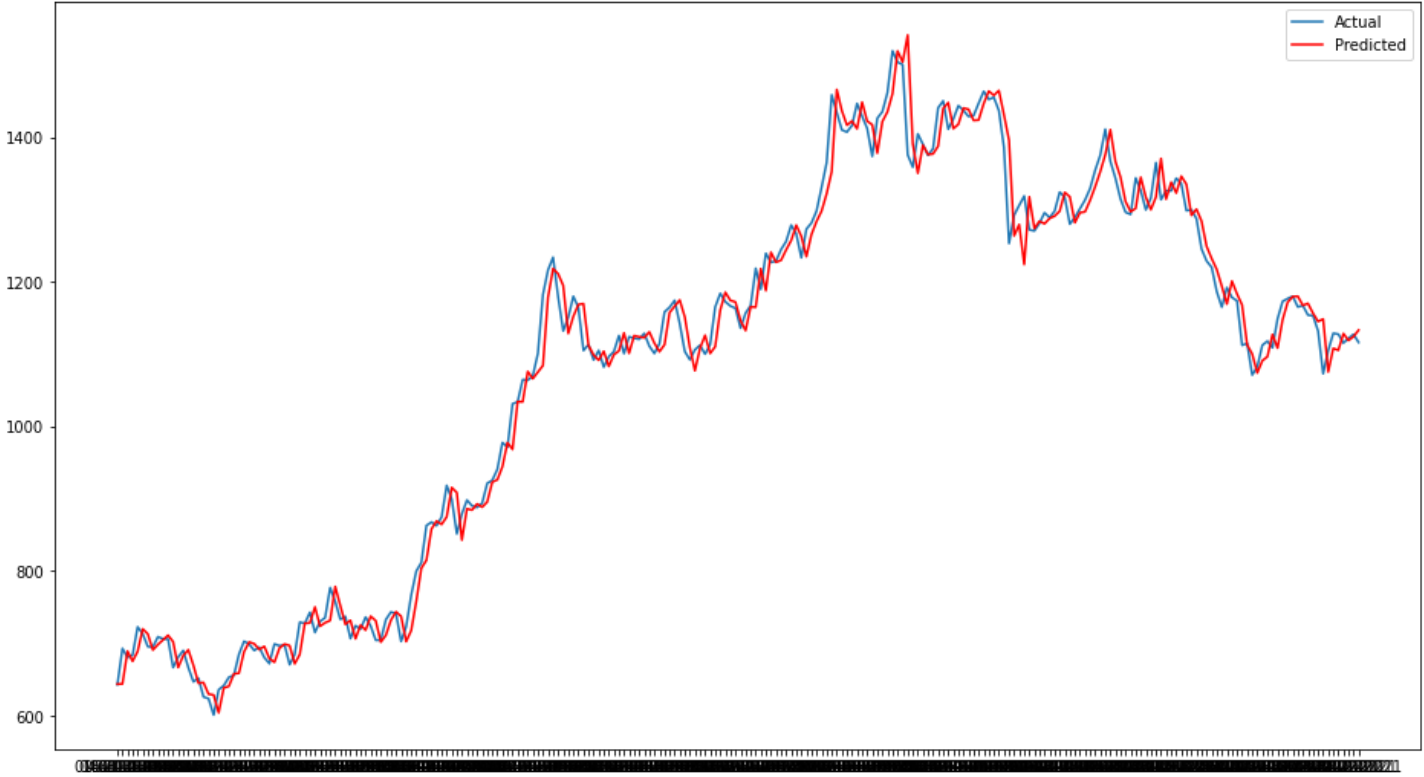

## 2. HINDALCO

## LINEAR REGRESSION

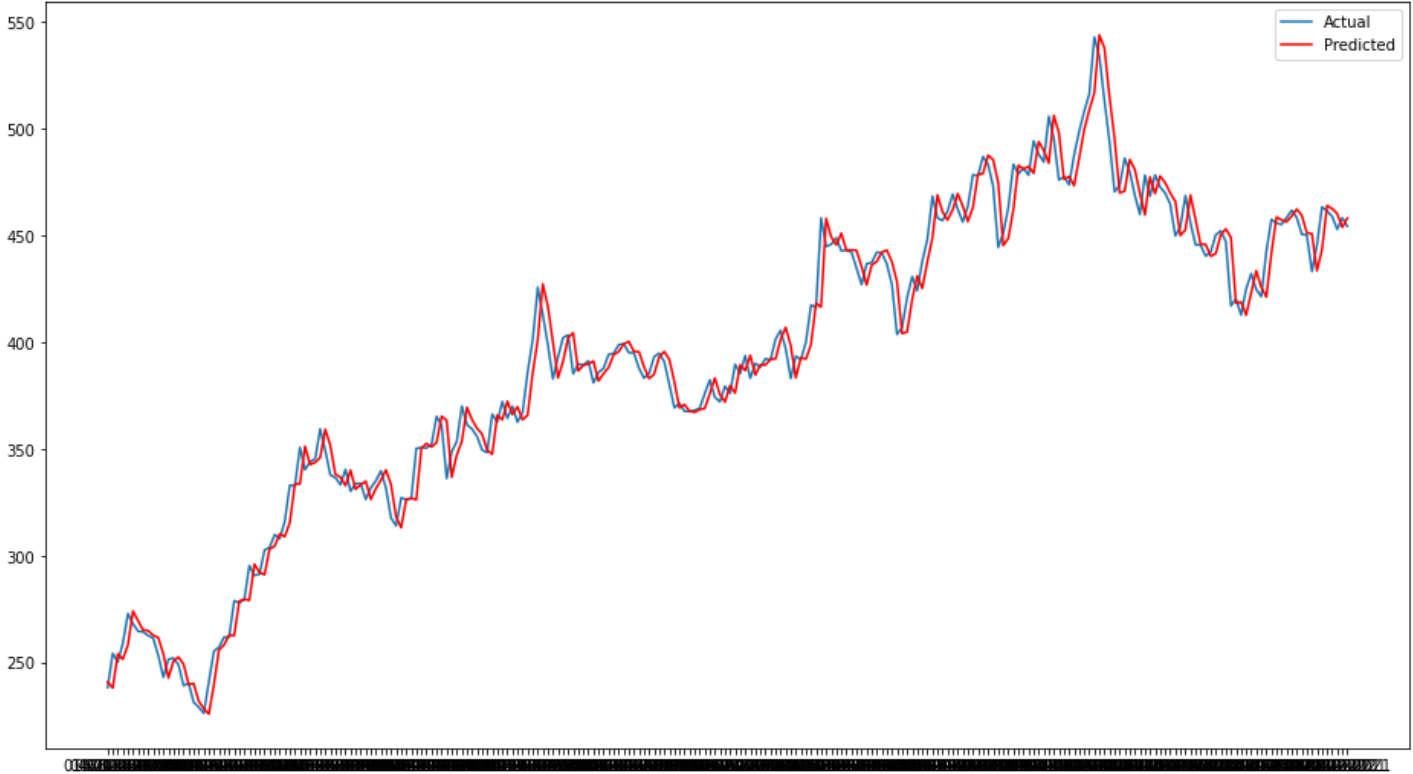

## RANDOM FOREST

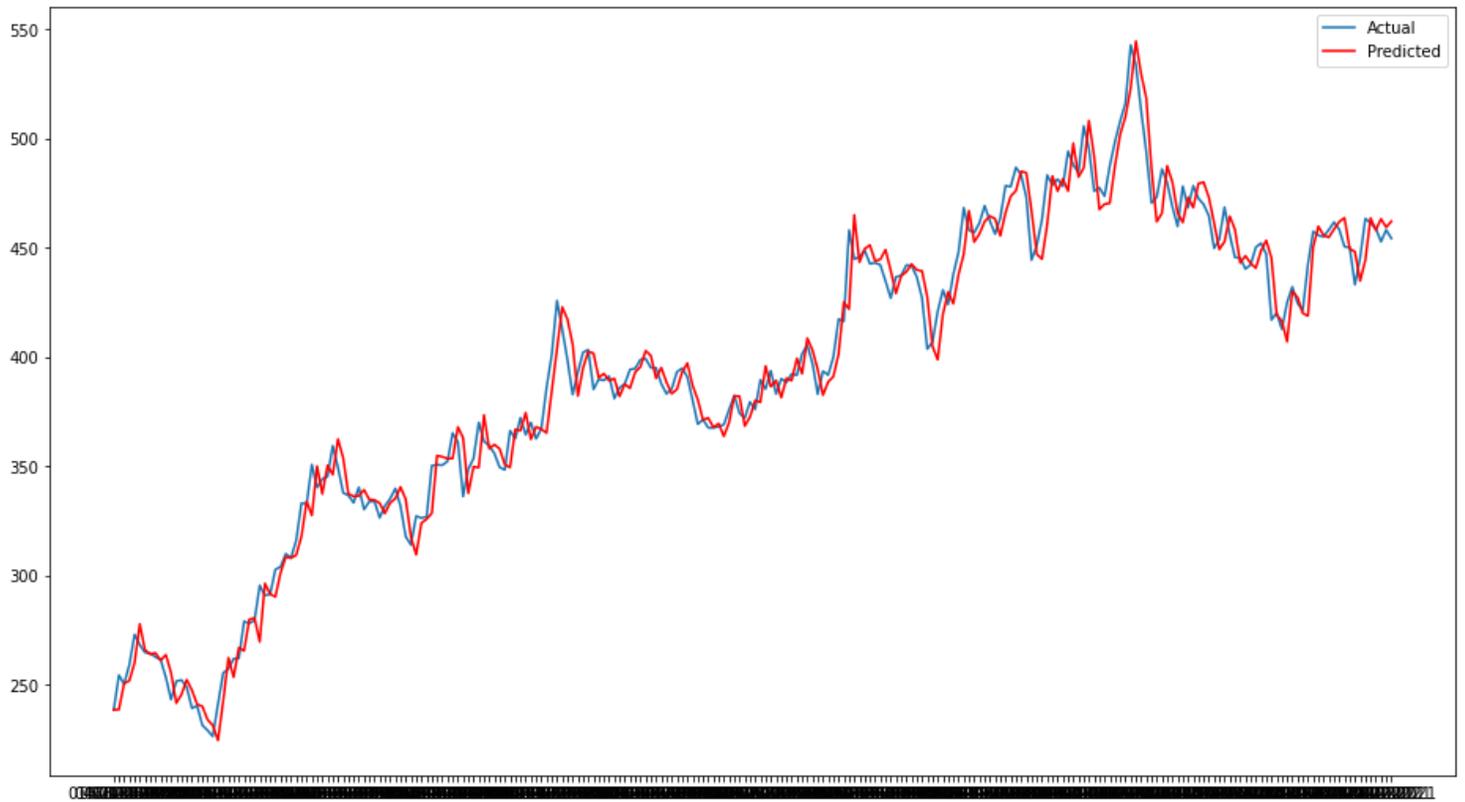

## GRADIENT BOOST

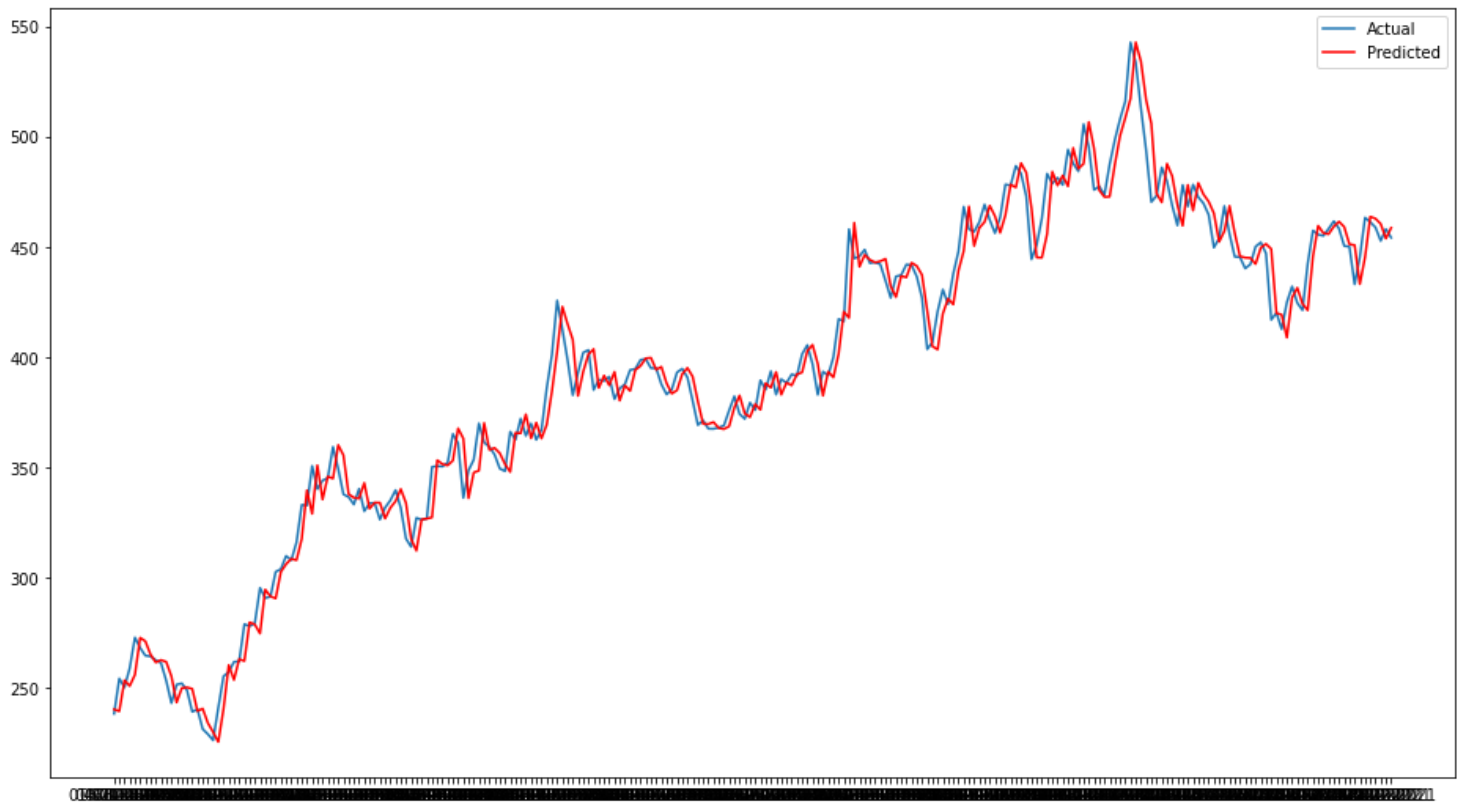

| STOCK NAME | LINEAR REGRESSION | RANDOM FOREST | GRADIENT BOOST |
|---|---|---|---|
| Tata Steel | 29.28 | 30.05 | 30.76 |
| Hindalco | 10.71 | 10.61 | 10.78 |
| JSW Steel | 14.60 | 15.90 | 15.96 |
| VEDL | 8.20 | 8.80 | 8.90 |

# IT SECTOR

1. TCS

LINEAR REGRESSION

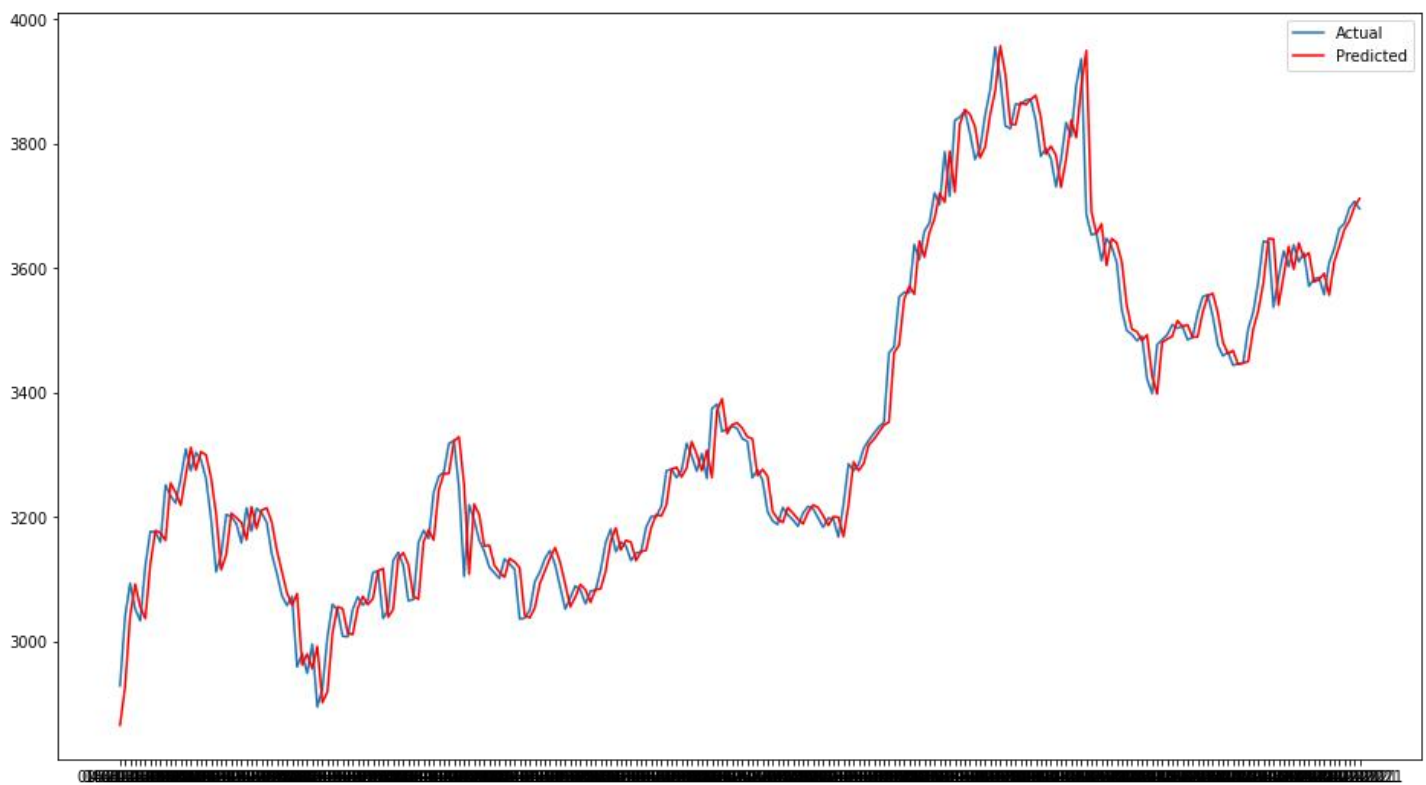

## RANDOM FOREST

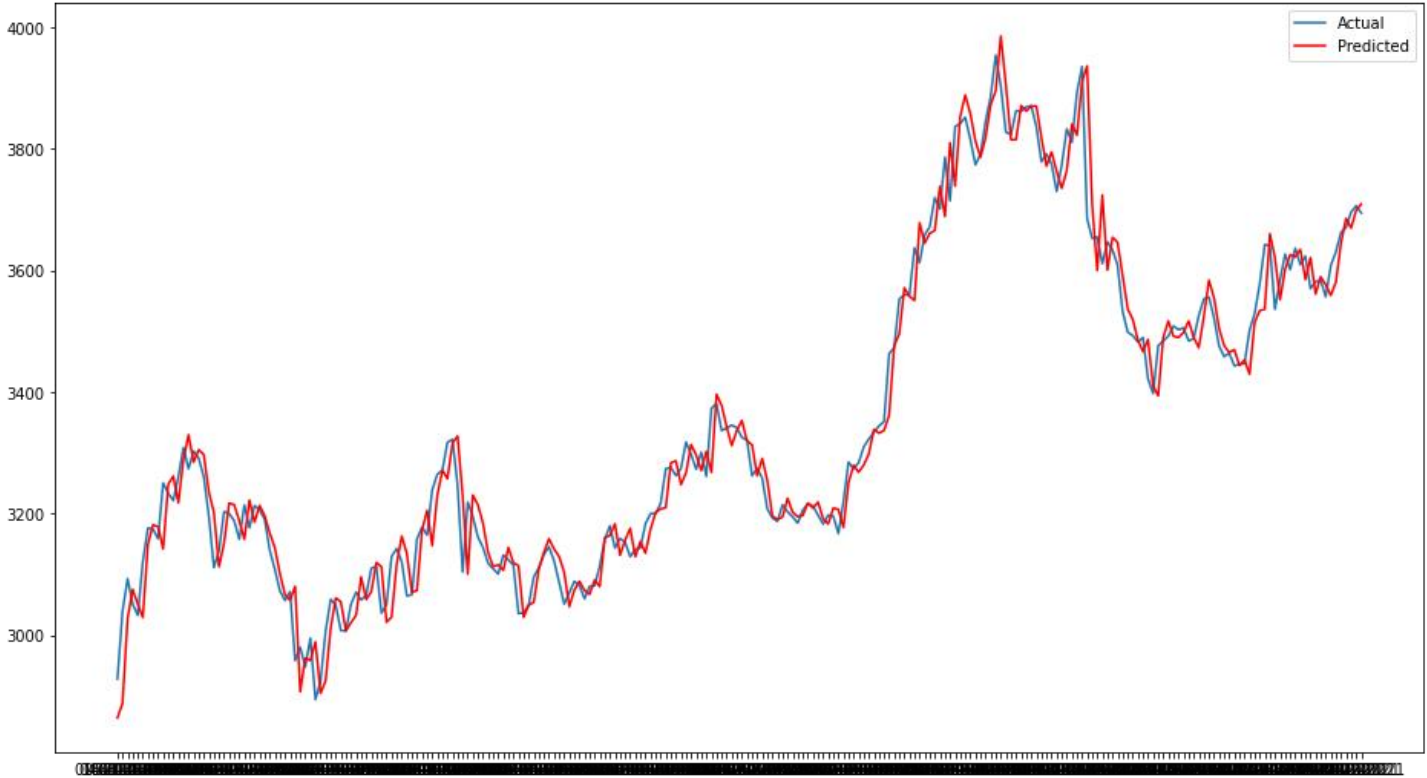

## GRADIENT BOOST

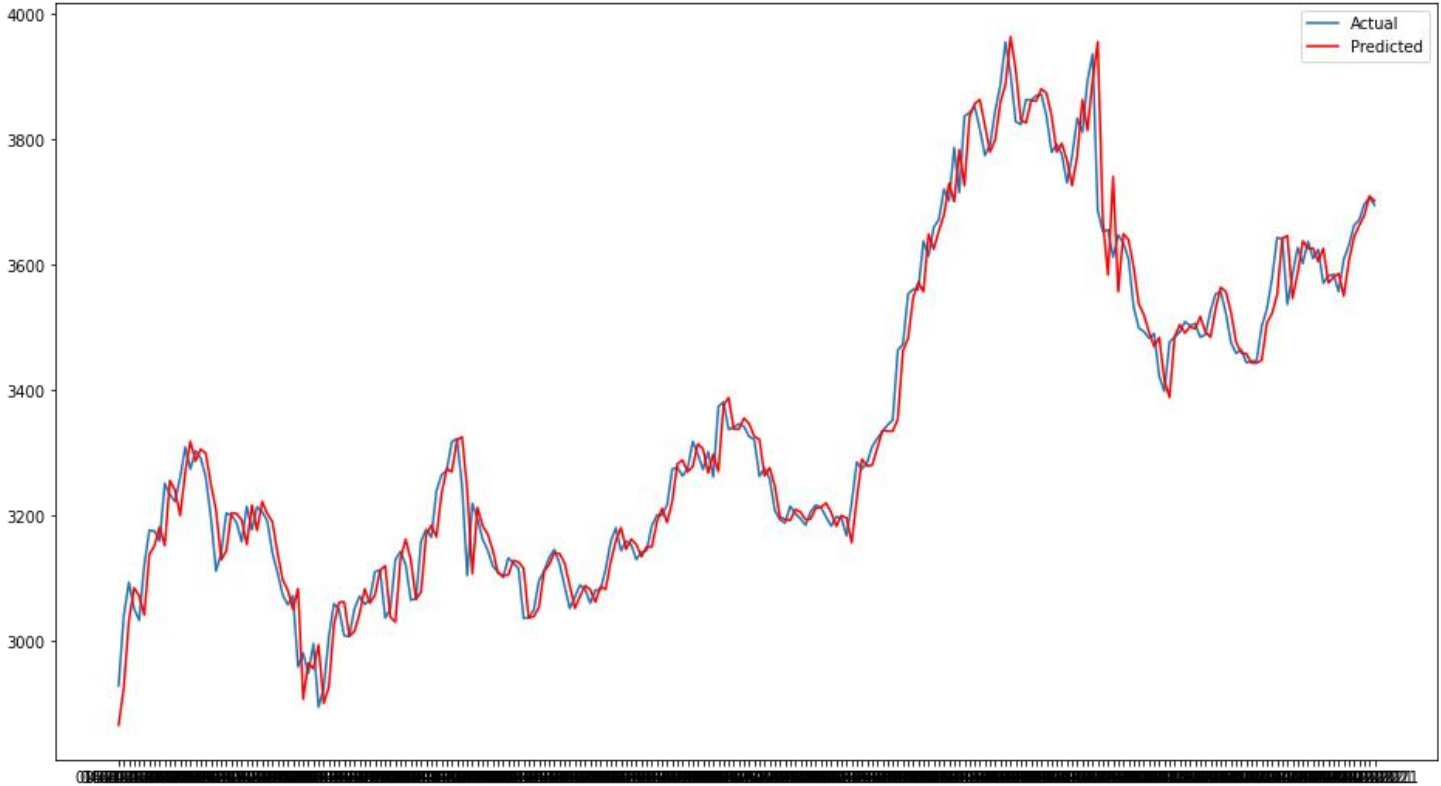

2. WIPRO

LINEAR REGRESSION

# RANDOM FOREST

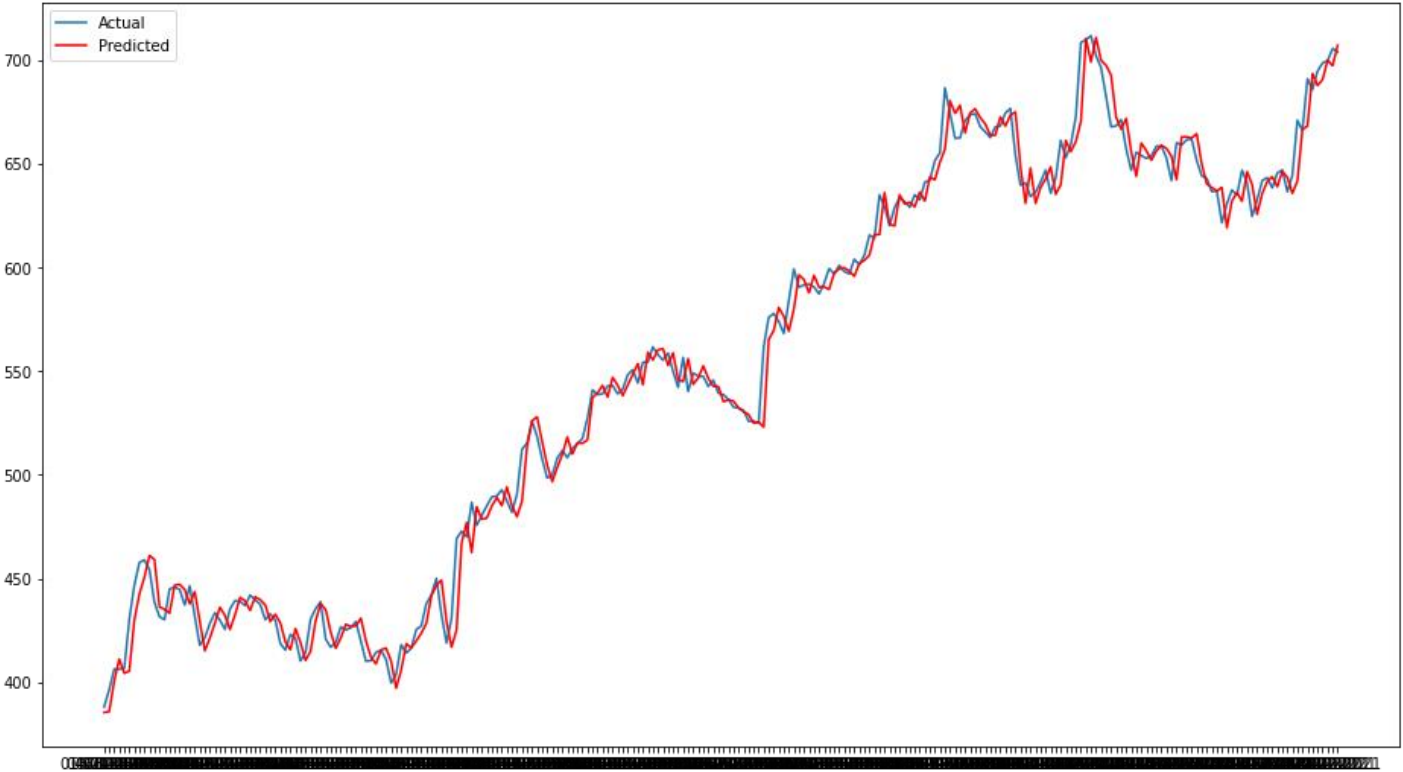

# GRADIENT BOOST

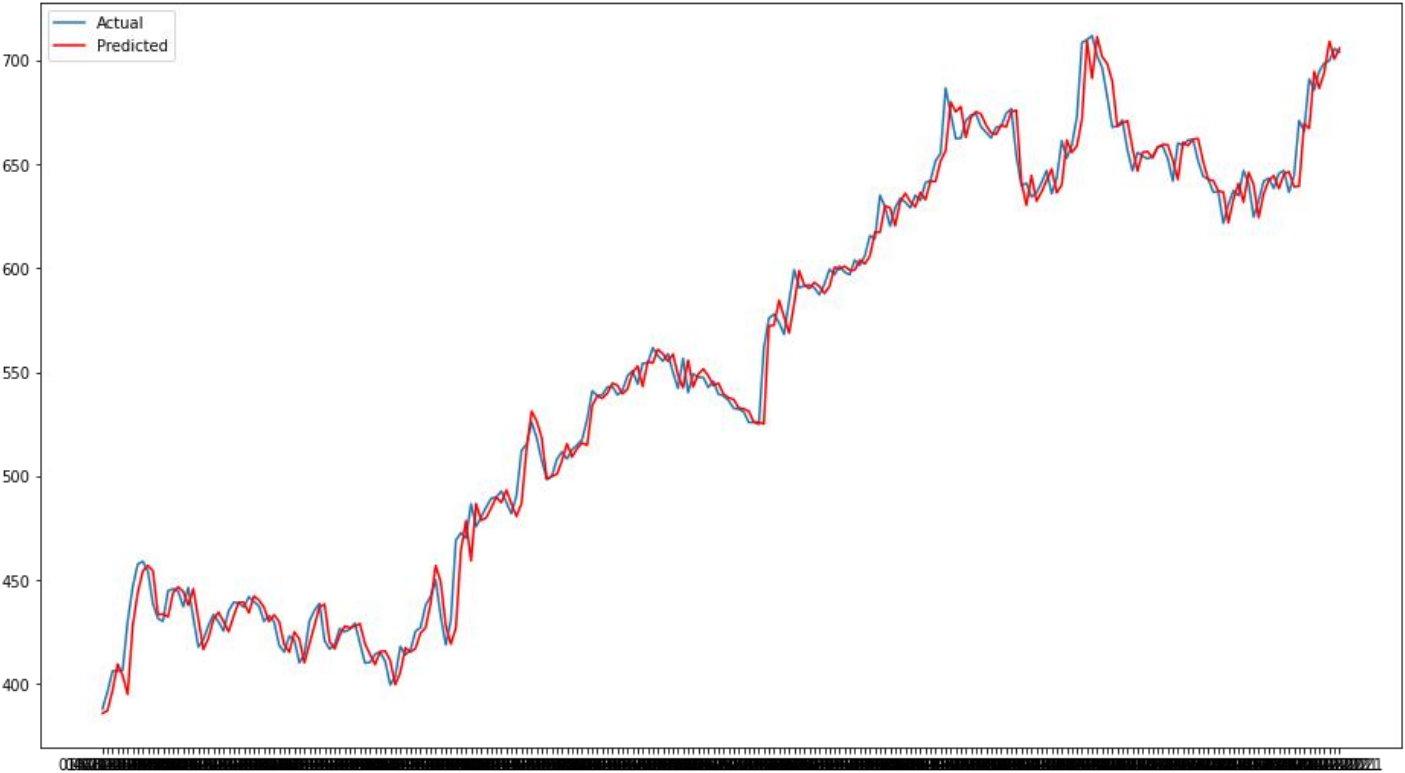

| STOCK NAME | LINEAR REGRESSION | RANDOM FOREST | GRADIENT BOOST |
|---|---|---|---|
| TCS | 45.73 | 46.94 | 46.87 |
| Infy | 20.69 | 23.58 | 22.77 |
| Wipro | 9.26 | 9.78 | 9.85 |
| Hcl tech | 18.55 | 20.01 | 18.97 |

# BANKING SECTOR

1. HDFC BANK

LINEAR REGRESSION

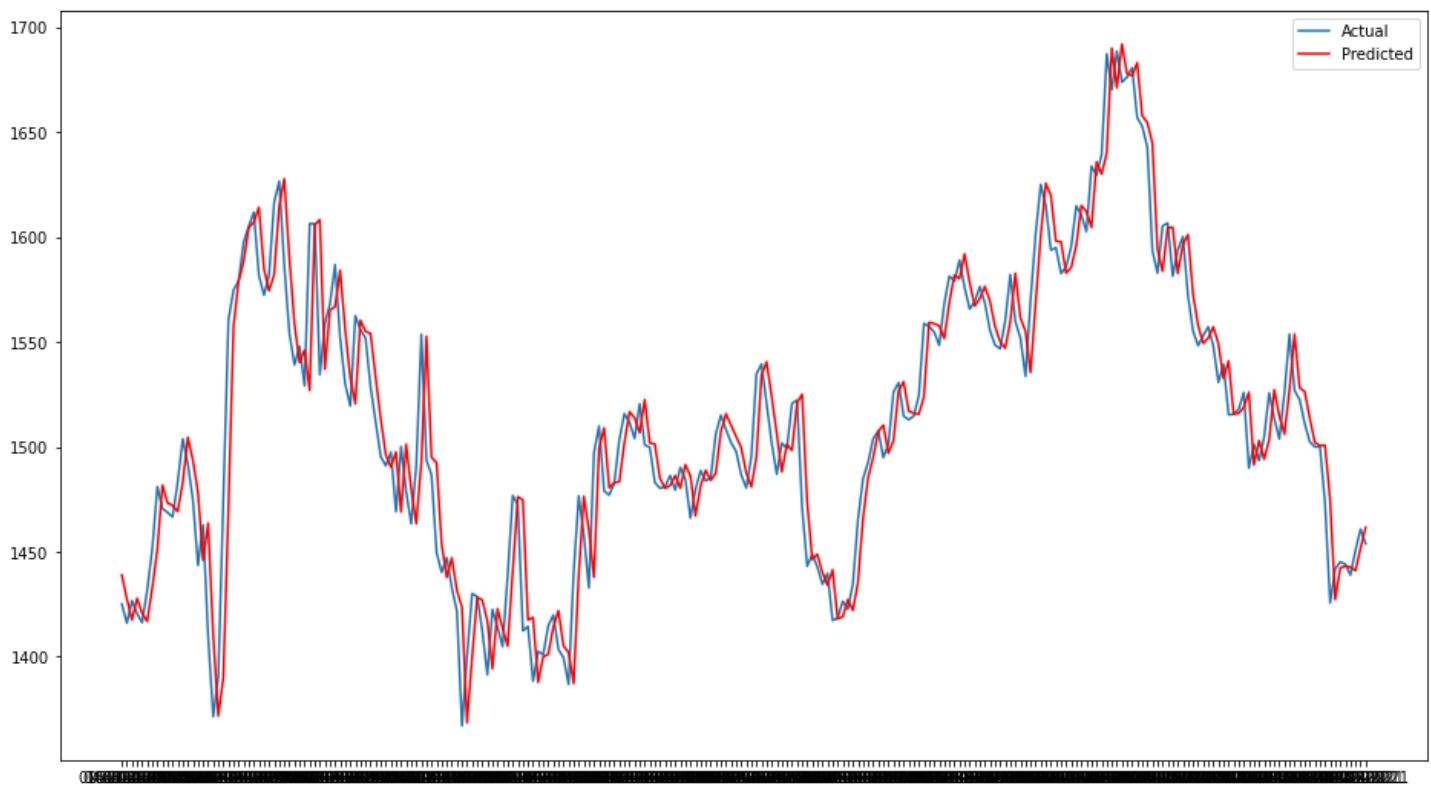

## RANDOM FOREST

## GRADIENT BOOST

## 2. ICICI BANK

## LINEAR REGRESSION

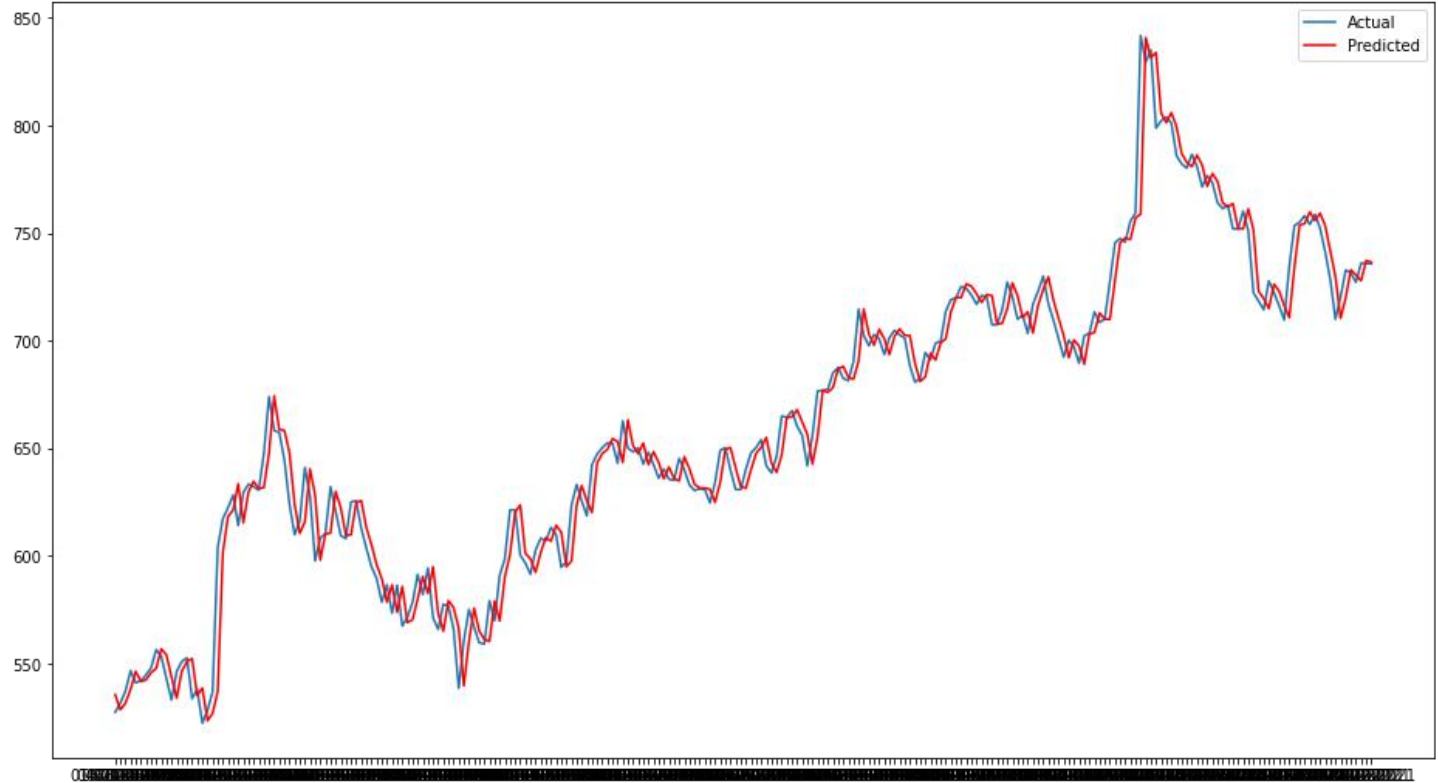

## RANDOM FOREST

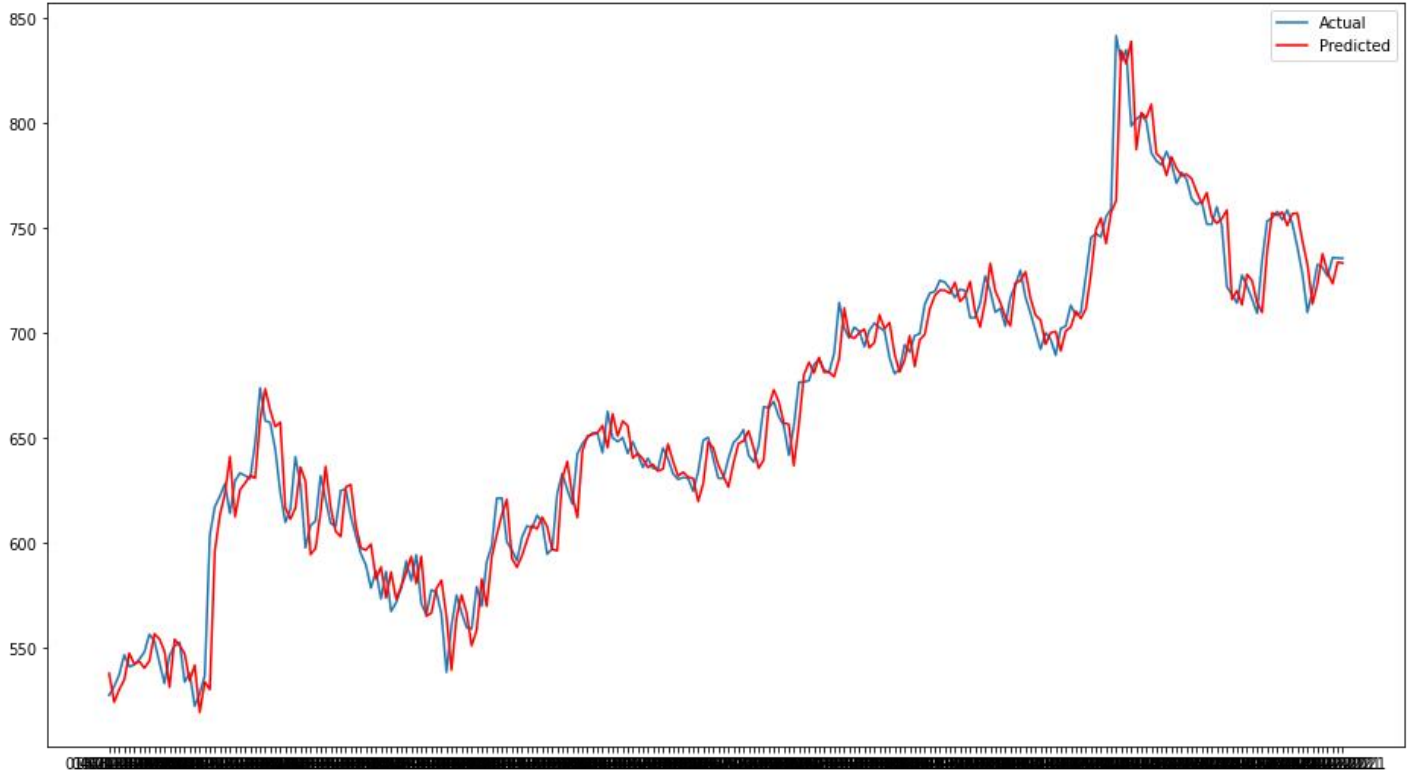

## GRADIENT BOOST

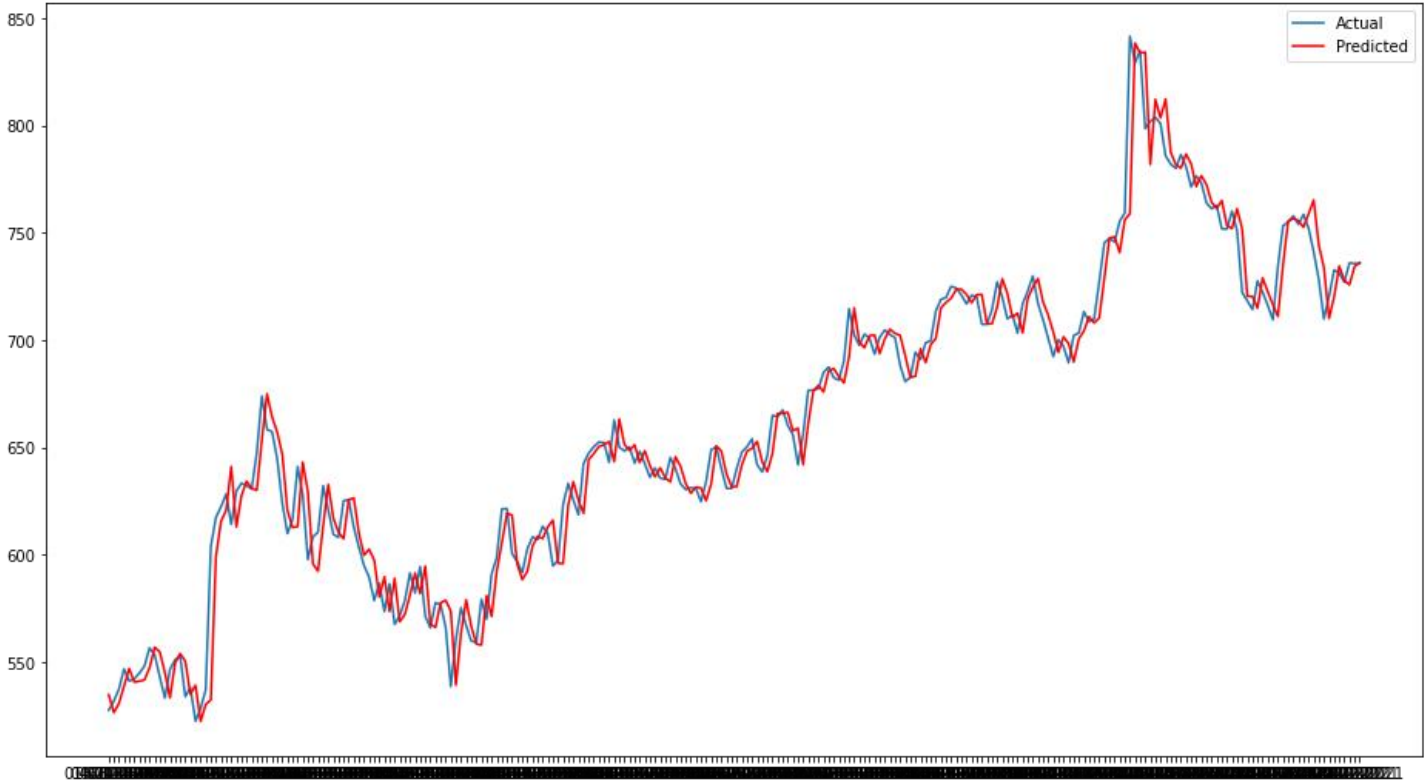

| STOCK NAME | LINEAR REGRESSION | RANDOM FOREST | GRADIENT BOOST |
|---|---|---|---|
| HDFC Bank | 23.09 | 22.85 | 23.17 |
| ICICI Bank | 12.72 | 13.53 | 13.31 |
| Kotak Bank | 33.74 | 35.18 | 34.54 |
| Axis Bank | 13.62 | 14.25 | 13.65 |

# FMCG SECTOR

1. HINDUSTAN UNILEVER

LINEAR REGRESSION

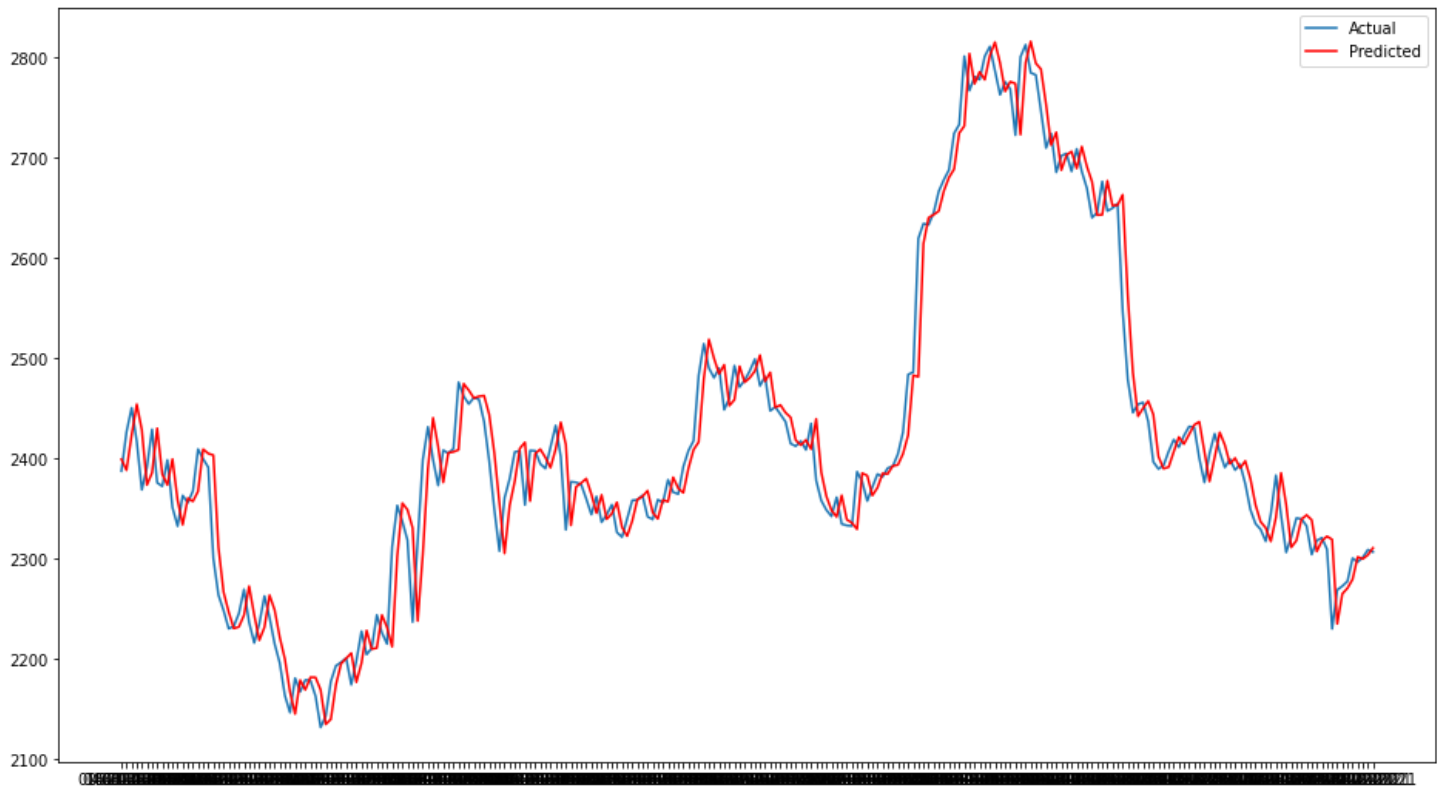

## RANDOM FOREST

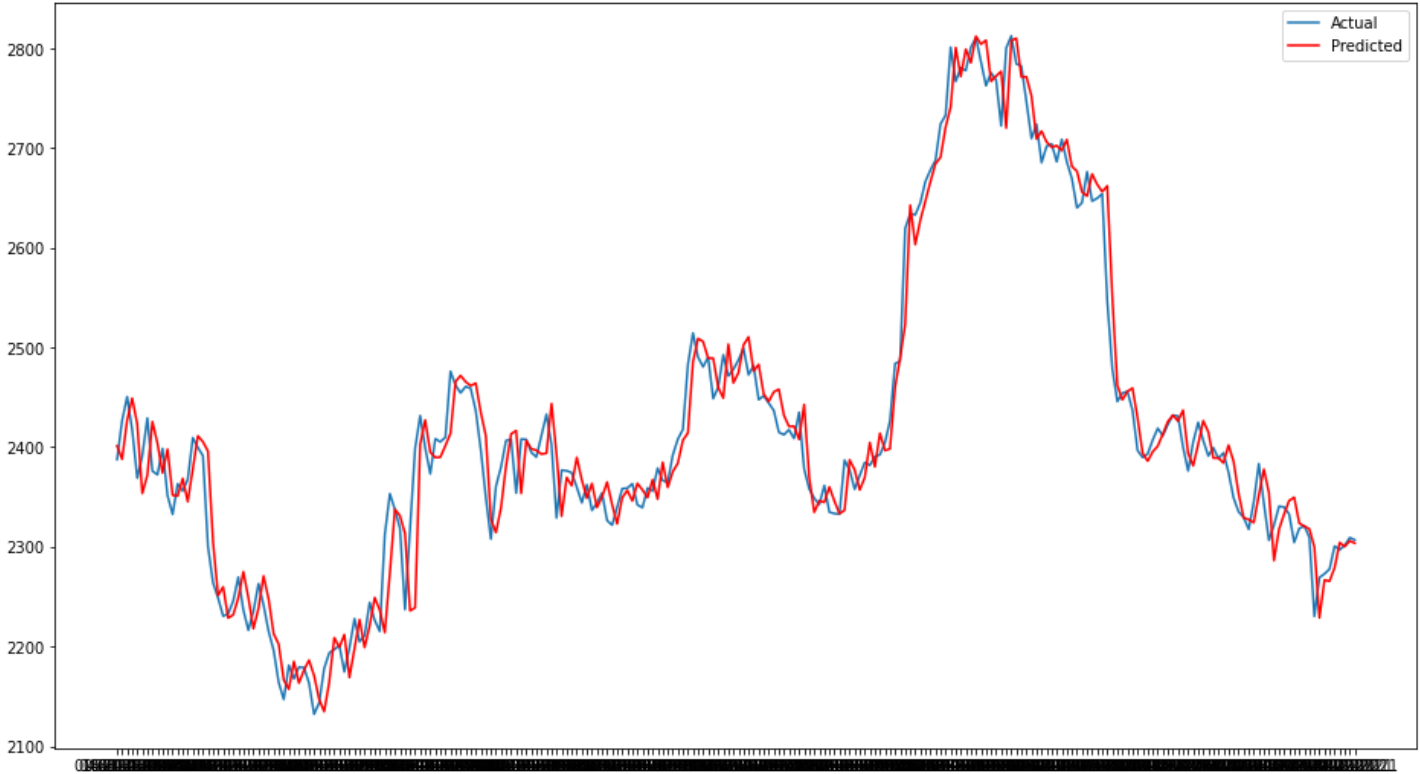

## GRADIENT BOOST

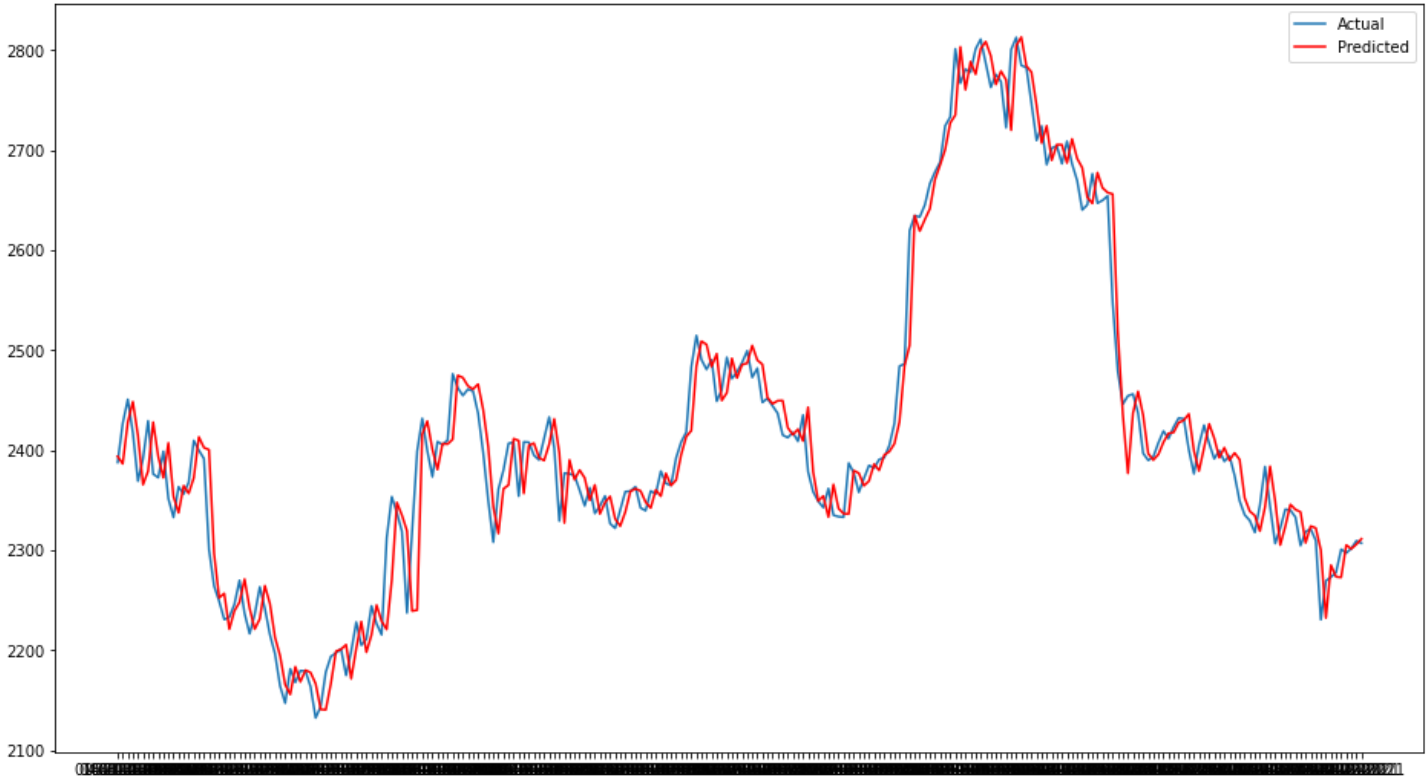

2. ITC
LINEAR REGRESSION

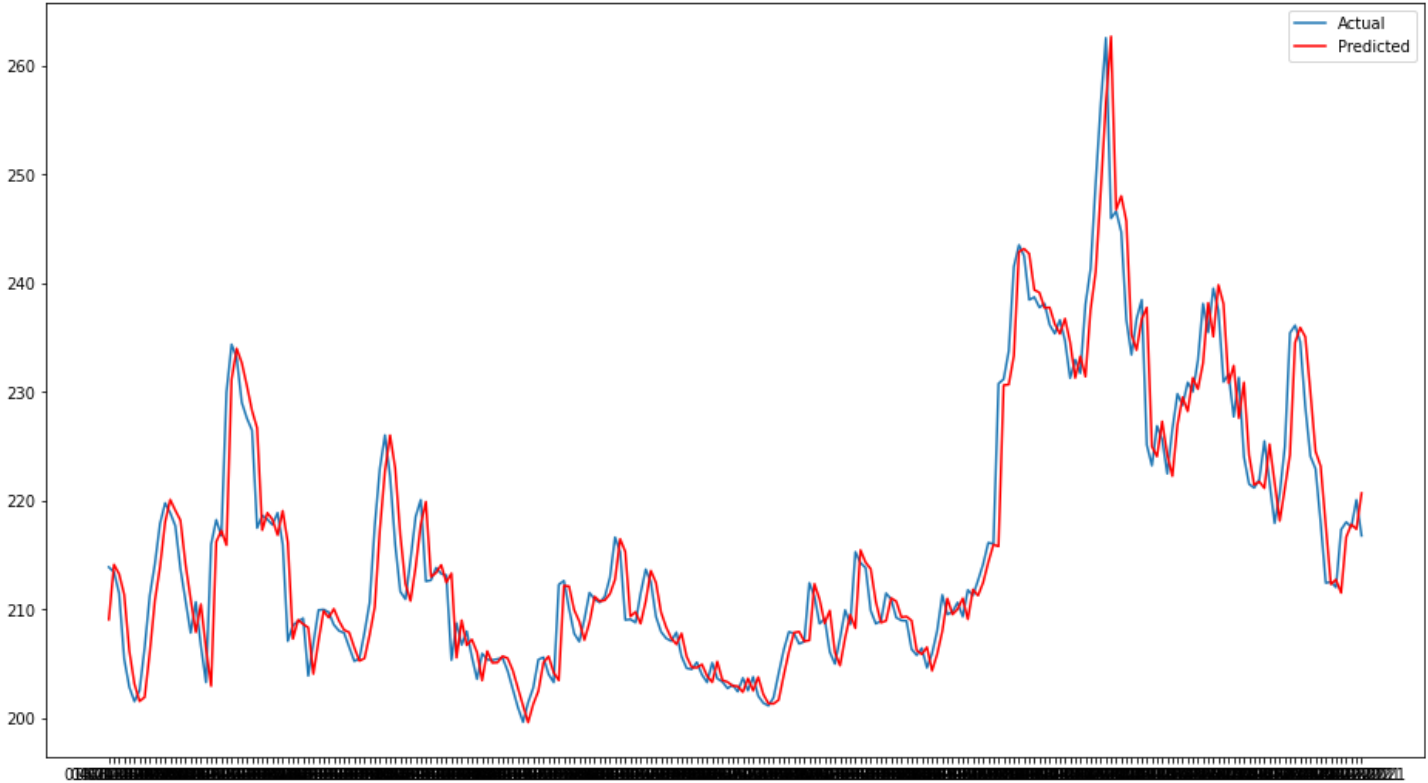

RANDOM FOREST

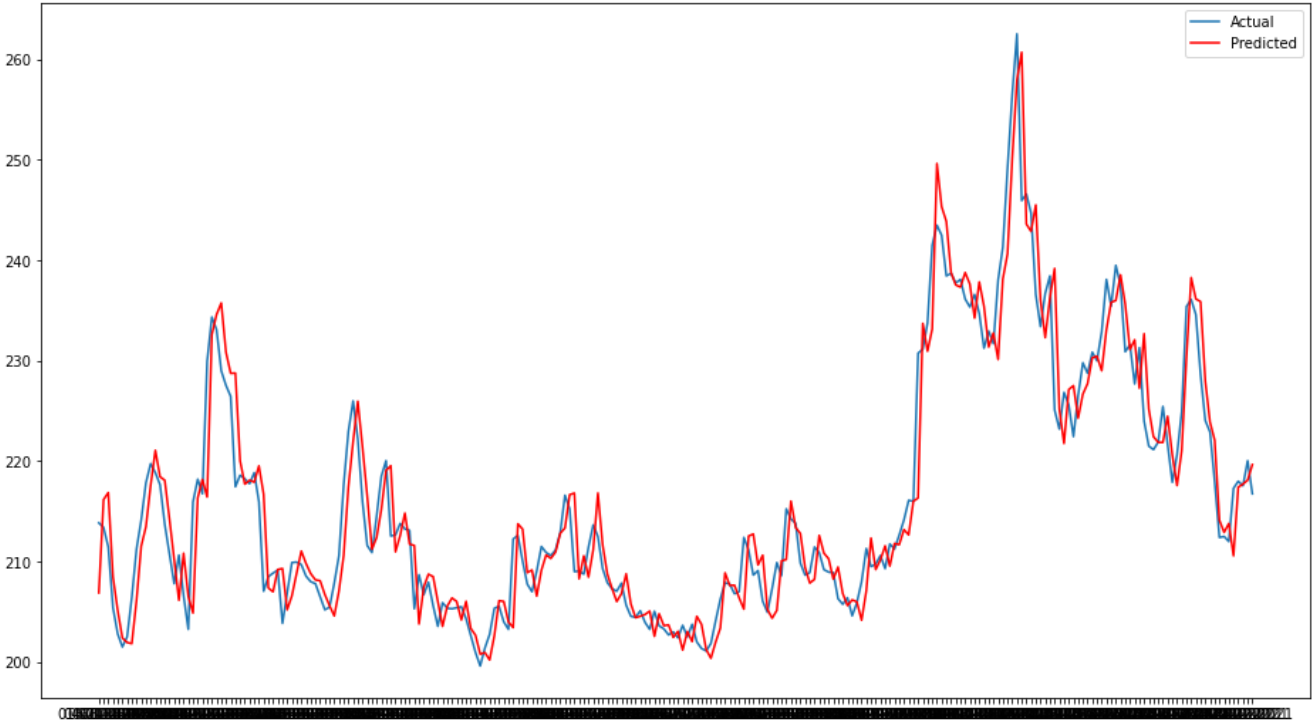

# GRADIENT BOOST

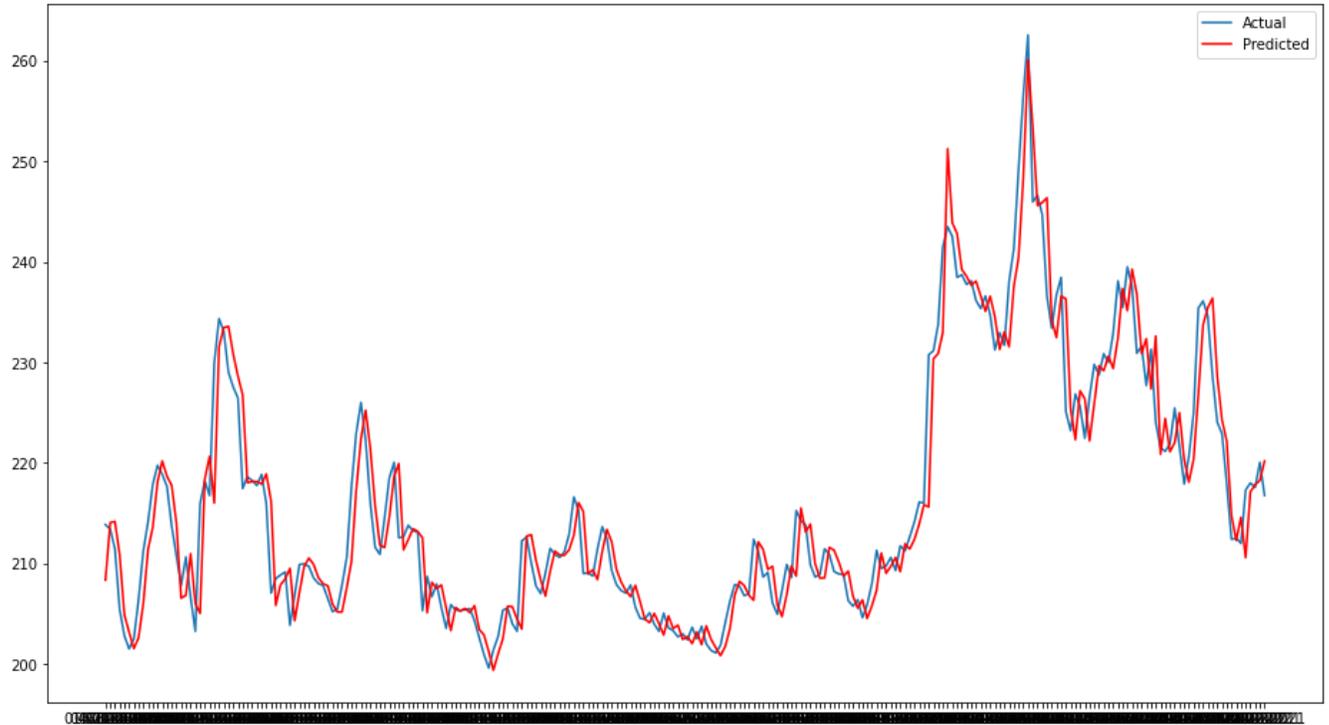

| STOCK NAME | LINEAR REGRESSION | RANDOM FOREST | GRADIENT BOOST |
|---|---|---|---|
| HUL | 10.51 | 11.31 | 10.86 |
| ITC | 3.76 | 3.80 | 3.60 |
| Nestle | 211.62 | 215.30 | 211.01 |
| Tata Consumer | 10.50 | 11.31 | 10.86 |

# AUTO SECTOR

1. TATA MOTORS

LINEAR REGRESSION

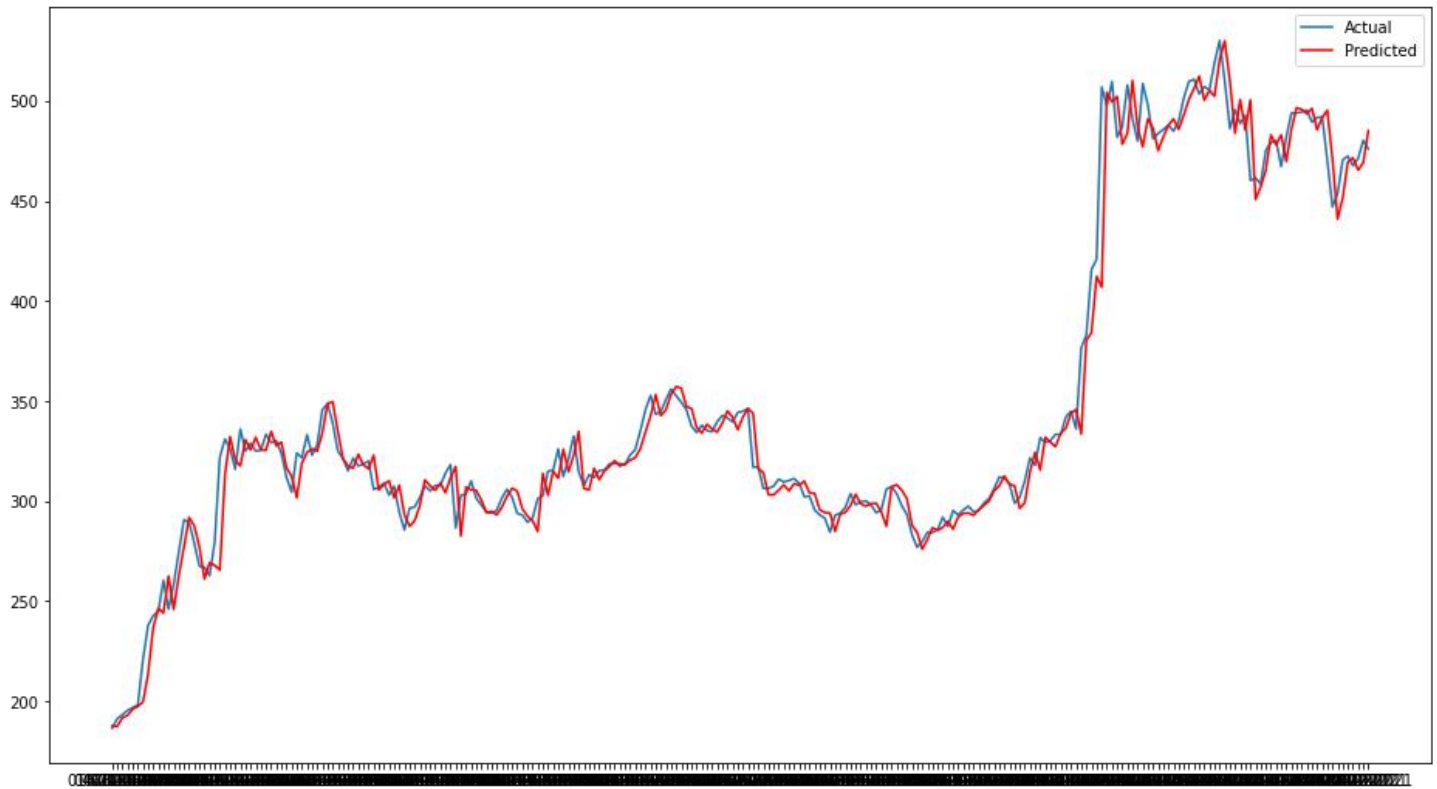

## RANDOM FOREST

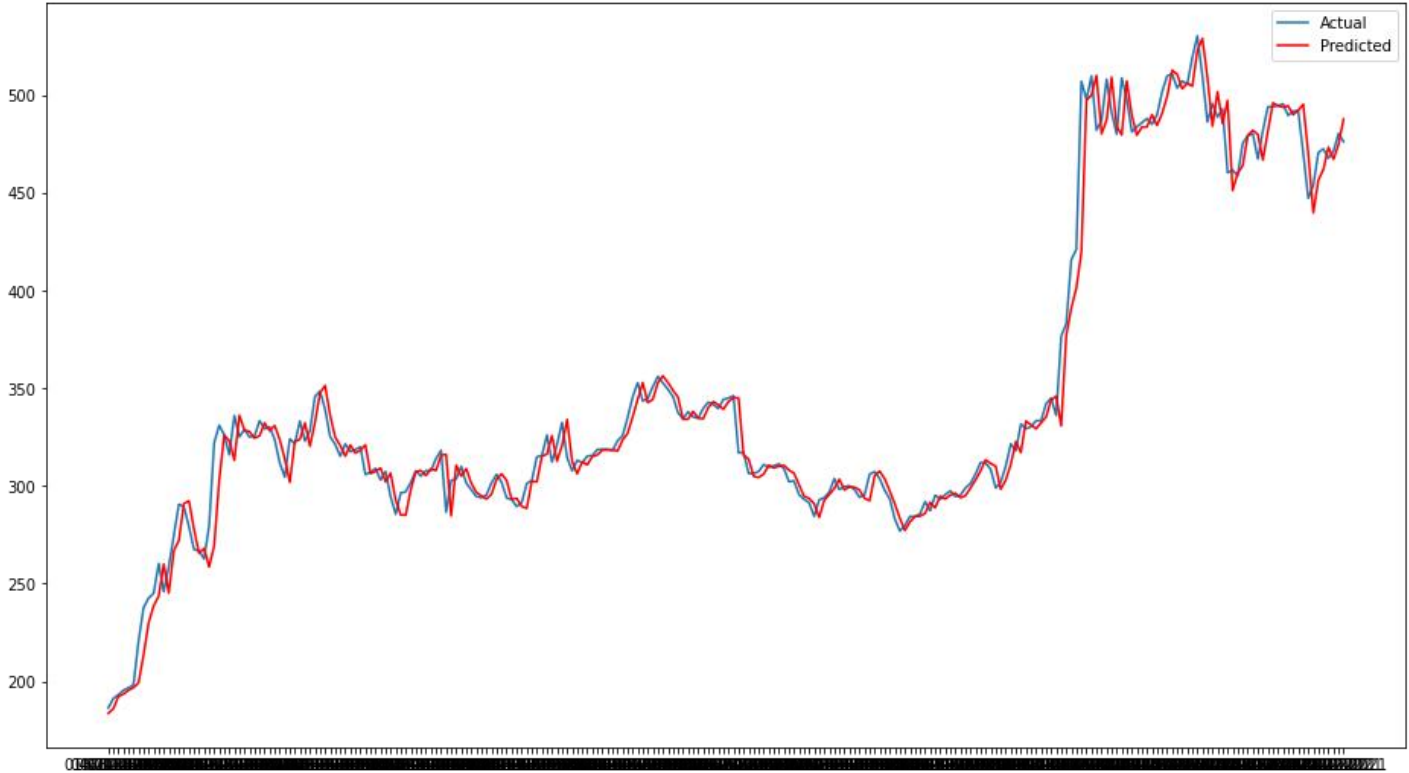

## GRADIENT BOOST

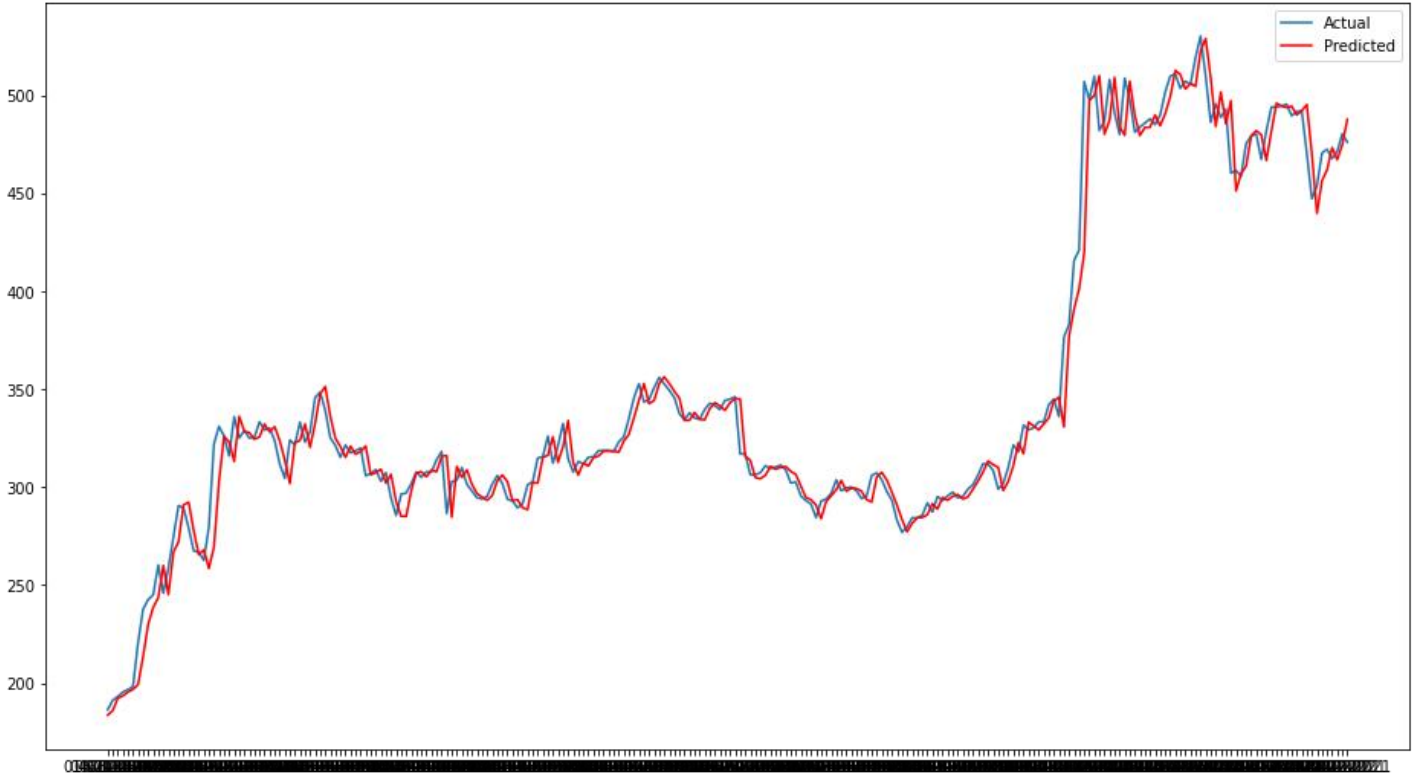

## 2. MARUTI

### LINEAR REGRESSION

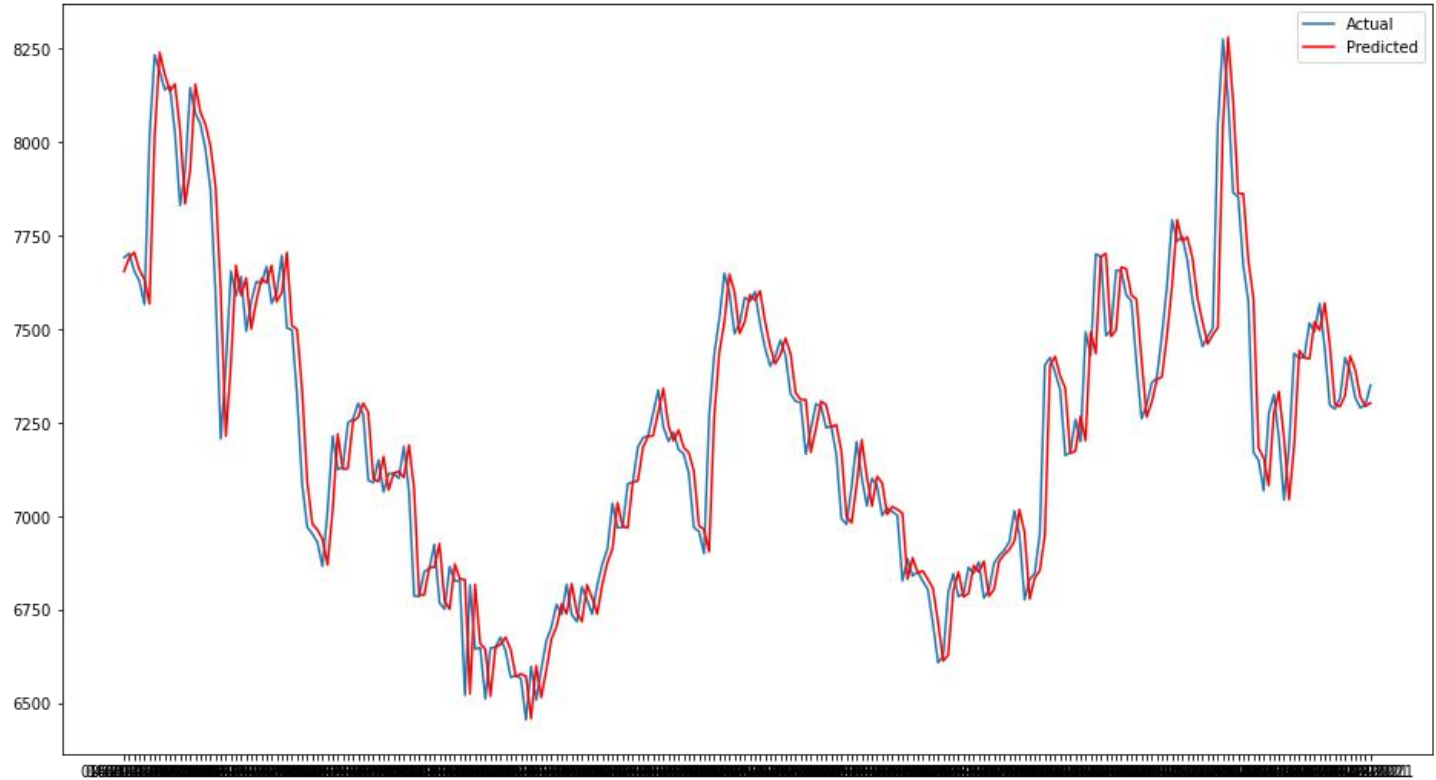

### RANDOM FOREST

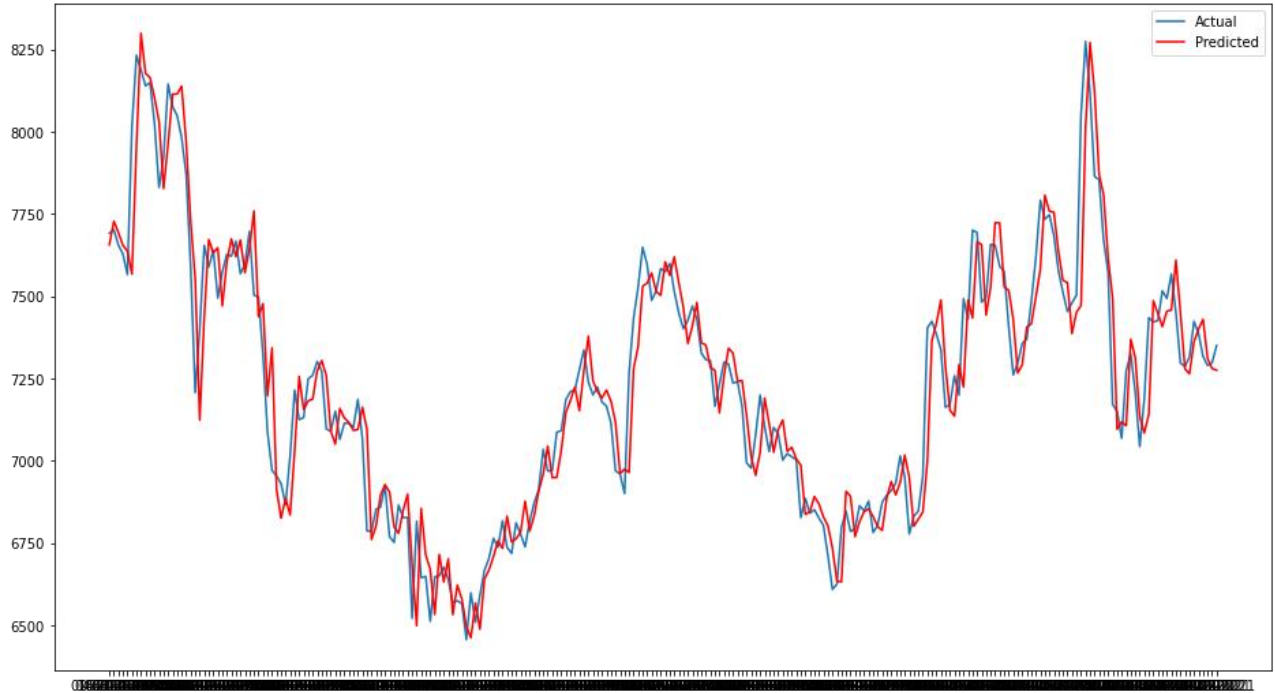

# GRADIENT BOOST

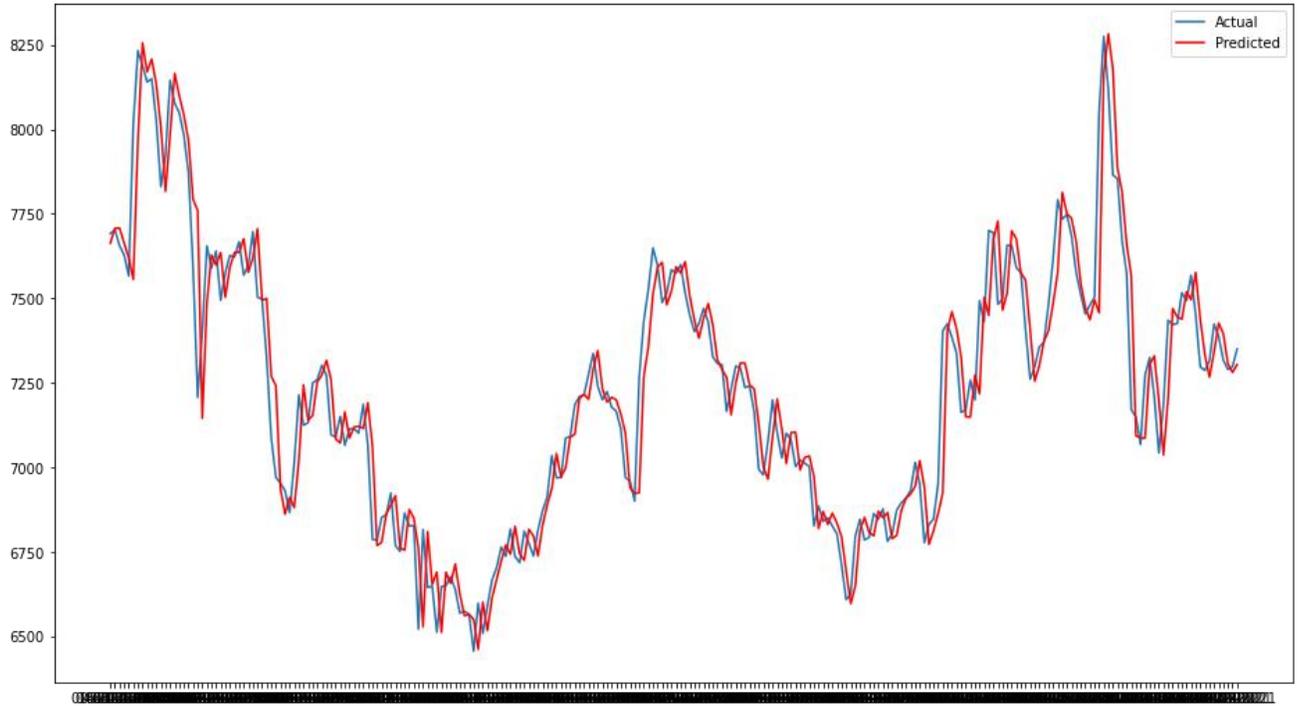

| STOCK NAME | LINEAR REGRESSION | RANDOM FOREST | GRADIENT BOOST |
|---|---|---|---|
| Tata Motors | 11.11 | 12.36 | 11.94 |
| Maruti | 123.58 | 123.08 | 123.95 |
| Bajaj Auto | 59.37 | 59.84 | 61.28 |
| M&M | 15.80 | 17.36 | 16.56 |

# PHARMA SECTOR

1. SUN PHARMA

LINEAR REGRESSION

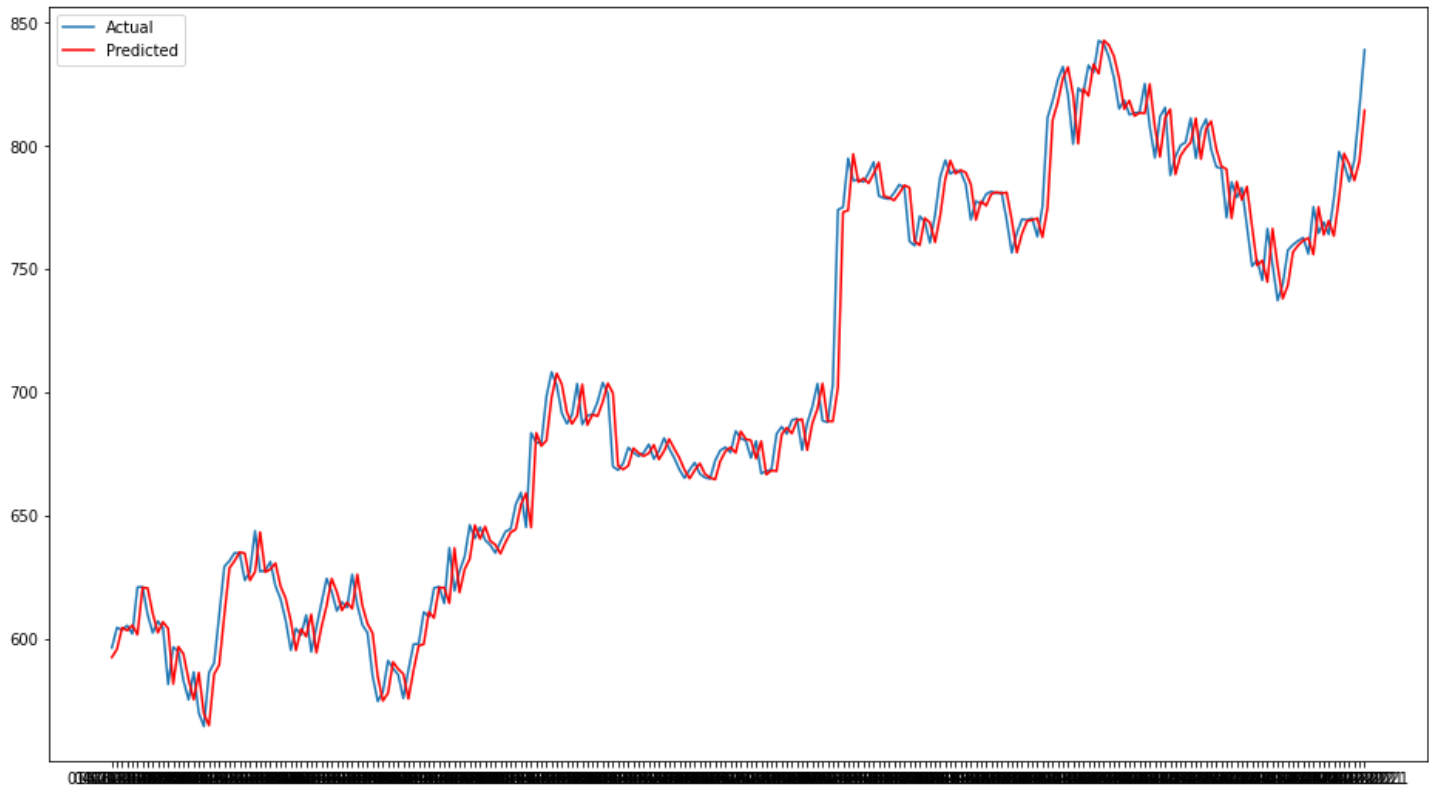

## RANDOM FOREST

## GRADIENT BOOST

## 2. CIPLA

### LINEAR REGRESSION

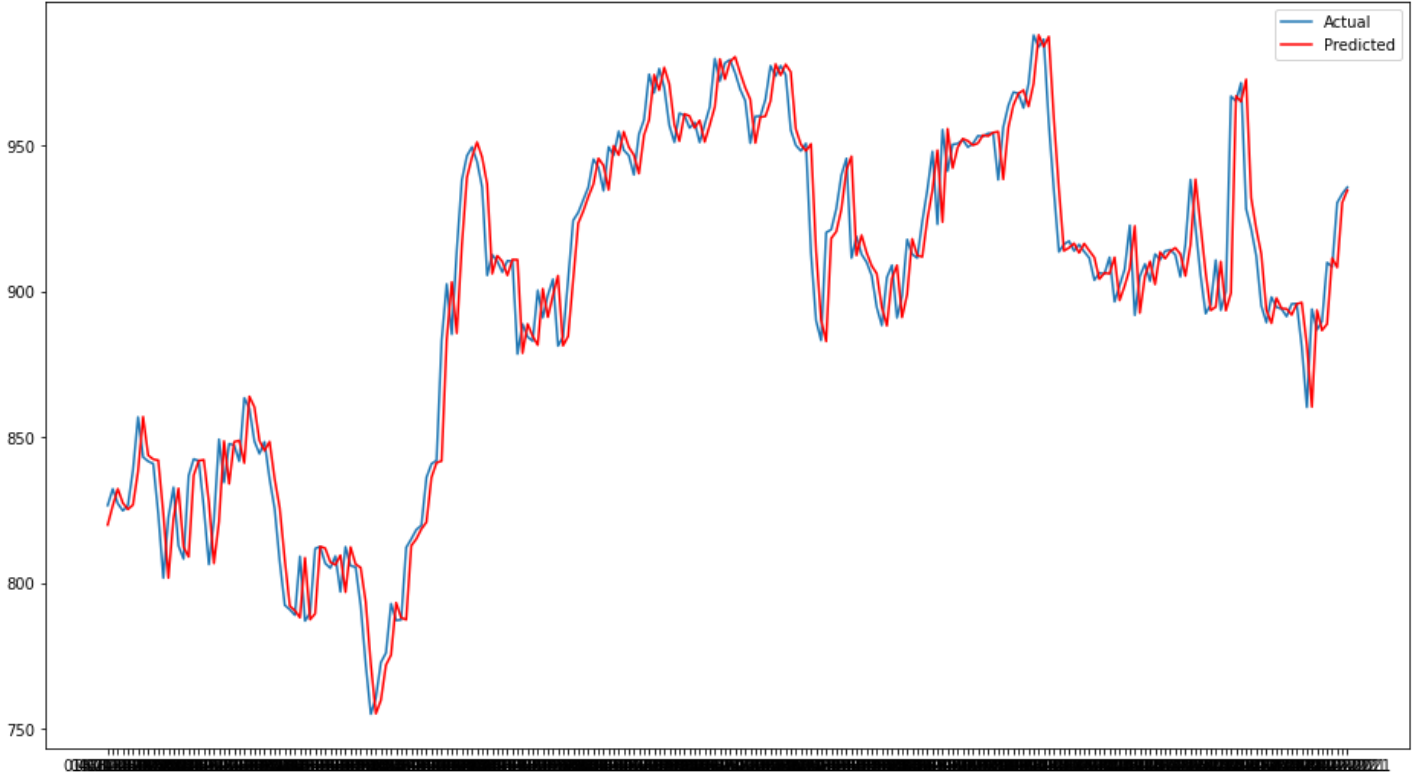

### RANDOM FOREST

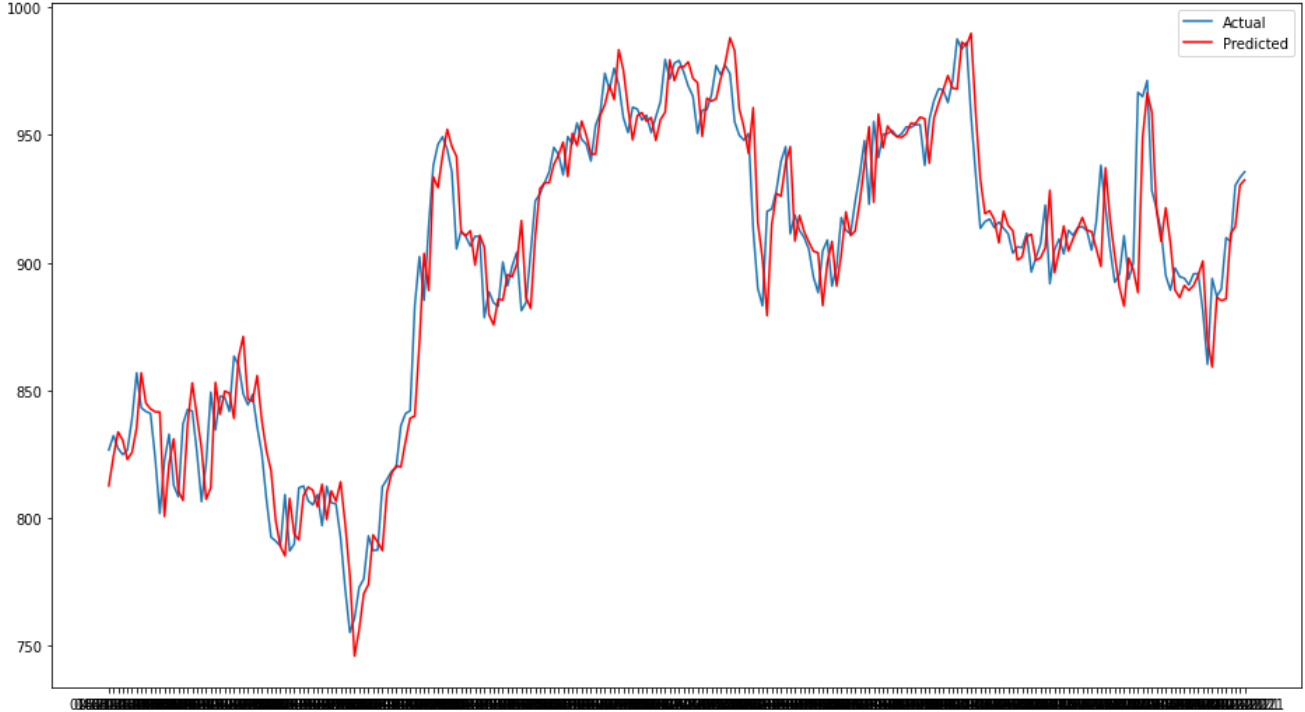

GRADIENT BOOST

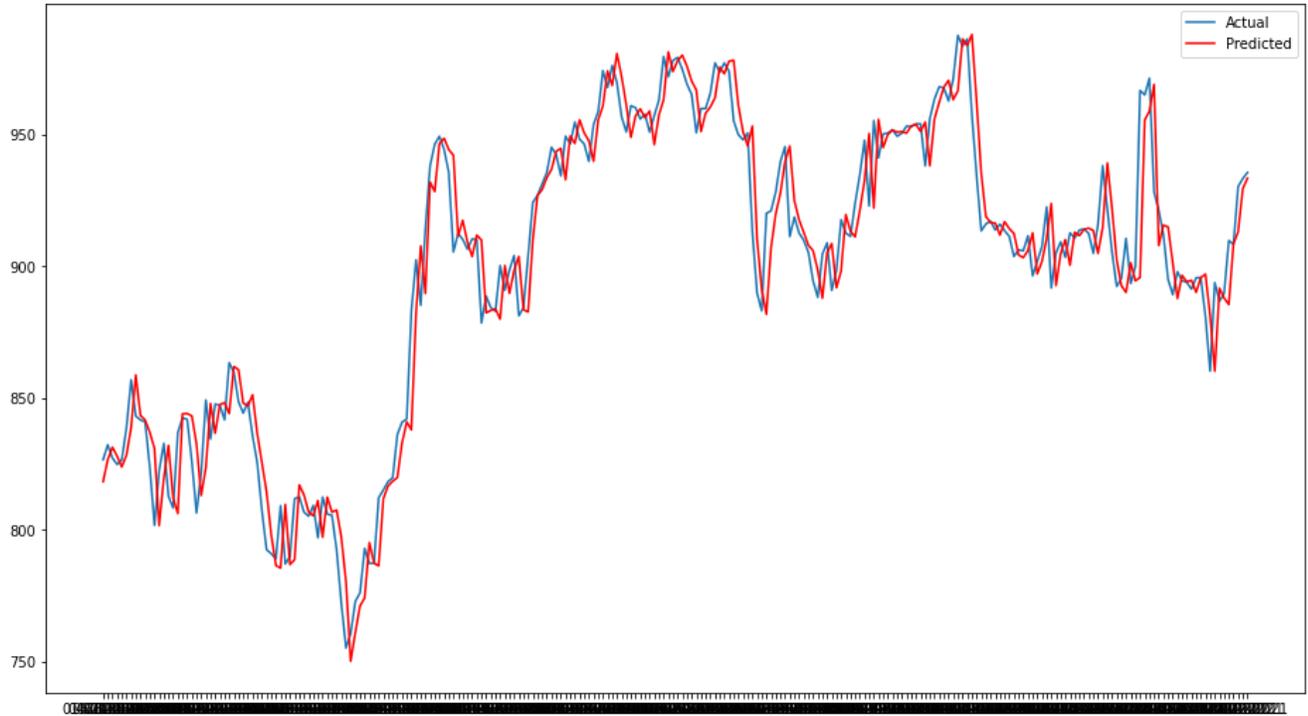

| STOCK NAME | LINEAR REGRESSION | RANDOM FOREST | GRADIENT BOOST |
|---|---|---|---|
| Sun Pharma | 11.44 | 12.31 | 12.54 |
| Cipla | 13.96 | 14.67 | 14.58 |
| Divislab | 68.49 | 73.12 | 71.71 |
| Drreddy | 75.78 | 78.15 | 77.26 |

# CLASSIFICATION REPORTS

## AUC VALUES OF DIFFERENT STOCKS

## METAL SECTOR:

| STOCK NAME | RANDOM FOREST | XG BOOST | GAUSSIAN NB | LOGISTIC REGRESSION | KNN |
|---|---|---|---|---|---|
| Tata Steel | 0.7822 | 0.7837 | 0.7097 | 0.6468 | 0.5068 |
| Hindalco | 0.7555 | 0.7714 | 0.6915 | 0.6177 | 0.5044 |
| JSW Steel | 0.7889 | 0.8033 | 0.7540 | 0.5987 | 0.5109 |
| Vedanta | 0.8126 | 0.8245 | 0.7700 | 0.7145 | 0.5644 |

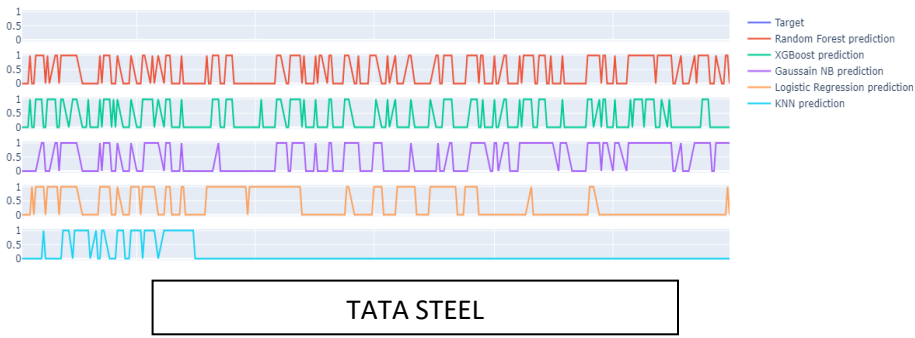

TATA STEEL

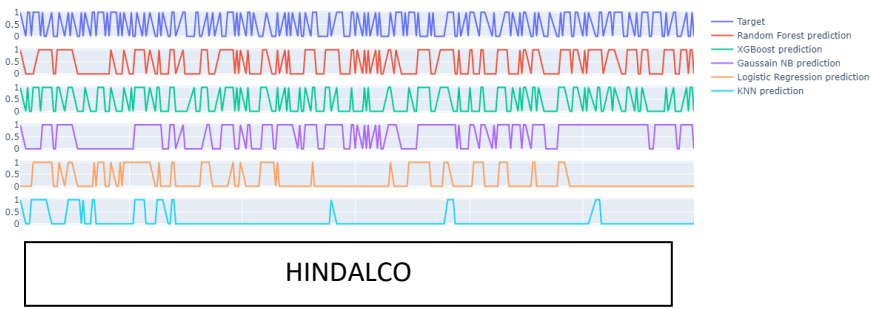

HINDALCO

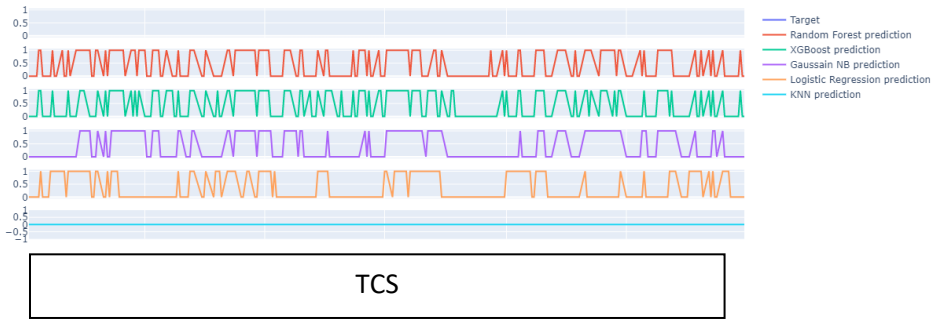

TCS

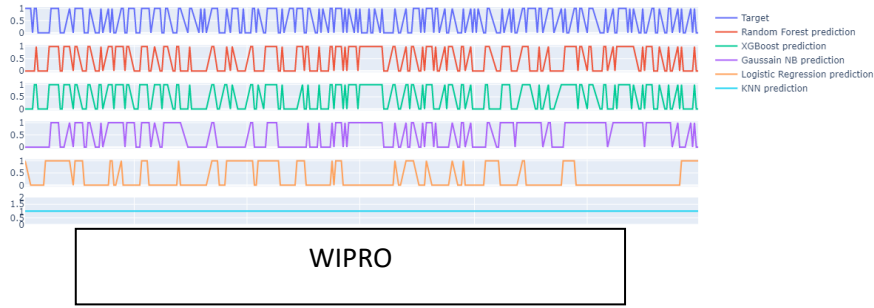

WIPRO

| STOCK NAME | RANDOM FOREST | XG BOOST | GAUSSIAN NB | LOGISTIC REGRESSION | KNN |
|---|---|---|---|---|---|
| TCS | 0.8015 | 0.7822 | 0.7329 | 0.6040 | 0.5000 |
| Infy | 0.7811 | 0.7350 | 0.7525 | 0.6225 | 0.5000 |
| Wipro | 0.8128 | 0.7923 | 0.7603 | 0.5931 | 0.5000 |
| Hcltech | 0.7851 | 0.7758 | 0.7328 | 0.6356 | 0.5000 |

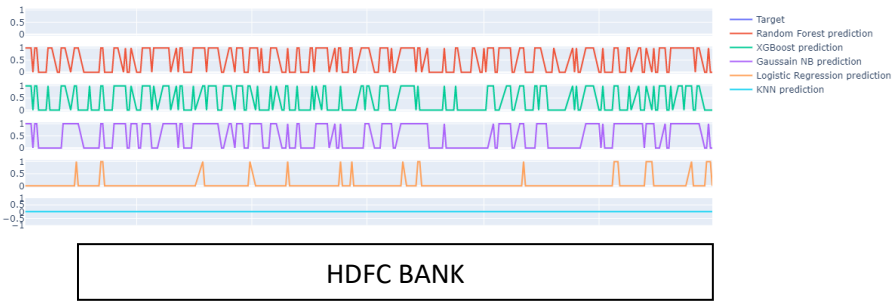

HDFC BANK

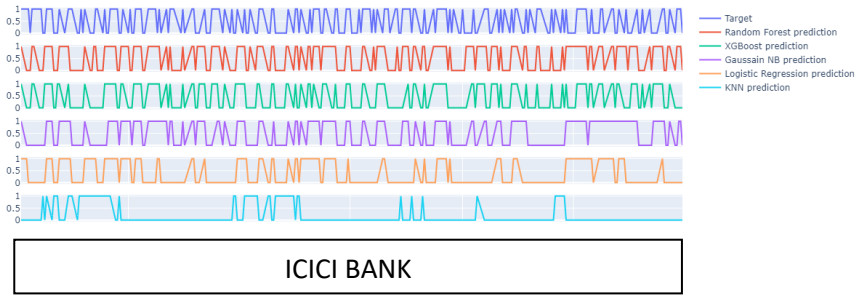

ICICI BANK

| STOCK NAME | RANDOM FOREST | XG BOOST | GAUSSIAN NB | LOGISTIC REGRESSION | KNN |
|---|---|---|---|---|---|
| HDFC Bank | 0.8333 | 0.8136 | 0.7912 | 0.6822 | 0.5687 |
| ICICI Bank | 0.8036 | 0.7951 | 0.7692 | 0.6944 | 0.4326 |
| Kotak Bank | 0.8217 | 0.8076 | 0.7674 | 0.6828 | 0.5325 |
| Axis Bank | 0.8167 | 0.8231 | 0.7892 | 0.6950 | 0.5141 |

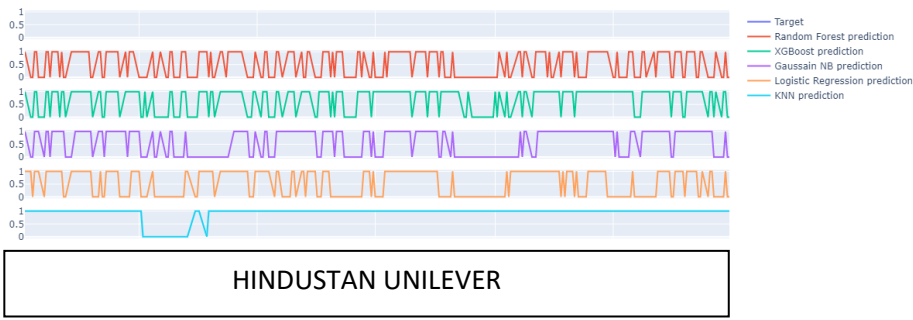

HINDUSTAN UNILEVER

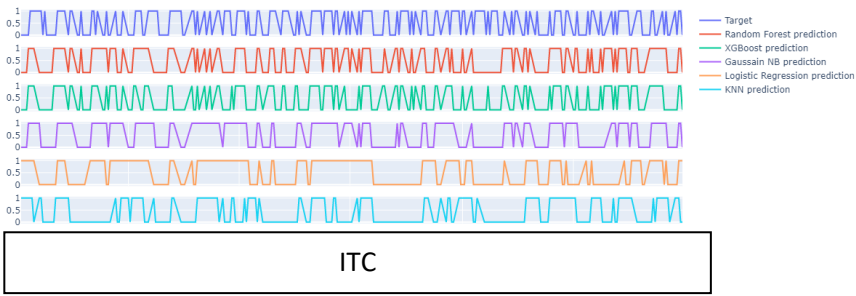

ITC

| STOCK NAME | RANDOM FOREST | XG BOOST | GAUSSIAN NB | LOGISTIC REGRESSION | KNN |
|---|---|---|---|---|---|
| HUL | 0.8090 | 0.7835 | 0.7389 | 0.6900 | 0.5168 |
| ITC | 0.8188 | 0.8019 | 0.7850 | 0.6791 | 0.5745 |
| Nestle | 0.8196 | 0.8022 | 0.6240 | 0.6039 | 0.4671 |
| Tata Consumer | 0.8157 | 0.8258 | 0.7365 | 0.6739 | 0.5000 |

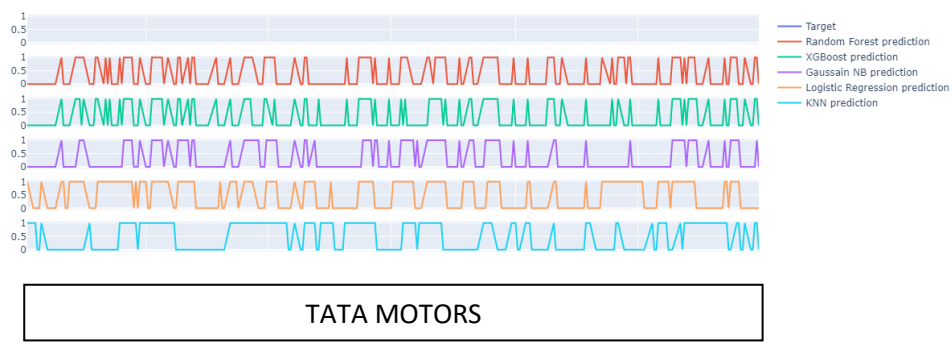

TATA MOTORS

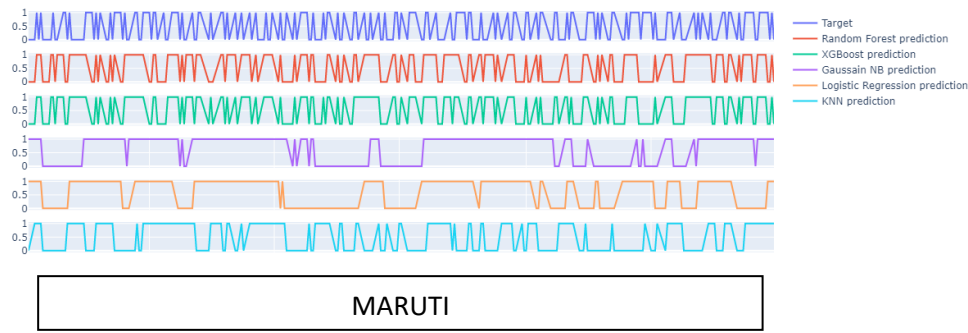

MARUTI

| STOCK NAME | RANDOM FOREST | XG BOOST | GAUSSIAN NB | LOGISTIC REGRESSION | KNN |
|---|---|---|---|---|---|
| Tata Motors | 0.7860 | 0.7761 | 0.7474 | 0.6973 | 0.5423 |
| Maruti | 0.7821 | 0.7787 | 0.6548 | 0.5843 | 0.4930 |
| Bajaj Auto | 0.7899 | 0.8104 | 0.7364 | 0.6017 | 0.4985 |
| M&M | 0.8225 | 0.8136 | 0.7863 | 0.6977 | 0.5437 |

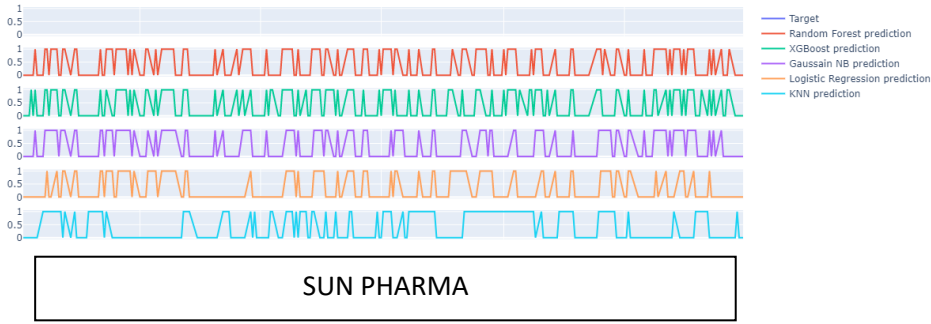

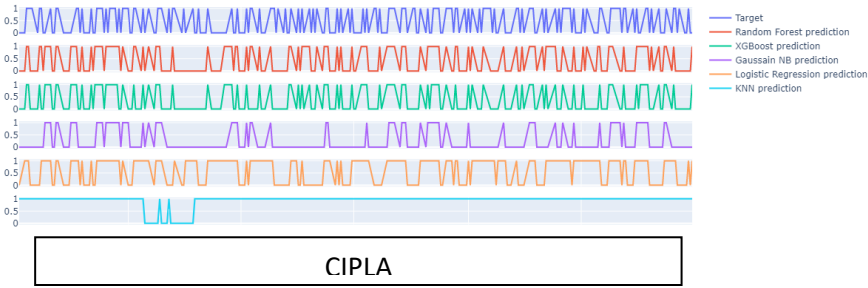

| STOCK NAME | RANDOM FOREST | XG BOOST | GAUSSIAN NB | LOGISTIC REGRESSION | KNN |
|---|---|---|---|---|---|
| Sun Pharma | 0.8184 | 0.8303 | 0.7785 | 0.7510 | 0.5039 |
| Cipla | 0.8212 | 0.8222 | 0.7828 | 0.6908 | 0.5029 |
| Divislab | 0.7944 | 0.8029 | 0.6825 | 0.5767 | 0.5074 |
| Drreddy | 0.7888 | 0.7909 | 0.7166 | 0.5893 | 0.5097 |

# CHAPTER-5

## DEEP LEARNING MODEL

Deep learning is a part of machine learning based on artificial neural networks. It teaches computer to work like that of humans. Deep learning is a key technology behind driverless cars, enabling them to recognize a stop sign, or to distinguish a pedestrian from a lamppost. It is the key to voice control in consumer devices like phones, tablets, TVs, and hands-free speakers. Deep learning is getting lots of attention lately and for good reason. Deep learning achieves recognition accuracy at higher levels than ever before. This helps consumer electronics meet user expectations, and it is crucial for safety-critical applications like driverless cars. Recent advances in deep learning have improved to the point where deep learning outperforms humans in some tasks like classifying objects in images. Deep learning requires large amounts of labeled data. Deep learning requires substantial computing power. High-performance GPUs have a parallel architecture that is efficient for deep learning. When combined with clusters or cloud computing, this enables development teams to reduce training time for a deep learning network from weeks to hours or less. The term "deep" refers to number of hidden layers in neural network. Deep learning network can have 100-150 hidden layers within it. Few applications of deep learning are  self driving cars, fraud detection, virtual assistants, visual recognition, pixel restoration, adding sounds to  silent movies, automatic game playing etc.

In this project we have worked on two deep learning techniques:-

❒ Long- and Short-Term Memory Network
❒ Convolutional Neural Networks

# Long and Short-Term Memory Network

Long Short-Term Memory (LSTM) networks are a type of recurrent neural network capable of learning order dependence in sequence prediction problems. This is a behavior required in complex problem domains like machine translation, speech recognition, and more. LSTMs are a complex area of deep learning. LSTM is an advanced RNN, which is sequential network that allows information to persist. It is capable of handling the vanishing gradient problem faced by RNN. A recurrent neural network is also known as RNN is used for persistent memory. LSTM was designed by Hochreiter & Schmidhuber. It tackled the problem of long-term dependencies of RNN in which the RNN cannot predict the word stored in the long-term memory but can give more accurate predictions from the recent information. As the gap length increases RNN does not give an efficient performance. LSTM can by default retain the information for a long period of time. It is used for processing, predicting, and classifying on the basis of time-series data. The recurrent neural network uses long short-term memory blocks to provide context for the way the program receives inputs and creates outputs. The long short-term memory block is a complex unit with various components such as weighted inputs, activation functions, inputs from previous blocks and eventual outputs. The unit is called a long short-term memory block because the program is using a structure founded on short-term memory processes to create longer-term memory.

Some applications of LSTM are:-

- ❖ Language Modelling

- ❖ Machine Translation

- ❖ Image Captioning

- ❖ Handwriting generation

- ❖ Chatbots

# Convolutional Neural Network

Convolutional Neural Network is a Deep Learning algorithm which can take in an input image, assign importance (learnable weights and biases) to various aspects/objects in the image and be able to differentiate one from the other. The pre-processing required in a CNN is much lower as compared to other classification algorithms. While in primitive methods filters are hand-engineered, with enough training, CNN have the ability to learn these filters/characteristics. The architecture of a CNN is analogous to that of the connectivity pattern of Neurons in the Human Brain and was inspired by the organization of the Visual Cortex. Individual neurons respond to stimuli only in a restricted region of the visual field known as the Receptive Field. A collection of such fields overlap to cover the entire visual area. CNN's were first developed and used around the 1980s. The most that a CNN could do at that time was recognizing handwritten digits. It was mostly used in the postal sectors to read zip codes, pin codes, etc. The important thing to remember about any deep learning model is that it requires a large amount of data to train and also requires a lot of computing resources. In the present work, we have used CNN to forecast the univariate time series data. CNN has two important processing layers, convolutional layers, and pooling layers. The convolutional layers read input using a filter by scanning across the input data field. The output of the convolutional layer is an interpretation of the input that is projected onto the filter map. The pooling layer takes the projections and reduces them to the most essential elements, using average pool or max pool. The convolution and pooling layers are repeated and the output of the final pooling layer is provided to one or more fully-connected layers that interpret what has been read.

The performance of the models is also studied by varying the number of filters. The objective of the same is to analyze how the increase or decrease in parameters would affect the accuracy of prediction and reduce/increase the time taken for computation. The loss function used is 'mean absolute error '(MAE), the optimizer used is Adam optimizer, and the activation function used is ReLU (Rectified Linear Unit).

# SECTOR WISE RESULT AND ANALYSIS

| STOCK NAME | PARAMETERS | MAE |
|---|---|---|
| Tata Steel | 793857 | 245.12 |
| Infy | 793857 | 425.63 |
| HDFC Bank | 793857 | 492.98 |
| ITC | 793857 | 136.41 |
| Maruti | 793857 | 3241.91 |
| Sun Pharma | 793857 | 336.29 |

# CHAPTER-6

# PORTFOLIO OPTIMIZATION

Earlier we have discussed the various statistical, econometric, machine learning, and deep learning models to predict stock prices of selected stocks from six sectors. This chapter discusses a systematic approach for building portfolios of stocks -minimum risk portfolio, optimal risk portfolio- from five sectors. Five stocks from each sector would be considered for portfolio building. Historical stock prices of five years that are from Jan 1, 2016, to Dec 31, 2020, would be used for building the portfolios. The performances of the portfolios are evaluated based on their returns after a period of one year. Several other attributes of the portfolios such as the risk, weights assigned to different stocks, and the correlation among the constituent stocks are also studied.

**Steps followed in portfolio design**

**Data acquisition of 5 stocks from each of the chosen five sectors**

As discussed in Chapter 2, using Yahoo Finance API, stock prices of five years i.e. from Jan 1, 2016, to Dec 31, 2020, are fetched for building the portfolio and from Jan 1, 2021, to Dec 31, 2021, for backtesting. Since the focus is on univariate analysis, 'close price' is the variable of interest.

**Computation of return and volatility**

The daily return and log return values of each stock of the sector are calculated. The daily return values are the percentage changes in the daily close values over successive days, while the log return values are the logarithms of the percentage changes in the daily close values. Using the daily volatility and the annual volatility of five stocks of each sector are computed. The daily volatility is defined as the standard deviation of daily return values. The annual volatility is yield by a multiplication factor of square root of 250. Annual volatility of a stock quantifies the risk associated with stock from point of view of an investor, as it indicates the amount of variability in price.

**Computation of covariance**

Once the volatilities and return of the stocks are computed, the covariances for the five stocks in each sector are calculated. Any pair exhibiting a high value of correlation coefficient indicates a strong association between them. A good portfolio aims to minimize the risk while maximizing the return. Risk minimization of a portfolio requires identifying stocks that have low correlation among themselves so that a higher diversity can be achieved. Hence, computation and analysis of the covariance of the stocks are of importance.

Based on various parameters we have built three types of portfolios:

OPTIMISED PORTFOLIO

EQUAL ALLOCATION PORTFOLIO

MINIMUM VOLATILITY PORTFOLIO

## OPTIMISED PORTFOLIO

The investors in the stock markets are usually not interested in the minimum risk portfolios as the return are usually less. In most cases, the investors are ready to incur risk if the associated return value is more. To compute the optimum risk portfolio, we use the metric Sharpe Ratio of a portfolio. The Sharpe Ratio of a portfolio is given below.

$Sharpe\ Ratio = (Rc - Rf) / σc$

Rc, Rf, and σc denote the return of the current portfolio, the risk-free portfolio, and the standard deviation of the current portfolio, respectively. Here, the risk-free portfolio is a portfolio with a volatility value of 1%. The optimum-risk portfolio is the one that maximizes the Sharpe Ratio for a set of stocks. This portfolio makes an optimization between the return and the risk of a portfolio. It yields a substantially higher return than the minimum risk portfolio, with a very nominal increase in the risk, and hence, maximizing the value of the Sharpe ratio.

## EQUAL ALLOCATION PORTFOLIO

Here, we proceed towards a deeper analysis of the historical prices of the five stocks in each of the five sectors. Here we construct a portfoliousing the twenty five stocks, with each stock carrying equal weight. Since there are five stocks in a sector each stock is assigned a weight of 0.04. Based on the training dataset and using an equal-weight portfolio, we compute the yearlyreturn and risk of each portfolio.

The yearly return and the yearly volatility of the equal-weight portfolio of each sector are computed using the training dataset. For this purpose, the mean of the yearly return values is derived using the resample function in Python with a parameter 'Y'. Yearly volatility values of the stocks in the equal-weight portfolio are derived by multiplying the daily volatility values by the square root of 250, assuming that there are 250 working days in a year for a stock exchange. The equal-weight portfolio of a sector gives us an idea about the overallprofitability and risk associated with each sector over the training period. However for future investments, their usefulness is very limited. Every stock in a portfolio does not contribute equally to its return and the risk.

## MINIMUM VOLATILITY PORTFOLIO

We build the minimum volatility portfolio for each sector using the records in its trainingdataset. The minimum volatility portfolio is characterized by its minimum variance. Thevariance of a portfolio is a metric computed using the variances of each stock in theportfolio as well as the covariance between each pair of stocks in the portfolio.

# ANALYSIS OF RETURN IN DIFFERENT TYPES OF PORTFOLIO

| STOCK NAME | OPTIMAL PORTFOLIO(%) | EQUAL ALLOCATION PORTFOLIO(%) | MINIMUM VOLATILITY PORTFOLIO(%) |
|---|---|---|---|
| AMOUNT SPENT FOR 1YR | 99950 | 99992 | 99982 |
| RETURN IN 1YR | 73287 | 70997 | 56751 |
| | | | |
| TATASTEEL | 7 | 6 | 8 |
| HINDALCO | 13 | 17 | 0 |
| JSW STEEL | 25 | 10 | 17 |
| VEDL | 42 | 25 | 16 |
| ADANI ENTERPRISE | 20 | 8 | 2 |
| MARUTI | 1 | 1 | 0 |
| TATAMOTORS | 11 | 22 | 1 |
| M&M | 2 | 6 | 3 |
| BAJAJ AUTO | 0 | 1 | 2 |
| EICHER MOTOR | 1 | 1 | 2 |
| HDFC BANK | 7 | 3 | 8 |
| ICICI BANK | 4 | 7 | 3 |
| AXIS BANK | 4 | 6 | 1 |
| KOTAK BANK | 3 | 2 | 1 |
| SBIN | 8 | 15 | 5 |
| INFY | 7 | 3 | 4 |
| TCS | 0 | 1 | 3 |
| HCL TECH | 4 | 4 | 8 |
| WIPRO | 16 | 11 | 27 |
| TECHM | 5 | 4 | 11 |
| ZEEL | 3 | 18 | 19 |
| PVR | 0 | 3 | 2 |
| TV18BRDCST | 12 | 130 | 24 |
| SUNTV | 8 | 8 | 2 |
| SAREGAMA | 15 | 48 | 29 |

# CHAPTER-7

## CONCLUSION

In this project, we have studied the performance of statistical, econometrical, machine learning and deep learning models in stock price prediction. Simple exponential smoothing, Holt Winter method, ARIMA is the statistical model studied whereas K Nearest Neighbor, Decision Tree, SVM, XGBoost and Random Forest are the machine learning models used for stock price prediction. Among the classification ML models, along with the above mentioned ones, logistic regression has also been studied. LSTM and CNN are the deep learning models used for stock price prediction. Performance of the deep learning models with increase or decrease of the parameters/ input values has also been analyzed. Sliding Window and Rolling Window are the validation method used for statistical, econometric and machine learning models. Since deep learning models give better results only if the training data is huge, traditional train slit method is used for training and testing. Some interesting patterns in day wise RMSE/ mean plots have been noticed while building machine learning and deep learning models. It is noticed that among the econometric models performed the best across the sectors, whereas Random Forest and XGBoost are at par with each other as far as ML classification or regression models are concerned. LSTM gave better results compared to CNN, though time taken for execution is much less for CNN.

We built three types of portfolio and analyzed return of different types of portfolio and then we studied NLP to do sentiment analysis and analyze impact of news on different stocks prices.

# REFERENCES

Sen, J. (2022) "A Forecasting Framework for the Indian Healthcare Sector Index", International Journal of Business Forecasting and Marketing Intelligence (IJBFMI), Art ID: IJBFMI-108504. (Accepted for Publication) DOI: 10.1504/IJBFMI.2022.10047095.

Sen, J. and Mehtab, S. (2021) "A Comparative Study of Optimum Risk Portfolio and Eigen Portfolio on the Indian Stock Market", International Journal of Business Forecasting and Marketing Intelligence (IJBFMI), Vol 7, No 2, pp. 143-193. Inderscience Publishers. DOI: 10.1504/IJBFMI.2021.10043037

Mehtab, S. and Sen, J. (2020) "A Time Series Analysis-Based Stock Price Prediction Using Machine Learning and Deep Learning Models", International Journal of Business Forecasting and Marketing Intelligence (IJBFMI), Inderscience Publishers, Vol 6, No 4, pp. 272-335. DOI: 10.1504/IJBFMI.2020.115691.

Sen, J., Mehtab, S. and Nath, G. (2020) "Stock Price Prediction Using Deep Learning Models", Lattice: The Machine Learning Journal, Vol 1, No 3, pp. 34-40, December 2020. DOI: 10.36227/techrxiv.16640197.v1

Sen, J. (2018) "Stock Composition of Mutual Funds and Fund Style: A Time Series Decomposition Approach towards Testing for Consistency", International Journal of Business Forecasting and Marketing Intelligence, Vol 4, No 3, pp. 235-292, 2018. DOI: 10.1504/IJBFMI.2018.092781.

Sen, J.and Datta Chaudhuri, T. (2018) "Understanding the Sectors of Indian Economy for Portfolio choice", International Journal of Business Forecasting and Marketing Intelligence, Vol 4, No 2, pp. 178 - 222, April 2018. DOI: 10.1504/IJBFMI.2018.090914.

Sen, J. (2017) "A Robust Analysis and Forecasting Framework for the Indian Mid Cap Sector Using Time Series Decomposition Approach", Journal of Insurance and Financial Management, Vol 3, No 4, pp. 1 - 32, September 2017. PDF DOI: 10.36227/techrxiv.15128901.v1

Sen, J. (2017) "A Time Series Analysis-Based Forecasting Approach for the Indian Realty Sector ", International Journal of Applied Economic Studies, Vol 5, No 4, pp 8 - 27, August 2017. PDF. DOI: 10.36227/techrxiv.16640212.v1.

Sen J. and Datta Chaudhuri, T. (2017) "A Predictive Analysis of the Indian FMCG Sector Using Time Series Decomposition-Based Approach", Journal of Economics Library, Vol 4, No 2, pp 206 - 226, June 2017.  http://dx.doi.org/10.1453/jel.v4i2.1282 PDF

Sen J. and Datta Chaudhuri, T. (2017), "A Time Series Analysis-Based Forecasting Framework for the Indian Healthcare Sector", Journal of Insurance and Financial Management, Vol 3, No 1 (2017), pp 66 - 94. PDF. DOI: 10.36227/techrxiv.16640221.v1.

Sen J. and Datta Chaudhuri (2016), "An Investigation of the Structural Characteristics of the Indian IT Sector and the Capital Goods Sector – An Application of the R Programming Language in Time Series Decomposition and Forecasting", Journal of Insurance and Financial Management, Vol 1, Issue 4, pp 68-132, June 2016. PDF. DOI:

10.36227/techrxiv.16640227.v1

Sen J. and Datta Chaudhuri (2016), "An Alternative Framework for Time Series Decomposition and Forecasting and its Relevance for Portfolio Choice – A Comparative Study of the Indian Consumer Durable and Small Cap Sectors", Journal of Economics Library, Istanbul, Turkey, Vol 3, Issue 2, pp 303 – 326, June 2016.  http://dx.doi.org/10.1453/jel.v3i2.787 PDF

Sen J. and Datta Chaudhuri (2015), "A Framework for Predictive Analysis of Stock Market Indices – A Study of the Indian Auto Sector, Calcutta Business School (CBS) Journal of Management Practices, Vol 2, Issue 2, pp 1- 20, December 2015. DOI: 10.13140/RG.2.1.2178.3448  PDF

Sen, J. (2022) "Designing Efficient Pair-Trading Strategies Using Cointegration for the Indian Stock Market" Proceedings of the IEEE ASIANCON'22, Pune, India, August, 2022.

Sen, J. and Dutta, A. (2022) "Design and Analysis of Optimized Portfolios for Selected Sectors of the Indian Stock Market", in Proceedings of the 2022 International Conference on Decision Aid Sciences and Applications (DASA), pp. 567-573, March 23-25, 2022, Chiangrai, Thailand. DOI: 10.1109/DASA54658.2022.9765289.

Sen, J. and Dutta, A. (2021), "A Comparative Study of Hierarchical Risk Parity Portfolio and Eigen Portfolio on the NIFTY 50 Stocks", In Proceedings of the 2nd International Conference on Computational Intelligence and Data Analytics (ICCIDA'21), January 8-9, 2021, Hyderabad, India. (In Press)

Sen, J., Dutta, A., Mondal, S., and Mehtab, S. (2021), "A Comparative Study of Portfolio Optimization Using Optimum Risk and Hierarchical Risk Parity Approaches", In Proceedings of the 8th International Conference on Business Analytics and Intelligence (ICBAI'21), December 20-22, 2021, IISc, Bangalore, India.

Bhowmick, H., Chatterjee, A., and Sen, J. (2021) "Comprehensive Movie Recommendation System", In Proceedings of the 8th International Conference on Business Analytics and Intelligence (ICBAI'21), December 20-22, 2021, IISc, Bangalore, India.

Sen, J. and Dutta, A. (2021), "Risk-Based Portfolio Optimization on Some Selected Sectors of the Indian Stock Market", In Proceedings of the 2nd International Conference on Big Data, Machine Learning and Applications (BigDML'21), December 19-20, 2021, NIT Silchar, Assam, India. (In Press)

Sen, J., Mehtab, S., and Dutta, A. (2021) "Precise Stock Price Prediction for Optimized Portfolio Design Using an LSTM Model", In Proceedings of the IEEE 19th OITS International Conference on Information Technology (OCIT'21), pp. 210-215, December 16-18, 2021, Bhubaneswar, India. DOI: 10.1109/OCIT53463.2021.00050.

Sen, J., Mondal, S., and Nath, G. (2021) "Robust Portfolio Design and Stock Price Prediction Using an Optimized LSTM Model", In Proceedings of the IEEE 18th India Council International Conference (INDICON'21), pp. 1-6, December 19-21, 2021, Guwahati, India. DOI: 10.1109/INDICON52576.2021.9691583.

Sen, J., Mehtab, S., Dutta, A., and Mondal, S. (2021) "Hierarchical Risk Parity and Minimum

Variance Portfolio Design on NIFTY 50 Stocks", In Proceedings of the IEEE International Conference on Decision Aid Sciences and Applications (DASA'21), pp. 668-675, December 7-8, 2021, Bahrain. DOI: 10.1109/DASA53625.2021.9681925.

Sen, J., Mondal, S. and Mehtab, S. (2021) "Portfolio Optimization on NIFTY Thematic Sector Stocks Using an LSTM Model", In Proceedings of the IEEE International Conference on Data Analytics for Business and Industry (ICDABI'21), pp. 364-369, Bahrain, October 25-26, 2021. DOI: 10.1109/ICDABI53623.2021.9655886.

Chatterjee, A., Bhowmick, H., and Sen, J.(2021) "Stock Price Prediction Using Time Series, Econometric, Machine Learning, and Deep Learning Models", In Proceedings of IEEE Mysore Sub Section International Conference (MysuruCon'21), October 24-25, 2021, pp. 289-296, Hassan, Karnataka, India. DOI: 10.1109/MysuruCon52639.2021.9641610.

Sen, J., Dutta, A., and Mehtab, S. (2021) "Stock Portfolio Optimization Using a Deep Learning LSTM Model", In Proceedings of IEEE Mysore Sub Section International Conference (MysuruCon), October 24-25, 2021, pp. 263-271, Hassan, Karnataka, India. DOI: 10.1109/MysuruCon52639.2021.9641662.

Sen, J., Mondal, S., and Mehtab, S. (2021) "Analysis of Sectoral Profitability of the Indian Stock Market Using an LSTM Regression Model", in Proceedings of the Deep Learning Developers' Conference (DLDC'21), September 24, 2021, Bangalore, India. DOI: 10.36227/techrxiv.17048579.v1.

Sen, J., Mehtab, S., and Dutta, A. (2021) "Volatility Modeling of Stocks from Selected Sectors of the Indian Economy Using GARCH", in Proceedings of the IEEE Asian Conference on Innovation in Technology (ASIANCON'21), pp 1-9, August 28-29, 2021, Pune, India. DOI: 10.1109/ASIANCON51346.2021.9544977. IEEE Xplore Link

Sen, J., Mehtab, S. (2021) "Accurate Stock Price Forecasting Using Robust and Optimized Deep Learning Models", in Proceedings of the IEEE International Conference on Intelligent Technologies (CONIT), pp. 1-9, June 25-27, 2021, Hubballi, India. DOI: 10.1109/CONIT51480.2021.9498565.

Sen, J., Dutta, A., and Mehtab, S. (2021) "Profitability Analysis in Stock Investment Using an LSTM-Based Deep Learning Model", in Proceedings of the IEEE 2nd International Conference for Emerging Technology (INCET), pp. 1-9, May 21-23, Belagavi, India. DOI: 10.1109/INCET51464.2021.9456385.

Mehtab, S. and Sen, J. (2022) "Analysis and Forecasting of Financial Time Series Using CNN and LSTM-Based Deep Learning Models", In: Sahoo J.P, Tripathy A.K., Mohanty M., Li KC., Nayak A.K. (eds) Advances in Distributed Computing and Machine Learning. Lecture Notes in Networks and Systems, vol 202, pp 405-423. Springer, Singapore. DOI: 10.1007/978-981-16-4807-6_39.

Mehtab, S. and Sen, J., and Dasgupta, S. (2020) "Robust Analysis of Stock Price Time Series Using CNN and LSTM-Based Deep Learning Models", in Proceedings of the IEEE 4th International Conference on Electronics, Communication and Aerospace Technology (ICECA), pp 1481-1486, November 5-7, 2020, Coimbatore, India. DOI: 10.1109/ICECA49313.2020.9297652.

Mehtab, S. and Sen, J. (2020), "Stock Price Prediction Using CNN and LSTM-Based Deep Learning Models", in Proceedings of the IEEE International Conference on Decision Aid Sciences and Applications (DASA), pp. 447-453, November 8-9, 2020, Sakheer, Bahrain. DOI: 10.1109/DASA51403.2020.9317207.

Mehtab, S. and Sen, J., and Dutta, A. (2021), and Abhishek Dutta, "Stock Price Prediction Using Machine Learning and LSTM-Based Deep Learning Models", Machine Learning and Metaheuristics Algorithms, and Applications (SoMMA), pp. 88-106, Communications in Computer and Information Science, Springer, Singapore. DOI: 10.1007/978-981-16-0419-5_8.

Mehtab, S. and Sen, J. (2020), "Stock Price Prediction Using Convolutional Neural Networks on a Multivariate Time Series", in Proceedings of the 3rd National Conference on Machine Learning and Artificial Intelligence (NCMLAI 2020), Lal Bahadur Shastri Institute of Management, New Delhi, INDIA, February 1, 2020. PDF DOI: 10.36227/techrxiv.15088734.v1.

Mehtab, S. and Sen, J. (2019), "A Robust Predictive Model for Stock Price Prediction Using Deep Learning and Natural Language Processing", in Proceedings of the 7th International Conference on Business Analytics and Intelligence (BAICONF, 2019), Indian Institute of Management, Bangalore, INDIA, December 5 – 7, 2019. DOI: 10.36227/techrxiv.15023361.v1.

Sen, J. (2018) "Stock Price Prediction Using Machine Learning and Deep Learning Frameworks", in Proceedings of the 6th International Conference on Business Analytics and Intelligence (ICBAI'2018), Indian Institute of Science, Bangalore, INDIA, December 20 – 21, 2018.

Sen J. and Datta Chaudhuri, T. (2017), "A Robust Predictive Model for Stock Price Forecasting", in Proceedings of the 5th International Conference on Business Analytics and Intelligence, Indian Institute of Management, Bangalore, December 11- 13, 2017, INDIA. DOI: 10.36227/techrxiv.16778611.v1

Mondal, S. and Sen, J. (2017) "A Framework of Predictive Analysis of Tourist Inflow in the Beaches of West Bengal: A Study of Digha-Mandarmoni Beach", in Proceedings of the First International Conference on Computational Intelligence, Communication and Business Analytics (CICBA-2017), Springer-Verlag, CCIS Series, pp. 161 - 176, March 24 - 25, 2017, Kolkata, INDIA. DOI: 10.1007/978-981-10-6427-2_14.

Sen J. and Datta Chaudhuri, T. (2016) "Decomposition of Time Series Data to Check Consistency between Fund Style and Actual Fund Composition of Mutual Funds", in Proceedings of the 4th International Conference on Business Analytics and Intelligence (ICBAI 2016), Indian Institute of Science, Bangalore, India, December 19 - 21, 2016. DOI: 10.13140/RG.2.2.33048.19206.

Sen J. and Datta Chaudhuri, T.(2016), "Decomposition of Time Series Data of Stock Markets and its Implications for Prediction – An Application for the Indian Auto Sector", in Proceedings of the 2nd National Conference on Advances in Business Research and Management Practices (ABRMP'2016), Kolkata, India, January 8-9, pp. 15 – 28, 2016. doi:10.13140/RG.2.1.3232.0241

Mehtab, S. and Sen, J. (2022) "Stock Price Prediction Using Deep Learning and Natural Language Processing", book chapter in: J. Sen and S. Mehtab (eds), Machine Learning in the Analysis and Forecasting of Financial Time Series, pp. 1-28, Cambridge Scholars Publishing, UK, May 2022. ISBN: 978-1-5275-8324-5.

Sen J. (2022) "Machine Learning and Deep Learning in Stock Price Prediction", book chapter in: J. Sen and S. Mehtab (eds), Machine Learning in the Analysis and Forecasting of Financial Time Series, pp. 29-67, Cambridge Scholars Publishing, UK, May 2022. ISBN: 978-1-5275-8324-5.

Sen, J. and Mehtab, S. (2022) "Stock Price Prediction Using Convolutional Neural Networks", book chapter in: J. Sen and S. Mehtab (eds), Machine Learning in the Analysis and Forecasting of Financial Time Series, pp. 68-101, Cambridge Scholars Publishing, UK, May 2022. ISBN: 978-1-5275-8324-5.

Dutta A. and Sen, J. (2022) "Robust Predictive Models for the Indian IT Sector using Machine Learning and Deep Learning", book chapter in: J. Sen and S. Mehtab (eds), Machine Learning in the Analysis and Forecasting of Financial Time Series, pp. 102-196, Cambridge Scholars Publishing, UK, May 2022. ISBN: 978-1-5275-8324-5.

Ashmita Paul and Jaydip Sen, "A Causality Analysis between the Indian Information Technology Sector Index and the DJIA Index", book chapter in: J. Sen and S. Mehtab (eds), Machine Learning in the Analysis and Forecasting of Financial Time Series, pp. 197-234, Cambridge Scholars Publishing, UK, May 2022. ISBN: 978-1-5275-8324-5.

Mehtab, S. and Sen, J. (2022) "Stock Price Prediction using Machine Learning and Deep Learning Algorithms and Models", book chapter in: J. Sen and S. Mehtab (eds), Machine Learning in the Analysis and Forecasting of Financial Time Series, pp. 235-303, Cambridge Scholars Publishing, UK, May 2022. ISBN: 978-1-5275-8324-5.

Sen J. (2022) "Analysis of Different Sectors of the Indian Economy for Robust Portfolio Construction", book chapter in: J. Sen and S. Mehtab (eds), Machine Learning in the Analysis and Forecasting of Financial Time Series, pp. 304-360, Cambridge Scholars Publishing, UK, May 2022. ISBN: 978-1-5275-8324-5.

Sen, J., Dutta, A. (2022), "Portfolio Optimization for the Indian Stock Market", in: Wang, J. (ed.) Encyclopedia of Data Science and Machine Learning, IGI Global, USA, August 2022. (In Press). DOI: 10.4018/978-1-7998-9220-5.

Sen, J. and Mehtab, S. (2022) "Long-and-Short-Term Memory (LSTM) Price Prediction – Architectures and Applications in Stock Price Prediction", Chapter 8, in: Singh, U., Murugesan, S., and Seth, A. (eds) Emerging Computing Paradigms – Principles, Advances, and Applications, Wiley, USA, August, 2022. (In Press)

Sen, J., Sen, R., and Dutta, A. (2021) "Machine Learning in Finance: Emerging Trends and Challenges", Introductory Chapter", in Machine Learning: Algorithms, Models, and Applications Editor: Jaydip Sen, IntechOpen, London, UK, December, 2021. DOI: 10.5772/intechopen.101120.

Sen, J. and Mehtab, S. (2021) "Design and Analysis of Robust Deep Learning Models for Stock Price Prediction", Book Chapter in Machine Learning – Algorithms, Models and Applications, Editor: Jaydip Sen. Chapter published in the book Machine Learning:

Algorithm, Models, and Application edited by Jaydip Sen and published online by IntechOpen, London, UK, in August 2021. DOI: 10.5772/intechopen.99982.

Sen J. (2018) "A Study of the Indian Metal Sector Using Time Series Decomposition-Based Approach", Book Chapter in Selected Studies on Economics and Finance, Editors: Selim Basar, A. Alkan Celik, and Turgut Bayramoglu, pp. 105 - 152, Cambridge Scholars Publishing, UK, March 2018.

Sen, J. and Mehtab, S. (2022) Machine Learning in the Analysis and Forecasting of Financial Time Series, Cambridge Scholars Publishing, Newcastle upon Tyne, UK, May 2022. ISBN: 978-1-5275-8324-5.

Sen, J. (2021) Machine Learning – Algorithms, Models, and Applications, InTechOpen Publishers, London, UK. ISBN: 978-1-83969-485-1. ISSN: 2633-1403.

Sen, J. and Datta Chaudhuri, T. (2017) Analysis and Forecasting of Financial Time Series Using R: Models and Applications, Scholar's Press, Germany. ISBN: 978-3-330-65386-3.